\documentclass[12pt]{ociamthesis}

\usepackage{amssymb,epsfig,cancel,comment}
\usepackage{setspace}
\usepackage{graphicx,feyn,feynarts}
\usepackage{pstricks}
\usepackage{color}
\usepackage[sumlimits]{amsmath}
\usepackage{appendix}

\usepackage{setspace}

\usepackage[font=md,captionskip=8pt]{subfig}

\def\to{\rightarrow}

\def\det{\mbox{det}}
\def\exp{\mbox{exp}}
\def\tr{\mbox{Tr}}
\def\hc{\mbox{h.c.}}

\newcommand{\cL}{{\cal L}}

\newcommand{\cO}{{\cal O}}

\newcommand{\ra}{\rightarrow}
\newcommand{\be}{\begin{equation}}
\newcommand{\ee}{\end{equation}}
\newcommand{\bea}{\begin{eqnarray}}
\newcommand{\eea}{\end{eqnarray}}

\def\be{\beta}

\def\frac#1#2{{\textstyle{{#1}\over {#2}}}}

\def\lsim{\mathrel{\rlap{\lower4pt\hbox{\hskip1pt$\sim$}}
    \raise1pt\hbox{$<$}}}
\def\gsim{\mathrel{\rlap{\lower4pt\hbox{\hskip1pt$\sim$}}
    \raise1pt\hbox{$>$}}}
\def\sqr#1#2{{\vcenter{\vbox{\hrule height.#2pt
         \hbox{\vrule width.#2pt height#1pt \kern#1pt
         \vrule width.#2pt}
         \hrle height.#2pt}}}}

\def\beq{\begin{equation}}
\def\eeq{\end{equation}}
\def\beqa{\begin{eqnarray}} 
\def\eeqa{\end{eqnarray}}

\def\laq{\raise 0.4 ex \hbox{$<$}\kern -0.8 em\lower 0.62 ex\hbox{$\sim$}}
\def\gaq{\raise 0.4 ex \hbox{$>$}\kern -0.7 em\lower 0.62 ex\hbox{$\sim$}}

\graphicspath{{./}{Figs/}{Plots/}}

\title{Naturalness of electroweak physics within minimal supergravity}
\author{Sebastian N. H. Cassel}
\college{St. John's College}
\degree{Doctor of Philosophy}
\degreedate{Trinity 2010}

\begin{document}

\baselineskip=18pt plus1pt

\setcounter{secnumdepth}{3}
\setcounter{tocdepth}{3}

\maketitle
\begin{acknowledgements}
I am extremely grateful to my supervisors, Prof. Graham Ross and Dr. Dumitru Ghilencea, who have provided generous support and insightful guidance. It has been an immense privilege to be able to work with such exceptionally knowledgeable people. I greatly appreciated the engaging discussions and challenging research propositions.

I would also like to thank the staff and students of the Oxford Theoretical Physics Department, for many stimulating conversations that helped develop my understanding of particle physics and more general areas of physics. These discussions commonly led to productive advances in my research, for which I am indebted.

The financial support of the UK Science and Technology Facilities Council through a postgraduate studentship, PPA/S/S/2006/04503, is gratefully acknowledged. I am also appreciative for the additional funding that allowed several visits to CERN in the past year, where part of this work was completed. For those trips, I thank the CERN Theory Department for their hospitality in accommodating me. The financial support of St. John's College and the Oxford Physics Department for attending summer schools and conferences is also gratefully acknowledged.

Finally, but of great significance, I thank my parents for all their support.

\end{acknowledgements}
\begin{abstract}

\begin{center}
\singlespacing
``Naturalness of electroweak physics within minimal supergravity"

\noindent
A thesis submitted for the degree of Doctor of Philosophy

\noindent
Sebastian Cassel, St. John's College

\noindent
Trinity 2010
\end{center}

\vspace{10mm}
Low energy supersymmetry is motivated by its use as a solution to the hierarchy problem of the electroweak scale. Having motivated this model with naturalness arguments, it is then necessary to check whether the experimentally allowed parameter space permits realisations of the model with low fine tuning. The scope of this thesis is a study of naturalness of the electroweak physics in the minimal supergravity model. The latest experimental constraints are applied, and the fine tuning is quantitatively evaluated for a scan across the parameter space. The fine tuning of the electroweak scale is evaluated at 2-loop order, and the fine tuning of the neutralino dark matter thermal relic energy density is also determined. The natural regions of the parameter space are identified and the associated phenomenology relevant for detection discussed. Naturalness limits are also found for the parameter space and spectrum. The minimum fine tuning found is 1 part in 9 when dark matter constraints are neglected, and 1 part in 15 when dark matter constraints are satisfied. For both cases, the minimum fine tuning is found for a Higgs mass of 115 GeV irrespective of whether the Higgs mass constraint is applied or not. The most natural spectrum includes light superpartner fermions, and heavy superpartner scalars. Minimal supergravity currently remains viable with respect to naturalness and a natural realisation may be discovered within the next couple of years.

\end{abstract}
\begin{romanpages}
\tableofcontents
\end{romanpages}

\doublespacing

\chapter{Introduction}

A theory of nature must account for the experimentally observed symmetries. These include the relativistic invariance of spacetime along with an independent set of symmetries, $SU(3)\times SU(2)\times U(1)$, of the fundamental matter. The Standard Model (SM) of particle physics has proved extremely successful in quantitatively matching experimental measurements, based on these local symmetries alone and the minimal particle content required to match observations.

There are however many problems with the Standard Model that motivate development of `beyond the Standard Model' (BSM) physics. An obvious flaw is the absence of gravitational interactions. Standard Model physics is based on quantum field theory, for which a consistent physical description of the non-renormalisable interactions and dynamics of gravity requires an infinite number of parameters. The unpredictability at ultraviolet energy scales is then interpreted as an indication that the model is incomplete or defective. The Standard Model also fails to explain neutrino masses or the existence of dark matter.


Hints for the existence of further BSM physics may be derived from the many apparently coincidental features of the Standard Model. For example, the SM gauge couplings seem to all converge at a high energy scale
on solving their renormalisation group equations. The quantum field theory anomalies of the quark sector cancel those of the independent lepton sector. The electric charges of all known matter is an integer multiple of a common quantised unit, leading to the electric charge of the electron to be exactly opposite that of the proton. Questions such as why there aren't more (or less) fundamental particles in nature, or why there are precisely three families of matter with identical sets of quantum numbers are also not answered by the Standard Model.


On examining the Standard Model, there are also various aesthetically unpleasant theoretical features. There are a large number of parameters which must be fixed from experiment. The values of these parameters vary over several orders of magnitude. Some parameters such as the degree of violation of the discrete charge and parity symmetries of the QCD sector are restricted to zero within experimental error without any theoretical justification. This leads to asking why the SM parameters take the values they do, and if there is a deeper reason to expect the required pattern of values to describe our universe.

The concept of naturalness has been a driving principle in recent decades for many proposals of BSM physics. The physical principles governing a model such as special relativity in an arbitrary quantum field theory permit a vast family of theories which seem equally viable before experimental validation. The philosophy of natural model building is that further physical principles must be identified in order to restrict the viable theories of nature. In a ``theory of everything," a desired quality is that the physics of our universe is fully explained with little need for calibration of its parameters, if any.

A measure of naturalness can be associated with the sensitivity of observable quantities to changes permitted in the underlying fundamental parameters. ``Un-natural" theories are then deemed to require large amounts of fine tuning of its parameters to match observations. The Standard Model suffers from such fine tuning problems. An alternative view to naturalness is to follow anthropic arguments and accept that arbitrary patterns just occur, or that a multiverse exists and conscious observers are necessarily only present in universes with appropriate physics leading to a selection bias. It is the implications of a principle of naturalness that is explored in this thesis.

With the Large Hadron Collider (LHC) beginning to open a new energy frontier and Fermilab still pushing up the luminosity frontier, we may soon be able to identify the physics responsible for electroweak symmetry breaking (EWSB) and/or discover BSM physics. In the Standard Model, the Higgs mechanism is used to trigger EWSB. The existence of the Higgs boson necessary for the Higgs mechanism remains the last component of the Standard Model unverified by experiment.

It emerges that from a theoretical perspective, the SM Higgs sector is responsible for its worst fine tuning problem, excepting the scale of the cosmological constant. Thus, if a naturalness principle is to be believed, identifying natural BSM theories with respect to the electroweak physics and its experimental signatures may lead to imminent discoveries of new physics beyond the Standard Model.

The subject of this thesis is an analysis of the degree of fine tuning present in a class of BSM theories with minimal supersymmetry broken through a restricted set of gravitational interactions. The most natural realisations of the BSM model are identified with indications for promising experimental search strategies discussed. In this introductory chapter, the formalism of supersymmetry is presented and notation defined. Supersymmetric Lagrangians are introduced with spontaneous supersymmetry breaking mechanisms and the structure of minimal supergravity (mSUGRA) reviewed. The hierarchy problem is then introduced and a quantitative measure of naturalness defined.

Chapter 2 presents the results of evaluation of the fine tuning of the electroweak scale for a scan of the allowed parameter space. A combination of analytic and numerical results are discussed, and the controlling features of fine tuning identified. Chapter 3 presents the results of fine tuning of the thermal relic density of the mSUGRA dark matter candidate with the relevant dark matter constraints applied. These results are compared to the fine tuning of the electroweak scale with the dark matter constraints then added to the results of Chapter 2. Chapter 4 discusses how extensions beyond the mSUGRA model may generate improved naturalness, and a model independent analysis is presented using an effective theory framework. Finally, the conclusions of this study are offered in Chapter 5.

The novel features of this work include analysis of the fine tuning of the electroweak scale at two loop order, a quantitative study of the effect of higher dimension operators on fine tuning, inclusion of the latest experimental constraints, and dependence of the fine tuning of electroweak physics on the composition of dark matter. The results of previous related work is also reviewed in the following chapters.






\section{Towards supersymmetry (SUSY)}

When building models of physics `beyond the Standard Model' (BSM), it is vital to understand the full class of theories into which the set of SM symmetries may be embedded. Before introducing supersymmetry, a brief review of the Standard Model symmetries is given where notation and conventions are defined.

\subsection{Poincar\'{e} symmetry}

In flat spacetime, the symmetry group associated with relativistic invariance is the Poincar\'{e} group. For a set of points in spacetime, $\{ x^\mu \}$, the Poincar\'{e} symmetry is defined by the quadratic form of any spacetime interval being invariant under linear co-ordinate transformations.
\begin{eqnarray}
(x-y)^2 &\equiv& \eta_{\mu\nu}^{} (x-y)^\mu (x-y)^\nu
\label{quadform}
\end{eqnarray}
where the metric of flat spacetime, $\eta = \mbox{diag} (1,-1,-1,-1)$. The general form of the co-ordinate transformation which satisfies this constraint is a combination of a Lorentz transform and translation.
\begin{eqnarray}
x^{\prime \mu} &=& \Lambda^{\mu}_{~\nu} \, x^{\nu} + a^{\mu}
\end{eqnarray}
for some constant 4-vector, $a^{\mu}$. A general Lorentz transform can be constructed from an $SO_+^{}(1,3)$ sub-group along with time and space reflections. The $SO_+^{}(1,3)$ group belongs to the set of real general linear 4 dimensional matrices subject to the constraints; $\mathbf{\Lambda^T \mathbf{\eta} \Lambda} = \mathbf{\eta}$, $\det \mathbf{\Lambda} = +1$,  $\Lambda^{0}_{~0} \geq +1$, where the last two constraints label the transformation as `proper' and `orthochronous' respectively. The subset of connected continuous Lorentz transformations are referred to as belonging to the restricted Lorentz group.

The generators of the Lorentz symmetry, $M_{\rho\sigma}^{}$, are defined by $\mathbf{\Lambda}=\exp \left( - \frac{i}{2} \omega^{\rho\sigma}  \mathbf{M}_{\rho\sigma}^{} \right)$
\begin{eqnarray}
\left( M_{\rho\sigma}^{} \right)^\mu_{~\nu} = i \left( \eta_{\sigma\nu}^{} \delta^\mu_{~\rho}-\eta_{\rho\nu}^{} \delta^\mu_{~\sigma} \right)
\end{eqnarray}
The generators can be decomposed into boosts, $K_i = -M_{0i}^{}$, and rotations, $J_i^{} = \frac12 \epsilon_{ijk}^{} M_{jk}^{}$ where $i,j,k \in \{ 1,2,3 \}$.

An equivalent method of expressing the above quadratic form exists in a complex 2-dimensional space. Introducing the following mapping operators, where $\{ \sigma_i^{} \}$ are the Pauli matrices,
\begin{eqnarray}
\sigma^\mu ~=~ \left( \mathbf{1}_2^{} , \{ \sigma_i^{} \} \right) 
\hspace{10mm}
\bar{\sigma}^\mu ~=~ \left( \mathbf{1}_2^{} , \{ - \sigma_i^{} \} \right)
\end{eqnarray}
spacetime 4-vectors can be mapped between the respective spaces according to:
\begin{eqnarray}
X ~=~ x_\mu^{} \sigma^\mu
\hspace{10mm}
x^\nu ~=~ \frac12 \tr \left[ X \bar{\sigma}^\nu \right]
\label{sl2cMap}
\end{eqnarray}
The quadratic form in eq~(\ref{quadform}) is then given by $\det \left(X-Y \right)$. This remains invariant under the following transformation.
\begin{eqnarray}
X^\prime = H X H^\dagger
\hspace{10mm}
\mbox{for some}~H\in SL(2,\mathbb{C})
\label{sl2cTrans}
\end{eqnarray}
$SL(2,\mathbb{C})$ is the group of 2-dimensional complex matrices with unit determinant. The above mappings lead to the relation, $\Lambda^{\mu}_{~\nu} (H) = \frac12 \tr \left[ \bar{\sigma}^\mu H \sigma_\nu H^\dagger \right]$. The groups $SO_+^{}(1,3)$ and $SL(2,\mathbb{C})$ are locally homomorphic, possessing homomorphic Lie algebras and so both describe the symmetries of proper orthochronous Lorentz transformations.


To classify the representations of the restricted Lorentz group, a common approach is to complexify the Lie algebra. There exists a 1:1 correspondence between the representations of real and complexified Lie algebras \cite{brocker} 
 making this technique useful. For the complex generators, $S_i^\pm = \left[ J_i^{} \pm i K_i^{} \right]/2$, the $S^\pm$ form independent $SU(2)$ sub-algebras. The complexified restricted Lie algebra is then homomorphic to $SU(2)\times SU(2)$ allowing its group representations to be classified by the eigenvalues, $\left( S_3^+ , S_3^- \right)$.

The $\left( S_3^+ , S_3^- \right)$ representation has a total spin of $S_3^+ + S_3^-$ and dimension $(2S_3^+ + 1).(2S_3^- + 1)$. Parity interchanges the eigenvalues, $S_3^\pm \to S_3^\mp$ and is obtained via Hermitian conjugation. By convention, the representation $\left( \frac12 ,0 \right)$ is referred to as left-handed, and $\left( 0 , \frac12 \right)$ as right-handed. Further examples of other representations are given in the following table.
\begin{table}[htdp]
\begin{center}
\begin{tabular}{|cl|} \hline
$\left( 0 ,0 \right)$ & spin-0 scalar \\
$\left( \frac12 ,\frac12 \right)$ & spin-1 4-vector \\
$\left( 1 ,0 \right)$ & spin-1 anti-symmetric self-dual tensor \\
$\left( 0 ,1 \right)$ & spin-1 anti-symmetric anti-self-dual tensor \\
$\left( 1 ,1 \right)$ & spin-2 symmetric traceless tensor \\
\hline
\end{tabular}
\end{center}
\label{reps}
\caption{Representations of the restricted Lorentz group}
\end{table}%

The possible dimensions of $SO_+^{}(1,3)$ representations restrict the spin of the states they can describe to integer values. However, the representations of $SL(2,\mathbb{C})$ also include half-integer spin states. $SL(2,\mathbb{C})$ is the spin group associated with the metric~$\eta$. The representations of a spin group are known as spinors. The fundamental spinors of $SL(2,\mathbb{C})$ are Weyl-type and can be used to construct higher spin states.
\begin{eqnarray}
\left( \frac12 ,0 \right) \otimes \left( 0, \frac12 \right) &=& \left( \frac12 ,\frac12 \right) \\
\left( \frac12 ,0 \right) \otimes \left(  \frac12 ,0 \right) &=& \left( 0 ,0 \right) \oplus \left( 1 ,0 \right)
\end{eqnarray}

The $\left( \frac12 ,\frac12 \right)$ state is equivalent to the fundamental representation of $SO_+^{}(1,3)$. To construct a parity invariant state with the same spin when $S_3^+ \neq S_3^-$, a direct sum of opposite parity states is required. For example, a Dirac spinor has the representation $\left( \frac12 ,0 \right) \oplus \left( 0, \frac12 \right)$.

The invariance under translations is included to extend the Lorentz symmetry to a Poincar\'{e} symmetry. This leads to a second Casimir operator of the symmetry group. In addition to spin, the rest mass associated with a Poincar\'{e} representation is then a necessary label.


\subsection{Fundamental spinors of $SL(2,\mathbb{C})$}


There exists four different transformation rules for fundamental representations of the $SL(2,\mathbb{C})$ group. For $H \in SL(2,\mathbb{C})$, these are given by;

\vspace{2mm}
\begin{tabular}{lcl}
$\bullet$\,~self representation & $\psi^{\prime}_A$ &$= H_A^{~B} ~ \psi_B^{}$ \\
$\bullet$\,~dual representation & $\psi^{\prime \, A}$ &$= \left(H^{-1 \, T} \right)^A_{~B} \psi^B_{}$ \\
$\bullet$\,~complex conjugate self representation & $\bar{\psi}^{\prime}_{\dot{A}}$ &$= \left(H^* \right)_{\dot{A}}^{~\dot{B}} \, \bar{\psi}_{\dot{B}}^{}$ \\
$\bullet$\,~complex conjugate dual representation \hspace{5mm}& $\bar{\psi}^{\prime \, \dot{A}}$ &$= \left(H^{*\, -1 \, T} \right)^{\dot{A}}_{~\dot{B}} \bar{\psi}^{\dot{B}}_{}$ 
\end{tabular}
\vspace{4mm}

\noindent
where the different representations are distinguished by upper and lower indices of dotted and undotted type. The indices can take values of $\{ 1,2 \}$.

However, the self and dual representations of $SL(2,\mathbb{C})$ are equivalent in the sense that there exists a matrix, $\epsilon \in SL(2,\mathbb{C})$ such that the transformation matrices are related by, $\epsilon H \epsilon^{-1} = H^{-1\, T}$. Up to a sign convention, this matrix is given by,
\begin{eqnarray}
\epsilon_{AB}^{} ~=& \left( \epsilon^{AB} \right)^{-1} &=~ -i\sigma_2^{} \nonumber\\
\epsilon^{AB}_{} ~=& \left( \epsilon_{AB}^{} \right)^{T} &=~ i\sigma_2^{}
\end{eqnarray}
where $\sigma_2^{}$ is the second Pauli matrix. The $\epsilon$ matrix acts like a metric to raise or lower indices of the $SL(2,\mathbb{C})$ space, $\psi^{A} = \epsilon^{AB} \psi_B^{}$, with the adopted notation.

Similary the complex conjugate self and dual representations are equivalent, where for $\bar{\epsilon} H^* \bar{\epsilon}^{-1} = H^{*\, -1\, T}$,
\begin{eqnarray}
\bar{\epsilon}^{\dot{A}\dot{B}}_{} ~=& \left( \bar{\epsilon}^{\dot{A}\dot{B}}_{} \right)^{-1} &=~ i\sigma_2^{} \nonumber\\
\bar{\epsilon}_{\dot{A}\dot{B}}^{} ~=& \left( \bar{\epsilon}^{\dot{A}\dot{B}}_{} \right)^{T} &=~ -i\sigma_2^{}
\end{eqnarray}
giving, $\bar{\psi}^{}_{\dot{A}} = \bar{\epsilon}_{\dot{A}\dot{B}}^{} \bar{\psi}^{\dot{B}}_{}$.

The complex conjugation operation does not generate equivalent representations. By considering a general transformation on a Dirac spinor (composed of both the direct sum of a left and a right-handed Weyl spinor) and its Hermitian conjugate, it follows that the relation,  $\psi^B = \bar{\psi}_{\dot{A}}^* \left(\bar{\sigma}^{0} \right)^{\dot{A} B}$ must hold \cite{Mueller} 
. The representation $\psi$ is referred to as a left-handed state, $\left( \frac12,0 \right)$, and $\bar{\psi}$ a right-handed state, $\left( 0,\frac12 \right)$.

Objects without free $SL(2,\mathbb{C})$ indices are Lorentz invariant. Where indices are suppressed, a summation convention is implied such that $\psi \phi \equiv \psi^A \phi_A^{}$, and $\bar{\psi} \bar{\phi} \equiv \bar{\psi}_{\dot{A}} \bar{\phi}^{\dot{A}}$. These definitions remove a sign ambiguity. It has been previously determined that these spinor fields have half-integer spin. By the spin-statistics theorem, it is then necessary to consider the spinor fields as Grassman variables leading to the relation, $\psi^A \phi_A^{} = -\phi_A^{} \psi^A ~\left( = \phi^A \psi_A^{} \right)$, and similarly for contraction of fields with dotted indices.

The Lorentz generators of $SL(2,\mathbb{C})$ in the spinor representations are for left- and right-handed states respectively:
\begin{eqnarray}
\sigma^{\mu\nu}  &=& \frac{i}{4} \left( \sigma^\mu \bar{\sigma}^\nu - \sigma^\nu \bar{\sigma}^\mu \right) \nonumber\\
\bar{\sigma}^{\mu\nu}  &=& \frac{i}{4} \left( \bar{\sigma}^\mu \sigma^\nu - \bar{\sigma}^\nu \sigma^\mu \right) 
\end{eqnarray}

To identify the type of $SL(2,\mathbb{C})$ indices required for the $\sigma^\mu$ and $\bar{\sigma}^\mu$ matrices, eqs~(\ref{sl2cMap}) and (\ref{sl2cTrans}) can be inspected, noting that the index structure of $SL(2,\mathbb{C})$ transformation matrices obeys $H_{A}^{~B}$ and $H^\dagger  = \left( H^* \right)^{\dot{B}}_{~\dot{A}}$. It then follows that $\left(  \sigma^\mu \right)_{A\dot{A}}^{}$ and $\left( \bar{\sigma}^\mu \right)^{\dot{A} A}$ are appropriate, with one dotted and one undotted index.


\subsection{Haag-$\cancel{\mbox{L}}$opusza\'{n}ski-Sohnius theorem}

The physics of particle interactions is governed by the S-matrix, as defined by:
\begin{eqnarray}
^{}_{\mbox{\tiny out}} \langle \mathbf{p}_1^{} \mathbf{p}_2^{} \, \ldots | \mathbf{k}_1^{} \mathbf{k}_2^{} \, \ldots \rangle^{}_{\mbox{\tiny in}} &\equiv&
 \langle \mathbf{p}_1^{} \mathbf{p}_2^{} \, \ldots |\, S \, | \mathbf{k}_1^{} \mathbf{k}_2^{} \, \ldots \rangle
\end{eqnarray}
where 
the {\it in} and {\it out} states are asymptotically defined in time by the tensor product of single particle states with definite momenta. 

Coleman and Mandula \cite{Coleman:1967ad} first proved that the most general connected symmetry of the S-matrix consistent with (i) relativistic invariance, (ii) the S-matrix being non-trivial and analytic, (iii) a unique vacuum, (iv) the presence of a finite number of massive particle states and (v) the Lie bracket of the symmetry algebra being a commutator, the group generators must decompose into a direct sum of the Poincar\'{e} generators, $P^\mu, M^{\mu\nu}$ and an arbitrary independent set of Lorentz scalar generators. The latter are known as ``internal" symmetry generators as they do not mix with the geometric symmetry generators of spacetime.

The Haag-$\cancel{\mbox{L}}$opusza\'{n}ski-Sohnius theorem \cite{Haag:1974qh} generalises the result of Coleman and Mandula by neglecting (v) above, finding the generalised symmetry with a $Z_2^{}$ graded Lie superalgebra where the odd generators that mix with the Poincar\'{e} generators transform as fundamental spinors under the restricted Lorentz group.

A $Z_2^{}$ graded algebra indicates that the generators can be decomposed into two groups, $L = L_0^{} \oplus L_1^{}$ and that for generators, $T_i^{} \in L_i^{}$, the Lie bracket satisfies $[ T_i^{}, T_j^{}] \in L_{\mbox{\tiny mod}[i+j,2]}^{}$. This property leads to the naming convention of generators in $L_0^{}$ being {\it{even}} and those in $L_1^{}$ being {\it{odd}}.

A $Z_2^{}$ graded superalgebra has a Lie bracket that satisfies the following symmetry
\begin{eqnarray}
\left[ T_i^{} , T_j^{} \right] &=& \left( -1 \right)^{i.j} \left[ T_j^{} , T_i^{} \right]
\end{eqnarray}
and the Jacobi identity becomes:
\begin{eqnarray}
\left( -1 \right)^{i.k} \left[ T_i^{}, \left[ T_j^{} , T_k^{} \right] \right] +
\left( -1 \right)^{j.i} \left[ T_j^{}, \left[ T_k^{} , T_i^{} \right] \right] +
\left( -1 \right)^{k.j} \left[ T_k^{}, \left[ T_i^{} , T_j^{} \right] \right] &=& 0
\end{eqnarray}
The Lie bracket is an anti-commutator when $(i,j)=(1,1)$, and a commutator otherwise.

The most general symmetry, referred to as supersymmetry, of the S-matrix within the given assumptions has the following algebra:
\begin{eqnarray}
\{ Q_A^{\,\alpha} , Q_B^{\,\beta} \} &=& \epsilon_{AB}^{} Z^{\alpha\beta} 
\hspace{15.5mm} 
\left[ Q_A^{\,\alpha}, P_\mu^{} \right]~~~=~ 0 \nonumber\\
\{ \bar{Q}_{\dot{A}}^{\,\alpha} , \bar{Q}_{\dot{B}}^{\,\beta} \} &=& \bar{\epsilon}_{\dot{A}\dot{B}}^{} \left( Z^\dagger \right)^{\alpha\beta}
\hspace{9.5mm} 
\left[ \bar{Q}_{\dot{A}}^{\,\alpha}, P_\mu^{} \right]~~~=~ 0  \nonumber\\
\{ Q_A^{\,\alpha} , \bar{Q}_{\dot{B}}^{\,\beta} \} &=& 2 \delta^{\alpha\beta} \sigma^\mu_{A\dot{B}} P_\mu^{}  
\hspace{10mm} 
\left[ Q_A^{\,\alpha}, M^{\mu\nu} \right]~=~ i \left( \sigma^{\mu\nu} \right)_A^{~B} Q_B^{\,\alpha} \nonumber\\
\left[ Q_A^{\,\alpha} , B_l^{} \right] ~&=& i S_l^{\alpha\beta} Q_A^{\,\beta}
\hspace{15.5mm}
\left[ \bar{Q}_{\dot{A}}^{\,\alpha} , B_l^{} \right] \hspace{5.5mm}=~ -i \left( S_l^{*} \right)^{\alpha\beta} \bar{Q}_{\dot{A}}^{\,\beta}
\label{alg}
\end{eqnarray}
where the central charges, $Z^{\alpha\beta}$, commute with all the generators and the Poincar\'{e} algebra is a subalgebra. The upper case Latin indices refer to $SL(2,\mathbb{C})$ indices, the $\mu,\nu$ Greek indices to Lorentz indices, and the $\alpha,\beta$ run from 1 to $N$ where $N$ identifies the ``number" of supersymmetries present. The lower case Latin indices associated with the Lorentz scalar generators, $B_l^{}$, run to some arbitrary number also, satisfying the algebra $\left[ B_j^{} , B_k^{}  \right] = i f_{jk}^{~~l} B_l^{}$, for some structure constants, $f_{jk}^{~~l}$.

The scalar generators associated with non-trivial $S_l^{}$ correspond to ``R-symmetries". For $N=1$ SUSY, the maximal R-symmetry is $U(1)$ and there is a null central charge. This R-symmetry leads to invariance under chiral rotations of the spinor generators. For internal symmetries, the associated $S_l^{}$ parameters must be zero to maintain independence from the generalised geometric symmetry.

The Coleman-Mandula theorem states that the even generators include a Lorentz vector, $P^\mu$ [in the $\left( \frac12 , \frac12 \right)$ representation], an antisymmetric tensor, $M^{\mu\nu}$ $\left[ \left( 1 , 0 \right) \oplus \left( 0 , 1 \right) \right]$ and a set of Lorentz scalars, $B_l^{}$ $\left[ \left( 0,0 \right) \right]$. From group theory arguments, it then becomes clear that the only odd generators that could mix with the Poincar\'{e} generators belong to the $\left( \frac12 , 0 \right)$ and $\left( 0 , \frac12  \right)$ representations.

To identify the superalgebra in eq~(\ref{alg}) as the most general form, consider the Lie bracket of two odd generators, $Q$, for an example in the required reasoning. The Lie bracket can only produce even generators due to the algebra grading and consideration of the direct product, $\left( \frac12 ,0 \right) \otimes \left( \frac12 ,0 \right) = \left( 0 ,0 \right) \oplus \left( 1 ,0 \right)$. By the Coleman-Mandula theorem, there are no even generators in the $\left( 1 ,0 \right)$ representation, but there may exist Lorentz scalar generators restricting the central charge, $Z^{\alpha\beta} = a_l^{\alpha\beta} B_l^{}$, for some co-efficient, $a_l^{\alpha\beta}$. The Jacobi identities further restrict the central charges to be anti-symmetric in their indices, with the $a_l^{\alpha\beta}$ co-efficients related to the $S_l^{}$ parameters, which are separately fixed to be Hermitian.

\section{Representations of the SUSY algebra}\label{sec:1:susyrep}

The Casimir operators of the superalgebra are given by
\begin{eqnarray}
P_\mu^{} P^\mu
~~~\mbox{and}~~~
 B_\mu^{} B^\mu P_\nu^{} P^\nu -  B_\mu^{} B^\nu P_\nu^{}  P^\mu
\end{eqnarray}
where $B_\mu^{}$ is a generalisation of the angular momentum operator, and $P_\mu^{}$ is the usual momentum operator. The first Casimir operator is common to the Poincar\'{e} algebra defining the rest mass of the representation, $m$. In the rest frame the second Casimir operator reduces to $m^2 \sum_{k=1}^{3} B_k^{} B_k^{}$, and the $B_k^{}$ commute with the spinorial generators. 
\begin{eqnarray}
B_\mu^{} &\equiv& \frac12 \epsilon_{\mu\nu\rho\sigma}^{} P^{\nu} M^{\rho\sigma} - \frac14 \bar{Q} \bar{\sigma}_\mu^{} Q \\
B_k^{\mbox{\scriptsize (rest frame)}} &=& m J_k^{} - \frac14 \bar{Q}\, \bar{\sigma}_k^{} Q
\hspace{15mm} (k \neq 0)
\end{eqnarray}
The commutation relations with the standard angular momentum operator obey:
\begin{eqnarray}
\left[ Q_A^\alpha , J_3^{} \right] ~=~ \frac12 \left( \sigma_3^{} \right)_A^{~B} Q_B^\alpha
\hspace{10mm}
\left[ \bar{Q}_{\dot{A}}^\alpha , J_3^{} \right] ~=~ - \frac12 \left( \sigma_3^{} \right)_{\dot{A}}^{~\dot{B}} \bar{Q}_{\dot{B}}^\alpha
\end{eqnarray}
The action of the generator $\bar{Q}_1^{}$ on a Poincar\'{e} state increases the spin eigenvalue, $j_3^{}$, by a half, and $Q_1^{}$ lowers the eigenvalue by a half (and vice versa for $\bar{Q}_2^{}, Q_2^{}$). Since a super-Poincar\'{e} representation contains multiple Poincar\'{e} representations of differing spin, SUSY representations are commonly referred to as a supermultiplets.


For a finite dimensional super-Poincar\'{e} representation, a Poincar\'{e} state within the supermuliplet can always be found for a given $b_3^{}$ value, such that when acted on by any left-handed spinorial operator, $Q_A^\alpha$, it is annihilated. This state is known as the Clifford vacuum, $|\Omega \rangle$. The additional states with the same $b_3^{}$ value can then be found through repeated applications of the $\bar{Q}_{\dot{A}}^\alpha$ operators:
\begin{eqnarray}
| m, b_3^{} ; j_3^{} \rangle  &=& |\Omega\rangle \nonumber \\
| m, b_3^{} ; j_3^{} + \frac12 \rangle &\propto& \bar{Q}_{1
}^{\alpha}  \, | \Omega \rangle \nonumber\\
| m, b_3^{} ; j_3^{} - \frac12 \rangle &\propto& \bar{Q}_{2
}^{\beta}  \, | \Omega \rangle \nonumber\\
| m, b_3^{} ; j_3^{} \rangle &\propto& \bar{Q}_{1
}^{\alpha} \, \bar{Q}_{2
}^{\beta} \, | \Omega \rangle
\label{susyladder}
\end{eqnarray}
and so on. This procedure must be repeated for each $b_3^{}$ component in order to construct a complete supermultiplet. Note that in the absence of spinorial charges, the usual spin multiplet is recovered on iterating through the $b_3^{}$ values. For $N=1$ SUSY, the anticommutation relations imply that there are no further states for a given $b_3^{}$ in a supermultiplet beyond those listed in eq~(\ref{susyladder}).

For massive supermultiplets, all the states in eq~(\ref{susyladder}) are physical as they have non-zero norm, allowing normalisation. Choosing the rest frame to test for non-zero norms, the anticommutation relations of type $\{ Q, \bar{Q} \}$ obey $\{ Q_A^{\,\alpha} , \bar{Q}_{\dot{B}}^{\,\beta} \} = 2m\,  \delta^{\alpha\beta} \delta_{A\dot{B}}$. There are then $2^{2N-1}$ distinct complex fermionic and bosonic degrees of freedom each, for $\alpha, \beta \in \{ 1,  \cdots, N \}$ with a given $b_3^{}$ value.

For massless supermultiplets where there is no rest frame, it is convenient to choose the frame with momentum aligned along the third spatial axis. The non-zero anticommutation relations of type $\{Q, \bar{Q}\}$ are $\{ Q_1^{\,\alpha} , \bar{Q}_{1}^{\,\beta} \} = 4E\, \delta^{\alpha\beta}$. The states with $\bar{Q}_2^{}$ acting on the Clifford vacuum thus have zero norm and are not physical states. This reduces both the fermionic and bosonic degrees of freedom to $2^{N-1}$ each for a given $b_3^{}$ value. Non-zero central charges (allowed when $N>1$) may also further reduce the number of non-zero norm states present for the massive and massless cases.

From now on, discussion will be restricted to the minimal $N=1$ SUSY. This is phenomenologically favourable as to construct matter states in the fundamental representation of a gauge symmetry, Lagrangians cannot be written for supermultiplets with spin 1 component states. This restricts the generalised angular momentum of the matter supermultiplet to zero, and $N \leq 2$. The $N=2$ and $b_3^{} \in \{0\}$ supermultiplet contains opposite parity fermions, but parity must be broken to realise the left-handed electroweak symmetry. $N=1$ SUSY allows for chiral supermultiplets and so any extended supersymmetry must break to $N=1$ SUSY above an energy scale where electroweak symmetry is considered to exist. The analysis within this thesis is concerned with the energy range where electroweak symmetry is appropriate, and so $N=1$ SUSY is now assumed.


\subsection{Superspace}

By introducing spinorial Grassman parameters, $\theta$ and $\bar{\theta}$, the SUSY algebra can be written in terms of commutation relations alone. For N=1 SUSY,
\begin{eqnarray}
\left[ \theta^A \, Q_A^{} , \bar{Q}_{\dot{B}}^{} \, \bar{\theta}^{\dot{B}} \right] &=& 2 \left( \theta \sigma^\mu \bar{\theta} \right) P_\mu^{}  
\end{eqnarray}
The usual technique for Lie groups of exponentiation can be used to find a representation for the symmetry transformation operator. Ignoring Lorentz transformations, this operator is given by:
\begin{eqnarray}
L (x^\mu,\theta,\bar{\theta}) &=& \exp \left[ -i x^\mu P_\mu^{} + i \theta Q + i \bar{\theta} \bar{Q} \right]
\end{eqnarray}
In addition to the usual spacetime co-ordinates, this formulation suggests that the new Grassman co-ordinates can be considered to generalise spacetime into a superspace.

Using the Baker-Campbell-Hausdorff formula, $e^A e^B = e^{A+B + \frac12 \left[ A,B \right] + \ldots}$, the action of two successive transformations is related to a single transformation by:
\begin{eqnarray}
L(0,\epsilon, \bar{\epsilon}) \, L(x^\mu,\theta,\bar{\theta}) &=& L(x^\mu + i\theta \sigma^\mu \bar{\epsilon} - i \epsilon \sigma^\mu \bar{\theta}, \theta+\epsilon, \bar{\theta} + \bar{\epsilon})
\end{eqnarray}

A field in the superspace, or superfield, can be expanded by a Taylor expansion along the Grassman co-ordinates to determine a general form of the superfield in terms of component fields that only depend on spacetime:
\begin{eqnarray}
\Phi (x, \theta ,\bar{\theta}) &=&
\phi(x) 
+ \sqrt2\,   \theta^A \psi_A^{} (x) + \sqrt2\, \bar{\theta}_{\dot{A}}^{} \bar{\chi}^{\dot{A}} (x) 
+ \theta \theta \, F(x) + \bar{\theta} \bar{\theta} \, G(x) 		\nonumber\\
&&~
+ \theta \sigma^\mu \bar{\theta} \, A_\mu^{} (x) 
+ \bar{\theta}\bar{\theta} \, \theta^A \lambda_A^{} (x) + \theta\theta \, \bar{\theta}_{\dot{A}}^{} \bar{\kappa}^{\dot{A}} (x) 
+ \theta \theta \, \bar{\theta}\bar{\theta} \, D(x) 
\label{superfield}
\end{eqnarray}
where under a SUSY transformation, $\Phi^\prime = L \Phi L^{-1}$.
This expansion is finite due to the symmetry properties of Grassman parameters, $\theta_A^{} \theta_B^{} = \frac12 \epsilon_{AB}^{} \, \theta \theta$. A product containing repeated Grassman numbers is zero. If the superfield has $SL(2,\mathbb{C})$ indices, then the component fields simply inherit these extra indices. To transform properly under the full super-Poincar\'{e} group, the component fields must be representations of the Poincar\'{e} group.

The product of an arbitrary number of superfields will also have a structure in terms of components that can be considered as a composite superfield. This composite superfield will then also transform correctly under SUSY transformations, a feature that will be useful for building Lagrangians.


The action of the symmetry generators on a superfield can be reproduced using differential operator representations. Considering an infinitessimal transformation, the momentum operator, $\hat{P}_{\mu}^{}$ is equivalent to $- i \partial_\mu^{}$. Similarly for the spinorial generators, the following identities apply:
\begin{eqnarray}
\hat{Q}_A^{} &=& - i\, \left( \partial_A^{} - i\, \sigma_{A\dot{B}}^\mu \bar{\theta}^{\dot{B}} \, \partial_\mu^{} \right)  \\
\hat{\bar{Q}}^{\dot{A}} &=& - i\, \left( \partial^{\dot{A}} - i \left( \bar{\sigma}^\mu \theta \right)^{\dot{A}}  \partial_\mu^{} \right) 
\end{eqnarray}
where $\partial_A^{} \equiv \partial / \partial \theta^A$ and $\partial^{\dot{A}} \equiv \partial / \partial \bar{\theta}_{\dot{A}}^{}$.

The effect of a SUSY transformation on the superfield, $\Phi$, will be to transform the component fields into a linear combination of the original fields and their first and second spacetime derivatives. It will generally only be the $\theta\theta \, \bar{\theta}\bar{\theta}$ term which remains invariant up to a total spacetime derivative term.

It is also useful to define covariant derivatives with respect to SUSY transformations, where by construction $\{ D_A^{}, \hat{Q}_B^{}\} = \{ D_A^{} , \hat{\bar{Q}}^{\dot{B}} \} = \{ \bar{D}_{\dot{A}}^{}, \hat{Q}_B^{}\} = \{ \bar{D}_{\dot{A}}^{},  \hat{\bar{Q}}^{\dot{B}} \} = 0$.
\begin{eqnarray}
D_A^{} &\equiv&  \partial_A^{} + i\, \sigma_{A\dot{B}}^\mu \bar{\theta}^{\dot{B}} \, \partial_\mu^{}
\\
\bar{D}^{\dot{A}} &\equiv&  \partial^{\dot{A}} + i \left( \bar{\sigma}^\mu \theta \right)^{\dot{A}} \, \partial_\mu^{}
\end{eqnarray}
These covariant derivatives anticommute with the symmetry transformation operators, and so superfields that transform correctly under symmetry operations can be constructed from the general superfield in eq~(\ref{superfield}) by repeated actions of these covariant derivatives.

Using the notation that $D^2 = D^A D_A^{}$ and $\bar{D}^2 = \bar{D}_{\dot{A}}^{} \bar{D}^{\dot{A}}$ and similarly for the Grassman parameters, it follows that $D_A^{} \, D^2\, \Phi = \bar{D}_{\dot{A}}^{} \, \bar{D}^2\, \Phi = 0$. This property can be used to introduce constrained superfields, $\Phi_L^{} = \bar{D}^2\, \Phi$ and $\Phi_R^{} = D^2\, \Phi$, where $\bar{D}\, \Phi_L^{} = 0$ and $D\, \Phi_R^{} = 0$. These constrained superfields are referred to as left and right-handed respectively by convention, or collectively as chiral. The application of constraints is important to identify the irreducible superfield representations as the degrees of freedom present in the general superfield do not match those identified in Section~\ref{sec:1:susyrep}.



A left-handed chiral superfield must generally be a function of $y^\mu \equiv x^\mu + i \theta \sigma^\mu \bar{\theta}$ and $\theta$ to satisfy the constraint, $\bar{D} \, \Phi_L^{} = 0$.
\begin{eqnarray}
\Phi_L^{} (y,\theta) &=& \phi_L^{} (y) + \sqrt{2}\, \theta \, \psi_L^{} (y) + \theta\theta \, F_L^{} (y) \\
\Phi_L^{} (x,\theta,\bar{\theta}) &=& \phi_L^{} (x) -i\theta \sigma^\mu \bar{\theta} \, \partial_\mu^{} \phi_L^{} (x) - \frac14\, \theta\theta \, \bar{\theta}\bar{\theta} \, \partial^\mu \partial_\mu \phi_L^{} (x) \nonumber\\
&&~ + \sqrt{2} \, \theta \, \psi_L^{} (x) + \frac{i}{\sqrt2} \, \theta\theta \, \partial_\mu \psi_L^{}  \, \sigma^\mu \bar{\theta} + \theta\theta \, F_L^{} (x) 
\label{leftx}
\end{eqnarray}
For the right-handed superfields, $\bar{y}^\mu \equiv x^\mu - i \theta \sigma^\mu \bar{\theta}$ is a convenient parameterisation giving, $\Phi_R^{} (\bar{y},\bar{\theta}) = \phi_R^{} (\bar{y}) + \sqrt{2}\, \bar{\theta} \, \psi_R^{}  (\bar{y}) + \bar{\theta}\bar{\theta} \, F_R^{} (\bar{y})$. A left handed superfield can be transformed into a right-handed superfield by Hermitian conjugation.

The $\theta\theta\bar{\theta}\bar{\theta}$ component is always a total spacetime derivative for a chiral superfield, but the $\theta\theta$ and $\bar{\theta}\bar{\theta}$ components also now remain invariant under SUSY transformations up to a total spacetime derivative. Products of superfields with equivalent chirality obey the same chirality constraint. A product of superfields with differing chirality does not obey either constraint, but the composite can still be considered as a general superfield for symmetry purposes.

Note that the degrees of freedom within the superfield have been reduced by the constraint. However, there still appears to be too many bosonic degrees of freedom in comparison with the discussion in Section~\ref{sec:1:susyrep}. It will be determined in the following subsection that a supersymmetric Lagrangian restricts the $F$ field to be non-dynamical. This field does remain vital though to maintain supersymmetric invariance and also to account for the difference in degrees of freedom when the dynamical fields are off-shell.


\section{Supersymmetric Lagrangians}\label{sec:1:susyLag}

Supersymmetry transformations change the component fields in the superspace co-ordinates. For physics to be invariant with respect to supersymmetry, the action must be invariant. This demands that any Lagrangian in terms of the component fields of a SUSY representation is invariant up to a total spacetime derivative, which vanishes for appropriate boundary conditions.

As observed in the previous subsection, this can be achieved by extracting the $\theta\theta\bar{\theta}\bar{\theta}$ component of a general superfield or the $\theta\theta$ ($\bar{\theta}\bar{\theta}$) component of a left (right) handed chiral superfield, or any product of such fields. Extraction of the components is possible using integration over the Grassman variables, where by definition,
\begin{eqnarray}
\int d^2 \theta \left[ f(x) + \theta^A \, g_A (x) + \theta \theta \,h (x) \right] ~\equiv~ h (x)
\label{gint1} \\
\int d^2 \bar{\theta} \left[ f(x) + \bar{\theta}_{\dot{A}}^{} \,g^{\dot{A}} (x) + \bar{\theta} \bar{\theta} \,h (x) \right] ~\equiv~ h (x) 
\label{gint2} \\
\int d^4 \theta \left[ f (x) +\, \cdots \,+  \theta \theta \bar{\theta} \bar{\theta} \, h(x) \right] ~\equiv~ h(x)
\label{gint3}
\end{eqnarray}
for arbitrary functions $f(x),g(x),h(x)$.

As a first step to a physical Lagrangian, consider the product of a left-handed chiral superfield and its Hermitian conjugate. Using eq~(\ref{leftx}) and Grassman identities, we find
\begin{eqnarray}
\int  d^4 \theta \left(\Phi_L^{} \right)^\dagger \Phi_L^{} &=& - \, \phi_L^* \, \partial_\mu \partial^\mu \phi_L^{} - i \bar{\psi}_L^{} \bar{\sigma}^\mu \partial_\mu^{} \psi_L^{} + F_L^* F_L^{}
\end{eqnarray}
which contains the standard kinetic terms for a complex scalar field, $\phi$, and a Weyl fermion, $\psi_L^{}$.

We can now associate mass dimensions with the fields and parameters. As the action is dimensionless in ``natural" units (where $\hbar = c = 1$), and the $d^4 x$ measure has mass dimension $-4$ (with $\partial_\mu$ having mass dimension 1), the scalar field, $\phi$, and chiral superfield, $\Phi$, must both have mass dimension 1. The Weyl fermion, $\psi$, has mass dimension $\frac32$ and the non-dynamical field, $F$, has mass dimension 2. We can also assign a mass dimension to the Grassman parameters with $\theta$ and $\bar{\theta}$ having mass dimension $-\frac12$ and the Grassman measure having opposite mass dimension for eqs~(\ref{gint1}) to (\ref{gint3}) to hold.

For a renormalisable model with a Hermitian Hamiltonian, a supersymmetric Lagrangian must then take the form,
\begin{eqnarray}
\int d^4 x \, d^4 \theta \left[  \left(\Phi_L^{} \right)^\dagger \Phi_L^{} + \left(  \bar{\theta}^2 \,W \left( \Phi_L^{} \right)  + \hc \right) \right]
\label{chiralLagrangian}
\end{eqnarray}
where $W$, known as the superpotential, is a polynomial in left-handed chiral superfields up to $O(\Phi_L^3)$ as it has mass dimension 3. Note that there can be no terms with derivatives of the component fields of a chiral superfield generated by Grassman integration of the superpotential, given eq~(\ref{leftx}). It will be shown that the superpotential is responsible for the interactions and mass of the chiral superfield component fields, hence the name.

To include gauge interactions and dynamics, it is necessary to find another type of constrained superfield with the appropriate degrees of freedom. For this, consider applying a reality condition to a general superfield, $V = V^\dagger$, restricting the vector component field, $A_\mu^{}$, in eq~(\ref{superfield}) to be real. $V$ is referred to as a vector superfield as it will contain a physical vector component but not because the superfield has any Lorentz indices (which it does not).

A superfield generalisation of a gauge transformation, or ``supergauge transformation", can be proposed to obey the following rules.
\begin{eqnarray}
e^{V^\prime} &=& e^{-i\Lambda^\dagger} \, e^V \, e^{i\Lambda} \\
e^{-V^\prime} &=& e^{-i\Lambda} \, e^{-V} \, e^{i\Lambda^\dagger} \\
\Phi^\prime_L &=& e^{-i\Lambda} ~ \Phi_L^{} \\
\Phi^\prime_R &=& \Phi_R^{} ~ e^{i\Lambda^\dagger} 
\end{eqnarray}
where $\Lambda$ is some arbitrary left-handed chiral superfield. From the Baker-Campbell-Hausdorff formula, the vector superfield transforms as follows.
\begin{eqnarray}
V^\prime &=& V + i \left( \Lambda - \Lambda^\dagger \right) +\frac{i}{2} \left[ V, \Lambda + \Lambda^\dagger \right] + O \left( \Lambda^2 \right)
\end{eqnarray}
Note that $V^\prime$ still satisfies the reality condition. The notation above implicitly assumes summation over the gauge group generators, $T^a$. In more explicit notation, let $V = 2g V^a T^a$ and  $\Lambda = 2g \Lambda^a T^a$. Using the notation that the scalar component of $\Lambda^a$ is $\xi^a$, the transformation of the $A_\mu^{}$ component follows:
\begin{eqnarray}
A^{a\,\prime}_\mu &=& A^a_\mu  +  \partial_\mu^{} \left( \mbox{Re} \left[ 2 \xi^a \right] \right) - g \, f^{abc} \, A^b_\mu \times \mbox{Re} \left[ 2 \xi^c \right] \, +\, O(\xi^2)
\end{eqnarray}
which does indeed give the usual gauge transformation for the vector component field, where $f^{abc}$ is the structure constant of the gauge algebra. The degrees of freedom in $\Lambda$ can be chosen to leave $\mbox{Re} \left[ \xi \right]$ arbitrary, but set others to cancel many of the component fields in $V$. This choice of supergauge is known as the Wess-Zumino gauge. In this gauge the non-zero components of the vector superfield are:
\begin{eqnarray}
V_{WZ}^a &=& \theta \sigma^\mu \bar{\theta} \, A_\mu^a (x) + \theta\theta \, \bar{\theta} \bar{\lambda}^a (x) + \bar{\theta}\bar{\theta} \, \theta \lambda^a (x) + \theta\theta \,\bar{\theta}\bar{\theta} \, D^a (x)
\end{eqnarray}
The action of a SUSY transformation will generally change the supergauge, but this supergauge choice is useful for identifying the number of degrees of freedom in the constrained superfield.

To construct a superfield that will allow for kinetic terms of the component fields, consider the following definitions,
\begin{eqnarray}
W_A^{} &\equiv& -\frac14 \,  \bar{D}^2 e^{-V} D_A^{} e^V
\label{Wa} \\
\bar{W}^{\dot{A}} &\equiv& -\frac14 \,  D^2 e^{V} D_A^{} e^{-V}
\label{barWa}
\end{eqnarray}
where under a supergauge transformation, $W^\prime_A = e^{-i\Lambda} \, W_A^{} \, e^{i\Lambda}$. Note also that due to the action of the contracted covariant derivatives, the $W$ superfields obey the constraints $\bar{D} \, W_A^{} = 0$ and $D \, \bar{W}^{\dot{A}} = 0$. The $W_A^{}$ ($\bar{W}^{\dot{A}}$) are thus left- (right-) handed chiral superfields with a $SL(2,\mathbb{C})$ index. Their $\theta\theta$ and $\bar{\theta}\bar{\theta}$ terms will be invariant under SUSY transformation up to a total spacetime derivative.

To identify the components of these chiral superfields, perform an expansion of the exponentials in eqs~(\ref{Wa}) and (\ref{barWa}).
\begin{eqnarray}
e^{-V} D_A^{} \, e^V &=& D_A^{} V +\frac12 \left[ D_A^{} V , V \right] + O(V^3)
\end{eqnarray}
For Abelian symmetries, the commutator is zero. In the Wess-Zumino gauge, the $O(V^3)$ corrections are zero, which gives the relation,
\begin{eqnarray}
W_A^a (y,\theta) &=& \lambda^a_A (y) +  2\, \theta_A^{} \, D^a (y)  - \left( \sigma^{\mu\nu} \theta \right)_A F^a_{\mu\nu} (y) \nonumber\\
&&~+ i \theta\theta \, \sigma^{\mu}_{A\dot{B}} \left[ \partial_\mu \bar{\lambda}^{a \dot{B}} (y) - g f^{abc} A_\mu^b \bar{\lambda}^{c \dot{B}} (y) \right]
\end{eqnarray}

To maintain Lorentz invariance, there must be no free $SL(2,\mathbb{C})$ indices in the Lagrangian. It follows that the only renormalisable term in the Lagrangian involving these chiral superfields must be composed of $W^{a \, A} W^a_A$ and its Hermitian conjugate.

For $A_\mu^{}$ to be identified as a gauge boson, the vector superfield must have mass dimension~0, permitting the exponentiation of the superfield for use in Lagrangians. Similary, the $W_A^{}, \bar{W}^{\dot{A}}$ must both have mass dimension $\frac32$.

A supersymmetric Lagrangian can now be written with dynamical matter and gauge fields. For an arbitrary set of matter chiral superfields, $\Phi_i^{}$, the Lagrangian density is given by:
\begin{eqnarray}
{\mathcal{L}} &=& \int d^4 \theta \left( \Phi_i^\dagger \, e^{2 g\, T^a V^a} \Phi_i^{} + 2\kappa \,  V^a + \left[  \bar{\theta}^2 \left( \frac14 \left[ 1+\frac{i\tilde{\theta}}{8\pi^2} \right] W^{a\, A} W_A^a + W \left(\Phi \right)  \right)+\hc \right] \right) \nonumber\\[12pt]
&=&
F_i^* F_i^{} - i \bar{\psi}_i^{} \, \bar{\sigma}^\mu \, D_\mu^{} \psi_i^{} - \phi_i^* \, D_\mu^{} D^\mu \phi_i^{} + 2g \, D^a  \left( \phi_i^* \, T^a \phi_i^{} \right) - \sqrt2 g \left(  \bar{\psi}_i^{} \bar{\lambda}^a  T^a \phi_i^{} + \hc \right) \nonumber\\
&& +\, 2 D^a D^a - i \bar{\lambda}^a \, \bar{\sigma}^\mu \partial_\mu^{} \lambda^a  - \frac14 \, F_{\mu\nu}^a F^{a\, \mu\nu} + \frac{\tilde{\theta}}{32\pi^2} F_{\mu\nu}^a \tilde{F}^{a \, \mu\nu} +2\kappa \, D^a
\nonumber\\
&& +\, \left[  \frac{\partial W (\phi)}{\partial \phi_i^{}} \, F_i^{} -  \frac12 \, \frac{\partial W(\phi)}{\partial \phi_i^{} \partial \phi_j^{}} \, \psi_i^{} \psi_j^{}  +\hc \right] 
\end{eqnarray}
where $\tilde{\theta}$ is not a Grassman variable, but a CP violating parameter for the gauge field dynamics. The use of Latin indices for the different chiral superfields' F-terms, $F_i^{}$, distinguish them from the gauge vector field strength, $F^a_{\mu\nu}$, which possess Greek indices. Similarly, the usual spacetime covariant derivative, $D_\mu^{}$ is separate to the vector superfield's D-term, $D^a$.

The non-dynamical fields, $F_i^{}$ and $D^a$, can be integrated out of the Lagrangian using their on-shell equations of motion. This proceedure generates the following Lagrangian which describes equivalent physics.
\begin{eqnarray}
{\mathcal{L}} &=& -\, i \bar{\psi}_i^{} \, \bar{\sigma}^\mu \, D_\mu^{} \psi_i^{} - \phi_i^* \, D_\mu^{} D^\mu \phi_i^{} 
- i \bar{\lambda}^a \, \bar{\sigma}^\mu \partial_\mu^{} \lambda^a   - \frac14 \, F_{\mu\nu}^a F^{a\, \mu\nu}
+ \frac{\tilde{\theta}}{32\pi^2} F_{\mu\nu}^a \tilde{F}^{a \, \mu\nu}
\nonumber\\
&& 
-\, \sqrt2 g \left(  \bar{\psi}_i^{} \bar{\lambda}^a  T^a \phi_i^{} + \hc \right) 
- \frac12 \left( g  \left( \phi_i^* \, T^a \phi_i^{} \right) + \kappa \right)^2
\nonumber\\
&& -\,  \left| \frac{\partial W (\phi)}{\partial \phi_i^{}} \right|^2 
- \left(  \frac12 \, \frac{\partial W(\phi)}{\partial \phi_i^{} \partial \phi_j^{}} \, \psi_i^{} \psi_j^{}  +\hc \right)
\label{genLag}
\end{eqnarray}

In the presence of multiple gauge fields, the exponent of the exponential is replaced by a sum over the gauge superfields. Note also that only $U(1)$ gauge symmetries will permit a $V^a$ term in the Lagrangian. A general superpotential will be of the form
\begin{eqnarray}
W (\Phi) = h^i \Phi_i^{} + \frac12 \, \mu^{ij} \Phi_i^{} \Phi_j^{} + \frac{1}{3!} \, Y^{ijk} \Phi_i^{} \Phi_j^{} \Phi_k^{} 
\label{superpot}
\end{eqnarray}
For a single matter chiral superfield the interactions present in eq~(\ref{genLag}) include
\begin{eqnarray}
\left| \frac{\partial W (\phi)}{\partial \phi_i^{}} \right|^2 &=&
 \mu^2 \phi^2 + Y \mu\,  \phi^3 + \frac{Y^2}{4} \phi^4 + h \left( h + 2\mu\, \phi + Y \phi^2 \right) \\
\frac12 \, \frac{\partial W(\phi)}{\partial \phi_i^{} \partial \phi_j^{}} \, \psi_i^{} \psi_j^{}  &=&
\frac12 \left( \mu\, \psi \psi + Y\, \psi \psi \phi \right)
\end{eqnarray}
where the scalar, $\phi$, and Weyl fermion, $\psi$, have degenerate masses and interactions controlled by only one further parameter. The mass degeneracy for degrees of freedom within a supermultiplet was expected from the previous arguments on SUSY algebra possessing the Casimir operator, $P_\mu P^\mu$. If this were not the case, the supersymmetry must have necessarily been broken.

In a general $N=1$ SUSY theory, a $U(1)$ R-symmetry is permitted. This involves chiral rotations of the Grassman parameters. The Grassman measure, $d^4 \theta$, is invariant under such an operation, as are $\Phi^\dagger \Phi$ terms. However, for invariance under the R-symmetry, the superpotential must have an R-charge of magnitude 2. R-charges must then be assigned to the matter chiral superfields and only terms with appropriate composite R-charge are allowed in the superpotential.

Noether currents can be associated with continuous symmetries of Lagrangians, and the derivation holds for invariance under transformations with a spinorial generator. Consider a component field transforming as $\phi^\prime = \phi + \epsilon^A \left( \Delta \phi \right)_A^{}$ for some infinitessimally small Grassman parameter, $\epsilon^A$, leaving the Lagrangian invariant up to some total spacetime derivative, $\mathcal{L}^\prime = \mathcal{L} + \epsilon^A \partial_\mu^{} \mathcal{J}_A^\mu$ for some $\mathcal{J}_A^\mu$. A conserved current is formed by the definition,
\begin{eqnarray}
j_A^\mu &=&  {\mathcal{J}}_A^\mu - \sum_i^{} \frac{\partial {\mathcal{L}}}{\partial \left( \partial_\mu^{} \phi_i^{} \right)} \left( \Delta \phi_i^{} \right)_A^{}
\end{eqnarray}
where $\partial_\mu \, j_A^\mu = 0$ and the sum is over all component fields. Similarly, an independent Noether current can be defined for transformations, $\phi^\prime = \phi + \bar{\epsilon}_{\dot{A}}^{} \left( \Delta \phi \right)^{\dot{A}}$. The SUSY Noether currents, or ``supercurrents", have three $SL(2,\mathbb{C})$ indices (as a Lorentz vector index, $\mu$, requires two $SL(2,\mathbb{C})$ indices), and have spin $\frac32$ each.

\section{Supersymmetry breaking}


The presence of supersymmetry forces a mass degeneracy between pairs of bosonic and fermionic degrees of freedom. As experimental limits exclude this possibility, then if SUSY is a symmetry of our universe it must be spontaneously broken in the phase that we observe. A sign of supersymmetry breaking is that the vacuum transforms non-trivially under SUSY transformations. For an infinitessimal transformation,
\begin{eqnarray}
| 0 \rangle^\prime &\equiv& | 0 \rangle + i \left( \epsilon^A Q_A^{} + \bar{\epsilon}_{\dot{A}}^{} \bar{Q}^{\dot{A}} \right) | 0 \rangle
\end{eqnarray}
and so for preservation of supersymmetry, $Q_A^{}$ and $\bar{Q}^{\dot{A}}$ must annihilate the vacuum. As the Hamiltonian can be written in terms of these spinorial generators using the SUSY algebra,
\begin{eqnarray}
\langle 0 | \hat{H} | 0 \rangle ~=~
\langle 0 | P^0 | 0 \rangle &=&
\frac14 \, \langle 0 |  \left( \{ Q_1^{}, \bar{Q}_{\dot{1}}^{} \} +  \{ Q_2^{}, \bar{Q}_{\dot{2}}^{} \} \right) | 0 \rangle
\end{eqnarray}
the energy of a vacuum invariant under (global) SUSY transformations must be zero. A theory with supersymmetry does not allow arbitrary shifts in the energy scale to set the vacuum to have zero energy.

The terms in the potential of the Lagrangian in eq~(\ref{genLag}) which only include scalar degrees of freedom are collectively known as the scalar potential. These can be expressed solely in terms of the auxiliary fields, which are non-dynamical.
\begin{eqnarray}
V_S^{} &=& \sum_i^{} \left| F_i^{} \right|^2 + 2 \, D^a D^a
\end{eqnarray}
As the gauge superfield $D$-terms are real, each term in the scalar potential is positive definite. For the vacuum to have non-zero energy and thus SUSY to be broken, one of the auxiliary fields must have a non-zero vacuum expectation value (vev). The terms in the potential with non-scalar fields must necessarily have null vevs to maintain Lorentz invariance.

To break supersymmetry whilst maintaining internal symmetries (in the presence of non-trivial interactions), multiple matter chiral superfields are required. To realise this, consider the case of a single matter superfield. If this solitary matter field is charged under a gauge group, quantum anomalies necessarily break the internal gauge symmetry, irrespective of whether supersymmetry is broken. The matter field is then excluded from being charged under the gauge group. If the matter field has mass and/or self interactions, the $F$-term vev is then always null, as the complex scalar field will acquire a vev in order to minimise the scalar potential.


The requirement for multiple matter fields is not a problem phenomenologically, but to illustrate SUSY breaking in a toy model, this must be realised. For an example of $F$-term SUSY breaking, consider the O'Raifearteigh model with the following superpotential
\begin{eqnarray}
W &=& m\, \Phi_2^{} \Phi_3^{} + \lambda\, \Phi_1^{} \left( \Phi_3^2 - \mu^2 \right)
\end{eqnarray}
and canonical diagonal kinetic terms. The scalar potential is given by,
\begin{eqnarray}
V_S^{} &=& 
\left| \lambda \left( \phi_3^2 - \mu^2 \right) \right|^2
+ \left| m \phi_3^{} \right|^2
+ \left| m \phi_2^{} + 2\lambda \phi_1^{} \phi_3^{} \right|^2 
\label{oRpot} \\
&=& \lambda^2 \mu^4 
+ \frac12
\left( \phi_3^* \hspace{2mm} \phi_3^{} \right)
\left( \begin{array}{cc} m^2 & -2\lambda^2 \mu^2 \\ -2\lambda^2 \mu^2 & m^2   
 \end{array}\right)
\left( \begin{array}{c} \phi_3^{}\\ \phi_3^*  \end{array}\right)
+ m^2 \left| \phi_2 \right|^2
\nonumber\\
&& +\,   \left( 2\lambda m\,  \phi_1^* \phi_2^{} \phi_3^{} +\hc \right) + 4 \lambda^2 \left| \phi_1^{} \phi_3^{} \right|^2 +\, \lambda^2 \left|  \phi_3 \right|^4 
\nonumber
\end{eqnarray}
It is then impossible for any choice of $\langle \phi_3^{} \rangle$ to obtain a vacuum solution with zero energy. Supersymmetry is broken spontaneously as although the Lagrangian is still invariant under SUSY transformations, the vacuum is not. The bosonic degrees of freedom have their degeneracy split with respect to the fermionic degrees of freedom. For example, when $m^2 \geq 2 \lambda^2 \mu^2$, $\langle \phi_{2,3}^{} \rangle = 0$ and the scalar mass spectrum includes $\sqrt{m^2 \pm 2\lambda^2\mu^2}$ whilst the non-zero fermion masses present are $m$. 


To demonstrate $D$-term SUSY breaking, consider the Fayet-Iliopoulos model which has a $U(1)$ gauge field with a non-zero $\kappa$ term in eq~(\ref{genLag}). There are also two matter chiral superfields, $\Phi_\pm^{}$ with respective $U(1)$ charges, $\pm q$. The superpotential can only contain the term, $W = m\, \Phi_+^{} \Phi_-^{}$. The Fayet-Iliopoulos model then contains the following equations of motion for the auxiliary fields,
\begin{eqnarray}
 F_\pm^*  &=& - m \phi_\mp^{}
\hspace{15mm}
 D  ~=~ - \frac12 \left[ \kappa + g q \left( \left|  \phi_+^{}  \right|^2 -  \left|  \phi_-^{}  \right|^2 \right) \right]
\end{eqnarray}
Note that it is impossible to simultaneously set $\langle F_\pm^{} \rangle = \langle D \rangle = 0$ for any non-zero $\kappa$ and so supersymmetry must be broken. If $m^2 > gq\kappa/2$, the vacuum is found at $\langle \phi_\pm^{} \rangle = 0$ which preserves the gauge symmetry and it is the $D$-term with a non-zero vev. The scalars again receive a mass splitting, $\sqrt{m^2 \pm g q\kappa / 2}$.  If $m^2 < gq\kappa/2$, both SUSY and the gauge symmetry are broken in this model.

The splitting of bosonic degrees of freedom is a common feature in spontaneously broken SUSY theories. For a tree level Lagrangian, the following sum rule is obeyed
\begin{eqnarray}
\mbox{Supertrace} ~\equiv~ \sum \left[ (-1)^{2J} (2J+1) \, m_J^2 \right] &=& -4 \sum g\, T^a \langle D^a \rangle
\end{eqnarray}
where the sum is over all superfields that may mix and $J$ is the spin of the component field.

After accounting for quantum corrections, the superpartner states of Standard Model fields should have been observed if the vev of an auxiliary field associated with a SM field is responsible for spontaneous SUSY breaking. This introduces the need of a hidden sector with no SM charges to be responsible for SUSY breaking in order to avoid the quantum corrected sum rule constraint. The breaking of supersymmetry would then be communicated through messenger states which couple to both the Standard Model and hidden sectors, and/or certainly through gravitational interactions. To avoid modelling the hidden sector physics, effective theories can be considered.

\section{Minimal supergravity}\label{mSUGRA}

Using the covariant derivative, $\bar{D}$, the action related to the Lagrangian in eq~(\ref{genLag}) can be equivalently expressed as follows,
\begin{eqnarray}
S &=& \int d^4x \, d^2 \theta  \left( - \frac{\bar{D}^2}{8} \,  \left[ \Phi^\dagger e^{V} \right]_i^{} \Phi_i^{} + W \left( \{ \Phi \} \right) + \frac14 \, W^{a \, A} W^a_{A}  \right) + \hc 
\end{eqnarray}
where the CP violating parameter, $\tilde{\theta}$, is dropped for simplicity, as is the $\kappa$ parameter that would lead to SUSY breaking.

For a theory to include gravity, non-renormalisable interactions must be present. The most general Lagrangian consistent with global supersymmetry is of the form:
\begin{eqnarray}
S_{\mbox{\scriptsize global}}^{} &=& \int d^4x \, d^2 \theta  \Big( - \frac{\bar{D}^2}{8} ~  \mathcal{K} \left( \left[ \Phi^\dagger e^{V} \right]_i^{},  \Phi_i^{} \right) \nonumber\\
&& \hspace{5mm}
+\,  W \left( \{ \Phi \} \right) + \frac14  \, f_{ab}^{} \left( \{ \Phi , W_B^c \} \right) W^{a \, A} W^b_{A}  \Big) + \hc
\hspace{10mm}
\end{eqnarray}
where the $\mathcal{K}$ function is real and includes terms with arbitrary positive powers of the chiral superfields and their Hermitian conjugates. The superpotential is now a polynomial of infinite order and the gauge kinetic function, $f_{ab}^{}$ is also a polynomial function of chiral superfields. To maintain the appropriate mass dimension of the action, non-renormlisable terms have co-efficients with negative mass dimension.


It is now stated without proof (see \cite{Wess:1992cp}), that for a theory with local supersymmetry, the general form of the action invariant under general superspace coordinate transformations is of the form:
\begin{eqnarray}
S_{\mbox{\scriptsize local}}^{} &=& \int \left( d^4x \, d^4 \theta \, E \right)  \Big( \frac38 \, M_P^{} \left( M_P^{} \, \bar{D}^2 - 8 \mathcal{R} \right) \, \exp \left[ -\frac13 \, K \left( \left[ \Phi^\dagger e^{V} \right]_i^{},  \Phi_i^{} \right) \right] \nonumber\\
&& \hspace{5mm}
+\, W \left( \Phi_i^{} \right) + \frac14 \, f_{ab}^{} \left( \Phi_i^{} , W_B^c  \right) W^{a \, A} W^b_{A}  \Big) + \hc
\end{eqnarray}
where $K \equiv -3 \log \left[ \mathcal{K} / (3 M_P^2) \right]$. The $E$ superfield is introduced to form an integral measure which is invariant under the coordinate transformations. This is the generalisation of the $\sqrt{-\mbox{det} g_{\mu\nu}^{}}$ factor in general relativity. The $\mathcal{R}$ superfield is also the generalisation of the Ricci scalar in general relativity.

The Hilbert action for a spin 2 field is obtained from the zeroth order term in the expansion of the exponential, and so the respective component field is identified as the graviton. The Rarita-Schwinger action for a spin $\frac32$ field is also found, with the field referred to as the gravitino being the superpartner of the graviton. For this reason, the presence of a local supersymmetry leads to the identification of theories with such symmetry as supergravity.

By analogy with gauge symmetries, we must expect a dynamical spin $\frac32$ field when promoting the supersymmetry from global to local. In gauge theories, an $A_\mu j^\mu$ term is introduced into the Lagrangian to maintain invariance up to a total spacetime derivative, where $j^\mu$ is the Noether current of the symmetry. For supersymmetry, the Noether current was found to be spin $\frac32$ in Section~\ref{sec:1:susyLag} and so a spin $\frac32$ field is necessary to couple to the current. When the gauge symmetry is spontaneously broken, the field coupling to the Noether current acquires a mass, and similarly the gravitino acquires a mass when SUSY is broken.

In order to identify a low energy effective theory that may be derived from supergravity, a minimal model is now considered. The K\"{a}hler potential, $K$ and superpotential are chosen to be of the form below with diagonal kinetic terms as a first step in avoiding flavour mixing 
\begin{eqnarray}
K &=&  \hat{K} \left( \Sigma_m^\dagger , \Sigma_n^{} \right) + M_P^{-2} \, \sum_i^{} \tilde{K}^{\left( i \right)} \left( \Sigma_m^\dagger, \Sigma_n^{} \right) \, \Phi_i^\dagger \Phi_i^{} \nonumber\\
&& \hspace{5mm}+ \, M_P^{-2} \, \sum_{i,j}^{} Z^{ij} \left( \Sigma_m^\dagger, \Sigma_n^{} \right) \, \Phi_i \Phi_j^{} + O\left( M_P^{-4} \right) 
\label{kahlerSugra} \\
W &=& \hat{W} \left(  \Sigma \right) 
+ \frac12 \, \mu^{ij} \left(  \Sigma  \right) \, \Phi_i^{} \Phi_j^{} + \frac{1}{3!} \, Y^{ijk} \left(  \Sigma  \right) \, \Phi_i^{} \Phi_j^{} \Phi_k^{}
\label{spotSugra}
\end{eqnarray} 
where $M_P^{}$ is the Planck mass, $\Phi_i^{}$ are some observable fields, and $\Sigma_n^{}$ are some hidden sector fields. The higher order terms in $\Phi$ are ignored as they will be suppressed by the Planck scale and so be negligible at low energy scales. However, with SUSY breaking occurring in the hidden sector the same approximation cannot be made for a small set of non-renormalisable interactions involving $\Sigma$.

The scalar potential of supergravity is given by
\begin{eqnarray}
V &=& - F_i^{} \, G^i_j \, \bar{F}^j - 3\, M_P^4 \, e^{-G/M_{P}^2} +  \frac12 \, D^a \, \left( \mbox{Re} f_{ab} \right)^{-1} D^b
\end{eqnarray}
where the $F_i^{}$ and $D^\alpha$ terms, and $G$ function are given by
\begin{eqnarray}
F_i^{} &=& 
M_{P}^{} \, e^{-G/(2M_P^2)} \left( G^{-1} \right)_i^j G_j^{}
+ \frac14 f^*_{ab, k} \left( G^{-1} \right)_i^k \bar{\lambda}^a \bar{\lambda}^b - \left( G^{-1} \right)_i^k G_k^{jl}\,  \psi_j^{} \psi_l^{} - \frac{1}{2 M_P^2} \, G_j^{} \, \bar{\psi}^j \psi_i^{}
\nonumber \\
D^a &=& G^i \left( T^a \right)_i^j \phi_j \nonumber\\
G &\equiv&  M_P^2 \left[ K - \log \frac{\left| W \left( \{ \Sigma \}, \{\Phi\} \right) \right|^2}{M_P^6} \right] 
\end{eqnarray}
using the notation that $G^i \equiv \partial G(\phi,\phi^*) / \partial \phi_i^{} $, and $G_j^{} \equiv \partial G(\phi,\phi^*) / \partial \phi_j^*$. The $\lambda^a$ are the gaugino fields and the component fields of $\Phi_j^{}$ include the scalar, $\phi_j^{}$, and Weyl fermion, $\psi_j^{}$.



On F-term SUSY breaking triggered by one of the hidden fields, the following terms are generated in the Lagrangian to leading order in $M_P^{-1}$:
\begin{eqnarray}
{\mathcal{L}}_{\mbox{\scriptsize soft}}^{} &=& - \left( m^2 \right)_i^j \, \phi_i^* \phi_j^{}  - \left[ \frac12 \, M_a^{} \lambda^a \lambda^a +  \frac12 \, \tilde{B}^{ij} \, \phi_i^{} \phi_j^{} + \frac{1}{3!} \, \tilde{A}^{ijk} \, \phi_i^{} \phi_j^{} \phi_k^{} + \hc \right]
\label{lsoft}
\end{eqnarray}
which when canonically normalised are given by:
\begin{eqnarray}
M_{a}^{} &=& \frac12\, \langle \mbox{Re} f_{a} \rangle^{-1} \langle F_k^{} \, \partial^k  f_{a}^{} \rangle   \\
\left(m^2 \right)_i^j &=& \delta_i^j \left[ m_{3/2}^2 +  V_0^{} / M_P^2 -  \langle F_n^{} \bar{F}^m  \, \partial^n \partial_m \log \tilde{K}^{(i)}  \rangle \right]   \\
\tilde{A}^{ijk} &=& Y^{ijk}_{\mbox{\tiny (phys.)}} ~ \langle  F_m^{}  \,  \partial^m \left( \hat{K} +  \log Y^{ijk} -  \log \left[ \tilde{K}^{(i)} \tilde{K}^{(j)} \tilde{K}^{(k)} \right] \right)  \rangle \\
\tilde{B}^{ij} &=& \mu^{ij}_{\mbox{\tiny (phys.)}} \, \langle F_m^{}  \,  \partial^m  \left( \hat{K} +  \log \mu^{ij} -  \log \left[ \tilde{K}^{(i)} \tilde{K}^{(j)} \right]  \right)  - m_{3/2}^{} \rangle
\end{eqnarray}
with the caveat that $O\left(Z^{ij} \right)$ terms have been omitted in $\tilde{B}^{ij}$, and the constraint $f_{ab}^{} = f_a^{} \delta_{ab}^{}$ is required to maintain gauge invariance of the observable fields. The gravitino mass, $m_{3/2}^{}$ and $V_0^{}$ parameter is given by
\begin{eqnarray}
m_{3/2}^{} &=& ~~M_P^{} \, e^{-\langle G \rangle / (2M_P^2)}  \\
V_0^{} &\equiv& - M_P^2 \, e^{-\langle G \rangle / M_P^2} \langle G^i \left( G^{-1} \right)_i^j G_j^{} + 3 \, M_P^2  \rangle
\end{eqnarray}

The choice of eqs~(\ref{kahlerSugra}) and (\ref{spotSugra}) has thus generated a certain pattern of terms in the effective Lagrangian that explicitly break supersymmetry. The $\tilde{K}^{(i)}$ parameters give the curvature of the metric in field space. If a common curvature is assumed for all observable fields, then universal soft scalar mass terms are generated.

The experimental constraints on flavour changing neutral currents places strong limits on the non-degeneracy of sfermion states with identical electroweak quantum numbers. The limits on third generation states are generally weaker though as, for example, the $B^0-\bar{B}^0$ mixing already has a large Standard Model contribution and significant theoretical errors. On running to the UV scale, it is important to obtain near universal soft masses amongst the generations for states with the same gauge quantum numbers. Further motivations will be provided in later sections for realising universality between other groups of states.

The procedure to canonically normalise leads to the $\mu^{ij}$ and $Y^{ijk}$ parameters present in the original superpotential of eq~(\ref{spotSugra}) not corresponding to the ``physical" parameters of the superpotential found after canonically normalising. The relation is not important for this discussion, but is mentioned for awareness. If the $\langle \partial^m \log Y^{ijk} \rangle$ term in $\tilde{A}^{ijk}$ is common to all observable fields, then the relation that $\tilde{A}^{ijk}$ is proportional to the physical Yukawa coupling, $Y^{ijk}$, is found. This also aids in eliminating the introduction of flavour changing processes.

Due to the proportionality of certain parameters, it is convenient to introduce new notation where $\tilde{B}^{ij} \equiv B^{ij} \, \mu^{ij}$, and $\tilde{A}^{ijk} \equiv A^{ijk} \, Y^{ijk}$, where the ``physical" label is now always used.


If the gauge kinetic function, $f_a^{} = \hat{f} + \left( \tilde{f}_a^{} / M_P^{} \right) \Sigma $, where the $F$-term of $\Sigma$ acquires a vev, the gaugino soft mass terms, $M_a^{}$ will be proportional to $\tilde{f}_a^{}$. In the minimal scenario, a common factor is assumed leading to universal gaugino masses.

Minimal supergravity (mSUGRA) is the model defined by assuming a structure in the non-renormalisable Lagrangian that leads to universal SUSY breaking parameters at tree level. These assumptions can be motivated as the following section will discuss, but it is not certain that nature will be realised in such a manner.

In addition to communicating SUSY breaking through the interactions of eq~(\ref{kahlerSugra}) and (\ref{spotSugra}), quantum anomalies associated with a local scaling of the superspace coordinates will generally break SUSY at loop level with a different pattern of contributions for the SUSY breaking parameters. If messenger states exist that couple to the hidden sector fields responsible for breaking SUSY and the observable SM fields, then further contributions to the SUSY breaking parameters will be present, again with a different pattern. It is assumed in mSUGRA that these contributions to the SUSY breaking parameters are negligible, and SUSY breaking is then parameterised at an appropriate energy scale by four universal parameters; $m_0^{}$, the soft scalar mass, $m_{1/2}^{}$, the soft gaugino mass, $A_0^{}$, the soft trilinear coupling, and $B_0^{}$, the soft bilinear parameter.

\section{Minimal Supersymmetric Standard Model}\label{MSSM}

The minimal set of superfields required to contain component fields that include a family of SM matter states and the Higgs field are listed in the Table~\ref{mssmRep}. There must then be three copies of the $Q,U,D,L,E$ left-handed chiral superfields to reproduce the three families, and vector superfields must be present to contain the gauge fields.
\begin{table}[h]
\center
\begin{tabular}[c]{|c||*{5}{c|}|c|c|}\hline
$$
& $Q_L^{\mbox{\textcolor{white}{t}}}$ &$\left(U_R \right)^{c}$& $ \left( D_R \right)^{c}$ 
& $L_L$ & $\left( E_R \right)^{c}$  
& $H_u^{}$ & $H_d^{}$
\\[2pt] \hline 
$SU(3)^{\mbox{\textcolor{white}{t}}}$
& $\mathbf{3}$ & $\mathbf{\bar{3}}$  & $\mathbf{\bar{3}}$ 
& $\mathbf{1}$ & $\mathbf{1}$ 
& $\mathbf{1}$ & $\mathbf{1}$
\\[2pt] \hline
$SU(2)^{\mbox{\textcolor{white}{t}}}$
& $\mathbf{2}$ & $\mathbf{1}$  & $\mathbf{1}$ 
& $\mathbf{2}$ & $\mathbf{1}$ 
& $\mathbf{2}$ & $\mathbf{2}$
\\[2pt] \hline
$U(1)_Y^{\mbox{\textcolor{white}{t}}}$
& $\frac16$ & $-\frac23$  & $\frac13$ 
& $-\frac12$ & $1$ 
& $\frac12$ & $-\frac12$
\\[2pt] \hline
\end{tabular}
\caption{Representations of MSSM particle content }
\label{mssmRep}
\end{table}

\noindent
$Q_L^{}$ contains the doublet of left-handed up and down quarks. The charge conjugate of $U_R^{}$ contains the charge conjugate of the right-handed up-type quark (which is left-handed), and similarly for $\left( D_R^{} \right)^c$. The $L_L^{}$ contains the doublet of left-handed electron and neutrino, and $E_R^{}$ is the right-handed electron.

The superpartners of the SM states have a naming convention that the partners of quarks and leptons are squarks, $\tilde{q}$, and sleptons, $\tilde{l}$. The partners of the gauge bosons are gauginos, $\tilde{g}$, and the partners of the Higgs bosons are Higgsinos, $\tilde{h}$.

In building the Minimal Supersymmetric Standard Model (MSSM), the only choice available to contain the SM matter is in chiral superfields. This is required since the fermion component field of a vector superfield must be in the adjoint representation of the gauge group for the Lagrangian of the vector component to respect gauge symmetry. None of the SM matter is in the adjoint representation of a gauge group. For the scalar Higgs, it is again a chiral superfield that is necessary for a scalar component field to be present.

The SM only contains a single Higgs field, but in constructing the MSSM, a second Higgs field is required. This is a consequence of the fermion component field affecting the triangular gauge anomaly. Without having two Higgs superfields with opposite hypercharge, the SM gauge symmetries would be broken through quantum loops. The holomorphicity restriction of the superpotential to only include superfields with the same chirality for SUSY invariance also leads to the requirement for a second Higgs superfield in order to give mass to both the up and down quark sector through the Higgs mechanism.

Having introduced the representations of the MSSM, a comment on gauge unification will now be offered (applying equally to the SM). One method of embedding the SM symmetries into a higher gauge group uses a $SU(5)$ symmetry. A selection of $SU(5)$ representations break down into $SU(3) \times SU(2) \times U(1)$ representations as follows:
\begin{eqnarray}
\mathbf{\bar{5}} &=& \left( \mathbf{\bar{3}}, \mathbf{1}, \frac13 \right) \oplus \left( \mathbf{1}, \mathbf{2}, -\frac12 \right) \\
\mathbf{10} &=& \left( \mathbf{3}, \mathbf{2}, \frac16 \right) \oplus \left( \mathbf{\bar{3}}, \mathbf{1}, -\frac23 \right) \oplus \left( \mathbf{1}, \mathbf{1}, 1 \right) \\
\mathbf{24} &=& \left( \mathbf{8}, \mathbf{1}, 0 \right) \oplus \left( \mathbf{1}, \mathbf{3}, 0 \right) \oplus \left( \mathbf{1}, \mathbf{1}, 0 \right) \oplus \left( \mathbf{3}, \mathbf{2}, -\frac56 \right) \oplus \left( \mathbf{\bar{3}}, \mathbf{2}, \frac56 \right) 
\end{eqnarray}
It can be seen that the matter fields can be grouped such that $ \{Q,U,E\} \in \mathbf{10}$, and $\{D,L)\} \in \mathbf{\bar{5}}$. The Higgs fields are contained in $\mathbf{5}$ and $\mathbf{\bar{5}}$, and the gauge fields in $\mathbf{24}$. The SM gauge symmetries could then be unified into $SU(5)$ at some high energy scale. Other embeddings are also possible.

The presence of gauge unification helps to justify some assumptions in minimal supergravity. Consider a non-renormalisable interaction, $W^A \, \Sigma \, W_A$ where the chiral superfield $\Sigma$ acquires an $F$-term vev. The $\Sigma$ superfield must be in a gauge representation that is the symmetric product of two adjoint representations in order to maintain the unified gauge symmetry. For $SU(5)$,
\begin{eqnarray}
\left( \mathbf{24} \times \mathbf{24}  \right)_s^{} &=& \mathbf{1} + \mathbf{24} + \mathbf{75} + \mathbf{200}
\end{eqnarray}
If $\Sigma$ is in the singlet representation, then the MSSM gauginos are forced to have universal soft masses at the scale of SUSY breaking. Soft masses for matter fields are similarly constrained when in a common multiplet of the unified gauge group. The assumptions leading to universality in mSUGRA may then be justified by further new physics. In general, there are 105 new real parameters required to describe the MSSM with explicit soft SUSY breaking terms. This is problematic for explorations of the parameter space and generally doesn't account for relations generated by UV completions with supersymmetry restored.


The MSSM superpotential can be constructed by writing down all interactions permitted by the $SU(3) \times SU(2) \times U(1)$ gauge symmetries.
\begin{eqnarray}
W_{\mbox{\tiny MSSM}}^{} &=&
\mu H_u^{} H_d^{} + Y_u^{} QU H_u^{} + Y_d^{} QD H_d^{} +Y_e^{} LE H_d^{} + W_{\mbox{\tiny RPV}}^{} \\
W_{\mbox{\tiny RPV}}^{} &\in& \{
L H_u^{} , LLE , LQD , UDD \}
\end{eqnarray}
where the Yukawa couplings, $Y_{u,d,e}^{}$ are $3\times3$ matrices in family space.

The interactions contained in $W_{\mbox{\tiny RPV}}^{}$ are allowed by the gauge symmetries but lead to phenomenological problems. This is where an R-symmetry can be helpful. A choice for the definition of the R-charge as $(-1)^{3(B-L)+2S}$, where $B, (L)$ is the baryon (lepton) number and $S$ is the spin of the component field is made. The presence of gaugino soft masses breaks the $U(1)_R^{}$ continuous symmetry allowed in $N=1$ SUSY to a discrete $\mathcal{Z}_2^{}$ parity. The R-parity violating (RPV) terms of the superpotential are then fully contained in $W_{\mbox{\tiny RPV}}^{}$. If this R-parity is realised, those terms are forbidden.

This choice of R-charges assigns a common charge to all of the component fields present in the SM, but an opposite charge to their superpartners. The presence of R-parity therefore makes the lightest supersymmetric particle (LSP) stable with respect to decay. If the LSP has no electric charge or colour, it becomes a candidate for dark matter. The Standard Model itself provides no such candidate.

However, it is not necessary to invoke R-parity to avoid the related phenomenological problems. Another candidate for a unified gauge group is $SO(10)$ that breaks to $SU(5)\times U(1)_\chi^{}$ with the $SU(5)$ breaking as mentioned before to the SM gauge groups. The necessary charge assignments of the $U(1)_\chi^{}$ forbid the terms in $W_{\mbox{\tiny RPV}}^{}$. The $SO(10)$ unified group also has the appealing feature that a whole family of matter is contained in its fundamental representation along with a SM gauge singlet that may be considered as a right handed neutrino. 

The phenomological problems alluded to from $W_{\mbox{\tiny RPV}}^{}$ include violation of lepton number $(LH_u^{}, LLE, LQD)$ and violation of baryon number $(UDD)$. The $UDD$ and $LQD$ interactions lead to tree level proton decay through squark mediation, $p \rightarrow \ell^+ \pi^0$, but the proton has an observed lifetime in excess of $10^{32}$ years. It is expected that gravitational interactions cause proton decay, and even Planck mass suppressed operators can produce too rapid proton decay. The presence of renormalisable interactions that lead to proton decay would therefore need extremely small coefficients or the squarks mediating decay must be very heavy in order to satisfy experimental bounds.  The following section will argue that both of these scenarios are ``un-natural" and so to maintain naturalness, R-parity will be assumed. Similar restrictions are found for the other RPV interactions.

\section{Hierarchy Problem}

The hierarchy problem of the Standard Model refers to the situation that the scale of electroweak symmetry breaking is $O(10^3\,\mbox{GeV})$ whereas gravity has an intrinsic scale set by the Planck Mass, $M_P^{} \sim O(10^{18}\,\mbox{GeV})$, hierarchically many orders of magnitude larger. As the Standard Model does not account for gravity, the model must be considered as an effective theory with an upper limit of validity at the energy scale where gravitational interactions become important.

To understand the influence of higher scale physics on low energy phenomenology, it is useful to consider renormalisable models with a hierarchical spectrum. The one loop corrections to scalar masses include terms of the form:
\begin{eqnarray}
\delta m^2 &\propto& i \lambda_j^{} \int \frac{d^4 k}{(2\pi)^4} \frac{1}{k^2 - m_j^2} 
\label{1lcor}
\end{eqnarray}
where $m_j^{}$ is of order the mass of the heaviest virtual particle within the loop of the given diagram, and $\lambda_j^{}$ is a coupling constant. The integral is divergent and must be regularised. For cutoff regularisation, the loop momentum is bounded by a regularisation mass scale, $\Lambda$, and for dimensional regularisation, evaluating the integral in $d~(= 4 -2\epsilon)$ dimensions produces a finite result. A regularisation mass scale is also required in dimensional regularisation to correct the mass dimensions of the Lagrangian parameters (eg $\lambda_j^{} \to \mu^{2\epsilon} \lambda_j^{}$). The value of the integral, $I$, in eq~\ref{1lcor} is then given by:
\begin{eqnarray}
I &=& \lim_{\Lambda \to \infty}^{} \left[ \frac{-i}{16\pi^2} \left( \Lambda^2 - m_j^2\, \log \frac{\Lambda^2 + m_j^2}{m_j^2} \right) \right] 
\\[10pt]
 &=& \,\, \lim_{\epsilon \to 0}^{} \, \left[ \frac{i m_j^2}{16\pi^2} \left( \frac{1}{\epsilon} + 1 - \gamma + \log (4\pi)   +  \log \frac{\mu^2}{m_j^2} + O(\epsilon)  \right)   \right]
\end{eqnarray}
for the cutoff and dimensional regularisation schemes respectively, where $\gamma$ is the Euler-Mascheroni constant.

To construct an effective theory, the degrees of freedom at the high mass scale, $M$, are removed from the Lagrangian. The contributions that this physics generates must be absorbed into the effective Lagrangian parameters of the reduced theory. The singularities are controlled through counterterms, but for both regularisation schemes there are $O(M^2 / 16\pi^2)$ contributions to scalar mass squared terms. The scalar masses then seem unprotected from acquiring masses of order close to the high energy physics scale.

In the effective theory of the Standard Model, the one loop corrections to the Higgs mass squared, $m_h^2$, is stated here with cutoff regularisation:
\begin{eqnarray}
\delta m_h^2 \approx \frac{3}{64\pi^2}  \left( 3g^2+g^{\prime 2} + 8\lambda -8 h_t^2 \right)  \Lambda^2 
\label{smHiggsLoop}
\end{eqnarray}
where $g\,(g^\prime)$ is the electroweak (hypercharge) gauge coupling, $\lambda$ is the quartic Higgs coupling $[(\lambda/4) h^4 \in \mathcal{L}]$, and $h_t^{}$ is the Yukawa coupling of the top quark. The cutoff here is restricted to be less than the Planck scale, as otherwise the effective theory framework breaks down.

The Higgs mass at tree level is given by, $m_h^2 = 2\lambda v^2$, where the Higs vev, $v = 246$\,GeV. In order to obtain a physical Higgs mass around the electroweak scale with new physics appearing at the Planck scale, the tree level mass squared then needs to be fine tuned to 1 part in $\sim10^{30}$ for cancellation of the loop corrections. This degree of tuning, controlled by the hierarchy of the electroweak and Planck scales, is considered extremely ``un-natural."

The reasoning for expecting the Higgs mass to be around the electroweak scale relies on demanding perturbativity and renormalisability below the Planck scale, and also for agreement with precision electroweak measurements. The 1-loop $\beta$ function for SM Higgs quartic coupling is,
\begin{eqnarray}
\beta_\lambda^{\mbox{\tiny SM}} \approx \frac{3}{8\pi^2} \left( 4\lambda^2 + 2 h_t^2 \lambda - h_t^2 \right)
\label{smBeta}
\end{eqnarray}
For $\lambda$ too large, a Landau pole will be encountered before the Planck scale. For $\lambda$ too small, the top Yukawa coupling can drive $\lambda$ negative before the Planck scale causing the Higgs potential to be unbounded from below at the renormalisable level. This bounds the Higgs mass from above and below which for a momentum cutoff at $O(M_P^{})$ permits a Higgs mass between around 130 and 180\,GeV \cite{pdg}. As the Higgs field contributes to loop corrections of electroweak physics, there are also indirect contraints from precision electroweak data giving an upper mass bound of 122\,(157)\,GeV at the 68\%\,(95\%) confidence level \cite{pdg}.

The above arguments hold for the Standard Model being valid up to the Planck scale, but for BSM models where new physics is encountered below the Planck scale, the contraints can be relaxed to permit a SM-like Higgs mass of up to $O(\mbox{TeV})$. A model independent argument for new physics close to the electroweak scale can also be found by demanding perturbative unitarity.


To determine the unitarity limit for a two body scattering process, it is useful to consider a partial wave expansion of the cross section.
\begin{eqnarray}
\sigma = \frac{16 \pi }{s} \sum_{\ell=0}^\infty \left( 2\ell+1 \right) | a_\ell^{} |^2
\end{eqnarray}
where $s$ is the centre of mass energy, $\ell$ is the angular momentum, and the partial wave amplitude satisfies, $| a_\ell^{} | \leq 1$. The unitarity limit is reached when $| a_\ell^{} | = 1$. Using the optical theorem, a general constraint is found, $|a_\ell^{}|^2 = \mbox{Im} (a_\ell^{})$, which given the structure of the partial wave amplitude implies that $|\mbox{Re} (a_\ell^{}) | \leq 1/2$.

Considering the case of elastic scattering of longitudinal electroweak gauge bosons, $W_L^\pm$, with momenta, $k_\pm^\mu \to \frac{\sqrt{s}}{2} \left( 1,0,0,\pm1 \right)$, and respective polarisation vectors, $\epsilon_\pm^\mu \to \frac{\sqrt{s}}{2m} \left( \pm1,0,0,1 \right)$, it follows that the contribution from the tree level diagram with a four point $W$ vertex to the $\ell = 0$ partial wave amplitude grows as $(s^2 / m_W^4)$. Adding tree level diagrams with photon and $Z$ mediation in the s- and t-channels, helps regularise the amplitude to grow as $(s / m_W^2)$. For perturbativity, a sufficiently accurate result should be obtainable by truncating the loop expansion. However, as the centre of mass energy is increased, the tree level amplitude will eventually violate unitarity. This signals a breakdown in the viability of the perturbative analysis.

Perturbative unitarity can be saved by a Higgs-like resonance, which modifies the tree level amplitude to behave as $(m_h^2 / m_W^2)$. An upper limit is then placed on this resonance, $m_h^{} < O(\mbox{TeV})$, such that the unitarity limit is not violated. The physics responsible for electroweak symmetry breaking is then expected from perturbativity arguments to be detectable below this $O(\mbox{TeV})$ scale. If perturbativity is not maintained, then composite electroweak boson states will form in order to obey unitarity of the scattering process. This situation will not be discussed further.

Solutions to the hierarchy problem through BSM physics are centred around modifying the loop corrections to the Higgs mass. It is useful to note the opposing contributions from the Higgs scalar quartic coupling and top Yukawa coupling in eq~(\ref{smHiggsLoop}). This follows from the spin statistics introducing a relative sign difference for scalar and fermion loops. To avoid fine tuning of the Higgs mass, the quartic coupling may be very close to cancelling the contributions from the other parameters in the coefficient of $\Lambda^2$. However, this scenario simply transfers the fine tuning onto the quartic/Yukawa coupling and will require a full loop expansion to confirm.

Supersymmetry provides a solution which greatly improves the degree of fine tuning present. As shown in previous sections, the degrees of freedom associated with SUSY representations come in pairs of bosonic and fermionic states. In addition, the bosonic couplings are related to the fermionic couplings with the precise relation required to cancel the quadratic dependence on the cutoff. 

For BSM models with SUSY, the dependence of scalar masses on the cutoff scale is logarithmic as found for fermion masses in the SM. The SM renormalisation group equations are then only valid up to the energy scale where superpartner states are accessible. In order for the hierarchy problem to be solved, superpartner states must exist around the electroweak scale. The non-observation of superpartner states has placed lower bounds on the masses of superpartner states that are beginning to  suggest the requirement of a small hierarchy between the electroweak scale and SUSY state masses. This is referred to as the little hierarchy problem. An analysis of the degree of fine tuning present on satisfying the experimental bounds for mSUGRA is the subject of this thesis.

Alternative solutions for the hierarchy problem include positing extra compact spacetime dimensions. This reduces the energy scale at which gravity must become important, and so reduces the hierarchy between the electroweak and scale of gravity. Another popular model is technicolour, which suggests that electroweak symmetry breaking has a dynamical origin similar to the spontaneous chiral symmetry breaking in QCD. The electroweak scale is set by dimensional transmutation on confinement of the technicolour charge. Such solutions will not be discussed further.

\section{Measure of naturalness}\label{measure}

The concept of naturalness is somewhat vague being an aesthetic quality. A definition proposed by 't Hooft is \cite{gth}:
\begin{quotation}
\begin{singlespace}
The naturalness criterion states that one such [dimensionless] parameter is allowed to be much smaller than unity only if setting it to zero increases the symmetry of the theory. If this does not happen, the theory is unnatural.
\end{singlespace}
\end{quotation}

This definition assigns parameters as being natural or unnatural, but does not give a quantitative measure of the degree of naturalness. The current use of the term naturalness in the particle physics field has developed into referring to the sensitivity of observable quantities to small changes of the fundamental parameters governing the respective physics.

To illustrate this identification of naturalness, a crude example would be walking into a room and observing a pencil standing on its point on a table. This situation requires fine tuning in its positioning (although much less fine tuning than present in the SM hierarchy problem) and so is considered to be in an unnatural state. It can be commented that we are able to control the positioning of the pencil, but we cannot change the Lagrangian parameters of a physical theory. However, if these parameters are actually governed by a more fundamental theory, then the analogy may be appropriate.

The sensitivity definition of naturalness characterises the presence of apparent theoretical accidents required for observed physics. The degree of naturalness is then associated with the magnitude of the sensitivity. It is common to consider tuning of more than 1 part in 100 as unnatural, but being an aesthetic criterion, the transition from a natural to an unnatural theory is not immediate.

A quantitative measure first introduced by Ellis {\it{et al}} \cite{Ellis:1986yg}, but commonly referred to as the Barbieri-Giudice measure \cite{r3} in the literature is defined by
\begin{eqnarray}
\Delta_p^{} = \left| \frac{p}{X} \frac{\partial X}{\partial p} \right|
\end{eqnarray}
where $X$ is some observable quantity, and $p$ is some input parameter.

For the case that the observable is a linear combination of the parameters, this measure extracts the contribution from parameter $p$ and normalises with respect to the magnitude of the observable. If there exist large cancellations in order to generate the observable, then $\Delta$ is large and the observable can roughly be considered to require a fine tuning of 1 part in $\Delta$ with respect to the associated parameter. If each parameter contributes a similar amount, then $\Delta \sim 1/n$ where $n$ is the number of parameters. This is the natural limit.

In general, observables are not linear combinations of fundamental parameters. The above measure has the useful features that it is dimensionless by construction and is invariant under rescaling of the observable or parameter.

The above measure does not always agree with the qualitative expectations of naturalness though. For example, the mass of the proton is extremely sensitive under this measure to small changes in the strong gauge coupling constant at a high energy scale. However, dimensional transmutation is proposed to be a very natural way of generating a mass scale. Some care must then be taken in interpreting the results of such a measure.

Anderson and Casta\~{n}o suggested a modification to the naturalness measure \cite{Anderson:1994dz}, such that the ratio of $\Delta_p^{}$ to some appropriate average $\langle \Delta_p^{} \rangle$ over the parameter space is considered. A suitable probability distribution must then be assigned across the parameter space for the averaging process. This prior is not well-defined and introduces further ambiguity. It has been common to restrict the averaging integral over parameter space to the experimentally viable region. For any given point in parameter space, the output of this fine tuning measure then depends on the current experimental status, which makes comparison with previous or future calculations non-trivial.

This Anderson-Casta\~{n}o measure will find reduced values as the experimental limits are strengthened, or after discovery as more precise measurements are made. This measure is thus more a description of likelihood than absolute sensitivity to fundamental parameters. As the latter interpretation has been associated with naturalness here, the Anderson-Casta\~{n}o modification is not used for the analysis.



An alternative modification to the Barbieri-Giudice measure was proposed by Ciafaloni and Strumia \cite{Ciafaloni:1996zh} specifically for experimentally observed parameters. This involves weighting the Barbieri-Giudice measure by $(\sigma_p^{} / p)$, where $\sigma_p^{}$ is the experimental uncertainty in its value. The motivation is for identification of a probability with the inverse of the measure that the observed physics is described by the set of input parameters (within experimental errors). The weighting rescales the measure to account for possible variations in the measured input parameter. This fine tuning measure shares some of the problems of the Anderson-Casta\~{n}o measure in interpretation of its results, and comparison with other work. However, when the parameters being tested involve a mixture of measured and unmeasured parameters, more importance is then attributed to the unverified part of the model.

Beyond the choice of quantitative measure, there are also ambiguities in how to define the overall fine tuning of a given model. The option adopted for the following analysis is to take the maximum $\Delta_p^{}$ over all choices for the parameter, $p$. This choice returns the worst sensitivity with respect to any parameter. Another option commonly employed is to consider $\sqrt{\sum \Delta_p^2}$. Returning to the example of an observable that is a linear combination of underlying parameters, if only two terms involve a large cancellation, then the difference from the previous option is a factor of $\sqrt2$.

These two choices of a maximum, or power sum of the individual $\Delta_p^{}$, do not distinguish between few or many parameters requiring fine tuning. One may also consider the product of $\Delta_p^{}$ to test for such cases. However, as the lower limit of $\Delta_p^{}$ is zero and not unity the results of such a measure can be misleading. This section mentioned various limitations with all discussed fine tuning measures, and none are perfect for an indication of naturalness. An awareness of these limitations must be maintained when analysing the results of the following chapters.

\chapter{Fine tuning of electroweak scale within minimal supergravity}

The discussion in the introductory chapter provides phenomenological and theoretical motivations for extending the Standard Model (SM) symmetry to include minimal supersymmetry (SUSY). The Minimal Supersymmetric Standard Model (MSSM) must have some mechanism to spontaneously break SUSY in order to agree with the observations of our universe. However, as argued in the introduction, phenomenological constraints force SUSY breaking to occur in some hidden sector. The communication of SUSY breaking is here assumed to be mediated via minimal supergravity (mSUGRA) interactions.

One of the main motivations for low energy SUSY is its use as a solution to the hierarchy problem. Having used naturalness to motivate the introduction of SUSY, it then must be checked that natural theories are experimentally viable. The measure can also be used to identify regions of parameter space that have an acceptable degree of fine tuning, and so determine promising search strategies for discovery of such physics. If naturalness is abandoned, then any new physics not excluded by experiment may be equally likely to be realised, which leads to immense freedom in model building.

Section~\ref{hpot} presents the physics of the Higgs potential and determines the one and two loop leading log corrections to the Higgs potential. Section~\ref{ftform} then evaluates an analytic formula for the fine tuning of the electroweak scale based on the former Higgs potential. Section~\ref{section3} presents results for fine tuning across the mSUGRA parameter space. The dependence on constraints, input parameters and observables is discussed, and the influence of the fine tuning with respect to individual parameters analysed. The natural regions of the parameter space are then identified and naturalness limits obtained for the parameter space and spectrum. Finally the phenomenology of the natural realisations are discussed and a summary given.


\section{Higgs potential}\label{hpot}

The introductory chapter identified the physics of the Higgs sector as contributing significantly to the degree of naturalness present in a model. It was also mentioned that this sector is soon to be tested by experiment and so is of topical interest. This section discusses the general structure for self interactions of the MSSM Higgs scalar fields at the electroweak scale, and will be referred to in following sections for explaining the behaviour of the naturalness measure.

Introducing an alternative notation to that presented in Section~\ref{MSSM}, with $H_1^{} = H_d^{}$ and $H_2^{} = H_u^{}$, the scalar potential involving Higgs field terms is parameterised by the following,
\begin{eqnarray} 
V&=&  m_1^2\,\,\vert H_1\vert^2
+  m_2^2\,\,\vert H_2\vert^2
- (m_3^2\,\,H_1 \cdot H_2+\hc )\nonumber\\[6pt]
 &&
 ~+~
\frac{1}{2}\,\lambda_1 \,\vert H_1\vert^4
+\frac{1}{2}\,\lambda_2 \,\vert H_2\vert^4
+\lambda_3 \,\vert H_1\vert^2\,\vert H_2\vert^2\,
+\lambda_4\,\vert H_1\cdot H_2 \vert^2\nonumber\\[5pt]
 &&
 ~+~
\bigg[\,
\frac{1}{2}\,\lambda_5\,\,(H_1\cdot  H_2)^2+\lambda_6\,\,\vert H_1\vert^2\, 
(H_1 \cdot H_2)+
\lambda_7\,\,\vert H_2 \vert^2\,(H_1 \cdot H_2)+\hc \bigg]
\hspace{10mm}
\label{2hdm}
\end{eqnarray}
as for a general two Higgs doublet model, and $H_1.H_2=H_1^0\,H_2^0-H_1^- H_2^+$.

For analysis of the scalar potential, it is convenient to introduce the following parameters,
\begin{eqnarray}
m^2 &=&
 m_1^2 \, \cos^2 \beta +  m_2^2
 \, \sin^2 \beta - m_3^2 \, \sin 2\beta\nonumber\\[3pt]
\lambda &=&\frac{ \lambda_1^{} }{2} \, \cos^4 \beta 
+ \frac{ \lambda_2^{} }{2} \,  \sin^4 \beta 
+ \frac{ \lambda_{345}^{} }{4} \, \sin^2 2\beta 
+ \sin 2\beta \left( \lambda_6^{} \cos^2 \beta 
+ \lambda_7^{} \sin^2 \beta \right)\label{ml2}
\end{eqnarray}
where $\lambda_{345}^{} = \lambda_3^{} + \lambda_4^{} + \lambda_5^{}$. On the Higgs field acquiring vevs, $\langle H_1^0 \rangle = v \sin \beta /\sqrt2$, and $\langle H_2^0 \rangle = v \cos \beta /\sqrt2$, a similar form to the single Higgs potential is found. The vacuum minimisation conditions provide the following constraints:
\bea
v^2=-\frac{m^2}{\lambda} ,\qquad 
2 \lambda\, \frac{\partial m^2}{\partial \beta}=m^2\, \frac{\partial
  \lambda}{\partial \beta}
\eea
or equivalently,
 \begin{eqnarray}
 \frac{2m_3^2}{\sin 2\beta} ~=~ m_1^2 + m_2^2 + \frac{v^2}{2} \left[
   \lambda_1^{} \, c^2_\beta + \lambda_2^{} \, s^2_\beta + 
 \lambda_{345}^{} + \left( \lambda_6^{}+ \lambda_7^{} \right)  s_{2\beta}^{} 
 +\lambda_6^{} \cot \beta + \lambda_7^{} \tan \beta \right] 
 \label{B0fix}
\nonumber \\
 m_1^2 - m_2^2 \tan^2 \beta ~=~ -\frac{v^2}{2} \left[ \cos^2 \beta
   \left( \lambda_1^{} - \lambda_2^{} \tan^4 \beta \right) +
  \sin 2\beta \left( \lambda_6^{} - \lambda_7^{} \tan^2 \beta \right) \right]
  \hspace{12mm}
 \label{mu0fix}
 \end{eqnarray}
 
 In the two Higgs doublet model, there are eight real degrees of freedom. Electroweak symmetry breaking will adopt three of these degrees of freedom in order to give mass to the electroweak gauge bosons. This leaves two CP even fields, $h^0, H^0$, a CP odd field, $A^0$ and two charged fields, $H^\pm$. Considering the case of real parameters, the CP eigenstate fields do not mix. The masses of the CP even fields are given below, using the notation of \cite{Davidson:2005cw}
 \begin{eqnarray}
 m_A^2 &=& \frac{2 m_3^2}{\sin 2\beta } -\frac{ v^2 }{2} 
\, \left( 2 \lambda_5^{} +  \lambda_{6}^{} \cot \beta + 
\lambda_7^{} \tan \beta  \right) \\
m_{h,H}^2 &=& \frac12 \Big[ \, m_A^2 + v^2 \left( 2 \lambda +
 \Lambda_5^{} \right) \pm \sqrt{ \left[ m_A^2 + v^2 \left( \Lambda_5^{}-
 2 \lambda \right) \right]^2 + 4 v^4 \, \Lambda_6^2\,} \, \Big]
\\[5pt]
\Lambda_5^{} &=&
\frac{s^2_{2\beta} }{4} \, \left( \lambda_1^{} + 
\lambda_2^{} -2 \lambda_{345}^{} \right) + \lambda_5^{}
 - \frac{s_{4\beta}^{}}{2} \,(\lambda_6^{}-\lambda_7^{})
\nonumber\\[3pt]
\Lambda_6^{} &=&  \frac{s_{2\beta}^{} }{2} \left( \lambda_{3451}^{}
 \,c^2_{\beta}- \lambda_{3452}^{} \,s^2_{\beta} \right) 
+ \frac{c_{2\beta}^{}}{2}   \left( \lambda_6^{} + \lambda_7^{} \right)
+ \frac{c_{4\beta}^{}}{2}   \left( \lambda_6^{} - 
\lambda_7^{} \right) \nonumber
\end{eqnarray}
where $\lambda_{345j}^{} = \lambda_{345}^{}-\lambda_j^{}$, $(s_\beta, c_\beta) = (\sin \beta , \cos \beta)$, and $h^0$ is conventionally lighter than $H^0$. For large splitting (large $m_A^{}$), $m_h^2$ reaches an upper limit of $2\lambda v^2$ (which tends to $\lambda_2^{} v^2$ for large $\tan \beta$). In order to allow a large $h^0$ mass, a large effective quartic coupling is then required. The masses of the $A^0, H^0, H^\pm$ are all similar in this limit, with the $h^0$ remaining relatively light. A decoupling limit is thus obtained where the light Higgs is then SM-like.

For the MSSM, the only tree level contributions to the quartic couplings come from the electroweak D-terms. The effective quartic coupling is then, $\lambda = (g^2/8) \cos^2 2 \beta$, and the individual $\lambda_j^{}$ are given by:
\bea
\lambda_1=\lambda_2=\frac14 \,(g_1^2+g_2^2),\,\,\,\,\,
\lambda_3=\frac14\, (g_2^2-g_1^2), \,\,\,\,\,
\lambda_4=-\frac12\, g_2^2,\,\,\,\,\,
\lambda_{5,6,7}=0
\eea
where $g_1^{} \, (g_2^{})$ is the hypercharge (weak) gauge coupling and $g^2 \equiv g_1^2 + g_2^2$. 

Given that $m_Z^2 = g^2 \, v^2 / 4$, the upper limit to the MSSM light Higgs is $m_Z^2 \, \cos^2 2 \beta$ at tree level. For SM-like couplings (found in the large $m_A^{}$ limit), this scenario is experimentally ruled out. It is then necessary to rely on the loop corrections in order to possibly escape this constraint through the increase of the effective quartic coupling.


\subsection{1 Loop Leading Log (1LLL) Terms}

Using the MSSM renormalisation group equations \cite{Martin:1993zk}, but neglecting the mass scales of the SM and superpartner states, the scalar potential remains of the form,
\begin{eqnarray}\label{v0}
V^{(0)} &=&  
\bar{m}_1^2 \left| H_1^{} \right|^2 + \bar{m}_2^2 \left| H_2^{}
 \right|^2 - \bar{m}_3^2 \left( H_1^{} H_2^{} + \hc \right)
+ \frac{g^2}{8} \left( \left| H_1^{} \right|^2 - \left| H_2^{}
 \right|^2 \right)^2
\end{eqnarray}
However, once low energy scales are reached, the masses of the MSSM states significantly affect the running. This can be approximately accounted for by running down to the mass scale of a given state, integrating that state out of the Lagrangian, and then running below the mass scale using the updated renormalisation group equations. The effect of including the threshold dependence introduces corrections to the scalar potential which are accounted for by addition of the Coleman-Weinberg potential \cite{Coleman:1973jx}.

The 1 loop Coleman-Weinberg potential has the form,
\begin{eqnarray}
V^{(1)} &=& \frac{1}{64 \pi^2} \, \sum_k^{} \, 
 (-1)^{2 J_k^{}} \, (2 J_k^{}+1)\, g_k^{} 
\, m_k^4 \left( \log \frac{m_k^2}{Q^2} - \frac32 \right)
\label{cwpotential}
\end{eqnarray}
where $m_k^{}$ is the field dependent mass, defined as the second derivative of the potential with respect to the given field, evaluated for the vacuum state.
The degeneracy factor, $g_k^{}$, is 6 for (s)quarks, and $J_k^{}$ is the particle spin. All parameters in eq~(\ref{cwpotential}) are evaluated at the scale Q using the RGEs which ignore the particle masses. The field dependent squark masses are (neglecting $O\! \left( g^4 \right)$ terms): 
\medskip
\begin{eqnarray}
m_{\tilde{t}_{1,2}^{}}^2 &\approx& M_S^2 + h_t^2  \left| H_2^{} \right|^2  
+ \frac{g^2}{8} \left( \left| H_1^{} \right|^2 - \left| H_2^{} \right|^2  \right)
\mp  h_t^{} 
 \left| A_t^{} H_2^{} - \mu H_1^* \right|
\\
m_{\tilde{b}_{1,2}^{}}^2 &\approx& M_S^2 + h_b^2  \left| H_1^{} \right|^2  
+ \frac{g^2}{8} \left( \left| H_2^{} \right|^2 - \left| H_1^{} \right|^2  \right)
\mp h_b^{} 
\left| A_b^{} H_1^{} - \mu H_2^* \right|
\end{eqnarray}
and where $m_{Q,U,D}^{} (M_S^{}) = M_S^{}$ is assumed. Here, $M_S^{}$
is  the soft SUSY breaking squark mass evaluated at the squark mass scale.

One can expand the non-linear (logarithmic) field dependence in $V^{(1)}$ in
inverse powers of $1/M_S$ in order to
find the dominant threshold corrections, which 
 come from the third generation squarks:
\medskip
\begin{eqnarray}
V^{(1)}_{\tilde{t}_{1,2}^{}} &\approx& \frac{3}{16\pi^2} \Bigg[ 
 t \left( h_t^4 \left|H_2^{}  \right|^4 + 2h_t^2 \, M_S^2 \left|
 H_2^{} 
 \right|^2 + h_t^2 \left| A_t^{} H_2^{} - \mu H_1^* \right|^2 \right)
\nonumber \\
&& \hspace{15mm}
+~ h_t^4 ~ \frac{\left| A_t^{} H_2^{} - \mu H_1^* \right|^2}{M_S^2}
 \left( \left| H_2^{} \right|^2 - \frac{\left| A_t^{} H_2^{} - \mu
 H_1^* 
\right|^2}{12 M_S^2} \right)\nonumber\\
&& \hspace{5mm}
+~ \frac{g^2}{8} \left( \left| H_1^{} \right|^2 - \left| H_2^{} \right|^2 \right)
\left( 2t \, h_t^2 \left| H_2^{} \right|^2 + 2 M_S^2 \left( t -1
 \right) + 
\frac{\left| A_t^{} H_2^{} - \mu H_1^* \right|^2}{M_S^2}  \right)
\Bigg]\qquad
\end{eqnarray}
\begin{eqnarray}
V^{(1)}_{\tilde{b}_{1,2}^{}} &\approx& \frac{3}{16\pi^2} \Bigg[ 
 t \left( h_b^4 \left|H_1^{}  \right|^4 + 2h_b^2 \, M_S^2 \left|
 H_1^{}  
\right|^2 + h_b^2 \left| A_b^{} H_1^{} - \mu H_2^* \right|^2 \right)
\nonumber\\
&& \hspace{15mm}
+~ h_b^4 ~ \frac{\left| A_b^{} H_1^{} - \mu H_2^* \right|^2}{M_S^2} 
\left( \left| H_1^{} \right|^2 - \frac{\left| A_b^{} H_1^{} - \mu
 H_2^* 
\right|^2}{12 M_S^2} \right)
\nonumber\\
&& \hspace{5mm}
+~ \frac{g^2}{8} \left( \left| H_2^{} \right|^2 - \left| H_1^{} \right|^2 \right)
\left( 2t \, h_b^2 \left| H_1^{} \right|^2 + 2 M_S^2 \left( t -1
 \right) + 
\frac{\left| A_b^{} H_1^{} - \mu H_2^* \right|^2}{M_S^2}  \right)
\Bigg]\qquad\label{lam}
\end{eqnarray}
where 
\bea
t = \log (M_S^2/Q^2)
\eea

 When running below the EWSB scale,
the inclusion of higher dimensional terms (threshold corrections) lead to a
re-summation such that $M_S^{}$ is replaced by a mass scale related to
the physical particle masses \cite{Carena:1995wu}.
 For the results found using analytic formulae presented in this thesis,
the geometric mean of the stop masses is used in the place of $M_S^{}$.

The above equations are valid down to the top mass scale; below this
 scale threshold corrections from the top quark should also be included. The
 dominant effect of running below the top scale can be absorbed by
 setting $Q$ in the above equations as the ``running'' top mass
 evaluated at the scale $Q$ instead of the pole mass.

From eqs.(\ref{v0}) to (\ref{lam}) one obtains the 
parameters entering in the scalar potential (\ref{2hdm}), evaluated at
the scale Q (below $M_S$), in the one-loop leading log approximation:
\medskip
\begin{eqnarray}
m_1^2 &=& \bar{m}_1^2 - \frac{6h_b^2}{16 \pi^2} \, M_S^2  
+ \frac{3 }{16 \pi^2}  \left( 2 h_b^2 \, M_S^2 + h_b^2 A_b^2 
+ h_t^2 \mu^2 \right) t  \label{m1loop1}  \\[3pt]
m_2^2 &=&  \bar{m}_2^2 - \frac{6h_t^2}{16 \pi^2} \, M_S^2 
 + \frac{3 }{16 \pi^2}  \left( 2 h_t^2 \, M_S^2 + h_t^2 A_t^2
 + h_b^2 \mu^2 \right) t  \\[3pt]
m_3^2 &=& \bar{m}_3^2 + \frac{3 }{16 \pi^2} \left( h_t^2 A_t^{}
 + h_b^2 A_b^{} \right) \mu\, t 
 \label{lambda1}
\end{eqnarray}
\begin{eqnarray}
\lambda_1^{} &=& \frac{g^2}{4} \left( 1 +  \frac{3 \left( h_t^2\, 
\mu^2 - h_b^2\, A_b^2 \right)}{16 \pi^2 \,  M_S^2} \right) + 
\frac{3}{8 \pi^2} \left( 
 \frac{h_b^4 \, X_b^{}}{2} 
- \frac{h_t^4 \, \mu^4}{12 M_S^4} \right)
+ \frac{3h_b^2 }{8 \pi^2} \left( h_b^2 - \frac{g^2}{4} \right) t 
\hspace{10mm} \\[10pt]
\lambda_2^{} &=& \frac{g^2}{4} \left( 1 + \frac{3 \left( h_b^2\, 
\mu^2 - h_t^2\, A_t^2 \right) }{16 \pi^2 \,  M_S^2} \right) + 
\frac{3}{8 \pi^2} \left( 
 \frac{h_t^4 \, X_t^{}}{2} 
- \frac{h_b^4 \, \mu^4}{12 M_S^4} \right)
+ \frac{3h_t^2 }{8 \pi^2} \left( h_t^2 - \frac{g^2}{4} \right) t  
\eea
\bea
\lambda_{34}^{} &=& -\frac{g^2}{4}  \left( 1 +  \frac{ 3 h_t^2 
\left( \mu^2 - A_t^2 \right)  }{32\pi^2 \,  M_S^2} +  
\frac{ 3 h_b^2 \left( \mu^2 - A_b^2 \right) }{32\pi^2 \,  M_S^2} \right)
+ \frac{3 \left( h_t^2 + h_b^2 \right) }{16 \pi^2}\, 
\frac{g^2}{4}  \, t \nonumber \\[5pt]
&& \hspace{10mm}
+~ \frac{3 h_t^4}{16\pi^2} \left( \frac{ \mu^2 }{M_S^2} - 
\frac{  \mu^2 A_t^2 }{3 M_S^4} \right)
+ \frac{3 h_b^4}{16\pi^2} \left( \frac{ \mu^2 }{M_S^2} - 
\frac{  \mu^2 A_b^2 }{3 M_S^4} \right) \qquad \qquad  \quad
\eea
\bea
\lambda_5^{} &=& 
- \frac{3 h_t^4}{96\pi^2}   \frac{\mu^2 A_t^2}{M_S^4}
- \frac{3 h_b^4}{96\pi^2}  \frac{\mu^2 A_b^2}{M_S^4}  \\[10pt]
\lambda_6^{} &=& 
\frac{g^2}{4} \left( \frac{3 \mu \left( h_b^2 A_b^{} - h_t^2 A_t^{} 
\right)}{ 32 \pi^2 \, M_S^2} \right)
+ \frac{3 h_t^4}{96\pi^2} \, \frac{\mu^3 A_t^{}}{ M_S^4} 
+ \frac{3 h_b^4}{96\pi^2} \, \frac{\mu}{M_S^{}} 
\left( \frac{ A_b^3}{ M_S^3} - \frac{ 6 A_b^{}}{M_S^{}}  \right) 
\eea
\bea
\lambda_7^{} &=& 
\frac{g^2}{4} \left( \frac{3 \mu \left( h_t^2 A_t^{} - h_b^2 A_b^{} 
\right)}{ 32 \pi^2 \, M_S^2} \right)
+ \frac{3 h_b^4}{96\pi^2} \, \frac{\mu^3 A_b^{}}{ M_S^4} 
+ \frac{3 h_t^4}{96\pi^2} \, \frac{\mu}{M_S^{}} 
\left( \frac{ A_t^3}{ M_S^3} - \frac{ 6 A_t^{}}{M_S^{}}  \right)
\label{lambda7}
\end{eqnarray}

\medskip\noindent
These analytic results agree with \cite{Carena:1995bx} when ignoring
 the stop mixing corrections to the D-terms. 

\subsection{2 Loop Leading Log (2LLL) Terms}

The two-loop leading log (2LLL) Coleman-Weinberg potential 
can be found in the arXiv version of \cite{Espinosa:2000df}
to $O\! \left( g_3^2 \, h_t^4 , g_3^2 \, h_b^4 \right)$ and $O\!
\left( h_t^6, h_t^4 \, h_b^2, h_t^2 \, h_b^4, h_b^6 \right)$,  see
also \cite{Martin:2001vx,Martin:2002iu} for the general case.
 The method of
the previous section may be used to determine the 2LLL contributions
to the Higgs scalar potential, however here a similar
approach to that in
 \cite{Carena:1995wu} is used, to RG-improve the 1-loop result into a 2LLL result.
A step approximation is applied to the $\beta$-functions so that the
MSSM RG eqs are used between the GUT and stop mass scale, 
then the 2HDM SM RG eqs between the stop and top mass scales, and finally
the top is integrated out to reach the electroweak scale.

When setting the renormalisation scale in eqs~(\ref{m1loop1}) to
 (\ref{lambda7}) as $Q=M_S^{}$, the logarithmic terms are removed but
 the finite corrections from stop mixing remain. These results are
 then used as boundary conditions for the parameters at the scale
 $M_S^{}$. A series  expansion of the RG eqs can then be applied:
\medskip
\begin{eqnarray}
\lambda \left( Q \right) &\approx& \lambda \left( M_S^{} \right) 
- \beta_\lambda^{} \left( M_S^{} \right) \, t + \frac12 \, 
\beta^\prime_\lambda \left( M_S^{} \right) \, t^2 + O\! \left( t^3 \right) \\
&=& \lambda \left( M_S^{} \right) - \beta_\lambda^{} 
\left( Q \right) \, t - \frac12 \, \beta^\prime_\lambda 
\left( Q \right) \, t^2 + O\! \left( t^3 \right)
\label{rgimp}
\end{eqnarray}
where $\beta_p^{} = \partial p / \partial \log Q^2$.
Eventually, all parameters will be expressed at a scale $Q$ as 
in the Coleman-Weinberg potential approach. For a
 $\beta_{\lambda}^{}$-function of the form
 $b \, \lambda + c$, eq~(\ref{rgimp}) becomes
\medskip
\begin{eqnarray}
\lambda  &\approx& \lambda \left( M_S^{} \right) - 
t \left[ b \, \lambda \left( M_S^{} \right) + c  \right]
+ t^2 \left[ b \,c  - \frac12 \, \beta^{\prime}_{\lambda} + O\! 
\left( \lambda \right) \right]
\label{lambdarg}
\end{eqnarray}

\medskip\noindent
where the couplings are evaluated at the scale $Q$ unless stated
 otherwise. The $\beta$-functions for the 2HDM SM \cite{Haber:1993an}
 are  listed below, 
neglecting $O\! \left( h_\tau^2 \right)$ terms, and
 with the $\beta_{\lambda_i^{}}^{}$-functions also neglecting
 $O\! \left( g^4 ,  g^2  \lambda _i^{} , \lambda_i^2 \right)$ terms:
\medskip
\begin{eqnarray}
16 \pi^2 \, \beta_{m_{1}^2}^{} &=&
3 h_b^2 \, m_1^2  + O\left( g^2 m^2 \right)\nonumber  \\
16 \pi^2 \, \beta_{m_{2}^2}^{} &=&
3 h_t^2 m_2^2 + O\left( g^2 m^2 \right)\nonumber \\
16 \pi^2 \, \beta_{m_{3}^2}^{} &=&
\frac32 \left( h_t^2 +h_b^2 \right) m_3^2  + O\left( g^2 m^2 \right)
\end{eqnarray}
\begin{eqnarray}
16 \pi^2 \, \beta_{\lambda_1^{}}^{} &\approx&
6 h_b^2 \left( \lambda_1^{}  - h_b^2 \right)\nonumber  \\
16 \pi^2 \, \beta_{\lambda_2^{}}^{} &\approx&
6 h_t^2 \left( \lambda_2^{} - h_t^2 \right) \nonumber \\
16 \pi^2 \, \beta_{\lambda_3^{}}^{} &\approx&
3\lambda_3^{}  \left( h_t^2 + h_b^2 \right) 
-6 h_t^2 h_b^2 \nonumber \\
16 \pi^2 \, \beta_{\lambda_4^{}}^{} &\approx&
3\lambda_4^{} \left( h_t^2 + h_b^2 \right)  
+6 h_t^2 h_b^2  \nonumber \\
16 \pi^2 \, \beta_{\lambda_5^{}}^{} &\approx&
3 \lambda_5^{} \left( h_t^2 + h_b^2 \right) \nonumber \\
16 \pi^2 \, \beta_{\lambda_6^{}}^{} &\approx&
 \lambda_6^{} \left( \frac92 \, h_b^2 + \frac32 \, h_t^2 \right)
\nonumber \\
16 \pi^2 \, \beta_{\lambda_7^{}}^{} &\approx&
 \lambda_7^{} \left( \frac92 \, h_t^2 + \frac32 \, h_b^2 \right) 
\end{eqnarray}
and finally
\begin{eqnarray}
16 \pi^2 \, \beta_{h_t^2}^{} &\approx&
h_t^2 \left(
\frac92 \, h_t^2 + \frac12 \, h_b^2 - 8 g_3^2 
- \frac94 \, g_2^2 -\frac{17}{12} \, g_1^2 \right) \\
16 \pi^2 \, \beta_{h_b^2}^{} &\approx&
h_b^2 \left(
\frac92 \, h_b^2 + \frac12 \, h_t^2 + h_\tau^2
- 8 g_3^2 - \frac94 \, g_2^2 -\frac{5}{12} \, g_1^2 \right) 
\end{eqnarray}

\medskip
Using (\ref{lambdarg}), the analytic 2-loop results in
\cite{Carena:1995bx} are then recovered when the same level
 of approximation is considered. For example,
\medskip
\begin{eqnarray}
\lambda_2^{} &\approx& 
\left[ \lambda_2^{} \left( M_S^{} \right) 
-\lambda_2^{} \, a_2^{} \, t \right]  - b_2^{} \, t
+  \left[ \, a_2^{} \, b_2^{}  + \frac{3 h_t^2}{16\pi^2 }
 \left( 2 \beta_{h_t^2}^{} - \beta_{\lambda_2^{}}^{}  \right)
 + O\! \left(  \lambda \right)  \right]  t^2
\\[2pt]
&=& \left[ \lambda_2^{} \left( M_S^{} \right)
 -\lambda_2^{} \, \frac{6h_t^2}{16\pi^2}  \, t \right] 
+ \frac{3 h_t^4}{8\pi^2} \left[ t + \frac{1}{16 \pi^2}    
\left( \frac32 \, h_t^2 + \frac12 \, h_b^2 - 8 g_3^2 \right)  t^2 \right]
\end{eqnarray}

\medskip\noindent
The couplings entering  in the expression of $\lambda_2^{} \left(
M_S^{} \right)$ are re-expressed in terms of their values
 at the scale $Q$, (with a
 logarithmic correction which compensates for the running below $M_S^{}$):
\medskip
\begin{eqnarray}
h_t^4 (M_S^{} ) &=& h_t^4 \left( 1 + \frac{t}{16\pi^2} \left( 9  h_t^2
+  h_b^2 - 16 g_3^2 \right) + O\! \left( g^2 \,t, t^2 \right) \right)
\label{YukRunT} 
\\[3pt]
h_b^4 (M_S^{} ) &=& h_b^4 \left( 1 + \frac{t}{16\pi^2} 
\left( 9  h_b^2 + 
 h_t^2 - 16 g_3^2 \right) + O\! \left( g^2 \, t, t^2 \right) \right)
\label{YukRunB}
\end{eqnarray}

\medskip\noindent
This
leads to the following expression,
 in agreement with \cite{Carena:1995bx},
 when the stop mixing contributions to the D-terms in the 
potential are again neglected (which is not assumed for the analytic results presented in the following sections):
\medskip
\begin{eqnarray}
\lambda_2^{} &\approx&
\frac{g^2}{4} \left( 1- \frac{3 h_t^2}{8 \pi^2} \, t \right) 
-  \frac{3 h_b^4}{96 \pi^2} \, \frac{\mu^4}{M_S^4} \left[ 1 + 
\frac{t}{16\pi^2} \left( 9h_b^2-5 h_t^2 - 16 g_3^2 \right) \right]
\nonumber \\[10pt]
&& +~ \frac{3 h_t^4}{8\pi^2} \left[ t +\frac{X_t^{}}{2} + 
\frac{t}{16 \pi^2} \left( \frac{3 h_t^2}{2} + \frac{h_b^2}{2} 
- 8 g_3^2 \right) \left( X_t^{} + t \right) \right] 
\label{lambda22lll}
\end{eqnarray}
where the following notation has been introduced,
\begin{eqnarray}
X_{t,b}^{} &=& \frac{2 A_{t,b}^2}{M_S^2} \left( 1- 
\frac{A_{t,b}^2}{12 M_S^2} \right)
\end{eqnarray}

\medskip\noindent
Note that these results assume that the CP odd Higgs mass is not
decoupled. If this is the case, the usual SM $\beta$-functions
should be used. The effective quartic coupling at the EW scale when $m_A^{} 
\lesssim M_S^{}$ is given by:
\medskip
\begin{eqnarray}
\lambda &\approx& \frac{g^2}{8} \cos^2 2\beta 
\Big[ 1 - \frac{3}{16 \pi^2} 
 \left( h_b^2+h_t^2+ \left( h_b^2 - h_t^2 \right) 
\sec 2\beta \right) t \Big]
\nonumber \\[4pt]
&&
+~ \frac{3 \, h_t^4 }{16\pi^2}\, \sin^4 \beta \bigg[ t+
 \frac{\tilde{X}_t^{}}{2} \, +\frac{1}{16 \pi^2} \left( \frac{3\,
 h_t^2}{2} + \frac{h_b^2}{2} - 8 g_3^2
 \right) \left( \tilde{X}_t^{} \, t+t^2  \right) + \delta_1^{} \bigg]
\nonumber \\[4pt]
&&
+~ \frac{3 \, h_b^4 }{16\pi^2}\, \cos^4 \beta \bigg[ t+
 \frac{\tilde{X}_b^{}}{2} +\frac{1}{16 \pi^2} \left( \frac{3\,
 h_b^2}{2} + \frac{h_t^2}{2} - 8 g_3^2
 \right) \left( \tilde{X}_b^{} \, t+t^2  \right) + \delta_2^{} \bigg]
 \label{lambda2lll}
\end{eqnarray}

\medskip\noindent
with the following notation:
\medskip
\begin{eqnarray}
\delta_1^{} &=&  \frac{3t \left( h_b^2 - h_t^2 \right)}{16 \pi^2}  ~
 \frac{\tilde{A}_t^{} \, \mu \cot \beta}{M_S^2} \, \left( 1-
 \frac{\tilde{A}_t^2 }{6 M_S^2}
 \right)  \\
\delta_2^{} &=&  \frac{3t \left( h_t^2 - h_b^2 \right) }{16 \pi^2} ~
\frac{\tilde{A}_b^{} \, \mu \tan \beta}{M_S^2} \, \left( 1-
\frac{\tilde{A}_b^2 }{6 M_S^2}
 \right)
\end{eqnarray}

\medskip\noindent
where $\tilde{X}_{t,b}^{}$ is defined as $X_{t,b}^{} (A_{t,b}^{} \to 
\tilde{A}_{t,b}^{} )$ with
\medskip
\begin{eqnarray}
\tilde{A}_t^{} &=& A_t^{} - \mu \cot \beta \nonumber\\
\tilde{A}_b^{} &=& A_b^{} - \mu \tan \beta
\end{eqnarray}

A similar but distinct result is obtained when $m_A^{} \sim M_S^{}$ 
(notably no $\delta_i^{}$ terms and a different dependence on $\tan
\beta$ and the mixed Yukawa couplings). The threshold corrections are
dependent on where the CP odd Higgs decouples. The same procedure is
applied to determine the 2LLL threshold corrections to the mass
terms. The results using these analytic formulae for the loop corrections assume that the CP odd Higgs has not been decoupled above the scale $Q = m_Z^{}$. This is inevitably wrong, but the errors involved are found to be insignificant given the other approximations made.

\section{Analytic formula for fine tuning}\label{ftform}

The fine tuning measure chosen for this analysis is the Barbieri-Giudice measure introduced in Section~\ref{measure}
\begin{equation} 
\Delta = \max \Delta_p^{},
\qquad \Delta _{p} = \left| \frac{\partial \ln
  v^{2}}{\partial \ln p} \right| \label{ftdef} 
\end{equation}
where the observable is taken as the electroweak vev squared. This is not actually an observable, but this approach is nearly equivalent to taking an electroweak gauge boson mass as the observable, where equivalency is broken by small quantum corrections to the self energy and gauge coupling running. It will turn out the approximations applied for calculation of the fine tuning generate more uncertainty than these choices of ``observable" introduces.

The fundamental parameters, $\{p\}$, are considered to consist of the soft SUSY breaking parameters and the Higgs bilinear parameter within the superpotential, $\{ m_0^2, m_{1/2}^2, A_0^2, B_0^2, \mu_0^2 \}$. The dominant dimensionless parameters controlling the renormalisation group equations, the strong gauge coupling and top Yukawa coupling, $\{ \alpha_3^{}, h_t^{} \}$, were also tested for fine tuning. However, the latter results were not considered as genuine measures of naturalness and were not included when finding $\max \Delta_p^{}$ . For example, the influence of the top Yukawa coupling leads to radiative symmetry breaking which, as will be discussed later, produces similar sensitivities as found for the dependence of the strong coupling on the proton mass scale. The Barbieri-Guidice measure is then not considered appropriate for these dimensionless parameters.

As gravitational interactions mediate supersymmetry breaking in mSUGRA, the universal pattern of soft masses determined in Section~\ref{mSUGRA} is valid around the Planck scale. These are then considered to be the fundamental parameters for the fine tuning tests. One motivation for universality relied on gauge unification. However, if unification occurs then the running of the Lagrangian parameters between the unification scale and Planck scale is certainly modified with respect to the defined MSSM case. The apparent scale of gauge unification in the MSSM is $\sim2 \times 10^{16}$\,GeV.

The running of soft masses between the unification and Planck scale will be ignored in the analysis. The relatively small hierarchy of these energy scales, and a preserved universality within any gauge unified multiplet may make this a reasonable assumption. The physics below the gauge unification scale is then described by the MSSM, and the associated renormalisation group equations are used to determine the low energy physics and their sensitivity to variations of the high energy inputs.




Note the choice of testing a mass-dimension-squared observable to sensitivity in mass-dimension-squared parameters for the SUSY parameters. This may be considered appropriate, as in the tree level vacuum minimisation condition, $v^2 = m^2 / \lambda$. The effective quartic coupling does not depend on the SUSY parameters at tree level, and so the observable is a linear combination of the chosen input parameters. As argued in Section~\ref{measure}, the measure then extracts the sole term within this linear combination that depends on the tested input parameter, and normalises the result with respect to the observable's magnitude. If the mass dimensions selected were not equivalent, then the measure would be scaled by a constant factor.

In general, the loop corrected mass parameters and quartic couplings in the two Higgs doublet potential will depend on $\{ \beta, p, v \}$. The measure is then given by,
\begin{eqnarray}
\frac{p}{v^2} \frac{\partial v^2}{\partial p}
&=&
\frac{p}{v^2} \left[ 
\frac{m^2}{\lambda^2} \left( \frac{\partial \lambda}{\partial p} + \frac{\partial \lambda}{\partial \beta} \frac{\partial \beta}{\partial p} + \frac{\partial \lambda}{\partial v^2} \frac{\partial v^2}{\partial p} \right)
- \frac{1}{\lambda} \left( \frac{\partial m^2}{\partial p} + \frac{\partial m^2}{\partial \beta} \frac{\partial \beta}{\partial p} + \frac{\partial m^2}{\partial v^2} \frac{\partial v^2}{\partial p} \right)
\right]
\end{eqnarray}
where $\partial \beta / \partial p$ can be determined by differentiating the second minimisation condition:
\begin{eqnarray}
\frac{\partial \beta}{\partial p}
   &=&
 \frac{1}{z} \left[ \lambda \left( 2 \frac{\partial^2 m^2}{\partial p \, \partial \beta}  +
  v^2 \, \frac{\partial^2 \lambda}{\partial p \, \partial \beta} \right)
  - \frac{\partial \lambda}{\partial \beta} \left( \frac{\partial m^2}{\partial p} +
 v^2  \frac{\partial \lambda}{\partial p}   \right)
   \right]  \nonumber\\[7pt]
&& ~+ \frac{1}{z}
 \frac{\partial v^2}{\partial p} \left[ \lambda \left( 2 \frac{\partial^2 m^2}{\partial v^2 \, \partial \beta} +
  v^2 \, \frac{\partial^2 \lambda}{\partial v^2 \, \partial \beta} \right)
- \frac{\partial \lambda}{\partial \beta} \left( \frac{\partial m^2}{\partial v^2} +
 v^2  \frac{\partial \lambda}{\partial v^2} \right)
 \right] \\[3pt]
 z &=&
 \left[ 
 \lambda \left( 2\,  \frac{\partial^2 m^2}{\partial \beta^2}
 + v^2 \, \frac{\partial^2 \lambda}{\partial \beta^2} \right)
- \frac{v^2}{2} \left( \frac{\partial \lambda}{\partial \beta} \right)^2 
  \right]
\end{eqnarray}

Using the vacuum minimisation equations to replace terms, the following formula for the measure is obtained.
\begin{eqnarray}
\Delta_p^{}
&=& -\frac{p}{z}\,\frac{1}{1+r}\,\bigg[\bigg(2
\frac{\partial ^{2}m^{2}}{\partial 
\beta ^{2}}+v^{2}\frac{\partial ^{2}\lambda }{\partial \beta ^{2}}\bigg)
\bigg(\frac{\partial \lambda }{\partial p}+\frac{1}{v^{2}}\frac{\partial 
m^{2}}{\partial p}\bigg)+\frac{\partial m^{2}}{\partial \beta }\frac{
\partial ^{2}\lambda }{\partial \beta \partial p}-\frac{\partial \lambda }{
\partial \beta }\frac{\partial ^{2}m^{2}}{\partial \beta \partial p}
\bigg] 
\label{ft} \\[3pt]
r &=&
\frac{v^2}{z} \left[
\left( 2 \frac{\partial^2 m^2 }{\partial \beta^2} + v^2 \frac{\partial^2 \lambda}{\partial \beta^2}  \right)
\left(  \frac{\partial \lambda}{\partial v^2} + \frac{1}{v^2} \frac{\partial m^2}{\partial v^2}  \right)
+ \frac{\partial m^2}{ \partial \beta} \frac{\partial^2 \lambda}{ \partial \beta \partial v^2}
- \frac{\partial \lambda}{\partial \beta}  \frac{\partial^2 m^2}{ \partial \beta\partial v^2} \right]
\nonumber
\end{eqnarray}
This agrees with the analytic result by Casas {\it{et al}} \cite{Casas:2003jx} when ignoring the vev dependence of the Lagrangian parameters generated by loops (by setting $r=0$).

The impact of including the $r$ factor tends to reduce the fine tuning by $O(30\%)$, however it can also increase the fine tuning for certain points in the parameter space. In the following analysis that is concerned with the minimum fine tuning obtained from scans of the parameter space, the effect of ignoring $r$ simply scales the curves of {\it{minimum}} fine tuning found in the subsequent plots by this $O(30\%)$ factor. The following analytic results are presented for the case of $r=0$, which is equivalent to considering the superpartner mass scale fixed at which they need to be integrated out. The analytic results are not used here for determining high precision fine tuning values, but as an indication of the preferred regions of parameter space. The final results are determined numerically via a finite difference method within {\texttt{SOFTSUSY}}, and so do not use this analytic formula or its assumed approximations.


In the case of ignoring the parameter dependence of the quartic couplings (valid at tree level when the input parameter is not an electroweak gauge coupling),  the ``master formula" of \cite{Dimopoulos:1995mi} (see also \cite{Chankowski:1997zh,Chankowski:1998xv}) is also found,
\bea
\Delta_{p}&=&\frac{p}{(\tan^2\beta-1)\,m_Z^2}\,
\bigg\{
\frac{\partial  m_1^2}{\partial p}-\tan^2\beta\,\frac{\partial
  m_2}{\partial p}\nonumber\\[7pt]
&&\qquad\qquad
-\,\,\frac{\tan\beta}{\cos
  2\beta}\bigg[1+\frac{m_Z^2}{m_1^2+\tilde m_2^2}\bigg]
\bigg[2\frac{\partial m_3^2}{\partial p}-\sin
  2\beta\,\Big(\frac{\partial m_1^2}{\partial p}+\frac{\partial
     m_2^2}{\partial p}\Big)
\bigg]\bigg\}\label{gd}
\eea
This formula is sometimes used
as the starting point in analyses that evaluate
electroweak scale fine-tuning, by using in it the loop corrected soft masses. 
However, for accurate estimates, it is necessary to
take full account of radiative corrections.

In the limit of large $\tan \beta$, eq~(\ref{ft}) reduces to
\bea
\Delta_p^{} 
&=& 
\frac{
[ 4 \lambda_7\,(m_3^2)' - 4 \lambda_7' \,m_3^2 ] 
+ [ \lambda_2^\prime \, v^2 +  2 ( m_2^2)' ]  \, [ \lambda_{3452} 
+ 2 ( m_1^2-  m_2^2)/v^2 ]  }{2 \lambda_7^2 \,v^2 - \lambda_2 
\,[ \lambda_{3452}\,\,v^2 + 2\,( m_1^2- m_2^2) ] }
+{\mathcal O}(\cot\beta)
\\[3pt]
&\ra&
- \frac{1}{\lambda_2^{} \, v^2} \left[
 2\, (m_2^2)'  +  \lambda_2^\prime \, v^2  
  +
  4\, v^2
\left(
\frac{
   \lambda_7\,(m_3^2)' -  \lambda_7' \,m_3^2  
 }{  \lambda_{3452}\,\,v^2 + 2\,( m_1^2- m_2^2)  }
\right)
\right],\,\,\,
(\textrm{if}\,\,\vert\lambda_7\vert\ll\vert\lambda_{2}\vert,\vert
\lambda_{3452}\vert)\nonumber
\label{fttan}
\end{eqnarray}
where $x^\prime \equiv \partial x / \partial \ln p$ is the partial derivative of $x$ wrt $\ln p$.



In order to match physical observations, the vacuum minimisation conditions are used to set the magnitude of $\mu$, using the electroweak gauge boson mass as input. The sign of $\mu$ however remains a free parameter. It is also common to exchange one of the mSUGRA parameters, $B_0^{}$, by $\tan \beta$ in order to characterise the position in parameter space. Although, as $\tan \beta$ and $m_Z^{}$ are not fundamental parameters of the high energy Lagrangian, the fine tuning will still be evaluated with respect to variations in $B_0^{}$ and $\mu$.






\section{Fine tuning results from parameter space scan}\label{section3}

The following presents results of evaluating the fine tuning measure over an experimentally allowed region. The fine tuning was first always evaluated using the analytic formula of the previous section, including the dominant third generation supersymmetric threshold effects in the scalar potential. The evaluation of the analytic formula was completed using {\texttt{Mathematica}} with a routine optimised to run quickly. The results were then used to identify regions of parameter space that possessed small fine tuning.

For the cases where an analysis of the general behaviour of fine tuning was considered sufficient, the analytic results were then used. This included testing the importance of various approximations. For the cases where more precise results were desired, the fine tuning measure was re-evaluated using the numerical routines of {\texttt{SOFTSUSY 3.0.10}} \cite{Allanach:2001kg}. It proved computationally prohibitive to begin a search of the parameter space using the numerical routines. This approach would also have made analysis of the influencing factors much more challenging.


The mSUGRA parameter space scanned over was selected to satisfy the following constraints:

$\bullet$ radiative electroweak symmetry breaking (EWSB) is found

$\bullet$ non-tachyonic SUSY particle masses 

\hspace{10mm}
(to forbid colour and charge breaking (CCB) vacua)

$\bullet$ experimental constraints satisfied

$\bullet$ if stated in plots, consistency of $m_h$ with the LEPII SM-like bound (114.4 GeV)

\noindent
where the experimental constraints include the bounds on superpartner masses, electroweak precision data, $b$ physics constraints, and the anomalous magnetic moment of the muon, as detailed in Table~\ref{contable}.

\begin{table}[htdp]
\begin{center} 
\begin{tabular}{|c|c|} \hline
Constraint & Reference \\ \hline
SUSY particle masses & Routine in {\tt{MicrOMEGAs 2.2}},  
``MSSM/masslim.c" \\
$\delta a_\mu^{} < 366 \times 10^{-11}$ &  PDG 
(sys. and stat. $1\sigma$ errors added linearly) \\
$3.20  < 10^{4} ~ \mbox{Br}(b \to s \gamma) < 3.84$ & PDG 
(sys. and stat. $1\sigma$ errors added linearly)
\\
$\mbox{Br} (B_s^{} \to \mu \mu) < 1.8 \times 10^{-8}$ &  Particle Data Group (PDG) \\
$-0.0007< \delta \rho < 0.0012$ &  Particle Data Group  (PDG) \\ \hline
\end{tabular}
\end{center}
\caption{\small Constraints tested using {\texttt{MicrOMEGAs 2.2}}. 
 PDG: http://pdg.lbl.gov/.}
\label{contable}
\end{table}


Regarding the Higgs mass constraint, the values determined for a given point in parameter space with {\texttt{SOFTSUSY}} agree with that found using
{\texttt{SuSpect}} \cite{Djouadi:2002ze}  within $0.1$\,GeV, but can 
differ by $2$\,GeV \cite{Allanach:2004rh} 
from the value found using {\texttt{FeynHiggs} \cite{Hahn:2009zz}. The difference is a 3-loop effect caused by different renormalisation schemes. The value of the Higgs mass obtained from {\texttt{SOFTSUSY}} should then be considered to possess a theoretical uncertainty of $O(2\,$GeV). For the numerical results, the {\texttt{SOFTSUSY}} Higgs mass is used.


Fig~\ref{loopP} presents results from the analytic fine tuning measure, demonstrating generic features for the minimum fine tuning allowed within the parameter space region scanned, which are discussed in the following subsections.

\begin{figure}[t]
\center
\def\baselinestretch{1.1}
\includegraphics[width=12cm] {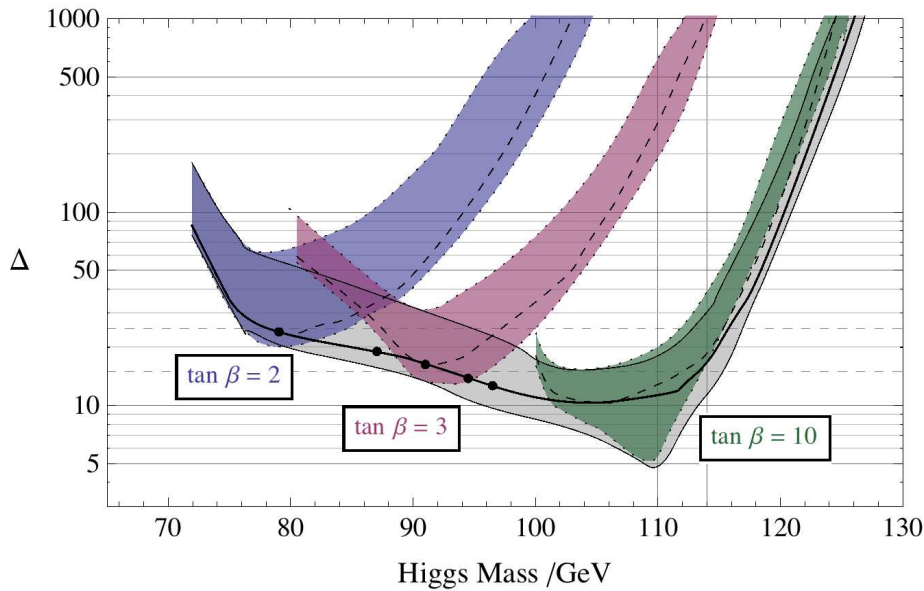} 
\caption{\small
Minimum (analytic) fine tuning versus 2LLL Higgs mass, showing the influence of
loop effects and $\tan \beta$ on fine tuning. All constraints listed
in Table~\ref{contable} are included. The upper and lower lines
associated with the coloured regions are the 1-loop without
thresholds for $\lambda$ and soft masses and ``full" 1-loop
results respectively (similarly for the grey region, for all
$\tan\beta$). The minimum 2-loop fine tuning  is found between these
two cases. The solid lines refer to the scan $2 \leq \tan \beta \leq
55$. The black points give the positions of minimum $\Delta$ 
for fixed $\tan \beta$ from 2 to 4 inclusive in steps of 0.5.}
\label{loopP}
\end{figure}

\subsection{Dependence on Higgs mass}

For any fixed $\tan \beta$, Fig~\ref{loopP} shows similar behaviour in the minimum fine tuning curves vs Higgs mass. There exists a global minimum at an intermediate Higgs mass, with rapidly increasing fine tuning as the Higgs mass is increased or decreased. It has been previously argued that if SUSY is a solution to the hierarchy problem then the superpartners that affect the Higgs potential (eg stops) are preferred to be light. The non-observation of superpartner states has introduced some tension by requiring heavy masses. The least fine tuning is then expected when the stops have masses just above the experimental limits. This corresponds to a certain value for the Higgs mass, for a given $\tan \beta$, which can be read from Fig~\ref{loopP}.

In order to get lighter Higgs mass, the $A_{t,b}^{}$ or $\mu$ parameters need to be large with an appropriate sense in order to reduce the loop contributions. However as the $\Delta$ measure grows like $A^2$ or $\mu^2$ (at least before leading log corrections, see Section~\ref{cmssm}), this leads to increased fine tuning.

On increasing the stop masses above their experimental limit, the Higgs mass is increased. Inverting the Higgs mass formula near the decoupling limit gives, $m_{\tilde{t}}^2 \propto \exp (m_h^2)$. As the stop-masses-squared are proportional to the soft SUSY breaking mass-squared terms at tree level, the $\Delta$ measure is then expected to roughly scale exponentially as the Higgs mass is increased.

\subsection{Dependence on $\tan \beta$}

Fig~\ref{loopP} also shows a clear dependence of minimum fine tuning on $\tan \beta$. As $\tan \beta$ is increased, the tree level Higgs mass is increased through an increase in the effective quartic coupling.  The analytic formula for fine tuning exhibits an inverse proportionality to this effective quartic coupling. The global minimum fine tuning for a given $\tan \beta$ is then expected to scale as $m_h^{-2}$ evaluated at this minimum, and this is roughly observed.

\subsection{Dependence on N-loop leading log corrections}

Similarly for the dependence on $\tan \beta$, the influence of the leading log corrections to the effective quartic coupling shifts both the Higgs mass and fine tuning. The 1LLL terms tend to increase the quartic coupling by a factor of $O(2)$, reducing the fine tuning by a similar factor and increasing the Higgs mass by the square root of the proportional change. This behaviour agrees with the results of \cite{Barbieri:1998uv} where a limited parameter space scan was completed for the experimental constraints as of 1998.

The inclusion of 2LLL corrections is seen to increase the fine tuning with respect to the 1LLL results. This again follows as the loop diagrams with strong gauge coupling dominate, and these lead to a reduction in the effective quartic coupling. The 2LLL corrections similarly decrease the Higgs mass.

As the change in fine tuning from 2LLL effects can introduce $O(2)$ corrections for a fixed Higgs mass, they are significant in evaluating naturalness of the parameter space. The 2LLL results are now used throughout the following analysis, and when numerically determining the fine tuning via {\texttt{SOFTSUSY}}, the 2-loop renormalisation group equations will be used.

The effect of going to 3-loop calculations generally introduces $O(1\%)$ changes to the Higgs mass. Having observed a correlation between minimum fine tuning and Higgs mass, it is then reasonable to expect that the fine tuning may vary by $O(2\%)$ at this level. This error is well within that generated by the various approximations, and so it is not currently worthwhile to include 3-loop corrections for evaluation of the fine tuning. The interpretation of such small differences in fine tuning is also not so meaningful.

\subsection{Dependence on superpartner mass limits}

Fig~\ref{chargcon} demonstrates the effects of varying the superpartner mass limits. This is useful to compare with previous fine tuning results when experimental constraints were weaker, and understand how stronger mass limits can affect the minimum fine tuning allowed.

\begin{figure}[t!]
\center
\def\baselinestretch{1.1}
\includegraphics[width=12cm] {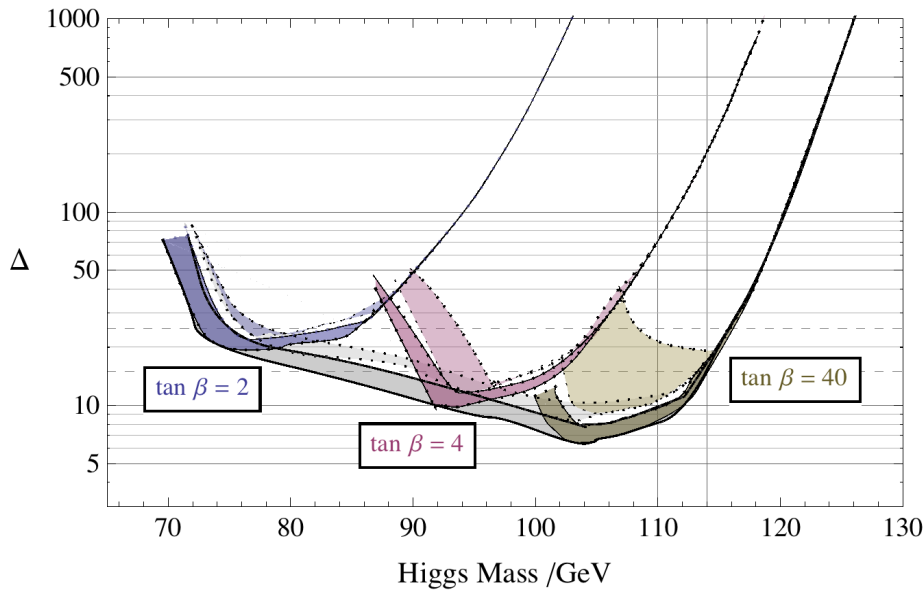} 
\caption{\small
Minimum (analytic) fine tuning vs Higgs mass, showing the influence of mass
constraints on fine tuning. The results are at 2-loop with the upper
shaded (coloured) areas connecting the case of only applying 
the SUSY spectrum constraints (lower line) to that with all 
constraints listed in Table~\ref{contable} (upper line).
The lower shaded (coloured) areas connect the
cases of only applying a chargino lower mass limit of 80 and 94 GeV
for the lower and upper lines respectively.
The results for the scan $2 \leq \tan \beta \leq 55$ are also shown
by the grey shaded area, 
with similar convention for upper/lower continuous lines delimiting it.
}\label{chargcon}
\end{figure}

It is found that with the current experimental limits, the light chargino mass constraint is almost wholly responsible for restriction on the minimum fine tuning found across the parameter space due to superpartner mass limits. This is a consequence of recent experiments, as when older experimental limits are used, the neutralino mass limit is found to be dominant. In both cases, it is masses controlled by the gaugino soft mass term, $m_{1/2}^{}$ that are found to be most significant for limiting fine tuning. The gluino mass limit is close to providing a matching constraint on fine tuning as the chargino limit, but is currently sub-dominant.

The current chargino mass limit varies according to the position of parameter space being tested, but for many cases, $m_{\chi^\pm}^{} > 94$\,GeV is roughly imposed. On decreasing the chargino mass limit, the minimum fine tuning is seen to reduce in Fig~\ref{chargcon} for $m_h^{} \lesssim 114$\,GeV, with little effect for heavier Higgs masses. On increasing the superpartner mass limits controlled by $m_{1/2}^{}$, there is a roughly parallel movement in the fine tuning vs Higgs mass space of the minimum fine tuning curve at low Higgs mass. At large Higgs masses, the exponential increase in fine tuning is still governed by large stop masses well above the experimental limits.

Note that this can be used as a test for naturalness. If a limit for fine tuning of 1 part in 100 ($\Delta = 100$) is demanded, then upper mass limits on superpartner states and Higgs mass can be obtained. If any state is observed in breach of these limits, then it can be concluded that a ``natural" mSUGRA theory cannot be realised. This may point to requiring an alternative theory to mSUGRA, or the abandonment of the principle of naturalness. The idea of setting of naturalness limits will be returned to when discussing the numerical fine tuning results.


\subsection{Dependence on $b\ra s\gamma$ constraint}

Figure~\ref{bsgcon} gives the impact of the $b\ra s\gamma$ constraint 
on $\Delta$. The lower limit of the $ b \to s \gamma$ constraint for a given
coloured area (fixed $\tan\beta$) restricts the
right hand edges of the plot, while the upper limit restricts
its left hand side. These curves also depend on the mass limits - these
 constraints are not fully independent.

\begin{figure}[t!]
\center
\def\baselinestretch{1.1}
\includegraphics[width=12cm]{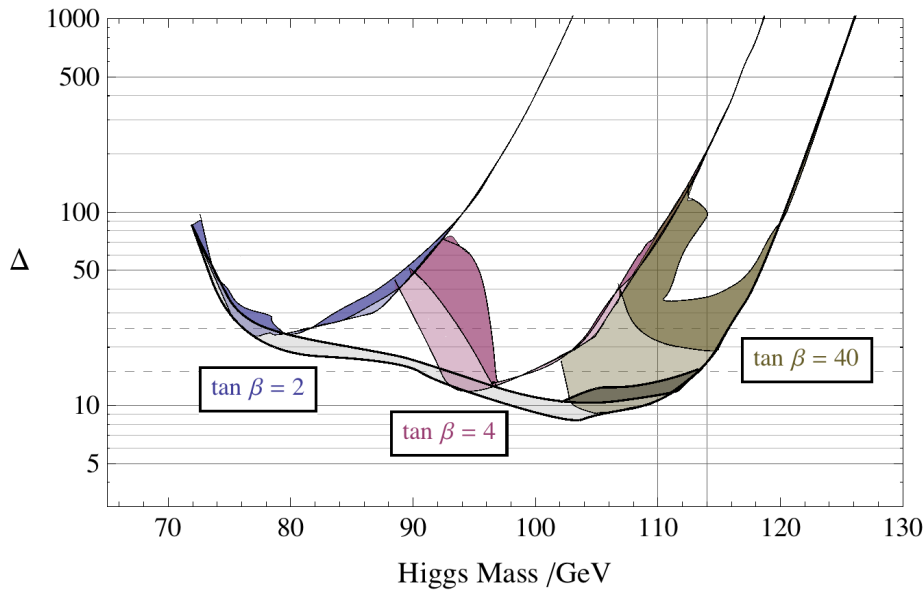} 
\caption{\small
Minimum (analytic) fine tuning vs Higgs mass, showing the influence of the $b \to
s \gamma$ constraint.  The results are at 2-loop with the lighter
shading connecting the case of only applying the SUSY spectrum
constraints (lower line) to that with also the $b \to s \gamma$
constraint listed in Table~\ref{contable} (upper line). This upper
edge of this shading is indistinguishable from the solid line which
includes all constraints in Table \ref{contable}. The darker shading
extends up to the minimum fine tuning limits for the stronger case, $3.52 <
10^{4}\, \mbox{Br}(b \to s \gamma) < 3.77$, with the other constraints
as given in Table~\ref{contable}. The results for the scan $2 \leq
\tan \beta \leq 55$ are also shown, between the two continuous 
and almost parallel lower curves.}\label{bsgcon}
\end{figure}

For the experimentally allowed area of $m_h>114.4$ GeV, the impact of
the $b\ra s\gamma$ constraint is rather small. The combination
of the SUSY mass limits and $b\ra s\gamma$ constraint currently
dominate the restriction on how small the fine tuning could be.

The other constraints listed in Table~\ref{contable} do eliminate further
mSUGRA points, but have a negligible 
effect on the fine tuning limits.
With the current mass limits, a change in the $\delta a_\mu^{}$ constraint by
factors of 2 or more does not noticeably affect these results.

Relaxing the $b\ra s\gamma$ constraint to be within the $3\sigma$ experimental error, or even neglecting the constraint totally only introduces $O(10\%)$ changes in the minimum fine tuning allowed for the full parameter space scan. This was verified with the numerical results also. However, for fixed large $\tan \beta$, the importance of this constraint remains significant.

\subsection{Dependence on the SUSY Lagrangian parameters}\label{cmssm}

The remaining analysis now includes a re-evaluation of the fine-tuning using {\texttt{SOFTSUSY}} in the regions of parameter space that were found to populate the low fine tuning points in each plot. The {\texttt{SOFTSUSY}} results were verified to exhibit the same generic features, with the quantitative differences attributed to a more robust evaluation of the renormalisation group equation solutions.

\begin{figure}[th!]
\center
\def\baselinestretch{1.1}
\hspace{-1mm}
\subfloat[Fine Tuning vs $\tan \beta$]{\includegraphics[width=7cm]{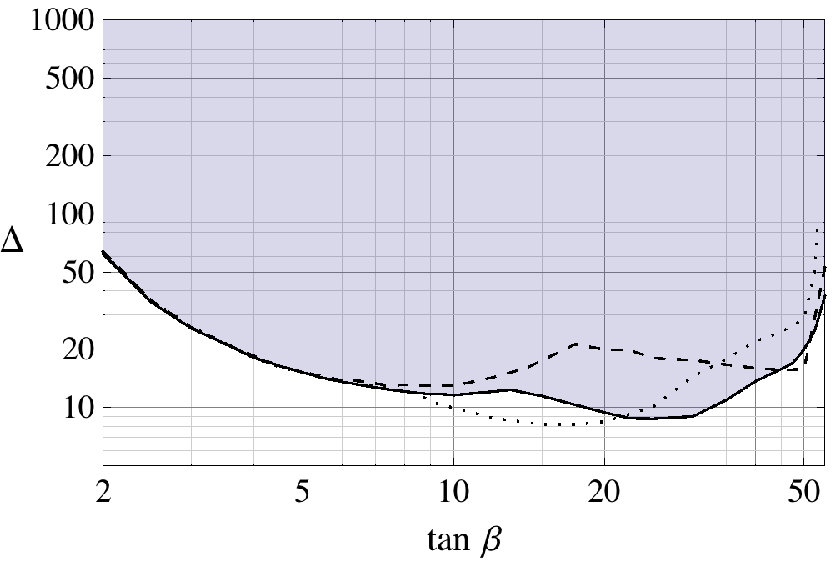}} 
\hspace{8mm}
\subfloat[Fine Tuning vs $\tan \beta$]{\includegraphics[width=7cm]{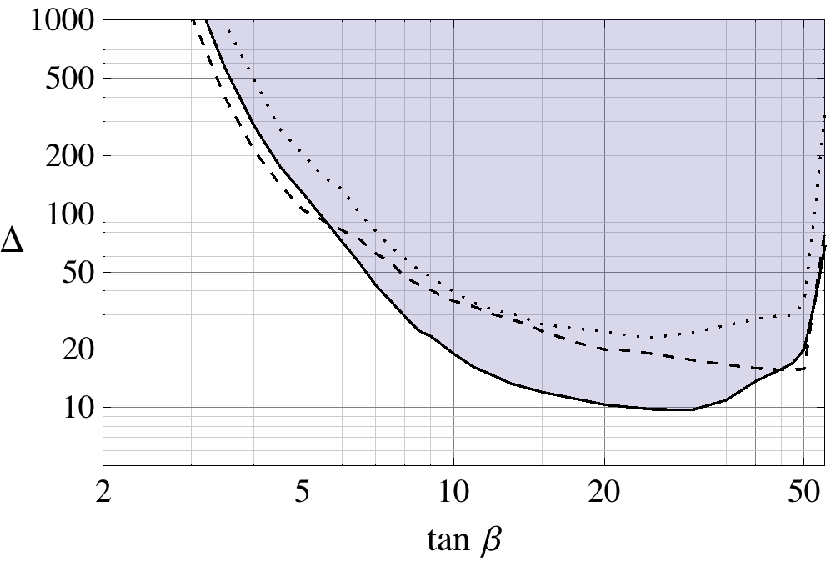}}
\hspace{4mm}
\subfloat[Fine Tuning vs
$A_0^{}$]{\includegraphics[width=7cm]{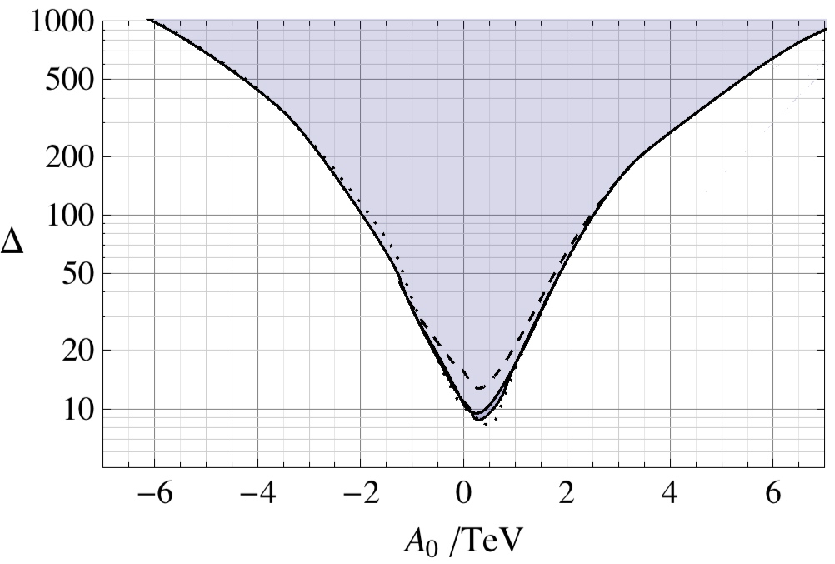}}
\hspace{8mm}
\subfloat[Fine Tuning vs $m_0^{}$]{\includegraphics[width=7cm]{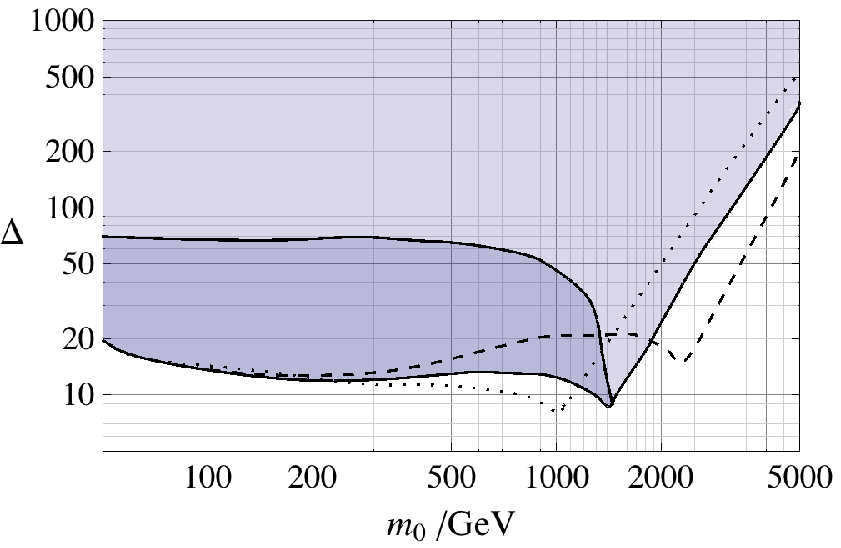}} 
\hspace{4mm}
\subfloat[Fine Tuning vs
$m_{1/2}^{}$]{\includegraphics[width=7cm]{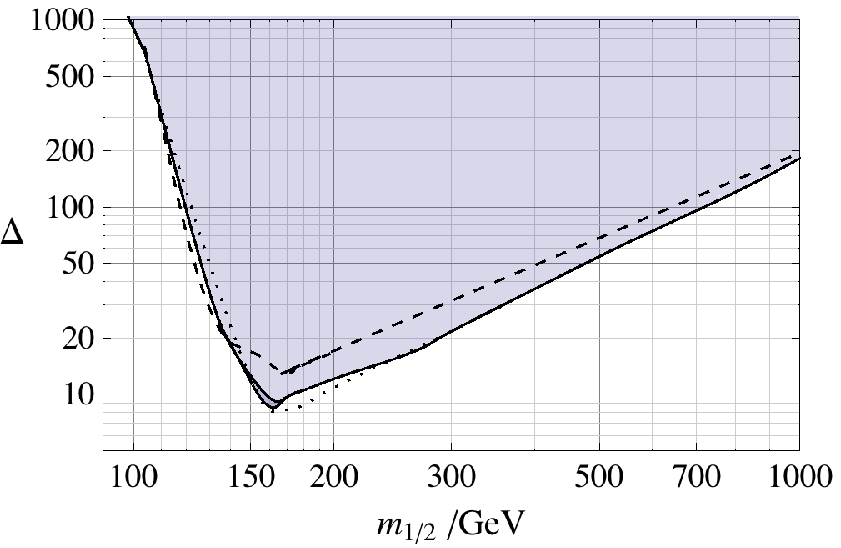}}
\hspace{8mm}
\subfloat[Fine Tuning vs $\mu$]{\includegraphics[width=7cm]{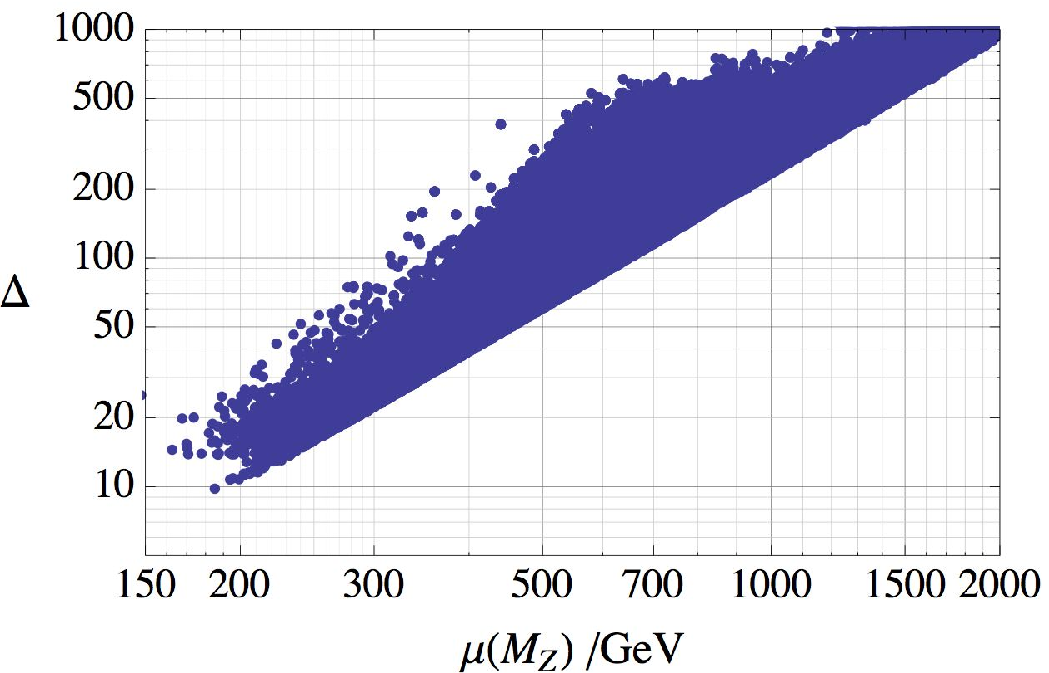}}
\caption{{\protect\small Dependence of minimum fine tuning on SUSY
parameters ($\protect\mu >0$). The solid, 
dashed and dotted lines are as explained in Fig~\protect\ref{2loop}. 
No bound on $m_h$ is applied in figure (a). 
In (c), (d), (e), the darker shaded regions are eliminated when
 $m_{h}^{{}}>114.4$~GeV is applied for the case with the central
 $\left( \protect\alpha _{3}^{{}},m_{t}^{{}}\right) $ values. 
In (b) and (f), $m_h > 114.4$\,GeV is applied, and the points in (f)
 are only for the central $\left( \protect\alpha
 _{3}^{{}},m_{t}^{{}}\right) $ values.}}\label{all}
\end{figure}

The results of this section identify which regions of parameter space permit small fine tuning of the electroweak scale, and will later be used to comment on the phenomenology of natural mSUGRA theories. In Figure~\ref{all} the dependence of the total fine tuning with respect to the Lagrangian parameters is plotted.

To understand the behaviour of the electroweak fine tuning, it is useful to consider one of the vacuum minimisation equations, which when ignoring quantum corrections to the quartic couplings simplifies to:
\begin{eqnarray}\label{mu0fixtree}
\frac{m_Z^2}{2} &=&  \frac{{m}_{1}^2-{m}_{2}^2 
\tan^2 \beta}{\tan^2 \beta -1}
\end{eqnarray}
%
\noindent
Using  2-loop RGE solutions which neglect MSSM state masses for $\tan \beta =10$,
\begin{eqnarray}
{m}_1^2 \,(m_Z^{}) &\approx& 0.99\, \mu_0^2 + 0.946\, m_0^2 
+ 0.331\, m_{1/2}^2 + 0.044\, A_0^{} \, m_{1/2}^{} - 0.013\, A_0^2
\hspace{10mm}
\label{m1uv}
\\
{m}_2^2 \,(m_Z^{}) &\approx& 0.99\, \mu_0^2 - 0.080\, m_0^2 
- 2.865\, m_{1/2}^2 + 0.445\, A_0^{} \, m_{1/2}^{} - 0.099\, A_0^2
\label{m2uv}
\end{eqnarray}


\noindent
The evaluation of $\Delta_{\mu_0^2}^{}$ for example, is roughly equal to $\left(2 \mu_0^2 / m_Z^2\right)$, when neglecting the leading log corrections. Small values for the Lagrangian parameters are thus preferred to obtain small fine tuning. The quadratic dependence on the respective Lagrangian parameter is a common feature, which is observed in the plots for large parameter values.

It is the large cancellation between the $\mu_0^2$ and $m_{1/2}^2$ terms
that is often  responsible for the large fine tuning (note however
that  this argument ignores the impact of quantum corrections to
quartic couplings, known to reduce the fine-tuning).
This leads to  the approximate relation
 $\Delta_{\mu_0^2}^{} \sim  \Delta_{m_{1/2}^2}^{}$. 
 The rise in fine tuning at small $m_{1/2}$ is a consequence of the constraints
such as the chargino mass limit.

The near flat distribution of minimum fine tuning in $m_0^{}$ is a
result of the coefficient of $m_0^{}$ in $m_2^{}$ being driven close
to zero. The fine tuning with respect to $m_0^{}$ then rarely
dominates, until we reach values of $m_h$ above the LEPII bound ($m_0$
at the edge of focus point region). The result of applying the Higgs
mass constraint also excludes a region with small $m_{1/2}$ at
$m_0^{}$ below 1.5~TeV.
 The focus point at $m_0^{} \sim 1.5$~TeV where the minimum
 of $m_{1/2}$ is possible, corresponds to the point 
where fine tuning is minimised. This only occurs for large $\tan \beta$,
and this available ``dip" in fine tuning in the mSUGRA space
disappears  as $\tan \beta$ is reduced.

Figure~\ref{all}(c) indicates that a small trilinear coupling 
$|A_0^{}| \lesssim 1$~TeV is preferred for the smallest
fine tuning. This follows from a similar argument for preferring
small $m_{1/2}$. Increasing $|A_0^{}|$ requires larger cancellations
with $\mu$ to set the electroweak scale. However, once the Higgs mass
constraint is applied, $A_0^{}$ is driven negative for small $\tan \beta$ in order to
maximise the stop mixing. The related increase in the minimum fine
tuning from being in this region of parameter space then
follows. This is important for small $\tan \beta$ where the
tree level Higgs mass is smallest.
 The sign structure of the UV
 parameter coefficients in $m_2^{}$ leads to a preference in a
 small, positive $A_0^{}$.

\subsection{Dependence on strong gauge coupling and top mass}

The large quantum corrections to the Higgs mass provides a sensitive dependence on the supersymmetric physics, and therefore fine tuning. We now return to examination of fine tuning in the Higgs mass space. Fig~\ref{2loop} demonstrates how the fine tuning of our observed physics varies according to the experimental input for the strong gauge coupling and top mass at the electroweak scale. Their parameters were varied within their $1\sigma$ experimental bounds to determine the combination that most strongly affects the fine tuning results.

\begin{figure}[t]
\center
\def\baselinestretch{1.1}
\includegraphics[width=10cm]{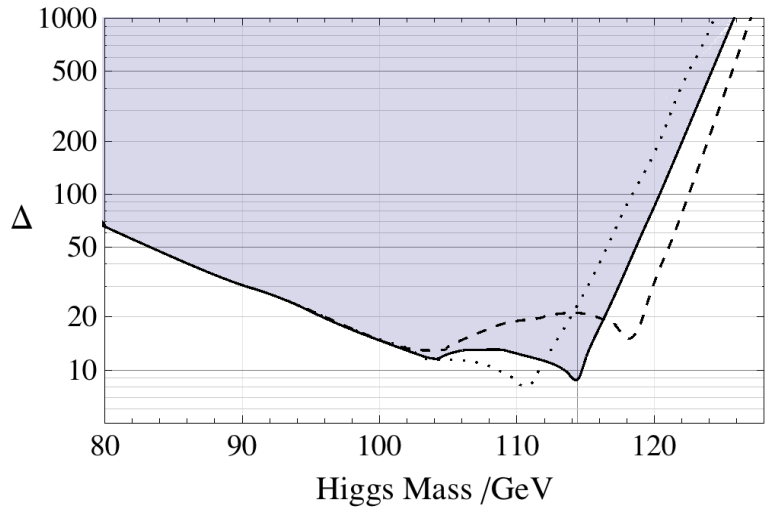} 
\caption{\small
Fine tuning vs Higgs mass, in a two-loop analysis.
 The data points are for $2 \leq \tan \beta \leq 55$.
The solid line is the minimum fine tuning with central values 
$(\alpha_3, m_t)=(0.1176,173.1$\,GeV).  The dashed line corresponds to
 $(\alpha_3, m_t)=(0.1156,174.4$\,GeV) and the dotted line to
$(0.1196,171.8$\,GeV), to account for  1$\sigma$ experimental errors in
$\alpha_3$ and top mass \cite{TeV}.
This is the ``worst'' case scenario, when such deviations combine such as
to give the largest variation of $\Delta$.
 The LEPII bound of $114.4$~GeV is indicated by a vertical line. 
Note the steep ($\approx$ exponential) increase of $\Delta$ 
on both sides of 
its minimum value situated near the LEPII bound.}\label{2loop}
\end{figure}

Note that this variation is philosophically different to testing sensitivity of observed physics to changes in the fundamental parameters. The variation found is not directly an indication of fine tuning, but a test of how accurate the fine tuning results can be considered to be.

Varying the gauge coupling and top mass influences the leading log corrections to the Lagrangian parameters. These affect both the Higgs mass and fine tuning to a similar order of magnitude. The experimental constraints are also affected, as the SUSY mass spectrum is generally modified. Also, the region of parameter space where radiative electroweak symmetry breaking occurs depends on these parameters.

An increase of $\alpha_3(m_Z)$ or reduction of
 $m_t(m_Z)$ by 1$\sigma$ have similar effects, which can be also
 understood from the relation between the mass of top evaluated 
 at $m_Z$ and at $m_t$. 
Keeping either $\alpha_3$ or $m_t$ fixed to its central value
and varying the other within
1$\sigma$ brings a curve situated half-way between the continuous line
and the corresponding dashed or dotted line.



\subsection{Individual fine tunings}

The individual contributions to $\Delta$ are shown in
Figure~\ref{dplot}. Below the LEPII bound, detailed calculations show that 
the minimal value of $\Delta$ is dominated by $\Delta_{\mu_0^2}$ 
and this increases rapidly for decreasing $m_h$.
For values of $m_h$ above the LEPII bound, $\Delta$ is  dominated 
by $\Delta_{m_0^2}$.
This happens at the edge of the focus point region.
 The transition from the dominant
$\Delta_{\mu_0^2}$ regime to the dominant $\Delta_{m_0^2}$ regime
occurs near the LEPII bound value, where the fine tuning happens to be minimised. This is a consequence of the present day experimental constraints.

\begin{figure}[t]
\center
\def\baselinestretch{1.1}
\includegraphics[width=11cm]{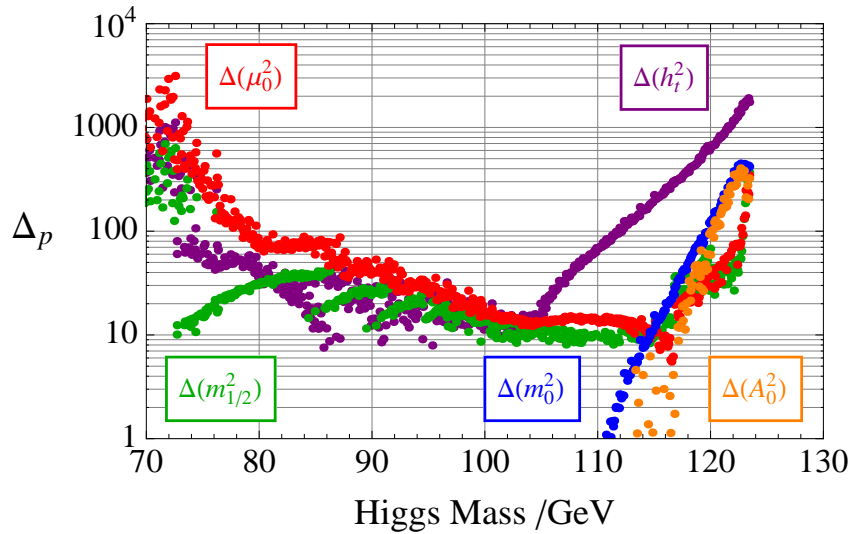}
\caption{\small
The plot displays the individual contributions $\Delta_p^{}$,
 to the minimum electroweak fine-tuning
$\Delta$ presented in Figure~\ref{2loop}. The largest of these for all
$m_h$ gives the curve presented in Figure~\ref{2loop}.
At low $m_h$, $\Delta_{\mu_0^2}$ (red) is dominant, while at large
$m_h$, $\Delta_{m_0^2}$ (blue) is dominant (with $\Delta_{A_0^2}$
reaching similar values near 120 GeV).
The transition between the two regions is happening at about 114.5
GeV. Note that in this plot the LEPII bound is not imposed at any time.
Although $\Delta_{h_t^2}$ (purple) is presented above for illustration, 
this contribution is always sub-dominant if using the  Ciafaloni-Strumia definition of fine-tuning \cite{Ciafaloni:1996zh}, for measured parameters. In all other plots, $\Delta_{h_t^2}$ is ignored for the calculation of $\Delta$ as it is not considered a genuine measure of naturalness, argued in the text.}\label{dplot}
\end{figure}

The large values of $\Delta_{h_t^2}^{}$ (using the Barbieri-Guidice measure) when $m_h^{} > 105$~GeV suggests an apparent large sensitivity to the top Yukawa coupling in the region argued to have small fine tuning in the other plots. 
The small fine tuning with respect to all other parameters is a consequence of being near the scalar focus point, where the sensitivity to the scalar soft mass parameter is reduced due to the renormalisation group equations.

To illustrate the presence of the scalar focus point, consider a moderate $\tan\beta$ scenario with $A_0^{} = 0$. The running of the soft mass terms is dominated by the top Yukawa coupling, where neglecting corrections from gauge interactions and subdominant Yukawa interactions, the 1-loop RGEs are:
\begin{eqnarray}
\frac{d}{dt}
\left(
\begin{array}{c} m_{H_2^{}}^2   \\ m_{\tilde{t}_R^{}}^2   \\ m_{\tilde{t}_L^{}}^2 \end{array}
\right)
&\approx&
\frac{h_t^2}{8\pi^2}
\left(
\begin{array}{ccc}
 3 & 3  & 3  \\
 2 & 2  & 2  \\
 1 & 1  & 1 
\end{array}
\right)
\left(
\begin{array}{c} m_{H_2^{}}^2   \\ m_{\tilde{t}_R^{}}^2   \\ m_{\tilde{t}_L^{}}^2 \end{array}
\right)
\end{eqnarray}
which leads to the following solution for the up-type Higgs soft mass term,
\begin{eqnarray}
m_{H_2^{}}^2 (t_1^{}) &\approx&
\frac{1}{2} 
\left[  
3\,e^{P} -1
+ (x_L^{} + x_R^{}) \left(  e^{P}  - 1 \right)
\right]
m_{H_2^{}}^2 (t_0^{})
\label{rgesol}
\end{eqnarray}
where $m_{\tilde{t}_{L,R}^{}}^2 (t_0^{}) \equiv \left( 1+x_{L,R}^{} \right) m_{H_2^{}}^2(t_0^{}) $, $P = \int_{t_0^{}}^{t_1^{}} \frac{6 h_t^2}{8\pi^2} \, d t$, and $t_i^{} = \log \left( Q_i^{} / Q_0^{} \right)$ for energy scale $Q_i^{}$. 
If the MSSM RGEs continued to arbitrarily low energy scales, there would always be a scale at which the up-type Higgs soft mass would be (approximately) independent of the high energy scale input. However, observables must be independent of renormalisation scale. In this discussion, the mention of a scalar focus point implicitly implies that this focus point scale occurs at the electroweak scale where the vacuum state is set. For this scenario, the electroweak physics is then insensitive to the fundamental scalar soft mass.

To avoid phenomenological problems in the MSSM, the high energy soft masses cannot be tachyonic. The top Yukawa coupling drives $m_{H_2^{}}^2$ to decreasing values as the energy scale is reduced below the gauge unification scale. Electroweak symmetry is then broken spontaneously by radiative corrections \cite{Ibanez:1982fr} as $m_{H_2^{}}^2$ becomes negative, perhaps requiring contributions neglected in eq~(\ref{rgesol}). It is the same radiative corrections that then control the positioning of the focus point scale.

In the mSUGRA scenario, $x_{L,R}^{} = 0$, and the scalar focus point for low to moderate $\tan\beta$ is found where $e^P \approx 1/3$. Higher loop corrections, thresholds and neglected contributions modify this relation, but the observed top mass generally places the focus point scale close to the electroweak scale with better matching at large $\tan\beta$.

Note that universality is not required for a focus point, as shifts of $x_{L,R}^{} \to x_{L,R}^{} + \epsilon_{L,R}^{}$ where $\epsilon_L^{} = - \epsilon_{R}^{}$ will maintain independence of the high energy value. The ratios must still be fixed with one underlying mass-scale parameter and not coincidental. Independent shifts in the ratios of other high energy scalar soft masses may also be found in the large $\tan \beta$ cases where the other third generation Yukawa couplings become significant \cite{Feng:1999zg}. However, for alternative mechanisms of mediating supersymmetry breaking such as gauge or anomaly mediation, the pattern of soft masses generally does not conform to that required for a focus point at the electroweak scale.

Returning to discuss the large apparent fine tuning to the top Yukawa coupling in Fig~\ref{dplot}, one has to be careful whether this large result is meaningful. In the example of the QCD scale, there was a large value for $\Delta$ with respect to the high energy gauge coupling input due to an exponential dependence for setting the confinement scale generated through running. Similarly here, there is an exponential dependence on a dimensionless parameter which controls the electroweak scale through radiative symmetry breaking.


For the following analysis, radiative EWSB is considered a natural mechanism. The large result of the $\Delta$ measure with respect to the top Yukawa coupling is then attributed to an inappropriate choice of fine tuning measure given how this parameter influences the electroweak scale, and ignored.
Anyway, once electroweak fine tuning is at a level of 1 part in 100, there are also problems such as the hierarchy of Yukawa couplings that should be addressed. Further new physics may provide reduced sensitivities to the true fundamental parameters.

\subsection{Naturalness limits}

As mentioned earlier, the fine tuning measure $\Delta$
can be used to  establish the remaining parameter space of mSUGRA
compatible with a solution to the hierarchy problem. Assuming that
$\Delta=100$ is the upper limit beyond which we consider that SUSY
failed to solve the hierarchy problem, the following naturalness bounds are obtained:
\begin{equation}\label{cmssmparameters}
\begin{tabular}{rlcrcl}
$m_{h}$ & $<~121$~\mbox{GeV} & \hspace{5mm} & $5.5~<$ & $\tan \beta $
  & 
$<~55
$ \\
$\mu $ & $<~680$~\mbox{GeV} &  & 120~\mbox{GeV}$~<$ & $m_{1/2}$ & $<~720$~
\mbox{GeV} \\
$m_{0}$ & $<~3.2$~\mbox{TeV} &  & $-2.0$~\mbox{TeV}$~<$ & $A_{0}$ & $<~2.5$~
\mbox{TeV}
\end{tabular}
\end{equation}

Note that the uncertainty in top mass and strong gauge coupling produces some variation in these limits. The Higgs mass limit varies at the $O(1\%)$ level, and the $A_0^{}, m_{1/2}^{}, \mu$ limits at the $O(10\%)$ level. The $m_0^{}$ limit is particularly uncertain though due to movement of the boundary where electroweak symmetry breaking occurs in the parameter space. The $m_0^{}$ limit ranges from 2.6 to 4.2~TeV on varying the Standard Model inputs within their 1$\sigma$ uncertainties.

The naturalness restriction on the mSUGRA parameter space also leads to a restriction on the MSSM superpartner spectrum. The results so far demonstrate that electroweak fine-tuning
has a strong sensitivity to parameters such
as  $\mu$, $m_{1/2}$, with a preference for lower values.
Regarding the $m_0$ dependence, $\Delta$ has a rather flat
dependence. 
The states that are dominantly controlled by the 
$\mu$, $m_{1/2}$  parameters are then the most important in determining
the naturalness of the proposed theory. These include the neutralinos,
charginos and the gluino states. 
\medskip
\begin{table}[ht]
\begin{center}
\begin{tabular}{|c||c|c|c|c||c|c||c|c||c|c|}
\hline
$\tilde{g}$ & $\chi_{1}^{0}$ & $\chi_{2}^{0}$ & $\chi_{3}^{0}$ & $
\chi_{4}^{0}$ & $\chi_{1}^{\pm}$ & $\chi_{2}^{\pm}$ & $\tilde{t}_{1}^{}$ & $
\tilde{t}_{2}^{}$ & $\tilde{b}_{1}^{}$ & $\tilde{b}_{2}^{}$ \\ \hline\hline
1720 & 305 & 550 & 660 & 665 & 550 & 670 & 2080 & 2660 & 2660 & 3140 \\ 
\hline
\end{tabular}
\end{center}
\caption{\small 
Upper mass limits on superpartners in GeV such that $\Delta<100$
remains possible. These limits apply individually on the respective state, independent of other states.}
\label{partlimit}
\end{table}
If any of these states have
masses in excess of those given in Table~\ref{partlimit}, it will
require less than 1\% tuning ($\Delta>100$) for the MSSM.
These upper mass limits scale approximately as
$\sqrt{\Delta_{\mbox{\tiny min}}^{}}$, so they may be adapted depending on
how much fine tuning the reader is willing to accept.

In the limit of optimum naturalness, the fine tuning of the electroweak scale is found to be minimised at $\Delta = 8.8$, for a Higgs mass of $114 \pm 2$ GeV. The uncertainty in Higgs mass is generated from the theoretical approximations, and the fine tuning may be considered accurate to $O(30\%)$ with the given definition. These results are calculated at the 2 loop level. As noted previously, the most natural Higgs mass for mSUGRA theories is fixed by the current bounds on the superpartner spectrum, and not directly the LEPII Higgs mass bound.



One may interpret the SUSY parameters
corresponding to the smallest fine tuned region as being the most likely, given our present
knowledge, and so it is of interest to compute the SUSY spectrum for natural
parameter choices as a benchmark for future searches. This is presented in
Table~\ref{favspec2} where it may be seen that there are very
heavy squarks and sleptons and lighter 
neutralinos, charginos and gluinos. This has  similarities
to the SPS2 scenario \cite{BA}. 

\vspace{3mm}
\begin{table}[th]
\begin{center}
\begin{tabular}{|c|c||c|c||c|c||c|c|}
\hline
$h^{0}$ & 114.5 & $\tilde{\chi}_1^0$ & 79 & $\tilde{b}_{1}^{}$ & 1147 & $
\tilde{u}_{L}^{}$ & 1444 \\[1pt] 
$H^{0}$ & 1264 & $\tilde{\chi}_2^0$ & 142 & $\tilde{b}_{2}^{}$ & 1369 & $
\tilde{u}_{R}^{}$ & 1446 \\[3pt] 
$H^\pm$ & 1267 & $\tilde{\chi}_3^0$ & 255 & $\tilde{\tau}_{1}^{}$ & 1328 & $
\tilde{d}_{L}^{}$ & 1448 \\[2pt] 
$A^0$ & 1264 & $\tilde{\chi}_4^0$ & 280 & $\tilde{\tau}_{2}^{}$ & 1368 & $
\tilde{d}_{R}^{}$ & 1446 \\[2pt] 
$\tilde{g}$ & 549 & $\tilde{\chi}_1^\pm$ & 142 & $\tilde{\mu}_L^{}$ & 1406 & 
$\tilde{s}_{L}^{}$ & 1448 \\[2pt] 
$\tilde{\nu}_{\tau}^{}$ & 1366 & $\tilde{\chi}_2^\pm$ & 280 & $\tilde{\mu}
_R^{}$ & 1406 & $\tilde{s}_{R}^{}$ & 1446 \\[2pt] 
$\tilde{\nu}_{\mu}^{}$ & 1404 & $\tilde{t}_{1}^{}$ & 873 & $\tilde{e}_L^{}$
& 1406 & $\tilde{c}_{L}^{}$ & 1444 \\[2pt] 
$\tilde{\nu}_{e}^{}$ & 1404 & $\tilde{t}_{2}^{}$ & 1158 & $\tilde{e}_R^{}$ & 
1406 & $\tilde{c}_{R}^{}$ & 1446 \\ \hline
\end{tabular}
\end{center}
\par
\caption{\small A favoured
 mSUGRA spectrum with $\Delta = 15$.
Masses are given in $GeV$.}
\label{favspec2}
\end{table}


\subsection{Predictions for SUSY searches at the LHC}\label{susylhc}

It is clear that there is still a wide range of parameters that needs to be
explored when testing mSUGRA. Will the LHC be able to cover the whole
range? To answer this, note that for a fine tuning measure
$\Delta<100$, one must be able to exclude the upper limits
of the mass parameters appearing in Table~\ref{partlimit}.
Of course the state
that affects fine tuning most is the Higgs scalar and one may see from
Figure~\ref{2loop}
 that establishing the bound $m_{h}>121$\,GeV will imply that 
$\Delta>100.$ However the least fine tuned region corresponds to the lightest
Higgs consistent with the LEPII bound and this is the region where the LHC
searches rely on the $h\rightarrow\gamma\gamma$ channel which has a small
cross section and will require some $30\,fb^{-1}$ at $\sqrt{s}=14$\,TeV to
explore. Given this, it is of interest to consider to what extent the direct
SUSY\ searches will probe the low fine tuned regions. 
Following the discussion in the previous section, the most significant
processes at the LHC will be those looking for gluinos, winos and
neutralinos.

Studies of SUSY\ at the LHC~$\cite{Baer:2009dn}$ have shown that the LHC
experiments have a sensitivity to gluinos of mass up to $1.9$~TeV for 
$\sqrt{s}=10$~TeV, and up to $2.4$~TeV for $\sqrt{s}=14$~TeV and luminosity $10fb^{-1}$. This corresponds to probing up to $\Delta =120,180$.

In a previous study of fine tuning \cite{Allanach:2000ii} for fixed $A_0^{} = 0$, $\tan \beta = 10, M_t^{} = 174~$GeV, a scan over the $(m_0^{}, m_{1/2}^{})$ space identified that values for $\Delta$ of up to 210 may be excluded, if mSUGRA physics remains unobserved at the LHC with $\sqrt{s}=14$~TeV and a luminosity of $10fb^{-1}$. The study included the dominant one loop corrections to the Higgs potential. Having identified the magnitude of errors associated with one loop fine tuning calculations in this work, and noting that fine tuning may be slightly reduced by varying $A_0^{}$ and $\tan \beta$, the limits on $\Delta$ found here are consistent with this previous work.

For the LHC to operate at $\sqrt{s}=10$~TeV and obtain a luminosity of $100\,pb^{-1}$, gluinos of up to 600~GeV may be discovered. However, if gluino masses are excluded to this level, $\Delta$ is only excluded to a value of 14. As has been discussed, charginos and neutralinos can be quite light, but their
signal events are difficult for LHC to extract from the background, owing in
part to a decreasing $M_{\widetilde{W}}-M_{\widetilde{Z}}$ mass gap as 
$\left\vert \mu \right\vert $
decreases~\cite{Barbieri:1991vk,Baer:2004qq}. An Atlas study
\cite{Vandelli:2007zza}  of the trilepton signal from
chargino-neutralino production found that $30fb^{-1}$ luminosity at
14 TeV is needed for a
 3$\sigma$ discovery significance
 for $M_2<300$~GeV and $\mu<250$~GeV \cite{CMS}.

\section{Summary}\label{summ2}

Supersymmetry was introduced to solve the electroweak hierarchy
problem and to avoid the large  fine-tuning in the
SM Higgs sector associated with the Planck or gauge unification scale
 when quantum corrections are included.
 While this hierarchy problem is solved
by TeV-scale supersymmetry, the non-observation, so far,
 of SUSY states means
that the MSSM has acquired some residual amount of fine-tuning 
related to  unnatural cancellations in the SUSY breaking sector.

The fine tuning measure $\Delta$  provides a quantitative
test of SUSY as a solution to the 
hierarchy problem and measures the  ``tension'' required to satisfy
the scalar potential minimum condition  $v^2\sim -m_{susy}^2/\lambda$, for  
a combination of soft masses $m_{susy}^2\sim$ TeV, with an effective quartic
coupling $\lambda$ remaining perturbative and $v\sim \cO(100)$ GeV.
Although the 
exact upper limit on the fine tuning beyond which a theory
fails to solve the hierarchy problem is debatable,
it is preferable for a given model to have a parameter space
configuration corresponding to the lowest value of $\Delta$. The fine tuning measure has been evaluated here at two-loop order with particular attention given to threshold
corrections and the $\tan\beta$ radiative dependence on 
the parameters. Such effects on  fine tuning were not fully considered in the past and turned out to reduce fine
tuning significantly.

The determination of the  fine tuning measure for mSUGRA included the theoretical constraints (radiative EWSB,
avoiding charge and colour breaking vacua), and also the 
experimental constraints (bounds on superpartner masses,
electroweak precision data, $b\ra s\,\gamma$, $B_s^{}\ra \mu\,\mu$ and
muon anomalous magnetic moment).

The analysis found that a fine tuning of 1 part in $O(10)$ remains consistent with the current experimental constraints. This is much more promising than the commonly reported limit of 1 part in $O(100)$ frequently repeated in the literature based on very primitive calculations. Moreover, these results indicate a preference for a Higgs mass at $\sim 115\,$GeV, which was found without application of the LEPII Higgs mass bound. This may be regarded as a theoretical prediction for the Higgs mass based on the principle of naturalness. Chapter~4 of this thesis will also discuss how non-minimal theories may improve naturalness beyond the degree found possible in mSUGRA.


The spectrum corresponding to the minimum value of the fine tuning shows similarities to the SPS2  scenario with light
neutralinos, charginos and gluinos (corresponding to light $\mu$,
$m_{1/2}$) and heavy squarks and sleptons corresponding to
large $m_0$, near the focus point limiting value \cite{Feng:2000bp,Chan:1997bi}. It provides the ``best" estimate for the SUSY spectrum given the present  experimental bounds.

Increasing $m_h$ above the minimum fine tuned value causes $\Delta$ 
to increase exponentially fast. One obtains $\Delta~\!=~\!100\,\,(1000)$ for a scalar mass
$m_h=121$ ($126$) GeV, respectively. Ultimately the question whether the SUSY solution to the hierarchy
problem has been experimentally tested relies on what value
of fine tuning represents the limit of acceptability.
Given a value, one can determine the range of parameter space 
that is still acceptable. For the case that the fine tuning measure
should satisfy $\Delta<100$, the corresponding 
ranges for the superpartners masses and mSUGRA parameters values have been determined to be 
relevant for SUSY searches.

\chapter{Fine tuning of dark matter within minimal supergravity}

In order to test a theory for naturalness, it is not sufficient to test whether just one observable, such as the electroweak scale, is fine tuned. One must test all observables. If any observable is found to require a large degree of fine tuning, then the theory must be deemed to be ``un-natural." Due to the electroweak scale being notably vulnerable to fine tuning issues, it is sensible to first test for naturalness of electroweak symmetry breaking (EWSB). Having found in the previous chapter fine tunings of less than 1 part in 10 consistent with current observations, attention is now turned to testing other physics within the mSUGRA model.

An exhaustive study of naturalness for any given model would investigate many different sectors of physics, such as flavour physics, QCD phenomenology and cosmology (inflation). For this thesis, the scope is restricted for consideration of the electroweak sector. As mentioned before, this is of particular interest for possible imminent discoveries. The phenomenology of other sectors are also not necessarily connected with electroweak physics, allowing this separation of analyses. However, as alluded to in earlier chapters, there are appealing motivations for unifying physics, and so unified models would need to check for naturalness in observables of all connected sectors.

In addition to motivating supersymmetry as a solution to the hierarchy problem, the presence of a dark matter candidate was argued. Evidence for dark matter is obtained from analysis of galactic rotation curves. The expected rotation from the observed luminous matter using standard gravity does not match the results found. The existence of diffuse halos of matter that do not interact significantly with light, ie ``dark" matter, are then required to match the observations.

Furthermore, the power spectrum of the cosmic microwave background exhibits peaks from which the ratio of luminous matter to total matter content of the universe can be inferred. The WMAP experiment \cite{wmap} determines this ratio to be $\sim 0.2$. The majority of matter is then non-luminous. Simulations of galaxy formation also suggest that the dominant contribution to dark matter is non-relativistic, or ``cold" dark matter.

The choice of R-parity within the MSSM stabilised the lightest supersymmetric particle (LSP) with respect to decay. The spectrum of mSUGRA is presented in Section~\ref{MSSMspectrum}, leading to identification of the LSP as having electroweak interactions (the gravitino is assumed heavy). Thus, as well as testing for naturalness of EWSB on introducing SUSY, one must also test the physics of this dark matter candidate to see if the electroweak sector is natural.

If R-parity is not realised, making the LSP unstable to decay, or the LSP is actually not related to the electroweak sector (such as an axino, gravitino, or a hidden sector state), then this analysis is not relevant. However, if the dark matter candidate considered here provides any thermal contribution to the relic energy density, the restriction on parameter space and naturalness constraints presented are applicable.

Section~\ref{RED} discusses how the thermal relic energy density of a stable particle can be calculated. 
Section~\ref{dmsrch} reviews the constraints obtained on dark matter physics from direct detection searches. Section~\ref{allowedSpace} presents the allowed parameter space found on imposing the dark matter constraints, and Section~\ref{dmResults} then presents the fine tuning of the relic density in the experimentally allowed parameter space.



\section{MSSM spectrum}\label{MSSMspectrum}

The spectrum of the superpartner states is controlled by the unknown effective soft SUSY breaking terms, and the renormalisation group equations (RGEs). For the case of mSUGRA, universal soft SUSY breaking parameters are fixed at the gauge unification scale. The MSSM renormalisation group equations \cite{Martin:1993zk} 
, which are dominated by the third generation couplings, then drive the stops and staus to be the lightest squarks and sleptons. 

As the 1-loop RGEs (see Appendix~\ref{appendix1}) obey the relation, $d \ln M_a^{} / dt = d \ln g_a^2 / dt  $, the gaugino soft mass, $M_a^{}$, run approximately equal to the gauge couplings squared, $g_a^2$. The universality of gaugino masses is broken on running away from the UV scale, but the ratio $M_1:M_2:M_3$ then roughly follows $\alpha_1^{} : \alpha_2^{} : \alpha_3^{}$. At the electroweak scale, this ratio is approximately given by $1:2:6$ (using an $SU(5)$ normalisation for hypercharge), and in terms of the mSUGRA soft mass parameter, $m_{1/2}^{}$, which sets the gaugino mass at the gauge unification scale, $M_1^{} (m_Z^{}) \sim 0.4 \, m_{1/2}^{}$. The ratio relations are modified by threshold corrections and higher loop effects.

The neutralino mass matrix in the $(\tilde{B}^0, \tilde{W}^0, \tilde{h}_1^0, \tilde{h}_2^0)$ basis is given by,
\begin{eqnarray}
\left( \begin{array}{cccc} 
M_1^{} & 0 & -m_Z^{} \, c_\beta^{} s_W^{} & ~~\,m_Z^{} \, s_\beta^{} s_W^{} \\
0 & M_2^{} & ~~\,m_Z^{} \, c_\beta^{} c_W^{} & -m_Z^{} \, s_\beta^{} c_W^{} \\
-m_Z^{} \, c_\beta^{} s_W^{} & ~~\,m_Z^{} \, c_\beta^{} c_W^{} & 0 & -\mu \\
~~\,m_Z^{} \, s_\beta^{} s_W^{} & -m_Z^{} \, s_\beta^{} c_W^{} & -\mu & 0
\end{array}\right)
\end{eqnarray}
where $s_\beta^{},\,(c_\beta^{}) = \sin \beta~(\cos \beta)$ and $s_W^{},\,(c_W^{}) = \sin \theta_W^{}~(\cos \theta_W^{})$. $\tilde{B}^0$ ($\tilde{W}^{}$) is the superpartner of the hypercharge (weak) gauge boson and known as the Bino (Wino), and $\tilde{h}_{1,2}^{}$ are the Higgsinos. Note that the gluino cannot mix with these states as colour is an unbroken non-Abelian symmetry. The chargino mass matrix in the $(\tilde{W}^\pm, \tilde{h}^\pm)$ basis is,
\begin{eqnarray}
\left( \begin{array}{cc} 
M_2^{} & \sqrt2 \, m_W^{} \sin \beta \\
\sqrt2 \, m_W^{} \cos \beta & \mu
\end{array}\right)
\end{eqnarray}

The magnitude of $\mu$ is fixed to reproduce the correct electroweak gauge boson masses. The vacuum minimisation equations of the Higgs potential tend to lead to the hierachy, $\mu \gtrsim M_2 > M_1$. However, in the region of parameter space where the scalar focus point exists, one finds $M_2 > \mu \gtrsim M_1$. In both cases, the lightest neutralino is lighter than the charginos. Neither the magnitude of $\mu$ nor $m_{1/2}^{}$ can be too small, as experimental limits on the superpartner masses forbid such cases. For mSUGRA models, these limits disallow $\mu \ll M_1^{}$.

For the common hierarchy, $\mu \gtrsim M_2 > M_1$, the lightest neutralino, $\tilde{\chi}_1^{}$, is dominantly Bino-like and $\tilde{\chi}_2^{}, \tilde{\chi}_{3,4}^{}$ are Wino-like and Higgsino-like respectively. Similarly, the light (heavy) chargino is Wino-like (Higgsino-like). The ratio of the two lightest neutralinos and gluino then also roughly follows the ratio, $\alpha_1^{} : \alpha_2^{} : \alpha_3^{}$.

For the scalar focus point region, $M_2 > \mu \gtrsim M_1$, there is significant mixing of the gaugino/Higgsino fields in the neutralino and chargino mass eigenstates. In the limit $M_1^{} \sim \mu$, and large $\tan \beta$, the lightest neutralino mass is approximately given by,
\begin{eqnarray}
m_{\tilde{\chi}_1^0} &\approx& \left[ 1-\frac{1}{\sqrt2} \left( \frac{m_Z^{} \, s_W^{}}{M_1^{}} \right) + O \left(\frac{m_Z^{} \, s_W^{}}{M_1^{}} \right)^2  \right] M_1^{}
\end{eqnarray}

The states $\{ \tilde{\chi}_1^\pm, \tilde{\chi}_{2,3}^0 \}$ are all $O(M_1^{})$, nearly degenerate with the lightest neutralino, and $\{ \tilde{\chi}_2^\pm, \tilde{\chi}_4^0 \}$ are Wino-like with mass $O(M_2^{} \approx 2 M_1^{})$. For the smallest experimentally allowed values of $M_1^{}$ in mSUGRA, the ratio of the gluino mass to the lightest neutralino increases from $\sim 6$ to $O(10)$. As the coloured gluino is one of the most promising particles for early detection at the LHC (assuming a light mass), this mass hierarchy will be referred to in the following results which plot the lightest neutralino mass. The mass ratios within the spectrum will be important to understand which regions of parameter space are detectable in the first runs of the LHC.

In order for the LSP to be a dark matter candidate, the regions of parameter space where the LSP is electrically charged (eg stau or stop) must be ruled out. The stop mass squared matrix is given by,
\begin{eqnarray}
\left( \begin{array}{cc} 
m_{\tilde{t}_L^{}}^{2} + m_t^2 + L_q^{} & m_t^{} \left( A_t^{} - \mu \cot \beta \right) \\
m_t^{} \left( A_t^{} - \mu \cot \beta \right) & m_{\tilde{t}_R^{}}^{2} + m_t^2 + R_q^{}
\end{array}\right)
\end{eqnarray}
where $L_q^{} = \left( \frac12 - \frac23 \sin^2 \theta_W^{} \right) m_Z^2 \cos 2\beta$, and 
$R_q^{} =  \frac23 \sin^2 \theta_W^{} m_Z^2 \cos 2\beta$.

There is a similar structure for sbottoms and staus (although with different $\tan \beta$ dependence and distinct $L_q^{}, R_q^{}$ definitions). It is usually the mass mixing induced by large $A_0 / m_0$ that leads to a small stop mass eigenstate. In the analysis of fine tuning for the electroweak scale, small $A_0^{}$ magnitudes were preferred on naturalness grounds, and so large sfermion mixing tends to require large fine tuning. As we are interested in determining the low fine tuned regions, this constraint is not problematic. Assuming that the gravitino is heavy, the lightest neutralino then is identified as the LSP. This particle will interact through the electroweak interactions.



\section{Thermal relic energy density}\label{RED}

In the Big-Bang scenario, an abundance of matter in the present day may result from an asymmetry between matter and antimatter. Alternatively, stable particles may leave thermal equilibrium with its annihilation products, ``freeze-out", to generate a relic density in the present day. The latter case is now discussed for its predictions of the relic density.
The freeze-out will occur roughly when the thermally
 averaged annihilation rate of this particle becomes slower than the
 expansion rate of the universe:
\begin{equation}
n \, \langle \sigma_{\mbox{\tiny ann}} \, v \rangle \sim H
\label{fro}
\end{equation}
 where $n$ is the number density of the particle freezing out, $\langle
 \sigma_{\mbox{\tiny ann}} \, v \rangle$ is the thermally averaged
 annihilation cross section weighted by the relative velocity, and $H$
 is the Hubble constant.
 
 The ensemble of frozen-out particles then
 expand isenthropically, allowing the present number of particles in a
 comoving volume to be calculated. The current contribution to the
 energy density of the universe can then also be determined. If the
 particles are
 non-relativistic at the time of freeze-out, the thermal average is given by:
\begin{eqnarray}
\langle \sigma_{\mbox{\tiny ann}} \, v \rangle &\approx& 
\frac{x^{3/2} }{2 \sqrt\pi} \int_0^\infty dv \, v^2 \,
 (\sigma_{\mbox{\tiny ann}} \, v) \, e^{-x v^2/4} 
\label{therm}
\end{eqnarray}
where $x \equiv m_{\mbox{\tiny DM}}^{} / T$ and $m_{\mbox{\tiny DM}}^{}$ is the mass of the stable particle (dark matter). The annihilation cross
 section can be calculated perturbatively, however for slow moving
 dark matter there can be important non-perturbative effects due to
 interactions before annihilation. This is known as the Sommerfeld effect. For further discussion of this effect, see
\cite{Cassel:2009wt} and the contained references. It turns out that for electroweak interactions the Sommerfeld effect is only important for precision calculations at a level beyond the error introduced from the following approximations.

 The temperature at which the freeze-out occurs is determined from
 eq~(\ref{fro}), and is parametrised by the value of $x$ at
 freeze-out \cite{Drees:2007kk}:
\begin{eqnarray}
x_F^{}   \approx \ln \left( \frac{\left[ 3.85 \times 10^{17} ~
 \mbox{\scriptsize GeV} \right] \, g_{\mbox{\tiny DM}}^{} \, m_{\mbox{\tiny DM}}^{}
 \langle \sigma_{\mbox{\tiny ann}} \, v
 \rangle_{x_F^{} } }{\sqrt{x_F^{} \, g_{*s}} } \right)
\label{xfreeze}
\end{eqnarray}
where $g_{\mbox{\tiny DM}}^{}$ is the number of degrees of freedom of
 the dark matter candidate. The $g_{*s}$ parameter is the effective
 number of relativistic degrees of freedom contributing to the entropy
 density of the universe. For $T \sim 300$ GeV, $g_{*s} = 106.75$
 (assuming only Standard Model particles contributing), and at $T = 1$
 GeV, $g_{*s} \sim 80$. The contribution to
 the energy density is then given by:
\begin{eqnarray}
\Omega_{\mbox{\tiny DM}}^{} h^2 &=& \frac{8.6 \times 10^{-11}
 \,\, \mbox{\scriptsize GeV}^{-2} }{\sqrt{g_{*s} (x_F)} \, J(x_F) }
\hspace{7mm}
\mbox{where} \hspace{2mm}
J(x_F) ~=~ \int_{x_F^{}}^\infty dx \, x^{-2} \langle
 \sigma_{\mbox{\tiny ann}} \, v \rangle
\label{omh}
\end{eqnarray}
 where the approximation that all annihilations cease after
 ``freeze-out" has not been applied. The ``annihilation integral,"
 $J(x_F^{})$, accounts for the reduction in particle number after
 ``freeze-out." The thermal relic will contribute to the non-baryonic
 matter energy density observed in the universe today. The 5-year data
 from WMAP \cite{wmap} suggests that $\Omega_{n.b.m.}^{} h^2 = 0.1099 \pm
 0.0062$, with the $1\sigma$ error stated.

 The thermal relic abundance increases as the annihilation cross
 section is reduced. The unitarity limit places an upper bound on dark matter masses with such thermal relic density at $O(10^2)$ TeV \cite{Griest:1989wd}. For dark matter with electroweak interactions, a mass of $O(\mbox{TeV})$ is required to obtain the observed energy density. Supersymmetry provides such a dark matter candidate as the lightest superpartner state, being stabilised by an R-parity. Note that the mass scale is the same as that argued for solving the hierarchy problem.


\section{Direct dark matter searches}\label{dmsrch}

In order to verify the existence of dark matter without inference from galactic rotation curves and cosmological studies, there are several types of experiment ongoing for this purpose. The experiments are either concerned with direct or indirect detection of dark matter. Indirect detection aims to observe the annihilation products of dark matter from astrophysical sources. The constraints from such experiments are currently subdominant to direct detection experiments for mSUGRA dark matter. Interpretation of the results also relies on astrophysical models and precise catalogs of background sources leading to large uncertainties. These types of experiment will not be discussed further.

Direct detection of dark matter may be acheived by measuring recoils of nuclear matter from collisions with dark matter passing through the experiment. From the rotation curve of our galaxy, the expected local dark matter density is $0.3$ GeV cm$^{-3}$ with an $O(2)$ factor uncertainty (see \cite{McCabe:2010zh} and contained references). The large uncertainty is a consequence of the difficulty determining the distance of our solar system from the galactic centre (our angular velocity is well known though).

Computational simulations suggest that the structure of observed galaxies is consistent with a smooth distribution of dark matter on the scales that are sampled by the direct detection experiments. The relative velocity, $v$, of the Earth to the dark matter halo is believed to be $220$\, km/s varying by $O(10)\,$km/s seasonally, and as direct dark matter detection experiments run over a timescale of years, this covers a $O(10^{-4}\,\mbox{Parsec})$ distance.

If the calculated mSUGRA dark matter candidate does not saturate the non-baryonic matter energy density as determined by WMAP, the missing energy density may be accounted for by an alternative dark matter state. The identity of alternative dark matter will not be discussed here, however such a situation affects how the constraints from direct detection dark matter experiments must be handled.

In the following analysis, the non-baryonic matter energy density is always assumed to match the result determined by WMAP. The local density of the mSUGRA dark matter, $\rho_\chi^{}$, is then assumed to scale proportionally as the fraction of mSUGRA dark matter to total dark matter present in the universe does. This is reasonable for (approximately) 
collisionless dark matter, which numerical simulations suggest is a property of dark matter.

The differential scattering rate per unit recoil energy and unit target mass is 
\begin{eqnarray}
\frac{\partial R}{\partial E_R^{}} &=& \frac{\rho_\chi}{2 \mu_{\chi \mbox{\tiny N}}^2 m_\chi^{}}  \int_{v_{\mbox{\tiny min}}}^{v_{\mbox{\tiny esc}}} \sigma \, F^2 (E_R^{}) \, \frac{f(v)}{|v|} ~d^3 v
\end{eqnarray}
where R is the number of recoil events measured, $E_R$ is the nuclear recoil energy, $\sigma$ is the elastic LSP-nucleon interaction cross section (for which an exclusion limit is to be found), $\mu_{\chi \mbox{\tiny N}}$ is the reduced mass of LSP-nucleon system, $m_\chi$ is the LSP mass, $F(q)$ is the nuclear form factor, 
$v_{\mbox{\tiny min}}$ is the minimum velocity required to give nuclear recoil energy $E_R$, $v_{\mbox{\tiny esc}}$ is the escape velocity of dark matter in our galaxy (boosted into Earth frame), and $f(v)$ is the velocity distribution of LSP in the local neighbourhood.

The non-observation of recoil events above the background rate places an upper limit on the LSP-nucleon cross section. The elastic scattering amplitudes can be separated into the contributions that are proportional to the spins of the scattering states (spin dependent), and those that are spin independent. As the two contributions do not interfere, separate limits can be placed on the spin dependent and spin independent cross sections.

It is the spin independent cross section limit that is the dominant constraint for neutralino dark matter. This may be expected as the spin independent amplitudes add coherently, whereas the spin dependent amplitudes are incoherent. In order to have large nuclear recoils, heavy nuclei are preferred for the experimental set up, and so the large number of nucleons lead to a relatively larger cross section for the spin independent case.

\begin{figure}[!th]
\center
\begin{tabular}{c} 
\subfloat[$\tan \beta \leq 45$]{\includegraphics[width=7cm]{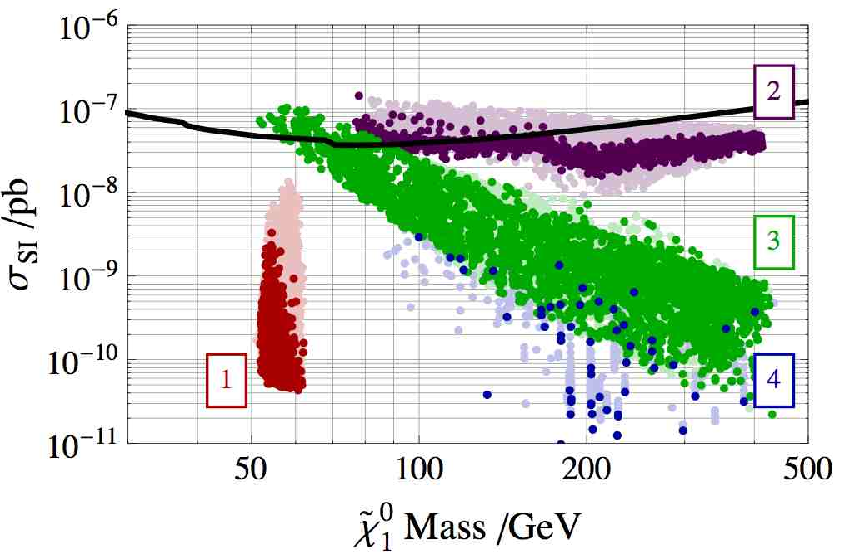}}
\hspace{4mm}
\subfloat[$\tan \beta \leq 45$, limit applied]{
\includegraphics[width=7cm]{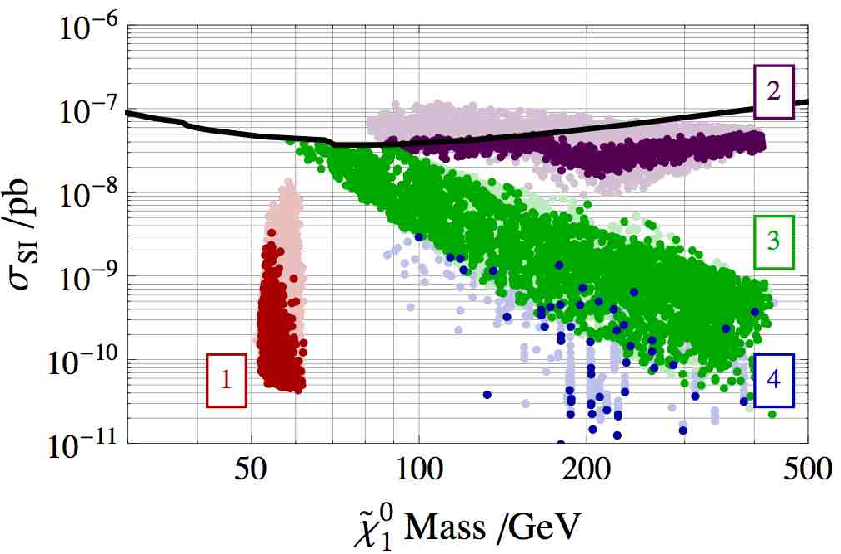}} 
\end{tabular}
\begin{tabular}{c} 
\subfloat[$50 \leq \tan \beta \leq 55$]{\includegraphics[width=7cm]{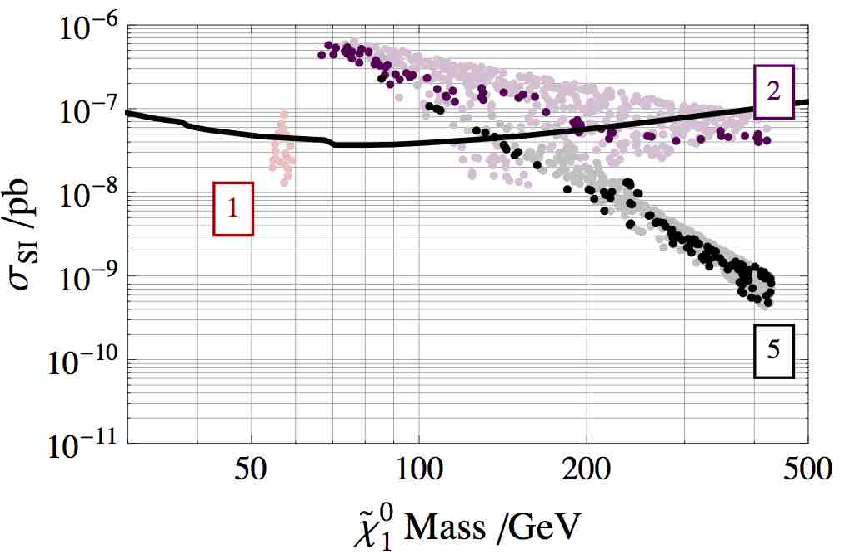}}
\hspace{4mm}
\subfloat[$50 \leq \tan \beta \leq 55$, limit applied]{
\includegraphics[width=7cm]{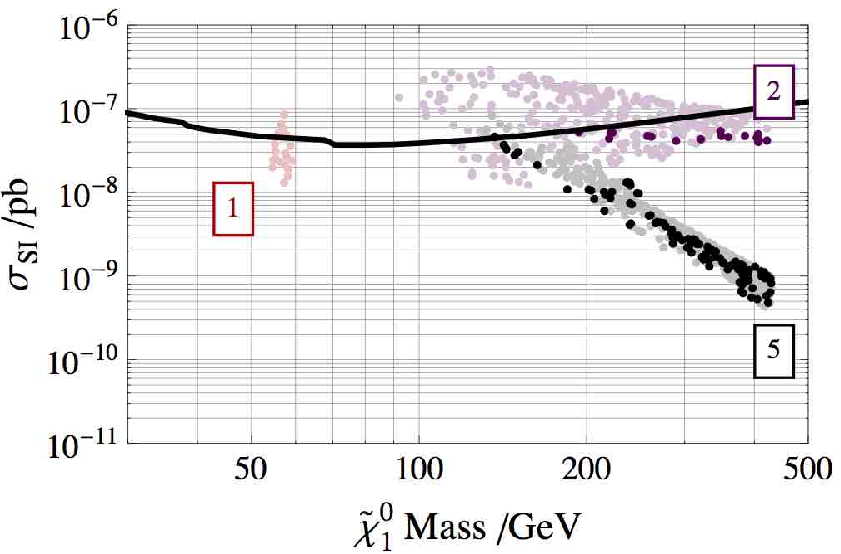}}
\end{tabular}
\caption{\protect\small 
Spin independent cross section for LSP-proton scattering, with an unrestricted Higgs mass. The solid line is the CDMS II limit, which is only applied in (b) and (d). The points satisfy the constraints in Table~\ref{contable} and have $\Omega h^2 < 0.1285$. Points with darker shading are within $3\sigma$ of the total non-baryonic matter relic density, $\Omega h^2 = 0.1099 \pm 3\times0.0062$. The labels distinguish between the dominant LSP annihilation mechanisms. The $h^0~(H^0, A^0)$ resonance is significant for region 1 (5) with points coloured red (black). Region 3 (4) realises stau (stop) co-annihilation with points coloured green (blue), and region 2, coloured purple, has increased Higgsino components.
}
\label{SInohiggs}
\end{figure}

\begin{figure}[!th]
\center
\begin{tabular}{c} 
\subfloat[$\tan \beta \leq 45$]{\includegraphics[width=7cm]{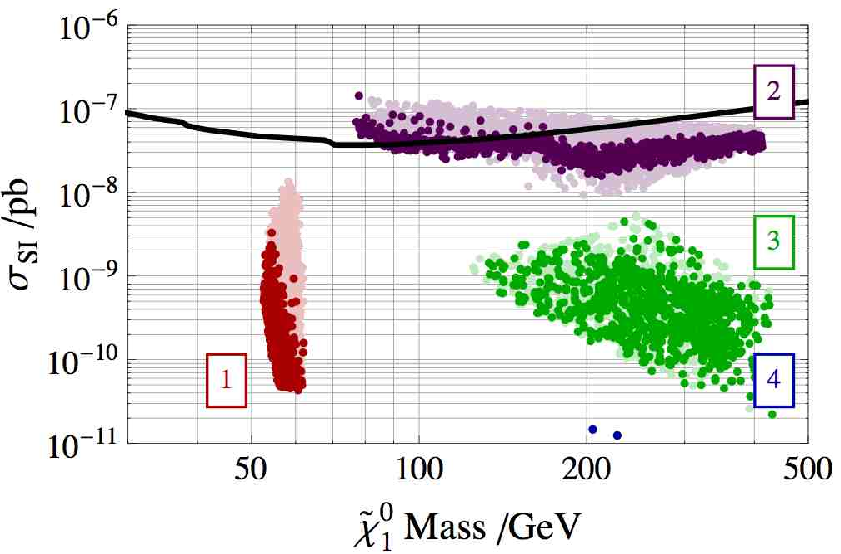}} 
\hspace{4mm}
\subfloat[$\tan \beta \leq 45$, limit applied]{
\includegraphics[width=7cm]{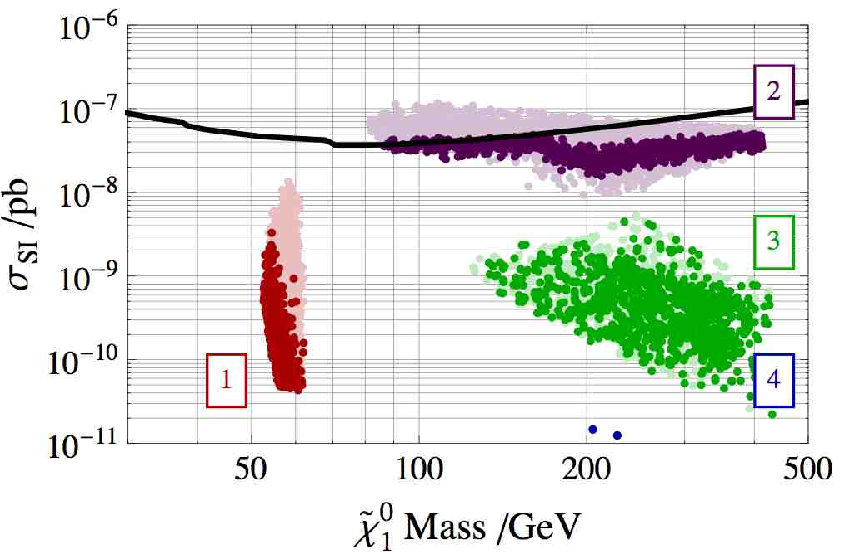}}
\end{tabular}
\begin{tabular}{c} 
\subfloat[$50 \leq \tan \beta \leq 55$]{\includegraphics[width=7cm]{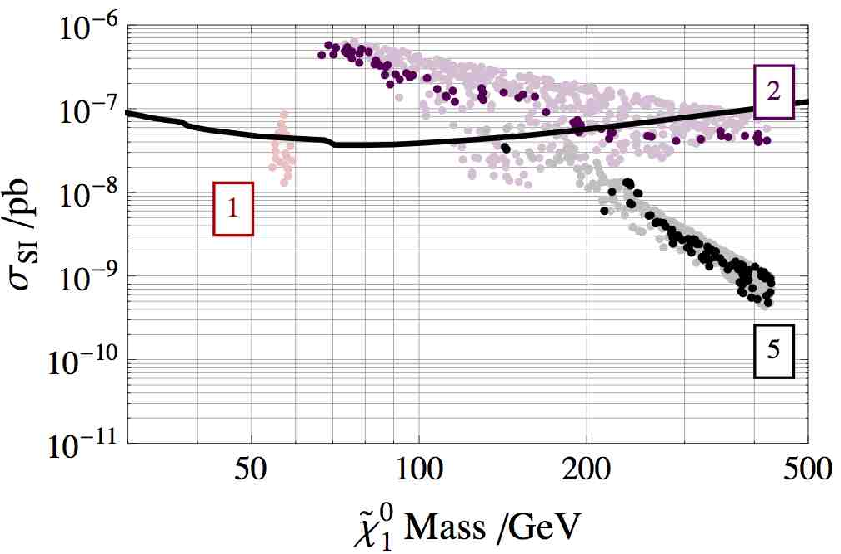}}
\hspace{4mm}
\subfloat[$50 \leq \tan \beta \leq 55$, limit applied]{
\includegraphics[width=7cm]{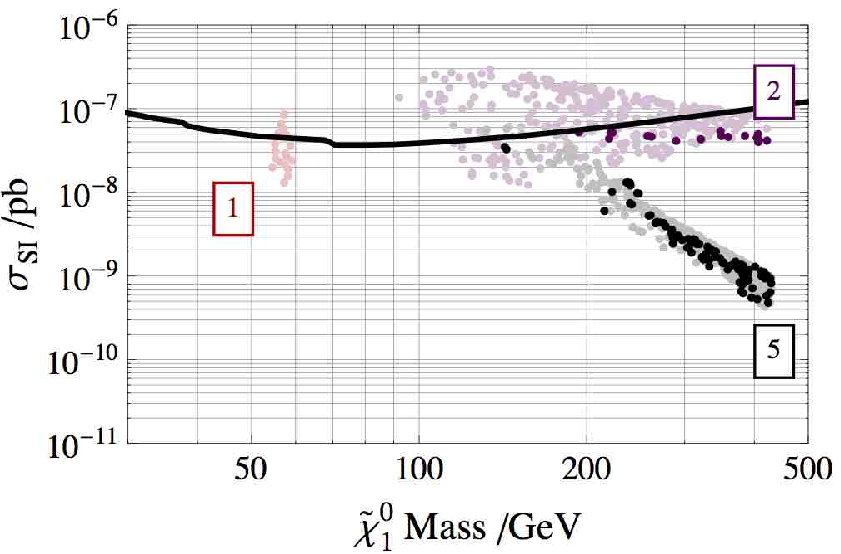}}
\end{tabular}
\caption{\protect\small 
Spin independent cross section for LSP-proton scattering, with $m_h > 114.4$~GeV. 
The solid line is the CDMS II limit, which is only applied in (b) and (d). The points satisfy the constraints in Table~\ref{contable} and have $\Omega h^2 < 0.1285$. Points with darker shading are within $3\sigma$ of the total non-baryonic matter relic density, $\Omega h^2 = 0.1099 \pm 3\times0.0062$ {\it (for all following plots also)}. The labels distinguish between the dominant LSP annihilation mechanisms. The $h^0~(H^0, A^0)$ resonance is significant for region 1 (5) with points coloured red (black). Region 3 (4) realises stau (stop) co-annihilation with points coloured green (blue), and region 2, coloured purple, has increased Higgsino components.
}
\label{SIhiggs}
\end{figure}

Figs~\ref{SInohiggs} and \ref{SIhiggs} show the experimental limits produced by the CDMS II experiment \cite{Ahmed:2009zw}. This is currently the strongest constraint on the LSP-nucleon scattering cross section for elastic dark matter. The points plotted correspond to points in the mSUGRA parameter space that satisfy the experimental limits listed in Table~\ref{contable} as well as the relic density constraint. 

The figures show that the direct dark matter constraint does exclude certain parts of the parameter space, whether the Higgs mass limit is applied or not. The phenomenology of the labelled regions is discussed in the following section. As future direct dark matter detection experiments will improve the reach by orders of magnitude in the next few years, there is competition with the collider experiments for discovery of supersymmetric states. 

The procedure of scaling the local LSP density by the relic density fraction means that a weaker limit is obtained for such points. This is the reason for finding some of the lighter shaded points (those that do not saturate the relic density) above the CDMS II line in subfigures (b) and (d), in which the constraint is appled - the line assumes that the LSP saturates the local dark matter.

The dominant astrophysical uncertainty is generated by the uncertainty in the local dark matter density, which is neglected for the limits produced by experimental groups. It is standard practice to assume the best fit values for the astrophysical variables, and so the experimental limits shown in the plots should be considered to vary by an $O(2)$ factor according to the astrophysical input.

\section{Phenomenology of allowed parameter space}\label{allowedSpace}

In addition to the constraints on the parameter space applied in the previous chapter (as listed in Table~\ref{contable}), the thermal relic density of the neutralino dark matter is now restricted such that, $\Omega h^2 < 0.1285$, for agreement with the WMAP 5 year results but allowing a 3$\sigma$ error deviation. The CDMS II limit is also applied as detailed in the previous section. The dark matter relic abundance and LSP-nucleon cross sections are calculated using {\texttt{MicrOMEGAs 2.2}} \cite{Belanger:2006is}.

The allowed parameter space of mSUGRA is strongly restricted by the condition of requiring a small dark matter relic density. This is a consequence of the neutralino LSP being dominantly Bino-like for most of the parameter space allowed before applying the dark matter constraint, as mentioned in Section~\ref{MSSMspectrum}. The small hypercharge coupling leads to a small annihilation cross section and, as Section~\ref{RED} demonstrated, the thermal relic density is inversely proportional to the annihilation cross section. 

The mechanisms for generating an increased annihilation cross section, and so a reduced relic density, distinguishes the allowed parts of parameter space that remain after applying the dark matter constraints. The universal structure of the soft SUSY breaking terms also leads to specific predictions of the mSUGRA spectrum that controls this phenomenology. In a more general supersymmetric model, the phenomenology is less restricted on application of a fixed number of constraints.

Mechanisms that can reduce the relic abundance include the presence of co-annihilation channels, where states nearly degenerate with the LSP interact with the LSP to produce final states without any superpartner states. If the parameters that lead to nearly degenerate states are distinct, this is expected to require fine tuning. Another mechanism is mixing with Higgsino states which is discussed further in Section~\ref{region2}.

The final mechanism to be discussed involves resonant annihilation. As the neutralino dark matter is non-relativistic at the time of freeze-out, the centre of mass energy for annihilation processes roughly obeys the relation, $\sqrt{s} \approx 2 m_{\chi_1^0}^{}$. If the $2 m_{\chi_1^0}^{}$ happens to be equal a bosonic state (such as a Higgs field), then the relic abundance will be depleted by the resonance. Again, this scenario suggests some degree of fine tuning is required.

The following subsections discuss the specific phenomenology of certain mechanisms that are found to lead to the relic density constraint being satisfied. The remaining allowed parameter space is separated into five regions, and the numbering is used to refer to each region in the plots of this chapter.

The case of a (left-handed) sneutrino LSP has been overlooked so far in the discussion. This is a viable dark matter candidate, but the large electroweak interactions lead to direct detection experiments excluding the candidate. The limits on superpartner masses and UV structure of mSUGRA also restricts the region of parameter space where the sneutrino is the LSP to a relatively small region, when the direct search limits are ignored. This situation will not be discussed further.


In past literature, the ``bulk" region has also been identified as consistent with the relic abundance. In this region, the Bino annihilation is increased by the exchange of light sleptons in the t and u-channel, which occurs at small $m_{1/2}^{}$ and small $m_0^{}$. Figs~\ref{pspace_noh} and \ref{pspace_h} demonstrate that this region is now excluded, which is a consequence of the improved experimental constraints on superpartner masses.

%
%

%




\begin{figure}[!th]
\center
\begin{tabular}{c} 
\subfloat[]{\includegraphics[width=7cm]{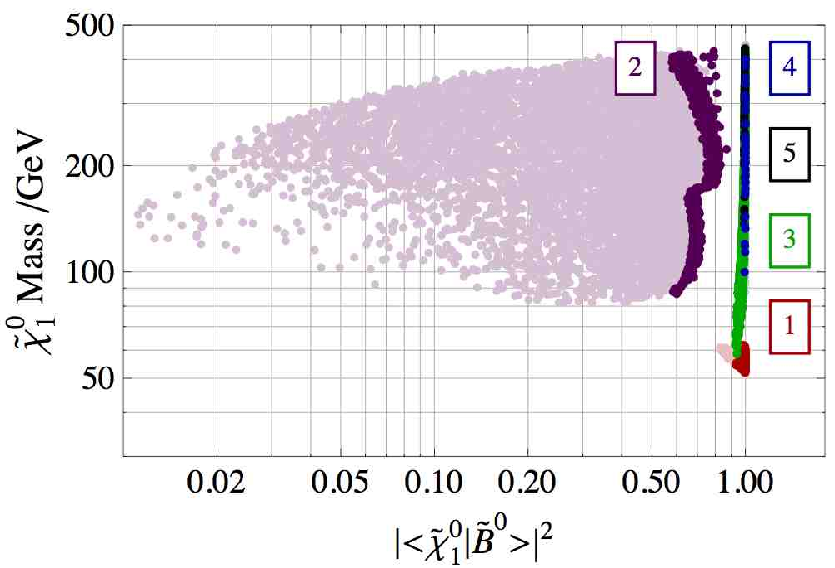}} 
\hspace{4mm}
\subfloat[]{
\includegraphics[width=7cm]{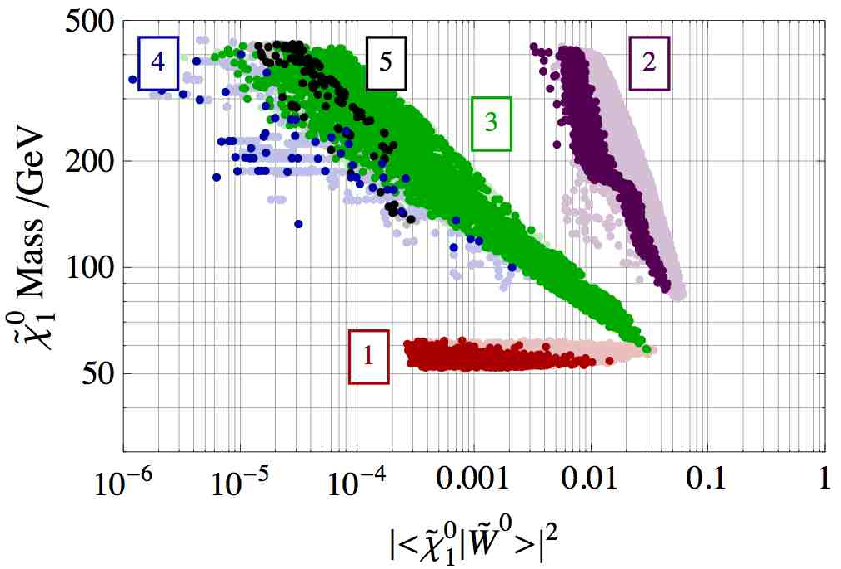}}
\end{tabular}
\begin{tabular}{c} 
\subfloat[]{\includegraphics[width=7cm]{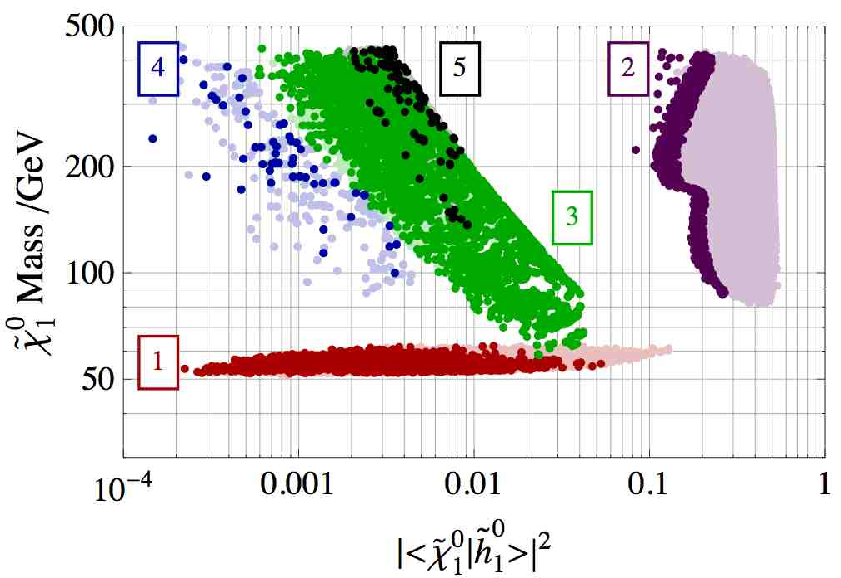}}
\hspace{4mm}
\subfloat[]{
\includegraphics[width=7cm]{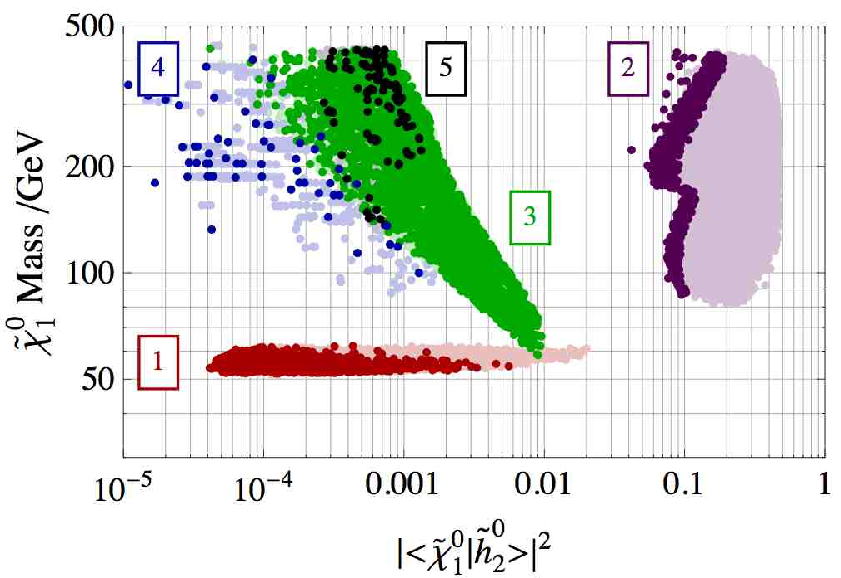}}
\end{tabular}
\caption{\protect\small 
Plot of LSP mass vs LSP composition in gaugino-Higgsino basis, with the Higgs mass unrestricted. The overlap with respect to the Bino is plotted in (a), the Wino in (b), the down-type Higgsino in (c) and the up-type Higgsino in (d). The labels distinguish between the dominant LSP annihilation mechanisms. The $h^0~(H^0, A^0)$ resonance is significant for region 1 (5) with points coloured red (black). Region 3 (4) realises stau (stop) co-annihilation with points coloured green (blue), and region 2, coloured purple, has increased Higgsino components.
}
\label{nmass_noh}
\end{figure}

\begin{figure}[!th]
\center
\begin{tabular}{c} 
\subfloat[]{\includegraphics[width=7cm]{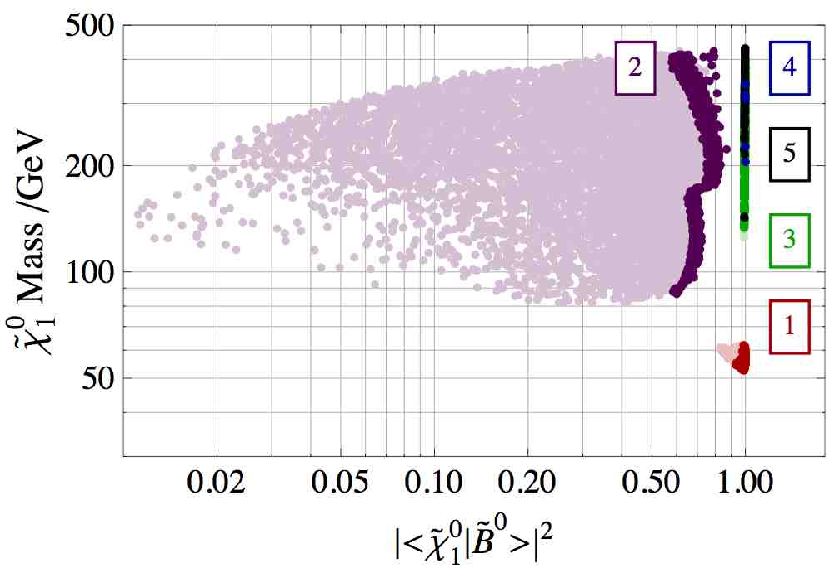}} 
\hspace{4mm}
\subfloat[]{
\includegraphics[width=7cm]{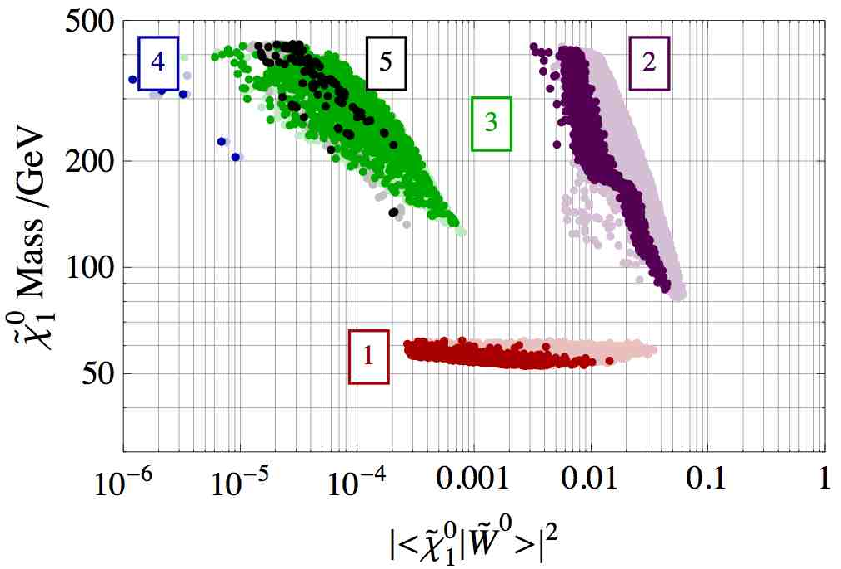}}
\end{tabular}
\begin{tabular}{c} 
\subfloat[]{\includegraphics[width=7cm]{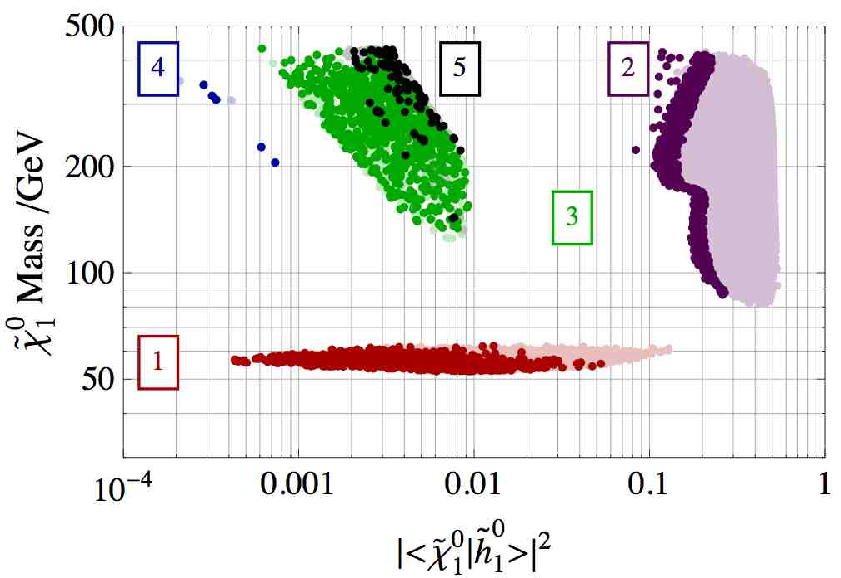}}
\hspace{4mm}
\subfloat[]{
\includegraphics[width=7cm]{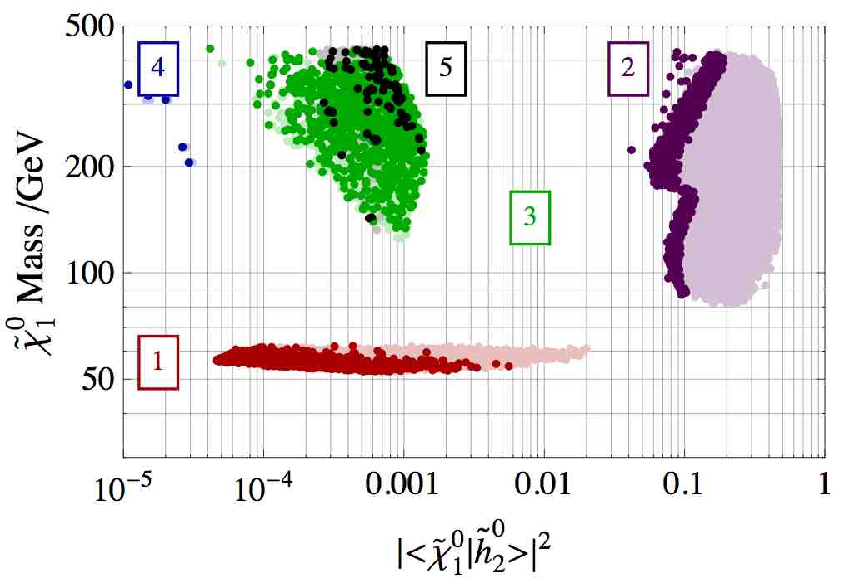}}
\end{tabular}
\caption{\protect\small 
Plot of LSP mass vs LSP composition in gaugino-Higgsino basis, with $m_h > 114.4$~GeV. The overlap with respect to the Bino is plotted in (a), the Wino in (b), the down-type Higgsino in (c) and the up-type Higgsino in (d). The labels distinguish between the dominant LSP annihilation mechanisms. The $h^0~(H^0, A^0)$ resonance is significant for region 1 (5) with points coloured red (black). Region 3 (4) realises stau (stop) co-annihilation with points coloured green (blue), and region 2, coloured purple, has increased Higgsino components.
}
\label{nmass_h}
\end{figure}

%
%
%
%

\subsection{Region 1: $h^0$ resonant annihilation}

The lower bound for lightest neutralino mass in the mSUGRA scenario is 46\,GeV for most of the parameter space. The centre of mass energy for LSP annihilation then obeys, $\sqrt{s} > 92\,$GeV. For all but the smallest values of $\tan \beta$, the light Higgs resonance will be accessible in some region of the parameter space allowed before dark matter constraints. The width of the $h^0$ resonance is $O(3\,$GeV), and so it may not require significant amounts of fine tuning. As for all the regions identified in this section, the degree of fine tuning will need to be tested.

Figs~\ref{nmass_noh} and \ref{nmass_h} demonstrate that only a small range of neutralino mass (50 to 62\,GeV) actually leads to a satisfactory relic abundance, and the LSP is indeed Bino-like. The Higgs mass constraint also does not considerably restrict this region of allowed parameter space. A light Higgs resonance at this energy scale will dominantly decay to $b\bar{b}$ products. It is then larger $\tan \beta$ that will lead to the large bottom Yukawa couplings to make the resonance most efficient. As the light Higgs is CP-even, the annihilation will dominantly proceed through p-wave annihilation. This is then less effective as other methods for reducing the relic density.

Figs~\ref{SInohiggs} and \ref{SIhiggs} show that the spin independent cross section for this region of parameter space is significantly below the current experimental limits so the most promising hope for early discovery is at colliders. The gluino is roughly six times heavier than the LSP in this region, and as the LSP (and gluino) here is only just above the current experimental limit, it will not require reaching the design centre of mass energies at colliders to generate these states in significant numbers.

Figs~\ref{pspace_noh} and \ref{pspace_h} indicate that this region is around the scalar focus point region, with a large $m_0^{}/m_{1/2}^{}$ ratio. The region also extends to large $A_0^{}$ parameters. This free variable can be used to introduce appropriate amounts of stop mixing to modify the Higgs quartic couplings and so shift the Higgs mass to be on resonance. As observed in the last chapter, this requires a sacrifice in fine tuning of the electroweak scale and so such limits for this region are not desirable on naturalness grounds.


\begin{figure}[!th]
\center
\begin{tabular}{c} 
\subfloat[]{\includegraphics[width=7cm]{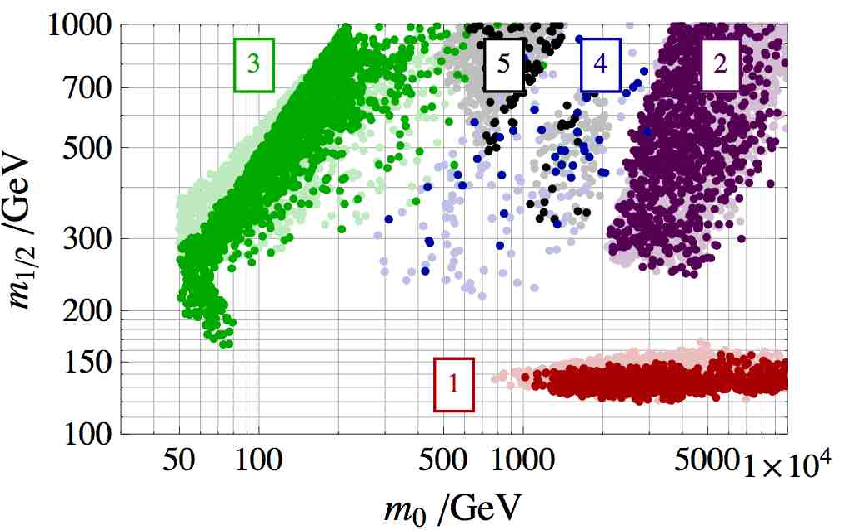}} 
\hspace{4mm}
\subfloat[]{
\includegraphics[width=7cm]{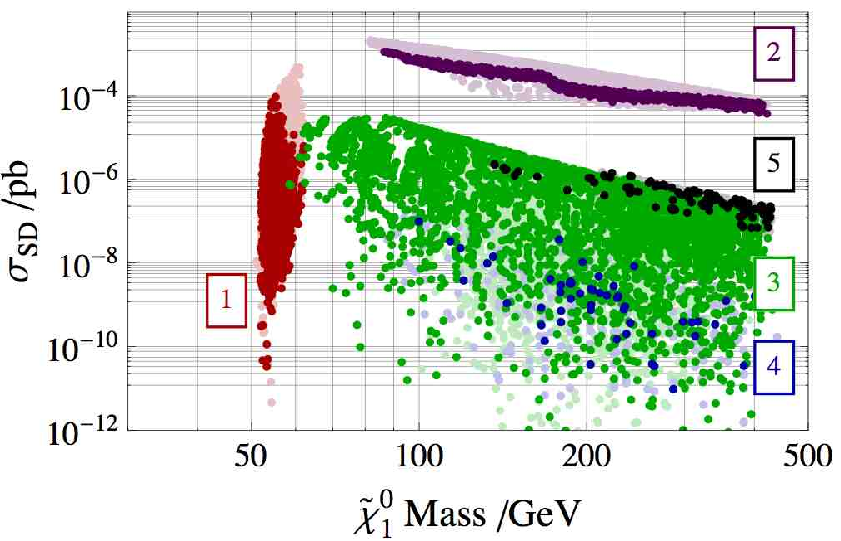}}
\end{tabular}
\begin{tabular}{c} 
\subfloat[]{\includegraphics[width=7cm]{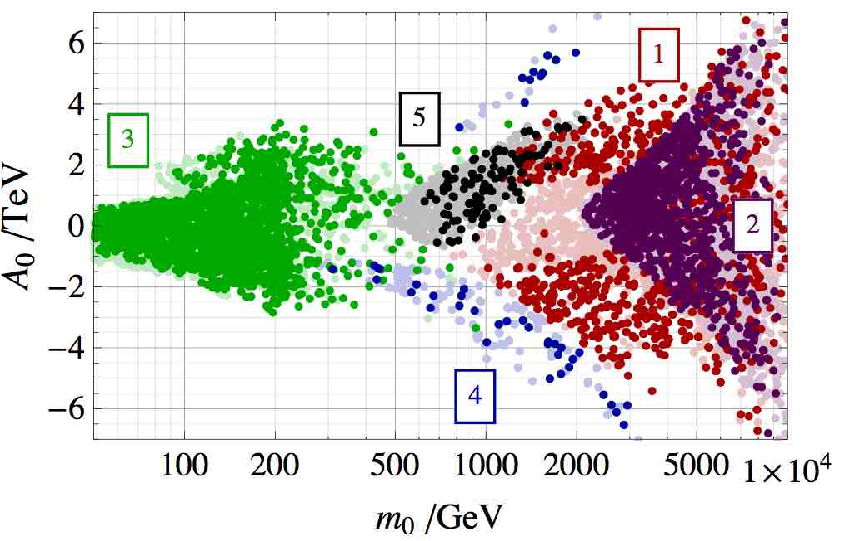}}
\hspace{4mm}
\subfloat[]{
\includegraphics[width=7cm]{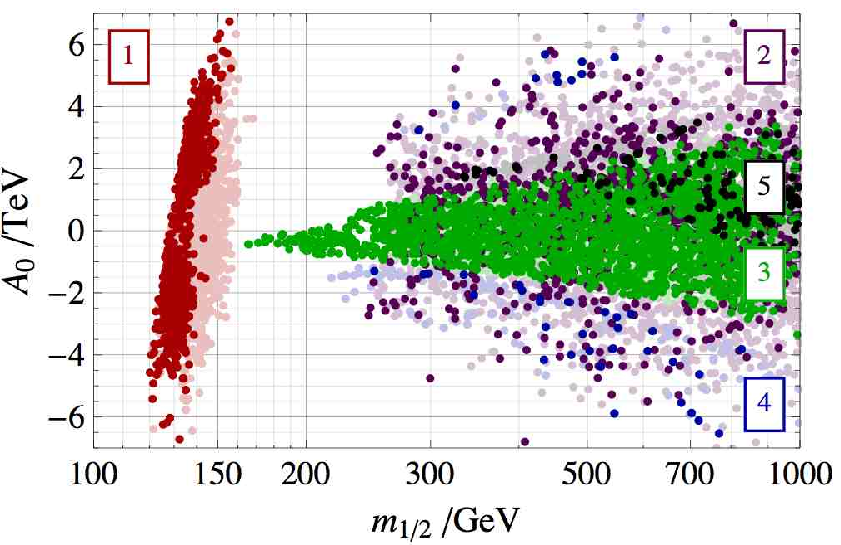}}
\end{tabular}
\begin{tabular}{c} 
\subfloat[]{\includegraphics[width=7cm]{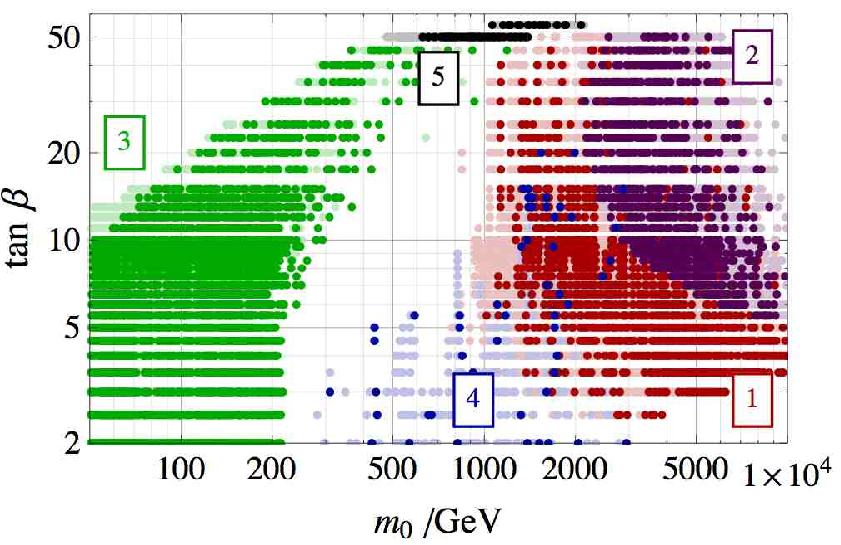}}
\hspace{4mm}
\subfloat[]{
\includegraphics[width=7cm]{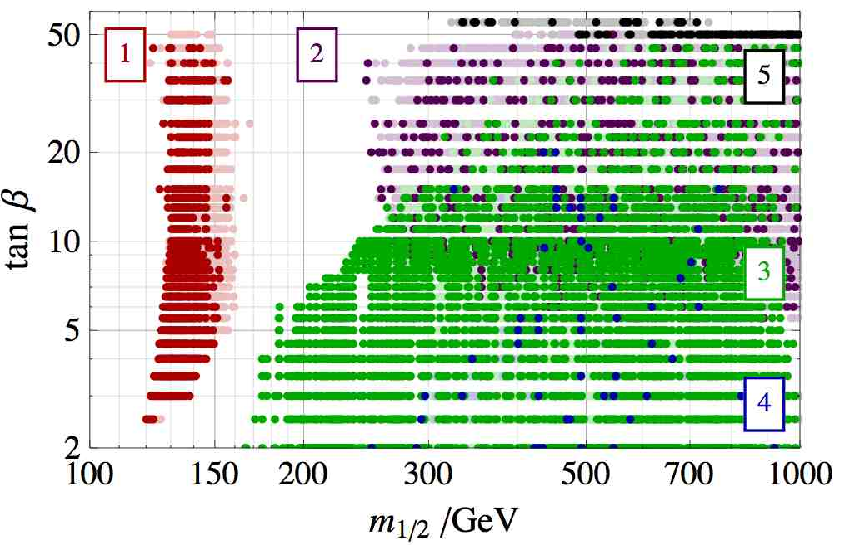}}
\end{tabular}
\caption{\protect\small 
Plots of the allowed mSUGRA parameter space subject to all constraints listed in Table~\ref{contable} and the dark matter constraints, and in (b) spin-dependent cross section vs LSP mass (where the experimental limit is off the top of the plot). The Higgs mass is unrestricted. The stepped choices of $\tan \beta$ is responsible for the lines of points in (e) and (f). The labels distinguish between the dominant LSP annihilation mechanisms. The $h^0~(H^0, A^0)$ resonance is significant for region 1 (5) with points coloured red (black). Region 3 (4) realises stau (stop) co-annihilation with points coloured green (blue), and region 2, coloured purple, has increased Higgsino components.
}
\label{pspace_noh}
\end{figure}

\begin{figure}[!th]
\center
\begin{tabular}{c} 
\subfloat[]{\includegraphics[width=7cm]{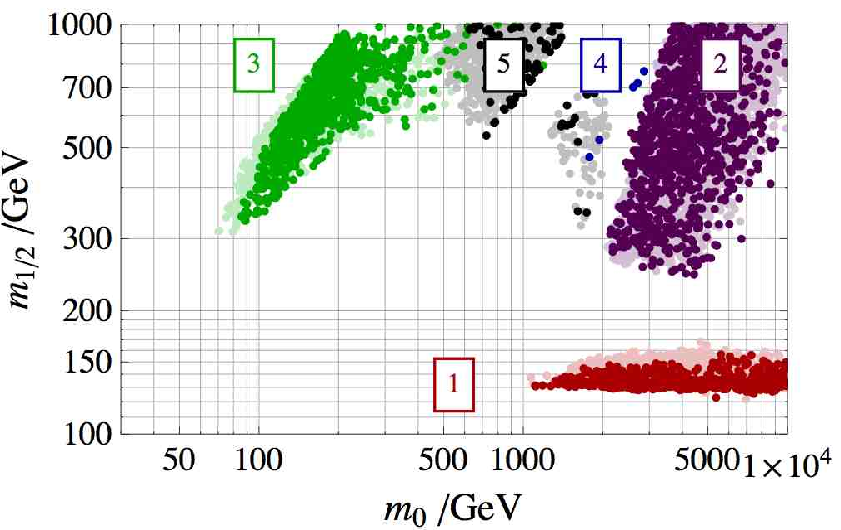}} 
\hspace{4mm}
\subfloat[]{
\includegraphics[width=7cm]{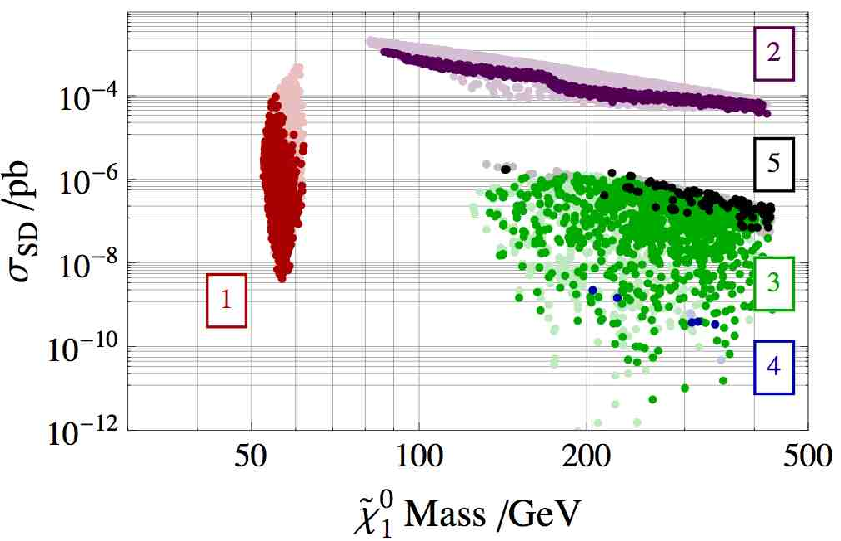}}
\end{tabular}
\begin{tabular}{c} 
\subfloat[]{\includegraphics[width=7cm]{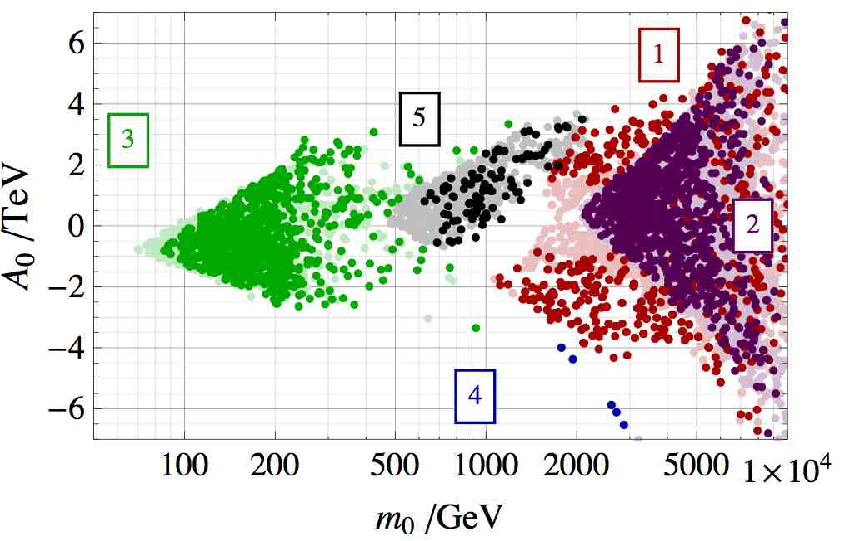}}
\hspace{4mm}
\subfloat[]{
\includegraphics[width=7cm]{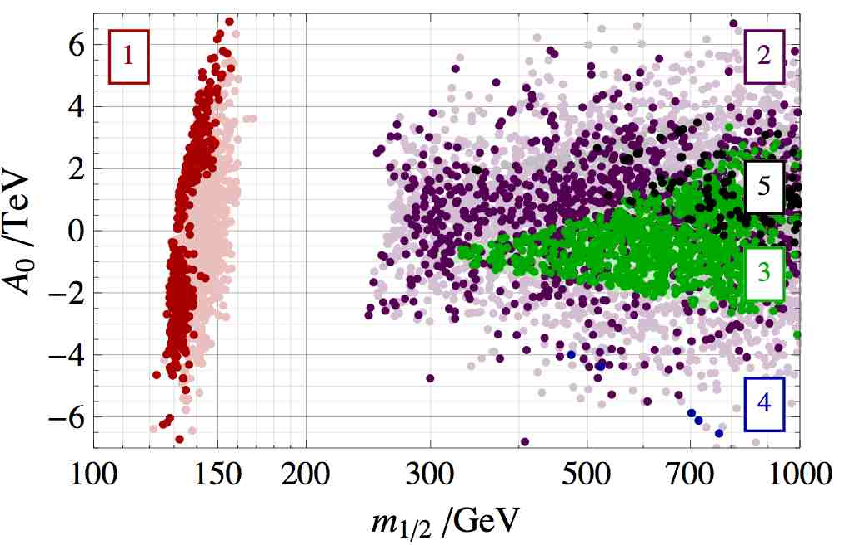}}
\end{tabular}
\begin{tabular}{c} 
\subfloat[]{\includegraphics[width=7cm]{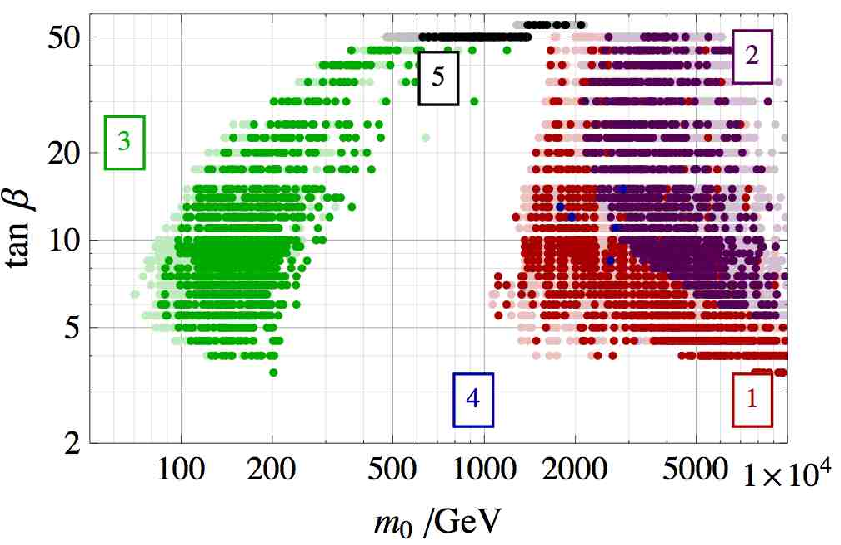}}
\hspace{4mm}
\subfloat[]{
\includegraphics[width=7cm]{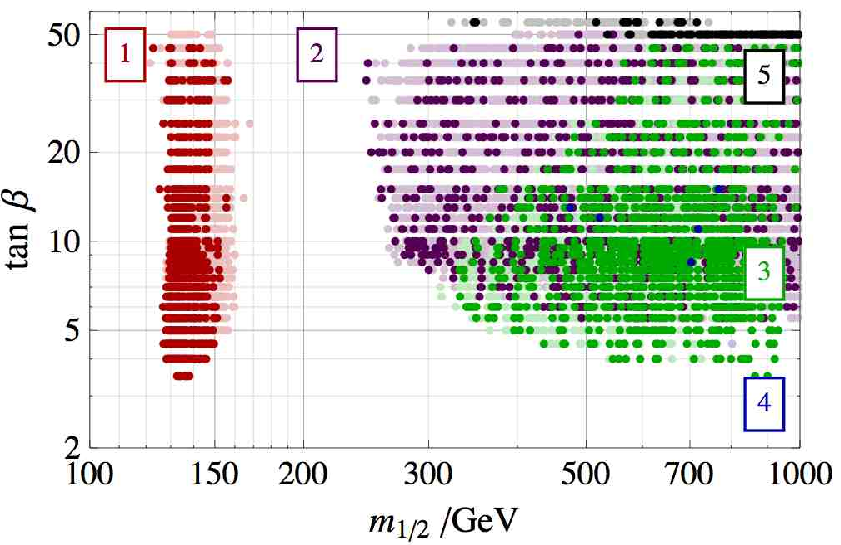}}
\end{tabular}
\caption{\protect\small 
Plots of the allowed mSUGRA parameter space subject to all constraints listed in Table~\ref{contable} and the dark matter constraints, and in (b) spin-dependent cross section vs LSP mass (where the experimental limit is off the top of the plot). The constraint, $m_h > 114.4$~GeV, is also applied.The stepped choices of $\tan \beta$ is responsible for the lines of points in (e) and (f). The labels distinguish between the dominant LSP annihilation mechanisms. The $h^0~(H^0, A^0)$ resonance is significant for region 1 (5) with points coloured red (black). Region 3 (4) realises stau (stop) co-annihilation with points coloured green (blue), and region 2, coloured purple, has increased Higgsino components.
}
\label{pspace_h}
\end{figure}

\subsection{Region 2: Bino-Higgsino mixing}\label{region2}

On the edge of parameter space where electroweak symmetry breaking is allowed, the SUSY $\mu$ parameter is driven towards zero by the Higgs potential vacuum minimisation conditions. Section~\ref{MSSMspectrum} indicated that the superpartner mass limits restrict the $\mu$ parameter from being much smaller than $M_1^{}$. In this region, there is a significant Higgsino component within the LSP.

The Higgsino can annihilate via a Higgsino t-channel to produce electroweak gauge bosons. The couplings are relatively large and so the annihilation is efficient. In fact, for pure Higgsino states, the relic abundance would be generally far below the WMAP result. An intermediate mixture of Bino and Higgsino is preferred for generating the observed relic abundance.

Section~\ref{MSSMspectrum} also mentioned that there is a near degeneracy in this region with the light chargino and all but the heaviest neutralino. There will therefore also be coannihilation channels from the $\tilde{\chi}_1^0 \, \tilde{\chi}_1^\pm \, W^\pm$ and $\tilde{\chi}_1^0 \, \tilde{\chi}_2^0 \, Z$ vertices.

Figs~\ref{nmass_noh} and \ref{nmass_h} demonstrate that in this region, the LSP is 60 to 75\% Bino, 10 to $\sim$20\% of each Higgsino type, and up to 5\% Wino when the non-baryonic matter relic density is saturated by the LSP contribution. In the case that the relic density is sub-saturating, the LSP may be half each Higgsino type (and $< 1\%$ Bino).

The Higgs mass constraint does not affect this region at all, but Figs~\ref{SInohiggs} and \ref{SIhiggs} show that the CDMS II constraint is restrictive for light neutralino masses. The large Higgsino components lead to large couplings with nucleons. It is this region of parameter space that the direct detection experiments are currently probing.




\subsection{Region 3: Stau co-annihilation}

The stau is driven relatively light in the region of parameter space where $m_{1/2}^{}/m_{0}^{}$ is large as observed in Figs~\ref{pspace_noh} and \ref{pspace_h}. This is expected, as the diagonal entries in the stau mass squared matrix are $O(m_0^2)$.

Figs~\ref{nmass_noh} and \ref{nmass_h} demonstrate that the LSP mass is correlated with the Wino and Higgsino components, being lighter for increased non-Bino components. This is just a consequence of the mSUGRA pattern for the neutralino mass matrix affecting the mixing as the magnitude of $m_{1/2}^{}$ is varied. The Higgs mass constraint signifcantly restricts this region of parameter space, as it is mostly found for low $\tan \beta$ values where the Higgs mass tends is light.

Figs~\ref{SInohiggs} and \ref{SIhiggs} show some restriction is provided by the CDMS II constraint when the Higgs mass is ignored. However, with the Higgs mass constraint, the points in this region are over an order of magnitude below the direct detection limits. The variation in the spin independent cross section for varying neutralino mass does have some kinematic origin, but the dominant influence is the varying Higgsino component which controls the effectiveness of coupling to nucleons.



\subsection{Region 4: Stop co-annihilation}

The stop co-annihilation region is in a similar region as the stau co-annihilation region. However, as indicated in Figs~\ref{pspace_noh} and \ref{pspace_h}, a large $A_0^{}$ parameter is necessary to induce large stop mixing. The mixing was less effective for staus due to the smaller tau Yukawa coupling.

Figs~\ref{nmass_noh} and \ref{nmass_h} demonstrate that the non-Bino components of the LSP are extremely small ($< 0.5\%$), and so the spin independent cross sections plotted in Figs~\ref{SInohiggs} and \ref{SIhiggs} are therefore relatively small. These points are also found for low to moderate $\tan \beta$, and so the Higgs mass constraint significantly restricts this region. At large $\tan \beta$ the off-diagonal terms in the stop mass matrix are reduced, and so there is less mixing.



\subsection{Region 5: $H^0, A^0$ resonant annihilation}

Similarly to region 1, where there is resonance with the light Higgs state, there exists a region of allowed parameter space where resonance with the heavy Higgs states aids in reducing the relic density. Due to the near degeneracy of these states and the $O(30\,$GeV) widths, it is not possible for this parameter space scan to resolve each resonance separately.

Figs~\ref{nmass_noh} and \ref{nmass_h} demonstrate that this region has a similar LSP composition to the stau co-annihilation region. The dependence of the spin independent cross section shown in Figs~\ref{SInohiggs} and \ref{SIhiggs} then also has a similar behaviour.

However, the structure of the mSUGRA mass spectrum requires large $m_{1/2}^{}$ and large $\tan \beta$ in order to find this resonance as indicated in Figs~\ref{pspace_noh} and \ref{pspace_h}. The annihilation via the CP odd $A^0$ can proceed via s-wave annihilation, but as for the $h^0$ resonance, the CP-even $H^0$ resonance must proceed via p-wave annihilation. A larger suppression is generally found when closer to the $A^0$ resonance.




\section{Fine tuning results from parameter space scan}\label{dmResults}

To quantify the fine tuning of the relic density, $\Delta^\Omega$, the Barbieri-Giudice measure is again used,
\bea
\Delta^\Omega=
\max\bigg\vert\frac{\partial \ln \Omega h^2}{\partial \ln q}
\bigg\vert
\eea
except that the parameters tested now include $q = \{ m_0^{}, m_{1/2}^{}, A_0^{}, \tan \beta  \}$. The Ciafaloni-Strumia measure was considered for the measured parameters, $ \{ m_Z^{}, m_t^{} \} $, but always found to be subdominant and so effectively ignored. The set of parameters are not all considered to be fundamental though. This compromise was necessary for using the numerical {\texttt{MicrOMEGAs}} package in order to determine the relic density. A routine for evaluation of fine tuning is also not built in the package as the fine tuning for the electroweak scale is within {\texttt{SOFTSUSY}}.

The procedure to evaluate the relic density fine tuning then used a finite difference method, checking the variation under small finite changes in the input parameters. Convergence tests were performed to confirm that the results were reliable to an $O(1\%)$ level. The theoretical uncertainty for most parts of the parameter space is expected to be at this level given comparisons with {\texttt{DarkSUSY}}. For earlier analyses of the fine tuning of the relic density, see \cite{Chankowski:1998za,Ellis:2007by} and the  references therein. A similar approach to that adopted here was used in those studies, although their scope did not cover a full parameter space scan, nor use the now available experimental constraints.

The following results demonstrate the degree of fine tunings present on application of the dark matter constraints. The results for fine tuning of the relic density and the electroweak scale are both presented, as well as plots to demonstrate the worst fine tuning present for a given point in the parameter space. For the latter, the worst fine tuning is defined to be max[$\Delta, \Delta^\Omega$], following a similar procedure as $\Delta$ being defined as max[$\{\Delta_p^{}\}$].

This procedure of combining fine tunings for different observables is not rigourously motivated. It was argued in the introductory chapter that the fine tuning of any one observable is not well defined, with many different approaches equally reasonable. One ambiguity was in choosing an appropriate parameterisation of the fundamental parameters for a given observable (dependent on its mass dimension). The ``worst" fine tuning is considered here for a straightforward interpretation of the results - the measure indicates the greatest degree of fine tuning necessary for either considered observable to assume its physical value.



\subsection{Dependence on dark matter constraints}

\begin{figure}[!th]
\center
\begin{tabular}{c} 
\subfloat[No direct search constraint]{\includegraphics[width=7cm]{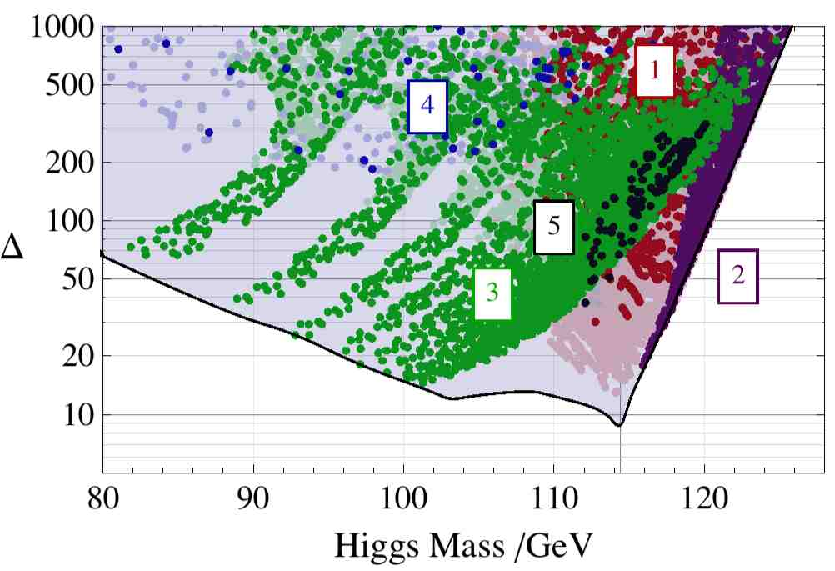}} 
\hspace{4mm}
\subfloat[CDMS II limit applied]{
\includegraphics[width=7cm]{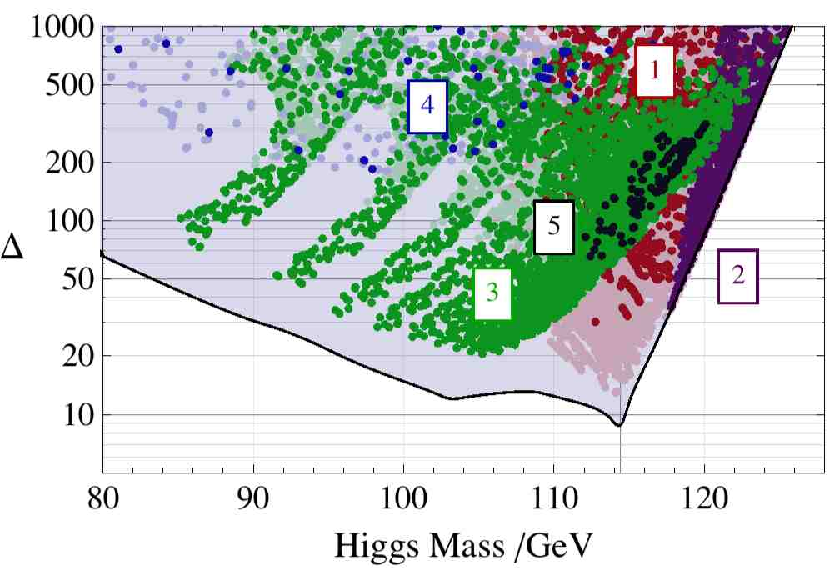}}
\end{tabular}
\begin{tabular}{c} 
\subfloat[No direct search constraint]{\includegraphics[width=7cm]{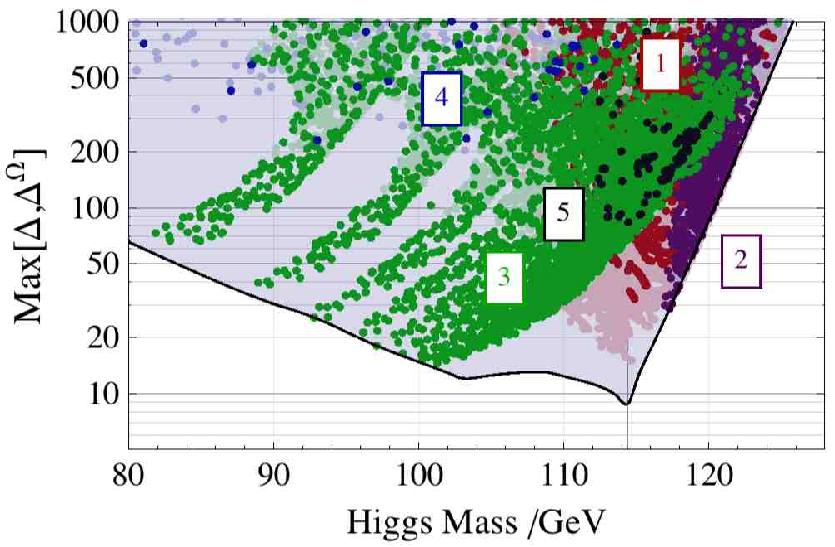}} 
\hspace{4mm}
\subfloat[CDMS II limit applied]{
\includegraphics[width=7cm]{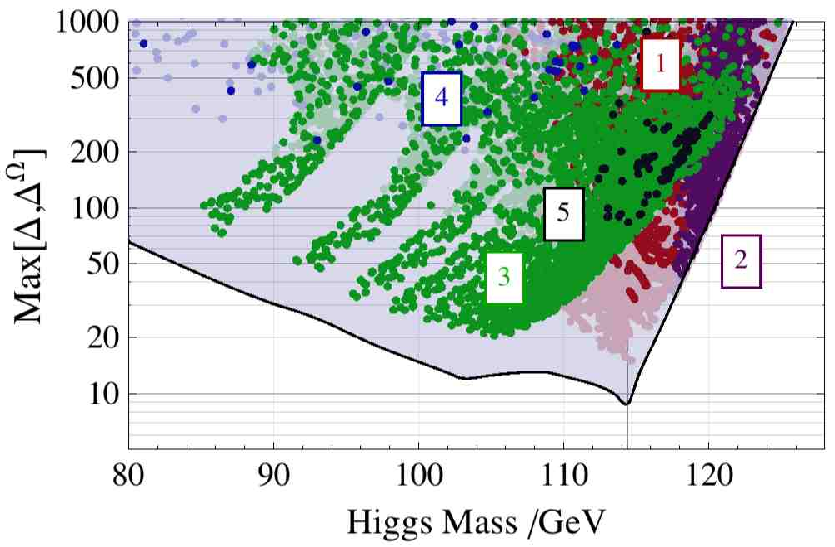}}
\end{tabular}
\caption{\protect\small 
Plots of fine tunings vs Higgs mass. The solid line and shading indicates the region allowed without any dark matter constraints. No direct detection experimental limits are applied in (a) and (c), but in (b) and (d) the CDMS II limit is applied for the points plotted. The fine tuning of the electroweak scale is plotted in (a) and (b), and the maximum fine tuning of the electroweak scale and relic density is plotted in (c) and (d). The labels distinguish between the dominant LSP annihilation mechanisms. The $h^0~(H^0, A^0)$ resonance is significant for region 1 (5) with points coloured red (black). Region 3 (4) realises stau (stop) co-annihilation with points coloured green (blue), and region 2, coloured purple, has increased Higgsino components.
}
\label{fthiggs}
\end{figure}

In order to compare with the results of the previous chapter, Fig~\ref{fthiggs} reproduces its results with the relic density constraint and subsequently the CDMS II constraint applied. The solid line and shaded area indicate the filling obtained without the dark matter constraint. The points (for which the dark matter constraint is applied) fill up only a fraction of that space.

Region 1 ($h^0$ resonance) is observed to occupy a region of low fine tuning around the LEP II Higgs mass bound. For points that sub-saturate the non-baryonic relic density, the smallest fine tuning is accessible. In fact, with the CDMS II constraint, the global minimum for the parameter space scan in the fine tuning of the electroweak scale and the overall fine tuning is found at the LEP II limit (just as was the case when ignoring the dark matter constraints). The impact of including the fine tuning of the relic density leads to slightly increased fine tuning in this limit and so both the electroweak scale and relic density are fine tuned to roughly 1 part in 15 at this minimum.

Region 2 (Bino-Higgsino mixing) is found to fill in the large Higgs mass region. The fine tuning of the relic density is greater than that for the electroweak scale at the lowest fine tuned points (low Higgs mass). However, these points are also excluded by the CDMS II constraint. For $m_h^{} > 118$\,GeV, the influence of dark constraints on the line of mininum fine tuning is negligible and the relic density fine tuning is not dominant in this limit except for the largest Higgs masses.

Region 3 (Stau coannihilation) seems to have strips of allowed space at low Higgs mass. This is in fact a result of the discrete stepping in $\tan \beta$ for the parameter space scan. A smoother scan across $\tan \beta$ fills in the gaps between the `strips.' It is observed that region 2 covers a wide range of Higgs mass. The relic density fine tuning does not noticeably affect the overall fine tuning. The impact of the CDMS II constraint is to eliminate points with low Higgs mass, at low $\tan \beta$. These points are far below the LEP II bound anyway.

Region 4 (Stop coannihilation) possesses large fine tuning of the electroweak scale, $\Delta \gtrsim 200$, which is frequently increased further by the relic density fine density at low $\tan \beta$. This was expected from the large $A_0^{}$ parameters required to generate the stop mixing.

Region 5 ($H^0, A^0$ resonance) also requires moderately large fine tuning, with the relic density fine tuning regularly dominating. The light Higgs mass ranges from just below the LEP II bound to 120\,GeV.


%




\subsection{Dependence on SUSY Lagrangian parameters}

The following results now always apply both the relic density and CDMS II constraints for the points plotted. The relic density fine tuning is also independently plotted for comparison and analysis. The plots of fine tuning for the electroweak scale and overall fine tuning indicate the region allowed when ignoring the dark matter constraints.

\begin{figure}[!th]
\center
\begin{tabular}{c} 
\subfloat[Higgs mass unrestricted]{\includegraphics[width=7cm]{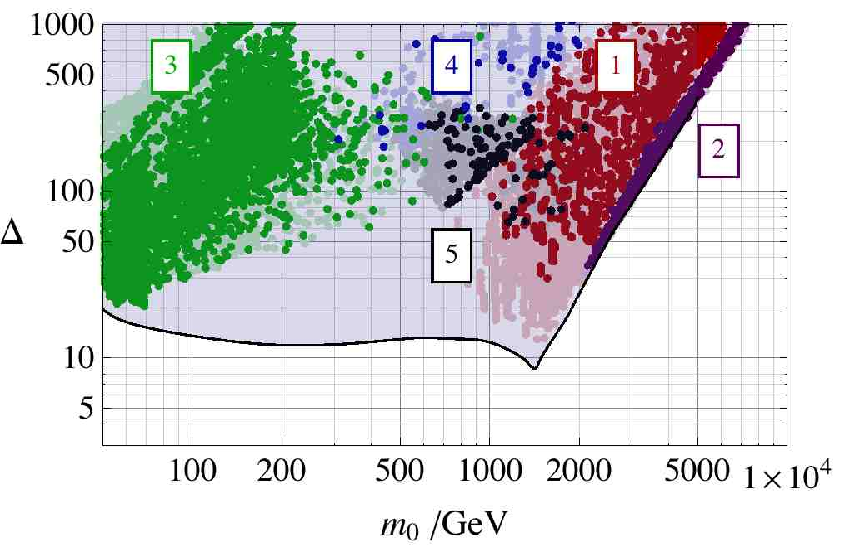}}
\hspace{4mm}
\subfloat[$m_h^{} > 114.4\,$GeV]{
\includegraphics[width=7cm]{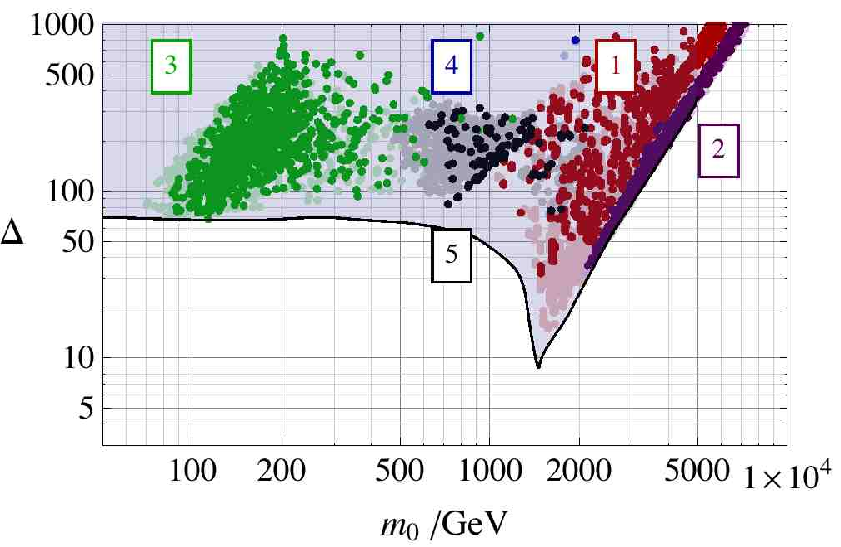}}
\end{tabular}
\begin{tabular}{c} 
\subfloat[Higgs mass unrestricted]{\includegraphics[width=7cm]{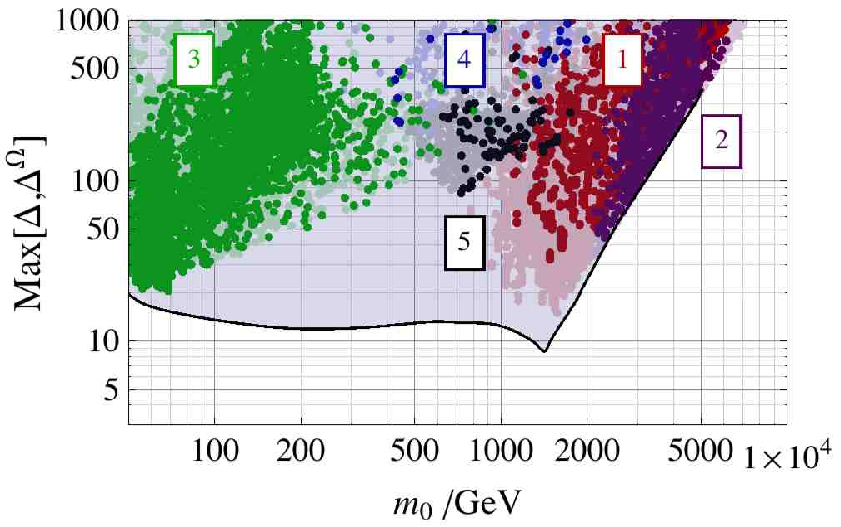}}
\hspace{4mm}
\subfloat[$m_h^{} > 114.4\,$GeV]{
\includegraphics[width=7cm]{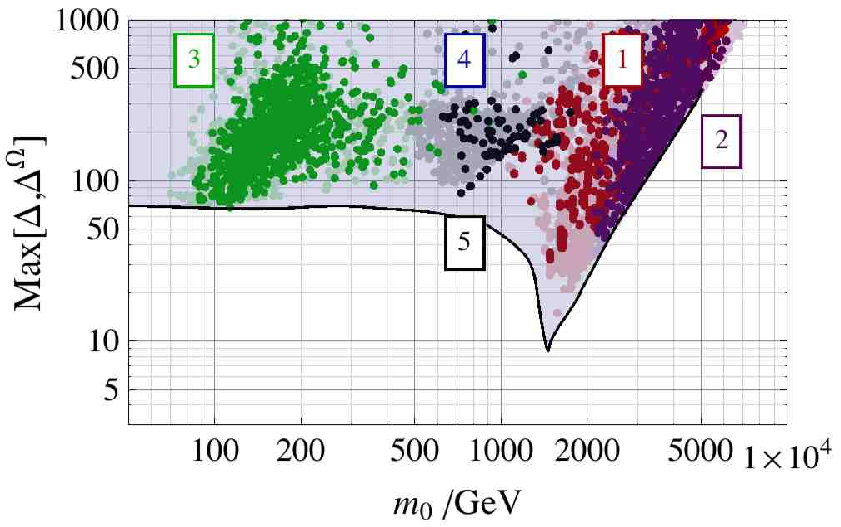}}
\end{tabular}
\begin{tabular}{c} 
\subfloat[Higgs mass unrestricted]{\includegraphics[width=7cm]{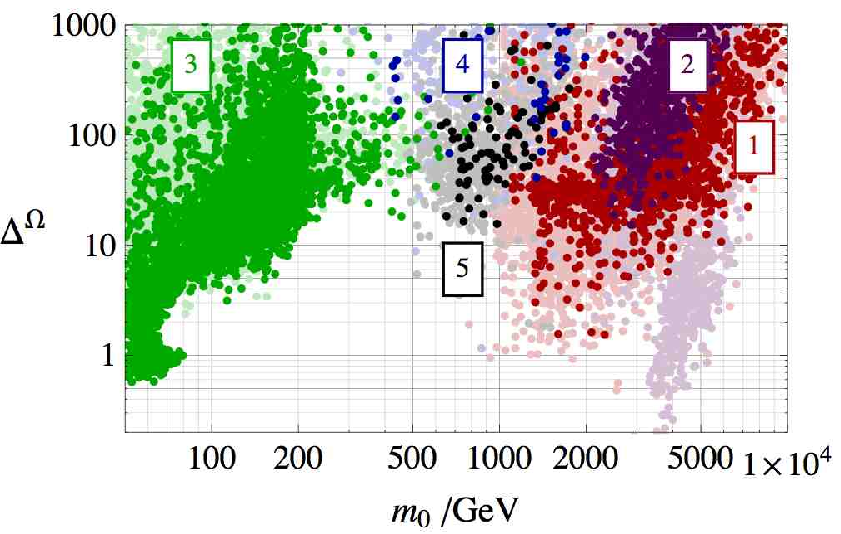}}
\hspace{4mm}
\subfloat[$m_h^{} > 114.4\,$GeV]{
\includegraphics[width=7cm]{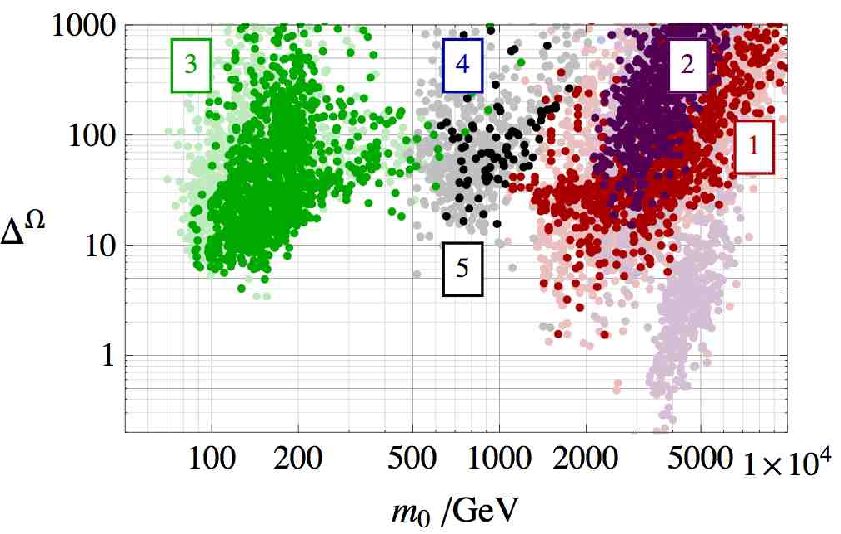}}
\end{tabular}
\caption{\protect\small 
Plots of fine tunings vs UV scalar soft mass, $m_0$. The fine tuning of the electroweak scale is plotted in (a) and (b), the fine tuning of the relic density in (e) and (f), and the maximum fine tuning between the two in (c) and (d). The labels distinguish between the dominant LSP annihilation mechanisms. The $h^0~(H^0, A^0)$ resonance is significant for region 1 (5) with points coloured red (black). Region 3 (4) realises stau (stop) co-annihilation with points coloured green (blue), and region 2, coloured purple, has increased Higgsino components.
}
\label{ftm0}
\end{figure}

Fig~\ref{ftm0} shows the dependence of fine tunings on the scalar soft mass. The impact of the dark matter constraints is to remove a large part of the parameter space with intermediate $m_{1/2}^{}$ with smallest fine tuning (when the Higgs mass is unrestricted).

Region 1 ($h^0$ resonance) is seen to populate the `dip' of fine tuning in $m_0^{}$ found on application of the LEP II bound. The relic density fine tuning also extends to very small values. Region 2 (Bino-Higgsino mixing) is seen to have a well-defined structure in the fine tuning of the electroweak scale, but the relic density fine tuning frequently raises the overall fine tuning significantly. Region 3 (Stau coannihilation) includes some very low relic density fine tuned points, but these are eliminated by the Higgs mass constraint. Region 4 (Stop coannihilation) and Region 5 ($H^0, A^0$ resonance) fill in the space between the other regions.

\begin{figure}[!th]
\center
\begin{tabular}{c} 
\subfloat[Higgs mass unrestricted]{\includegraphics[width=7cm]{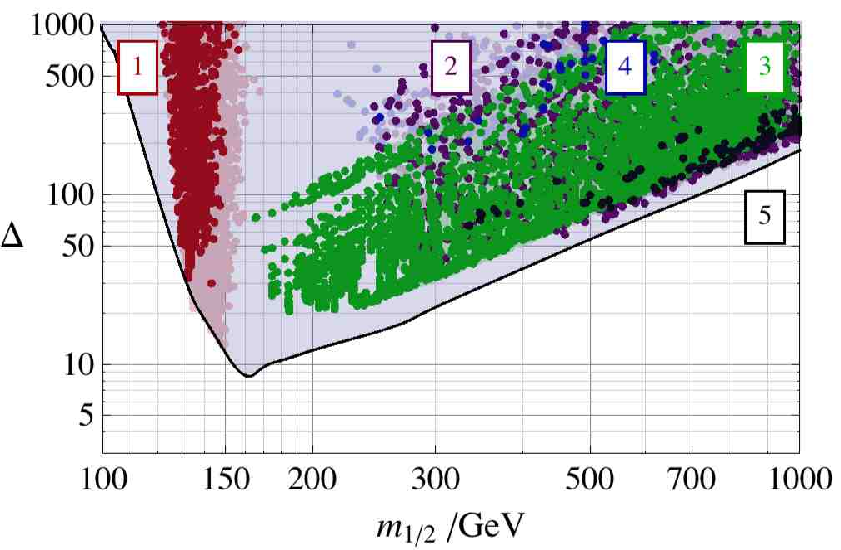}}
\hspace{4mm}
\subfloat[$m_h^{} > 114.4\,$GeV]{
\includegraphics[width=7cm]{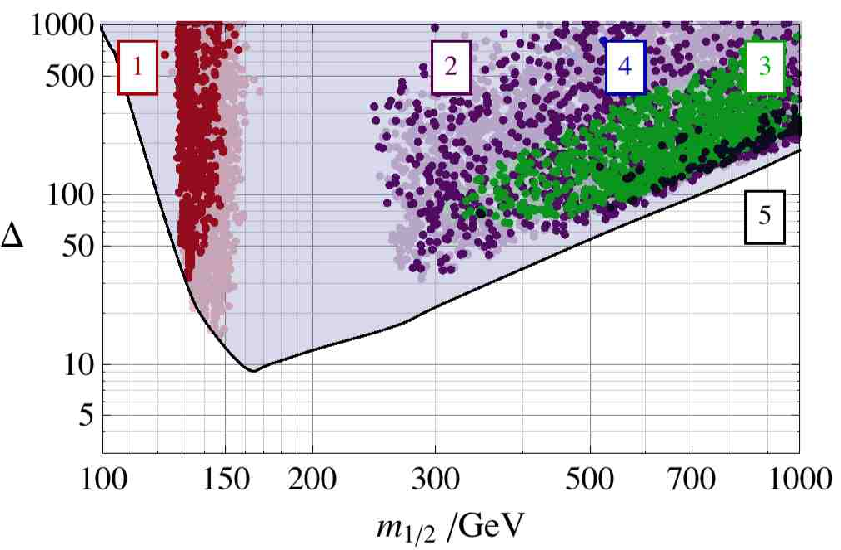}}
\end{tabular}
\begin{tabular}{c} 
\subfloat[Higgs mass unrestricted]{\includegraphics[width=7cm]{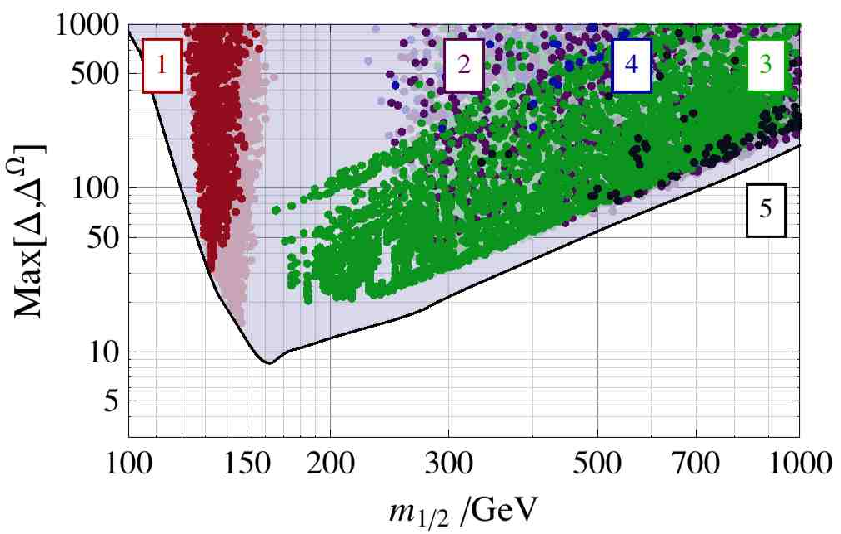}}
\hspace{4mm}
\subfloat[$m_h^{} > 114.4\,$GeV]{
\includegraphics[width=7cm]{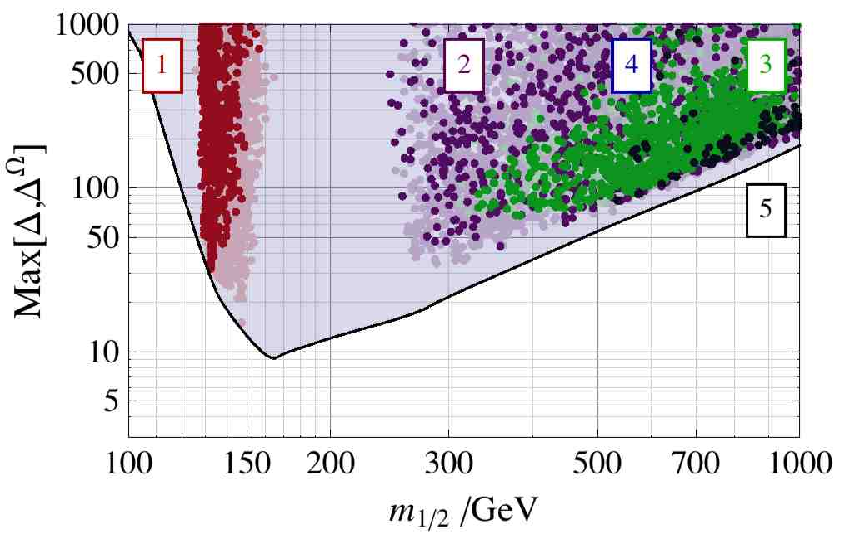}}
\end{tabular}
\begin{tabular}{c} 
\subfloat[Higgs mass unrestricted]{\includegraphics[width=7cm]{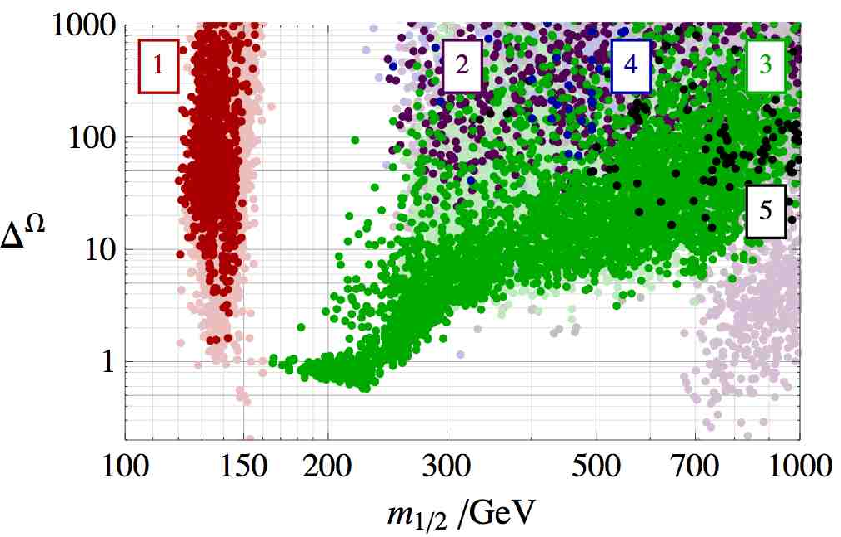}}
\hspace{4mm}
\subfloat[$m_h^{} > 114.4\,$GeV]{
\includegraphics[width=7cm]{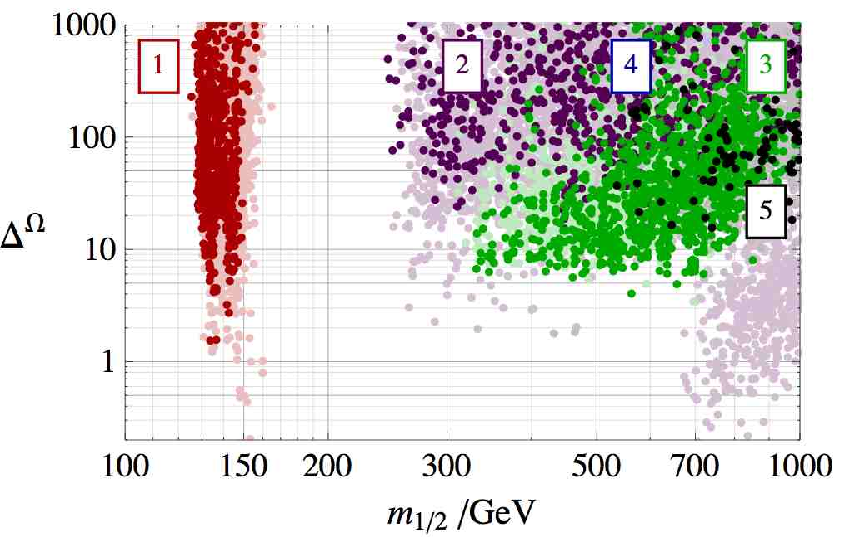}}
\end{tabular}
\caption{\protect\small 
Plots of fine tunings vs UV gaugino soft mass, $m_{12}$. The fine tuning of the electroweak scale is plotted in (a) and (b), the fine tuning of the relic density in (e) and (f), and the maximum fine tuning between the two in (c) and (d). The labels distinguish between the dominant LSP annihilation mechanisms. The $h^0~(H^0, A^0)$ resonance is significant for region 1 (5) with points coloured red (black). Region 3 (4) realises stau (stop) co-annihilation with points coloured green (blue), and region 2, coloured purple, has increased Higgsino components.
}
\label{ftm12}
\end{figure}

\begin{figure}[!th]
\center
\begin{tabular}{c} 
\subfloat[Higgs mass unrestricted]{\includegraphics[width=7cm]{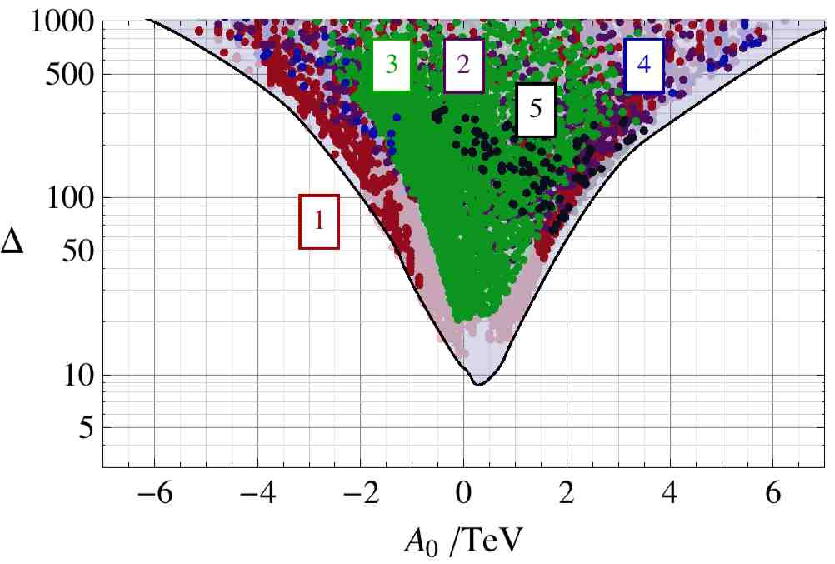}}
\hspace{4mm}
\subfloat[$m_h^{} > 114.4\,$GeV]{
\includegraphics[width=7cm]{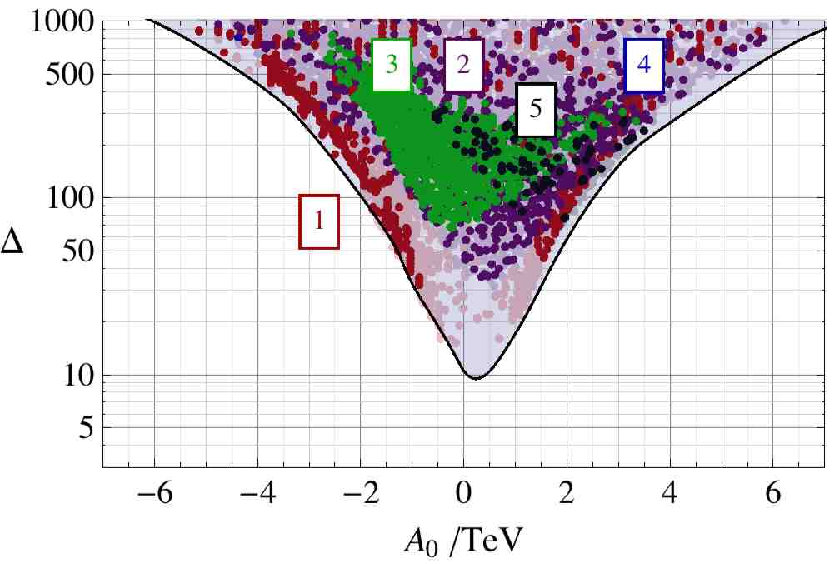}}
\end{tabular}
\begin{tabular}{c} 
\subfloat[Higgs mass unrestricted]{\includegraphics[width=7cm]{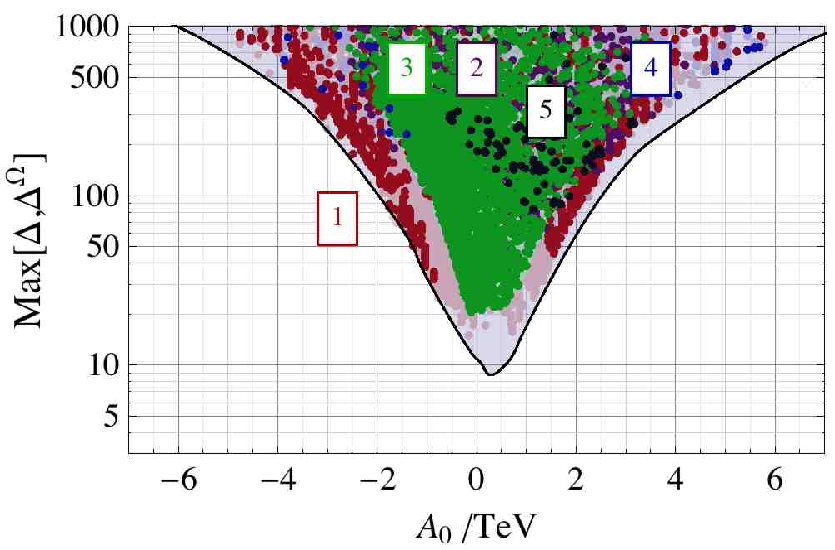}}
\hspace{4mm}
\subfloat[$m_h^{} > 114.4\,$GeV]{
\includegraphics[width=7cm]{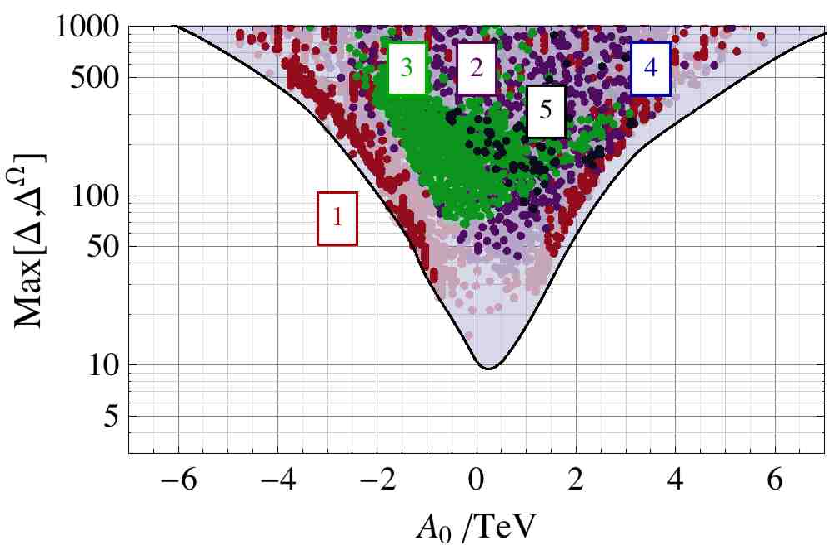}}
\end{tabular}
\begin{tabular}{c} 
\subfloat[Higgs mass unrestricted]{\includegraphics[width=7cm]{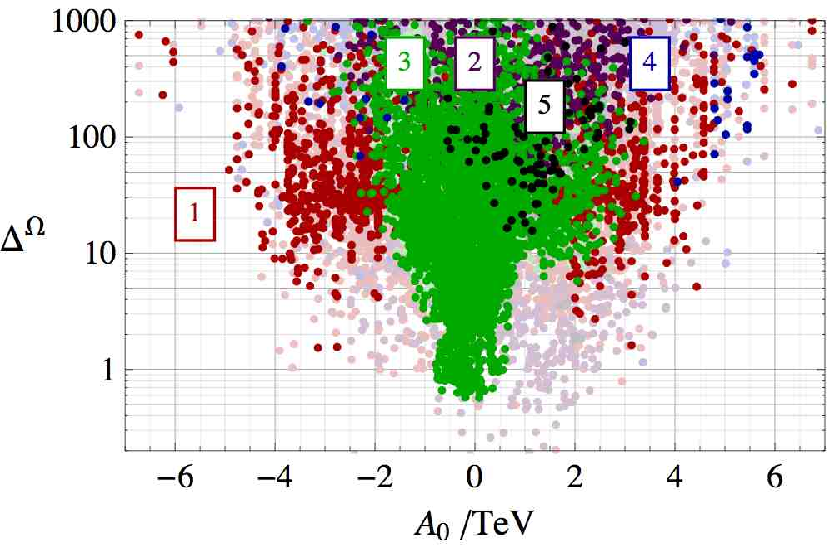}}
\hspace{4mm}
\subfloat[$m_h^{} > 114.4\,$GeV]{
\includegraphics[width=7cm]{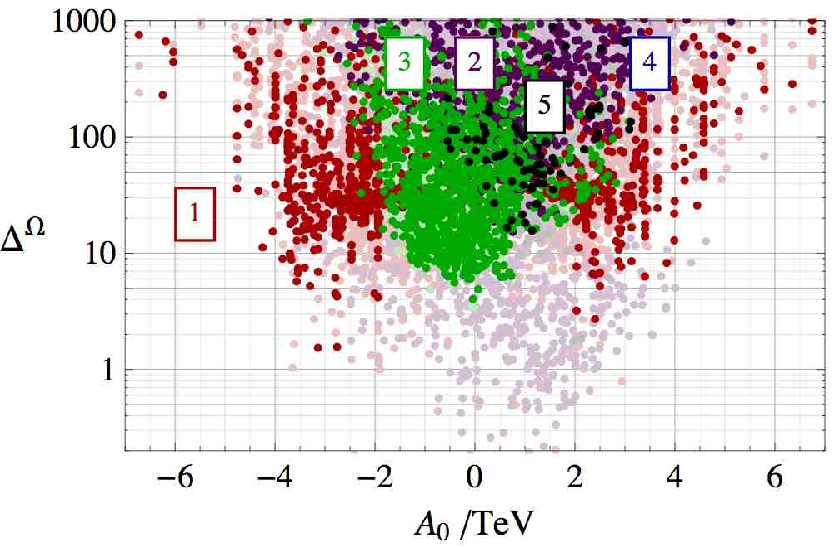}}
\end{tabular}
\caption{\protect\small 
Plots of fine tunings vs UV trilinear soft coupling, $A_0$. The fine tuning of the electroweak scale is plotted in (a) and (b), the fine tuning of the relic density in (e) and (f), and the maximum fine tuning between the two in (c) and (d). The labels distinguish between the dominant LSP annihilation mechanisms. The $h^0~(H^0, A^0)$ resonance is significant for region 1 (5) with points coloured red (black). Region 3 (4) realises stau (stop) co-annihilation with points coloured green (blue), and region 2, coloured purple, has increased Higgsino components.
}
\label{fta0}
\end{figure}

Fig~\ref{ftm12} shows the dependence of fine tunings on the gaugino soft mass. Region 1 ($h^0$ resonance) fills in a restricted range in $m_{1/2}$ as already noted from the previous plots of neutralino mass. Region 2 (Bino-Higgsino mixing) sets the minimum fine tuning for large $m_{1/2}^{}$, which is slightly raised from the case ignoring dark matter constraints. Region 3 (Stau coannihilation) is seen to be restricted to large $m_{1/2}^{}$ on application of the Higgs mass, which opens a gap in the allowed parameter space from 160 to 250\,GeV. The discrete lines visible within this region in plots (a) and (b) are a result of the discrete steps in $\tan \beta$, where larger $\tan \beta$ has reduced fine tuning. Region 4 (Stop coannihilation) has intermediate $m_{1/2}^{}$ and Region 5 ($H^0, A^0$ resonance) is on the low fine tuning edge at large $m_{1/2}^{}$.

Fig~\ref{fta0} shows the dependence of fine tunings on the trlinear soft parameter. Region 1 ($h^0$ resonance) has smallest fine tuning for non-zero $A_0^{}$, and fills the area just above the line of minimum fine tuning when ignoring the dark matter constraints. Region 2 (Bino-Higgsino mixing) prefers near zero $A_0^{}$, with the minimum edges becoming visible on application of the Higgs mass contraint (which doesn't affect this region, but excludes points previously overlaying the points of this region). Region 3 (Stau coannihilation) prefers near zero $A_0^{}$, but does not extend to the edges found without applying dark matter constraints. Region 4 (Stop coannihilation) requires large $A_0^{}$ in order to generate the stop mixing as previously argued, where the fine tunings are large. Region 5 ($H^0, A^0$ resonance) prefers $A_0^{} \sim 1\,$GeV for both fine tunings.

\subsection{Dependence on LSP composition}

Figs~\ref{ftbino} to \ref{fth2ino} plot the fine tunings vs the fraction of the LSP wavefunction given by the neutral gaugino and Higgsino states. Fig~\ref{ftbino} is not helpful for analysing the Bino-like regions, but for Region 2 (Bino-Higgsino mixing), the fine tuning of the electroweak scale is noticeably reduced as the Bino component is increased.

\begin{figure}[!th]
\center
\begin{tabular}{c} 
\subfloat[Higgs mass unrestricted]{\includegraphics[width=7cm]{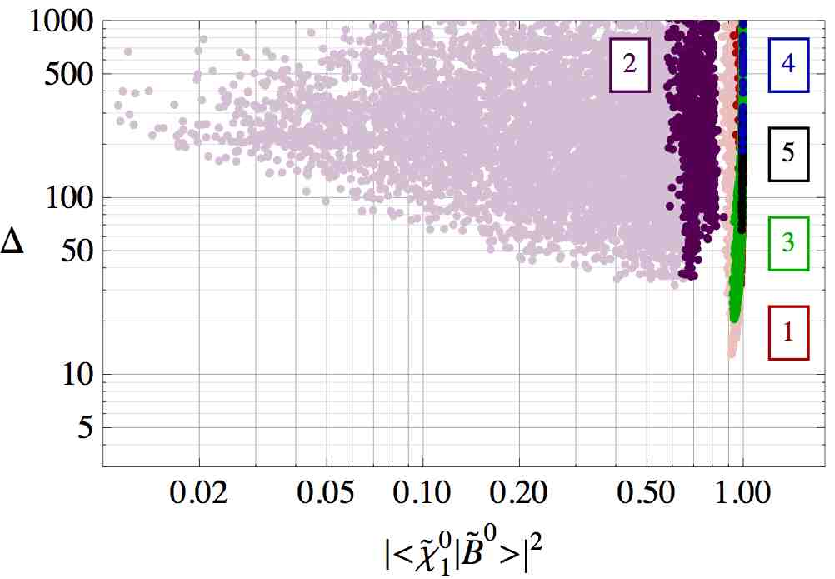}} 
\hspace{4mm}
\subfloat[$m_h^{} > 114.4\,$GeV]{
\includegraphics[width=7cm]{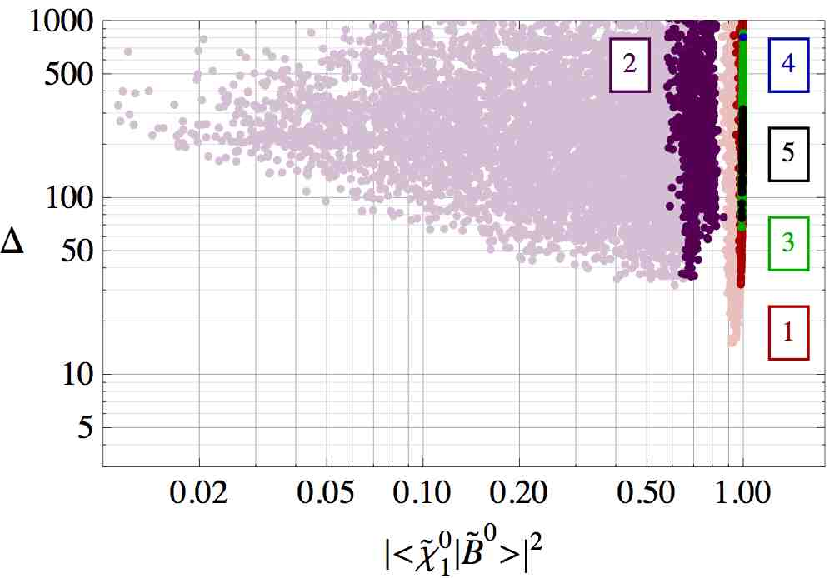}}
\end{tabular}
\begin{tabular}{c} 
\subfloat[Higgs mass unrestricted]{\includegraphics[width=7cm]{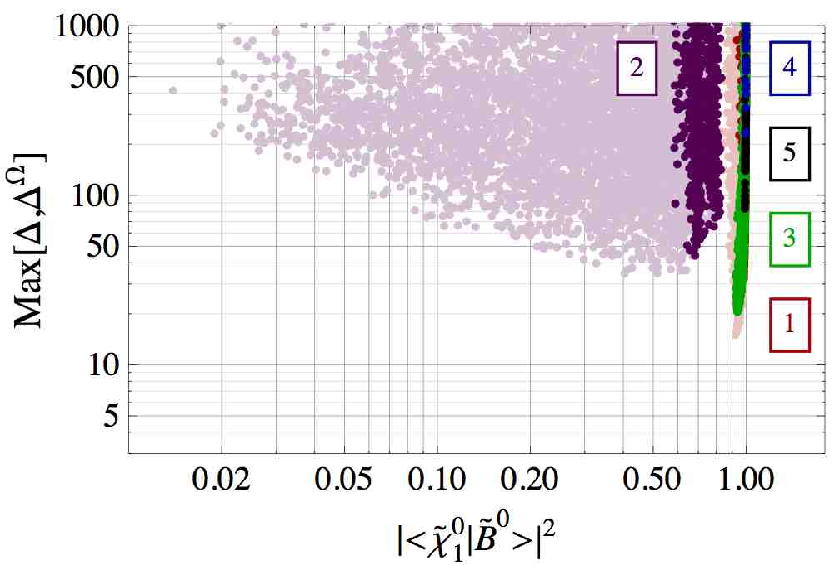}}
\hspace{4mm}
\subfloat[$m_h^{} > 114.4\,$GeV]{
\includegraphics[width=7cm]{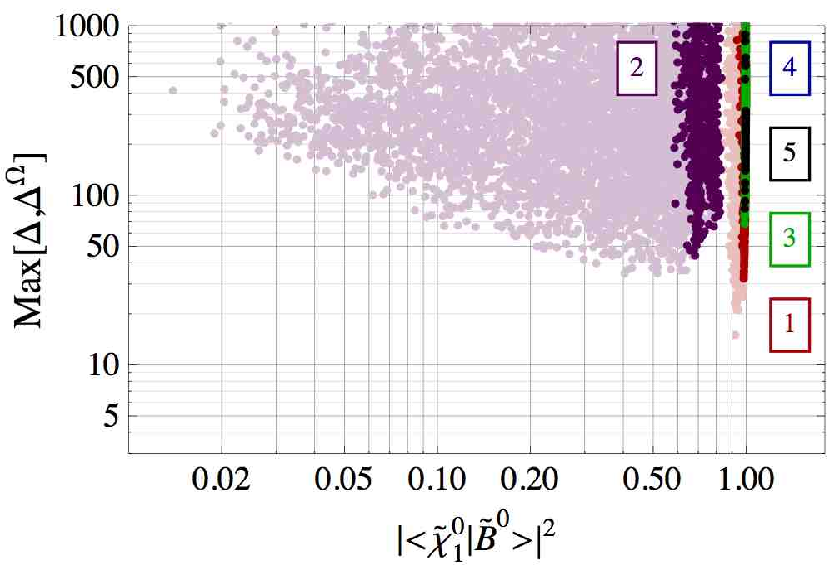}}
\end{tabular}
\begin{tabular}{c} 
\subfloat[Higgs mass unrestricted]{\includegraphics[width=7cm]{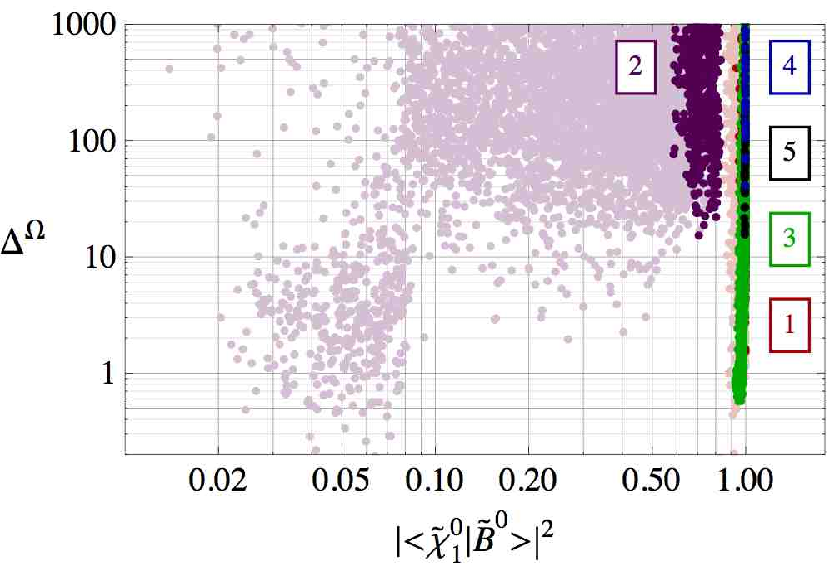}}
\hspace{4mm}
\subfloat[$m_h^{} > 114.4\,$GeV]{
\includegraphics[width=7cm]{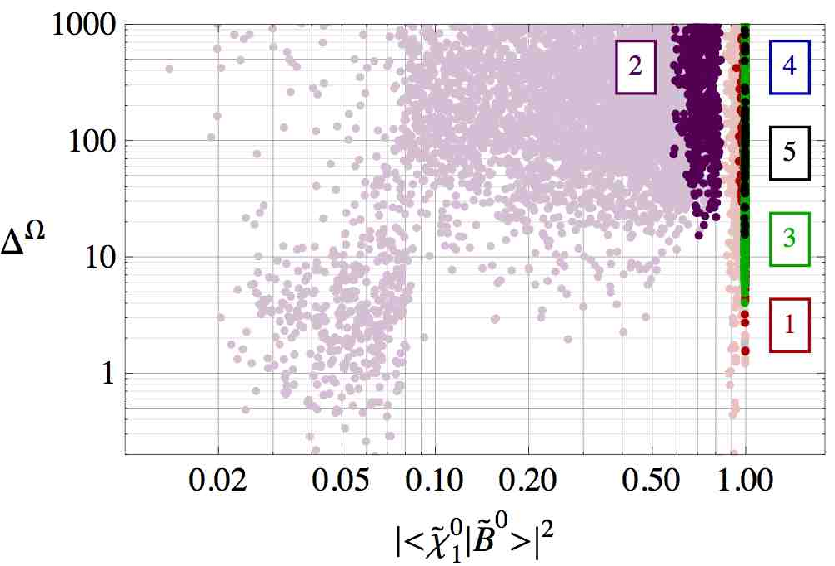}}
\end{tabular}
\caption{\protect\small 
Plots of fine tunings vs Bino overlap of LSP. The fine tuning of the electroweak scale is plotted in (a) and (b), the fine tuning of the relic density in (e) and (f), and the maximum fine tuning between the two in (c) and (d). The labels distinguish between the dominant LSP annihilation mechanisms. The $h^0~(H^0, A^0)$ resonance is significant for region 1 (5) with points coloured red (black). Region 3 (4) realises stau (stop) co-annihilation with points coloured green (blue), and region 2, coloured purple, has increased Higgsino components.
}
\label{ftbino}
\end{figure}

\pagebreak
\begin{figure}[!th]
\center
\begin{tabular}{c} 
\subfloat[Higgs mass unrestricted]{\includegraphics[width=7cm]{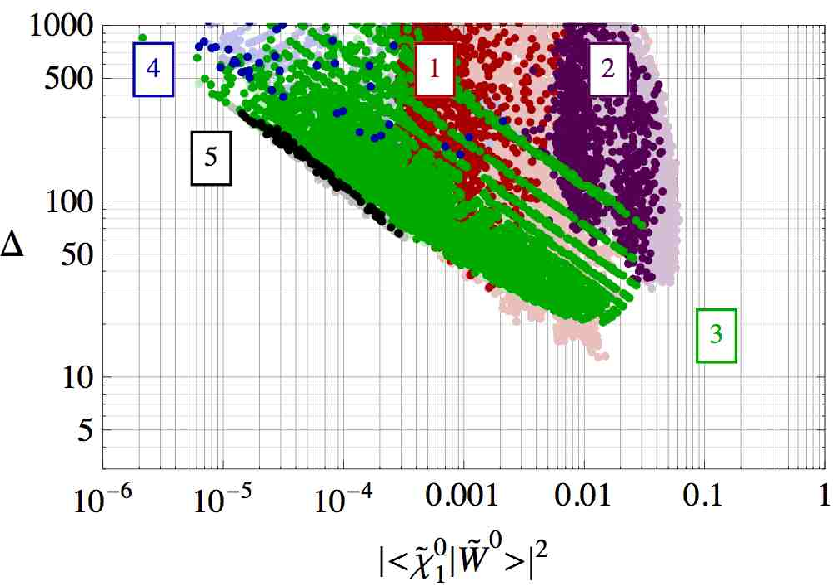}}
\hspace{4mm}
\subfloat[$m_h^{} > 114.4\,$GeV]{
\includegraphics[width=7cm]{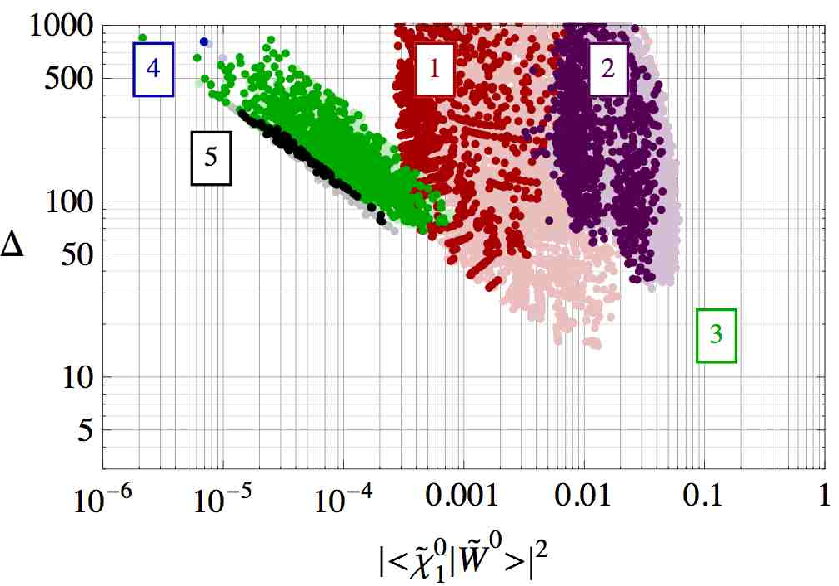}}
\end{tabular}
\begin{tabular}{c} 
\subfloat[Higgs mass unrestricted]{\includegraphics[width=7cm]{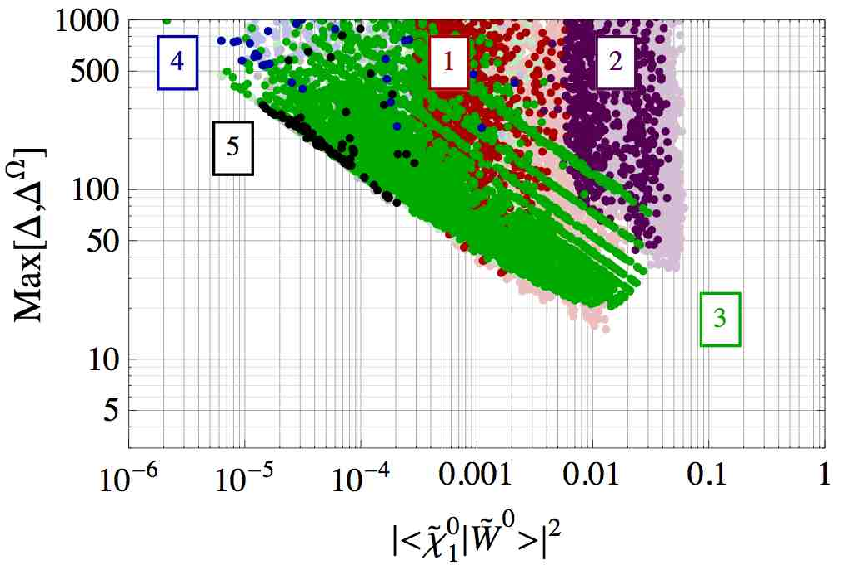}}
\hspace{4mm}
\subfloat[$m_h^{} > 114.4\,$GeV]{
\includegraphics[width=7cm]{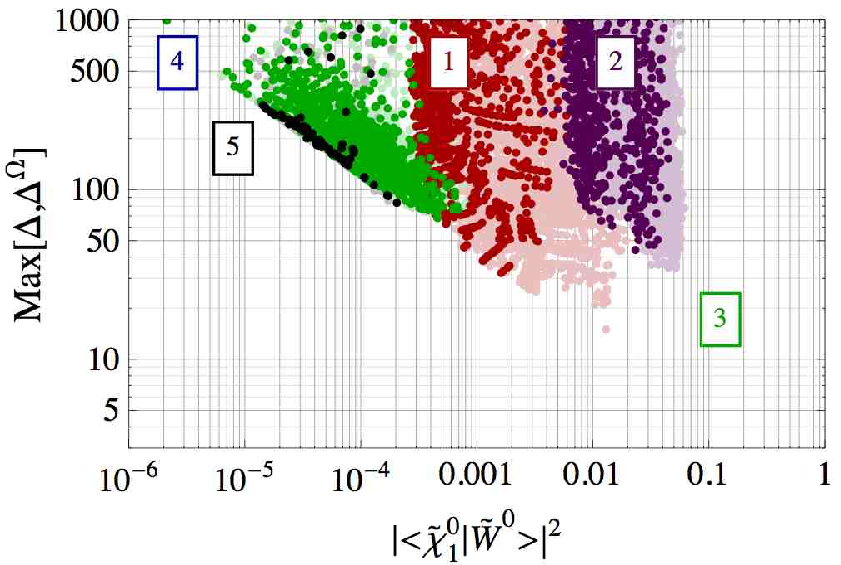}}
\end{tabular}
\begin{tabular}{c} 
\subfloat[Higgs mass unrestricted]{\includegraphics[width=7cm]{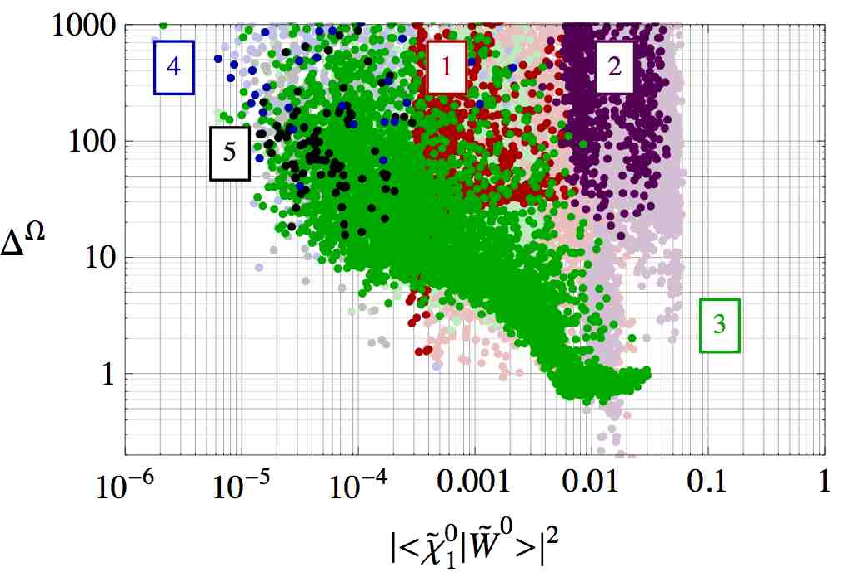}}
\hspace{4mm}
\subfloat[$m_h^{} > 114.4\,$GeV]{
\includegraphics[width=7cm]{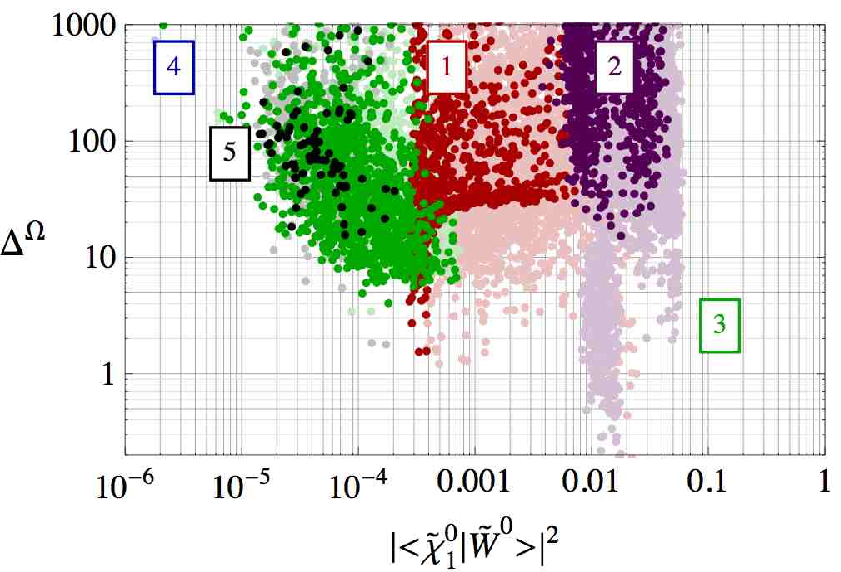}}
\end{tabular}
\caption{\protect\small 
Plots of fine tunings vs Wino overlap of LSP. The fine tuning of the electroweak scale is plotted in (a) and (b), the fine tuning of the relic density in (e) and (f), and the maximum fine tuning between the two in (c) and (d). The labels distinguish between the dominant LSP annihilation mechanisms. The $h^0~(H^0, A^0)$ resonance is significant for region 1 (5) with points coloured red (black). Region 3 (4) realises stau (stop) co-annihilation with points coloured green (blue), and region 2, coloured purple, has increased Higgsino components.
}
\label{ftwino}
\end{figure}

\pagebreak
\begin{figure}[!th]
\center
\begin{tabular}{c} 
\subfloat[Higgs mass unrestricted]{\includegraphics[width=7cm]{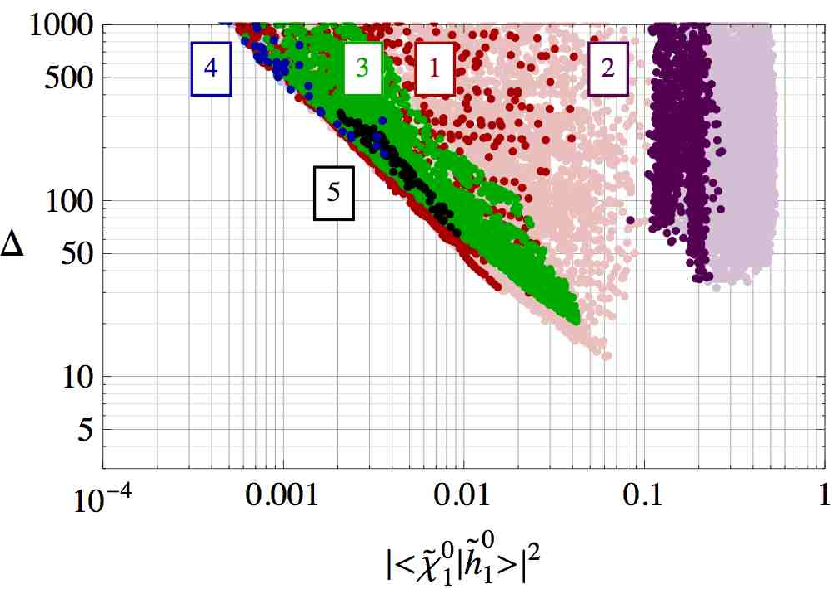}}
\hspace{4mm}
\subfloat[$m_h^{} > 114.4\,$GeV]{
\includegraphics[width=7cm]{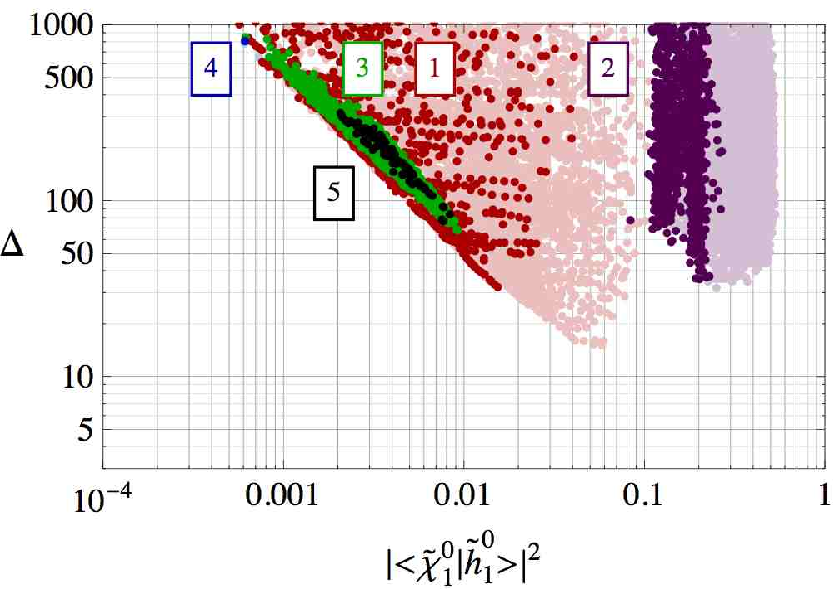}}
\end{tabular}
\begin{tabular}{c} 
\subfloat[Higgs mass unrestricted]{\includegraphics[width=7cm]{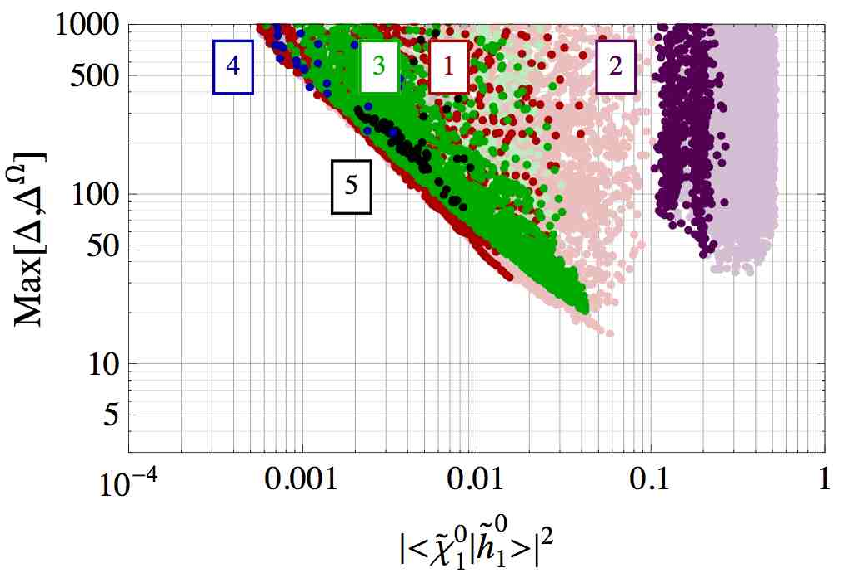}}
\hspace{4mm}
\subfloat[$m_h^{} > 114.4\,$GeV]{
\includegraphics[width=7cm]{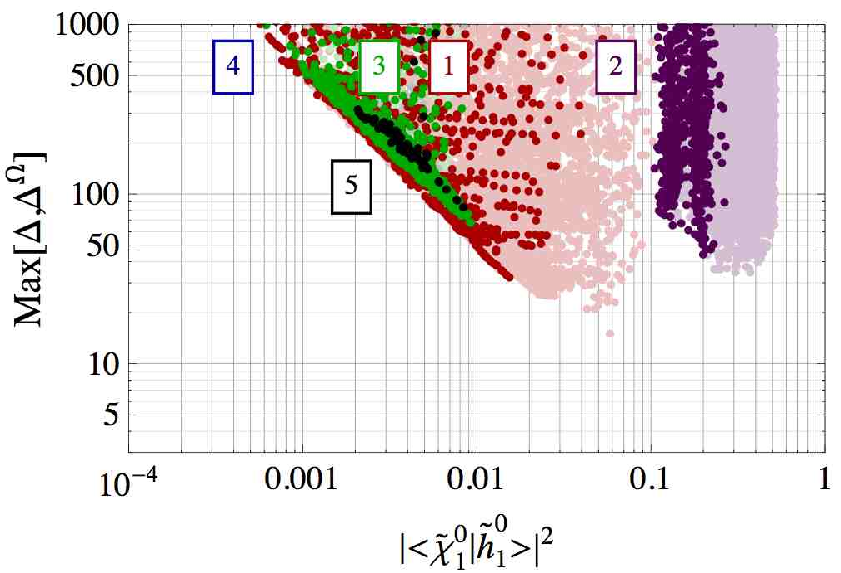}}
\end{tabular}
\begin{tabular}{c} 
\subfloat[Higgs mass unrestricted]{\includegraphics[width=7cm]{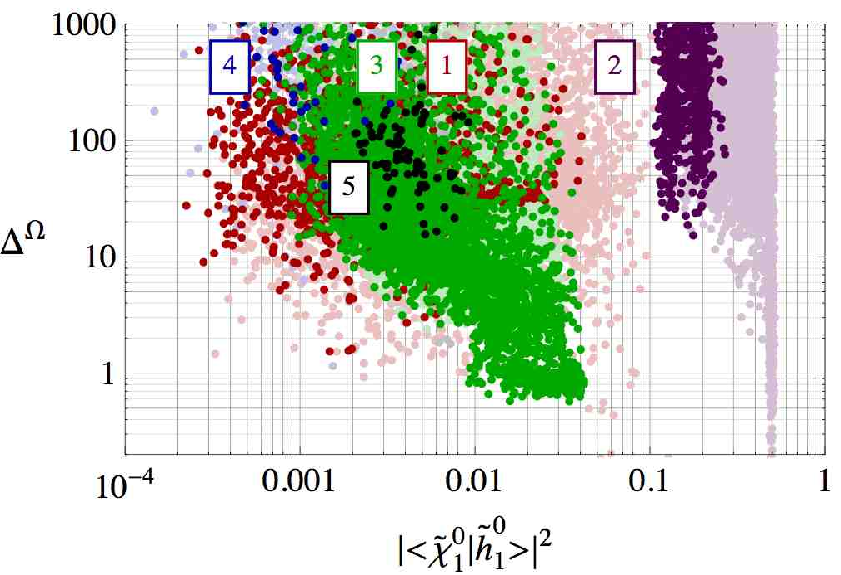}}
\hspace{4mm}
\subfloat[$m_h^{} > 114.4\,$GeV]{
\includegraphics[width=7cm]{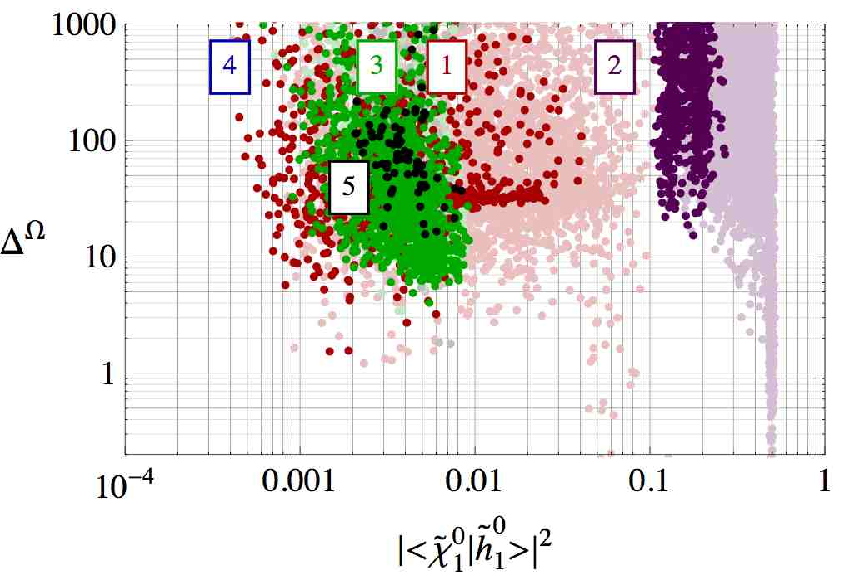}}
\end{tabular}
\caption{\protect\small 
Plots of fine tunings vs Higgsino (neutral down-type) overlap of LSP. The fine tuning of the electroweak scale is plotted in (a) and (b), the fine tuning of the relic density in (e) and (f), and the maximum fine tuning between the two in (c) and (d). The labels distinguish between the dominant LSP annihilation mechanisms. The $h^0~(H^0, A^0)$ resonance is significant for region 1 (5) with points coloured red (black). Region 3 (4) realises stau (stop) co-annihilation with points coloured green (blue), and region 2, coloured purple, has increased Higgsino components.
}
\label{fth1ino}
\end{figure}

\pagebreak
\begin{figure}[!th]
\center
\begin{tabular}{c} 
\subfloat[Higgs mass unrestricted]{\includegraphics[width=7cm]{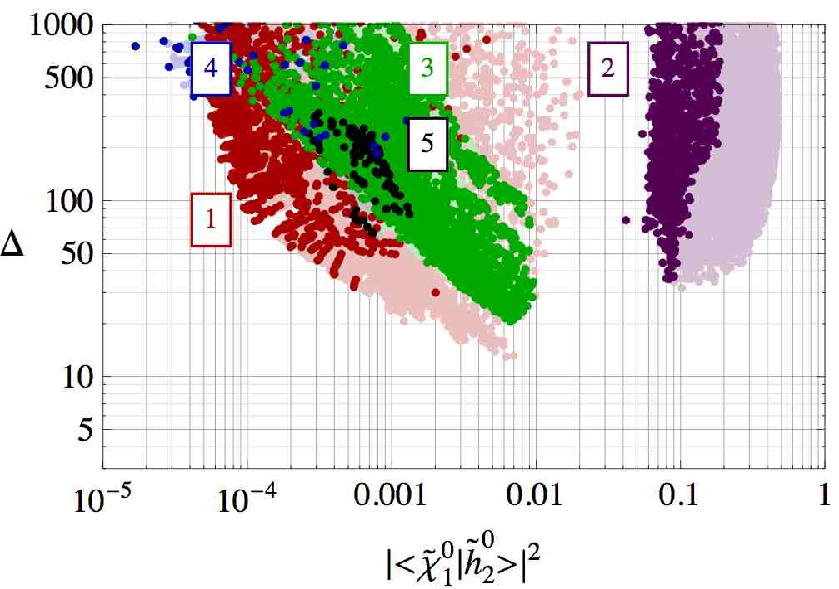}}
\hspace{4mm}
\subfloat[$m_h^{} > 114.4\,$GeV]{
\includegraphics[width=7cm]{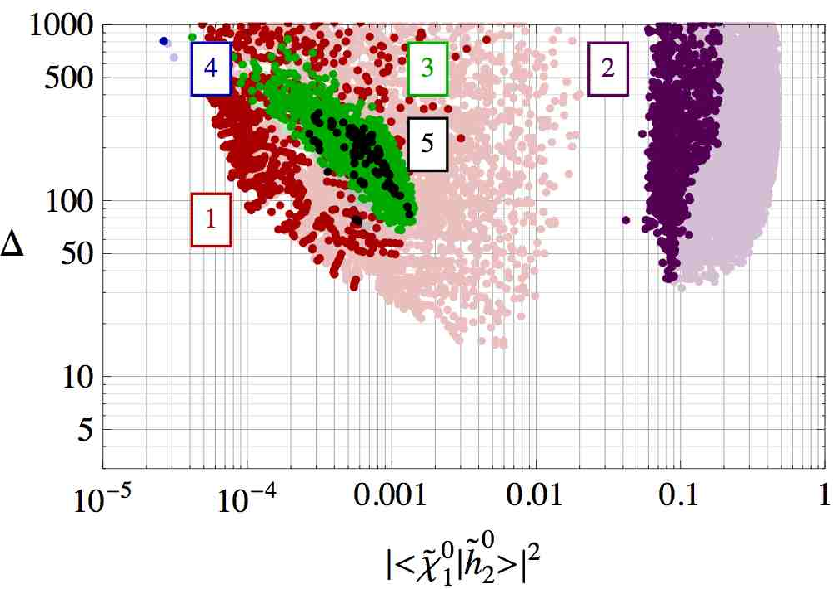}}
\end{tabular}
\begin{tabular}{c} 
\subfloat[Higgs mass unrestricted]{\includegraphics[width=7cm]{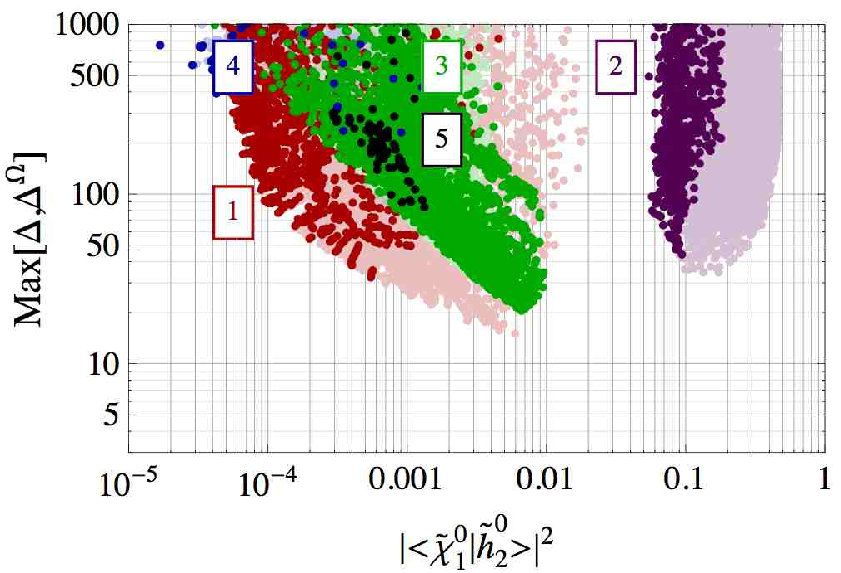}}
\hspace{4mm}
\subfloat[$m_h^{} > 114.4\,$GeV]{
\includegraphics[width=7cm]{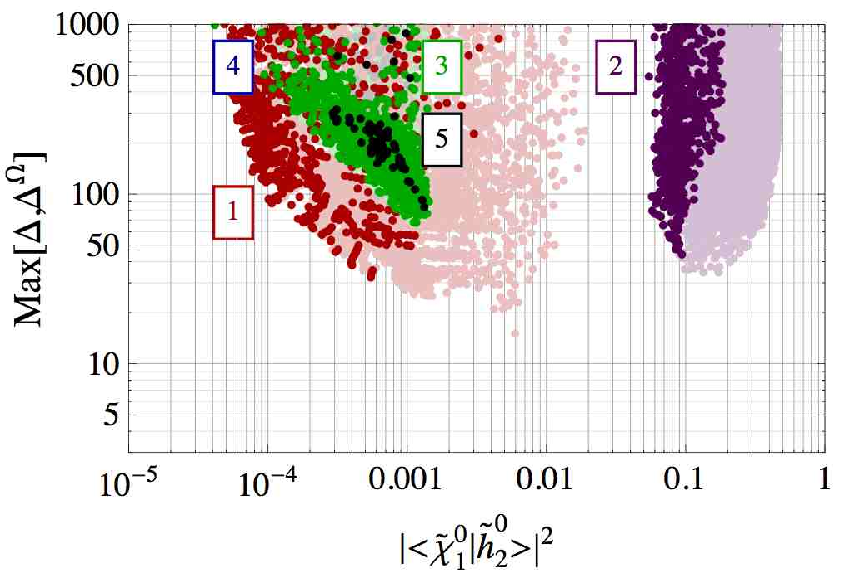}}
\end{tabular}
\begin{tabular}{c} 
\subfloat[Higgs mass unrestricted]{\includegraphics[width=7cm]{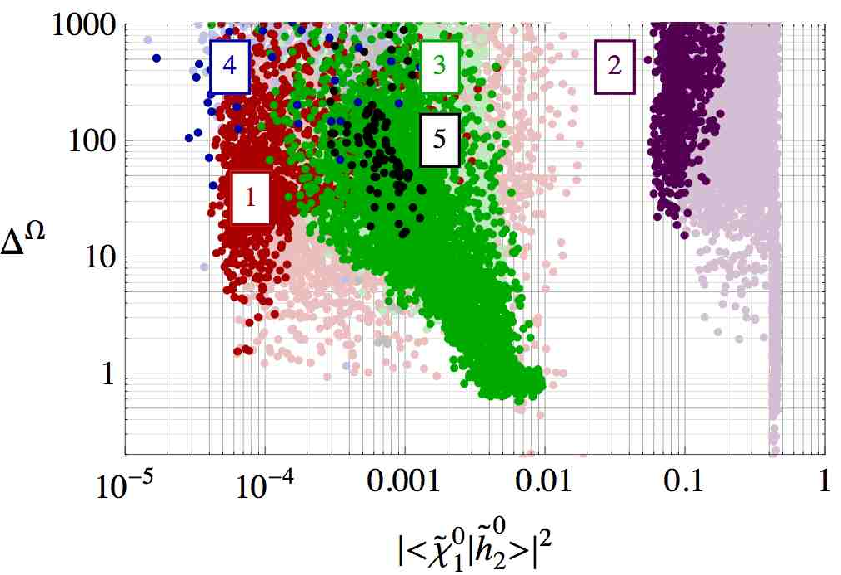}}
\hspace{4mm}
\subfloat[$m_h^{} > 114.4\,$GeV]{
\includegraphics[width=7cm]{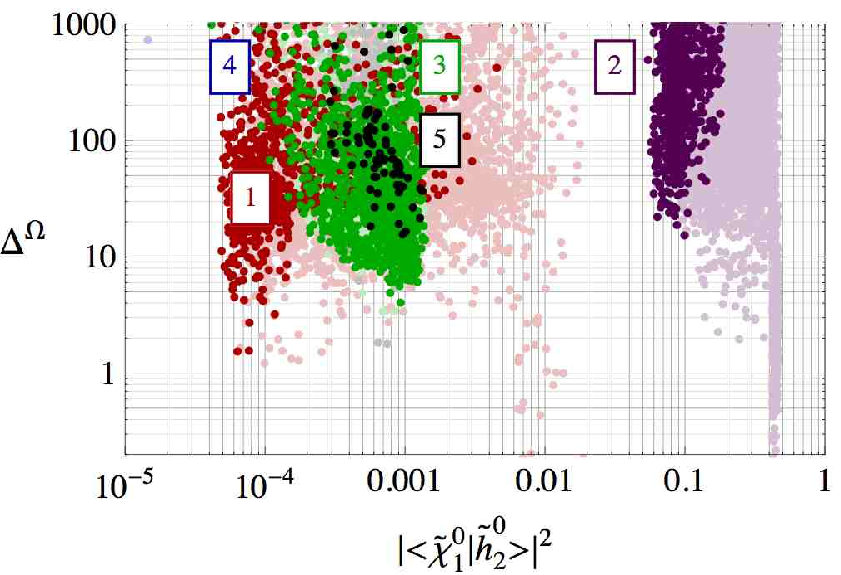}}
\end{tabular}
\caption{\protect\small 
Plots of fine tunings vs Higgsino (neutral up-type) overlap of LSP. The fine tuning of the electroweak scale is plotted in (a) and (b), the fine tuning of the relic density in (e) and (f), and the maximum fine tuning between the two in (c) and (d). The labels distinguish between the dominant LSP annihilation mechanisms. The $h^0~(H^0, A^0)$ resonance is significant for region 1 (5) with points coloured red (black). Region 3 (4) realises stau (stop) co-annihilation with points coloured green (blue), and region 2, coloured purple, has increased Higgsino components.
}
\label{fth2ino}
\end{figure}

\pagebreak
\noindent
This follows since for the points of smallest Bino component, the $m_{1/2}^{}$ is a factor of two larger than at the minimum of fine tuning. The heavier spectrum leads to increased fine tuning of the electroweak scale. Note that very natural relic densities can be achieved in the limit of small Bino component, even with the Higgs mass constraint.

Fig~\ref{ftwino} shows a strong dependence on the LSP's Wino composition against fine tuning. Fine tuning is reduced for a large Wino component. The strips observed in Region 3 (Stau coannihilation) is an artefact of the stepped $\tan \beta$ scan, with low $\tan \beta$ having larger fine tuning. The fine tuning dependence is similarly observed in Fig~\ref{fth1ino} and \ref{fth2ino} for the Higgsino components. For Region 1 ($h^0$ resonance), increased non-Bino components indicates moving into the focus point region, where the fine tuning is reached. For Regions 3 (Stau coannihilation), 4 (Stop coannihilation) and 5 ($H^0, A^0$ resonance), the smaller non-Bino components occur for large $m_{1/2}^{}$, where mixing is reduced. The heavy spectrum then leads to increased fine tunings.

\section{Summary}

The generation of a dark matter candidate for the mSUGRA model with electroweak interactions requires a consideration of the constraints from dark matter physics on the allowed parameter space of the model. As the dark matter candidate is mostly Bino-like in the majority of parameter space, with small annihilation cross sections, the requirement for a thermal relic density in agreement with the WMAP observations for the non-baryonic energy density strongly restricts the allowed parameter space.

The regions of remaining parameter space where increased annihilation cross sections were found could be distinguished according to the mechanism present that strengthened the annihilation. The mechanisms included resonance via the Higgs fields, $h^0, H^0, A^0$, coannihilation with the staus and stops and residing in the focus point region where significant Bino-Higgsino mixing is present.

The CDMS II constraint was found to further limit the allowed parameter, being most sensitive to the focus point region. The composition of the LSP was inspected, from which the phenomenology of the dark matter could be understood. The allowed mSUGRA parameter space was presented, demanding consistency with present day experimental limits, including the dark matter constraints, for a scan of all mSUGRA parameters.

The fine tunings were plotted and compared for determining the worst fine tuning of observables present in the model that are connected to the electroweak physics. The CDMS II constraint was found to be only significant below the LEP II Higgs limit. The minimum fine tuning with dark matter constraints was 1 part in 15 (for which relic density does not saturate the WMAP result and annihilation is via the $h^0$ resonance), with $m_h = 114.5\,$GeV. This is only a factor of two larger than that found when ignoring the dark matter constraints. In the case of requiring saturation of the non-baryonic matter by the neutralino LSP, and the Higgs mass constraint, the minimum fine tuning is 1 part in 20 with $m_h^{} = 115\,$GeV (again corresponding to annihilating close to the $h^0$ resonance). For points in the focus point, the minimum fine tuning of the electroweak scale is 1 part in 30, but for this point the relic density fine tuning is 1 part in 40, with $m_h^{} = 118\,$GeV. These results were found without application of the Higgs mass bound.

The dark matter constraints have slightly raised the minimum fine tuning of the electroweak scale present, but the naturalness limits ($\Delta < 100$) are not significantly changed for the allowed range for the mSUGRA parameters. The consideration of the fine tuning of the relic density is also frequently competitive with that of the electroweak scale for the lowest fine tuned points.

In the near future, different sections of the mSUGRA parameter space will be tested by direct dark matter searches and searches at the LHC. In the first run of the LHC which is due to end during winter 2011, an integrated luminosity of $1\,$fb$^{-1}$ will be collected at the centre of mass energy of $\sqrt{s} = 7\,$TeV. The gluino mass reach for the mSUGRA model during this period will extend to 620\,GeV \cite{Baer:2010tk}. This will be able to test the region where the LSP annihilates via the $h^0$ resonance (region 1) for which the spin independent cross for scattering of the LSP with a nucleon is up to five orders of magnitude below the current experimental limits. The gluino mass for region 1 points is $O(400\,$GeV), but the remaining labelled regions all have too heavy gluino masses for detection in the first run.

For discovery of the gluino, the kinematic edges observed in the distributions of its decay products such as charginos and neutralinos and the constrained ratios of the mSUGRA spectrum states can be used to demonstrate detection. However, the backgrounds are usually significant for the region 1 points and so rapid observation is not expected. A useful signal for heavier spectra is the presence of a strong dilepton peak from subsequent electroweak boson decay, but within region 1, the mass difference of kinematically accessible neutralinos and charginos is insufficient to produce on-shell electroweak bosons from cascade decays of the gluino.

In comparison, the direct dark matter detection experiments probe the region of mSUGRA parameter space close to the focus point where there is significant Bino-Higgsino mixing in the LSP (region 2). The points in this region have a gluino mass in excess of $750\,$GeV. The points with smallest fine tuning that satisfy the Higgs mass bound populate regions 1 and 2, and so these two complementary types of experiment provide separate tests of naturalness and either may be expected to soon discover natural SUSY physics.



\chapter{Improving naturalness through non-minimality}

The previous chapters have analysed the degree of fine tuning present in minimal supergravity with universal soft mass parameters. This was motivated to produce up to date and more precise results over a more complete parameter space scan than the analyses previously published. In addition, the minimal scenario was selected for its greater transparency in what features control the fine tuning measure. The motivations for low energy supersymmetry presented in the introduction did not however require a minimality condition. For an appreciation of how naturalness may be modified by non-minimal models, this chapter explores such scenarios. In particular, the features of models that allow smaller fine tuning with respect to mSUGRA will be identified.

One structure alluded to in previous discussion that may permit reduced fine tuning is alternative patterns in the soft mass parameters. For example, consider the separate contributions to the electroweak scale from the different soft terms in the MSSM. For $\tan \beta = 2.5$, \cite{Kane:1998im}:
\begin{eqnarray}
m_Z^2 /2 &=& -0.87 \left| \mu_0 \right|^2 + 3.6 \, M_3^2 - 0.12 \, M_2^2 + 0.007 \, M_1^2 \nonumber\\[2pt]
&& 
~ + 0.25 \, M_2^{} \, M_3^{} + 0.03 \, M_1^{} \, M_3^{} + 0.007 \, M_1^{} \, M_2^{} \nonumber\\[3pt]
&& 
~ -0.71 \, m_{H_2^{}}^2 + 0.19 \, m_{H_1^{}}^2 + 0.48 \left( m_Q^2 + m_U^2 \right) \nonumber\\[2pt]
&& 
~ -0.34 \, A_t^{} \, M_3^{} - 0.07 \, A_t^{} \, M_2^{} - 0.01 \, A_t^{} \, M_1^{} + 0.09 \, A_t^2
\end{eqnarray}
where all parameters in the above equation are defined at the UV scale of the theory. If all these parameters are independent, then fine tuning will be found to be large. However, if say the gaugino masses are fixed according to certain ratios which suppress their overall contribution to the electroweak scale, a gaugino focus point is found \cite{Kane:1998im, Horton:2009ed}. If a non-singlet unified gauge representation is responsible for SUSY breaking, non-universal but fixed ratio gaugino masses are generated that may lead to such a gaugino focus point.

An alternative approach for finding improved naturalness is through non-minimal particle content. This may relax the experimental constraints, and so open up less fine tuned points, help increase the annihilation cross section of the dark matter (or provide an alternative candidate) and/or increase the effective quartic coupling of the Higgs potential. The investigation of the analytical fine tuning results identified that increasing the effective quartic coupling significantly improves the fine tuning.

To improve naturalness with extra physics, complete model extensions can be considered, or if there is a scale separation between the MSSM and the new physics, an effective theory can be examined. The latter investigation is now considered for extensions of mSUGRA. Once the most efficient couplings for the higher dimension operators are ascertained, the models of new physics that can generate such operators will be inferred as being natural extensions.


An understanding of extensions of SUSY models that improve naturalness may become particularly useful if restrictive experimental limits force benchmark models such as mSUGRA into increasingly fine tuned regions (for example, if an exclusion limit of $m_h^{} > 121\,$GeV is found). The premise of natural supersymmetry would then suggest specific new degrees of freedom are necessary to preserve the viability of the principle of naturalness.

For model building beyond the MSSM, there are many directions that can be followed. The next-to-minimal supersymmetric model (NMSSM) \cite{Dermisek:2007yt} has 
one extra chiral singlet, but an arbitrary number of superfields could be added with appropriate quantum numbers and Lagrangian parameters to survive experimental constraints. In the following analysis, the model-independent approach of using an effective theory with higher dimensional operators \cite{Barbieri:1999tm,Piriz:1997id,Polonsky:2000zt,M0,Dine,Antoniadis:2008es,Antoniadis:2009rn} is adopted in order to avoid needing to specify a model.




\section{Higher dimensional operators of the MSSM Higgs sector} 
\label{s1} 
 
In this section the effective operators of dimension $d=5,6$ are listed that 
can be present in the Higgs sector consistent with the symmetries of the 
MSSM. These operators parametrise new physics beyond the MSSM and affect the 
Higgs scalar potential. Therefore they also affect the amount of fine tuning 
of the electroweak scale, as discussed in detail in the next section. The 
($R-$parity conserving) $d=5$ operators in the MSSM Higgs sector are: 
\medskip  
\begin{eqnarray} 
\mathcal{L}_{1} &=&\frac{1}{M_{\ast }}\int d^{2}\theta \,\lambda 
(S)\,(H_{1}\,H_{2})^{2},  \label{dimensionfive} \\[0.06in] 
\mathcal{L}_{2} &=&\frac{1}{M_{\ast }}\int d^{4}\theta \,\,%
\,A(S,S^{\dagger })D^{\alpha }\Big[B(S,S^{\dagger })\,H_{2}\,e^{-V_{1}}\Big]%
D_{\alpha }\Big[C(S,S^{\dagger })\,e^{V_{1}}\,H_{1}\Big] 
+\hc  
\end{eqnarray}

\medskip\noindent
where $S$ is the spurion field, $S=\theta \theta \,m_{0}$,
$A(S,S^{\dagger })$, $B(S,S^{\dagger })$, $C(S,S^{\dagger })$ are 
polynomials in $S,S^{\dagger }$ 
 and $m_{0}$ is the soft scalar mass term in the visible 
sector. As will be demonstrated 
in Section \ref{origin} the first operator can be generated, for example, by 
integrating out massive gauge singlets or $SU(2)$ triplets, while the second 
is easily generated by integrating out a pair of massive Higgs
doublets  \cite{Antoniadis:2008es}, all of mass of order $M_{\ast }$.

In \cite{Antoniadis:2008es,Antoniadis:2009rn}
 it was shown that by using general 
field redefinitions one can remove $\mathcal{L}_{2}$ from the action. The 
effect of this is an overall renormalisation of the soft terms and of the $%
\mu $ term. Since the fine tuning measure includes the fine tuning 
with respect to each of these soft operators separately, adding $\cL_{2}$
cannot reduce the overall fine tuning. For this reason 
only $\cL_{1}$ will be included in the discussion of fine tuning with $d=5$
operators. 

The $\mathcal{L}_{1}$ term generates the following contributions to the Higgs scalar potential on integrating out the auxiliary fields,
\begin{eqnarray}
V_1^{} &=& \zeta_1^{} h_1^{} h_2^{} \left( \left| h_1^{} \right|^2 + \left| h_2^{} \right|^2 \right) + \frac12 \,\zeta_2^{} \left(  h_1^{} h_2^{} \right)^2 + \hc
\end{eqnarray}
where $\zeta_1^{} = 2 \lambda \mu_0^{} / M_\ast$, and $\zeta_2^{} = - 2 \lambda^\prime m_0^{} / M_\ast$ given that $\lambda(S) \equiv \lambda + \lambda^\prime S$. In the two Higgs doublet notation, $\lambda_5^{} = \zeta_1^{}$ and $\lambda_{6,7}^{} = \zeta_2^{}$. At tree level in the MSSM, $\lambda_{5,6,7}^{} = 0$.

There are also $d=6$ operators that can be present in addition to the MSSM 
Higgs sector. These are suppressed relative to the $d=5$ operators by the 
factor $1/M_{\ast }$. However they may give contributions to the 
Higgs potential enhanced by $\tan \beta $  relative to the $d=5$
 so cannot be ignored at very large $\tan \beta $. The 
list of $d=6$ operators  is  (see also \cite{Piriz:1997id,Polonsky:2000zt}): 
\medskip  
\begin{eqnarray} 
\mathcal{O}_{i} &=&\frac{1}{M_{\ast }^{2}}\int d^{4}\theta \,\,\mathcal{Z}%
_{i}(S,S^{\dagger })\,\,(H_{i}^{\dagger }\,e^{V_{i}}\,H_{i})^{2},\qquad 
i=1,2.\qquad   \nonumber \\ 
\mathcal{O}_{3} &=&\frac{1}{M_{\ast }^{2}}\int d^{4}\theta \,\,\mathcal{Z}%
_{3}(S,S^{\dagger })\,\,(H_{1}^{\dagger 
}\,e^{V_{1}}\,H_{1})\,(H_{2}^{\dagger }\,e^{V_{2}}\,H_{2}),\qquad  
\label{dim6} 
\end{eqnarray}

\medskip\noindent
The above operators can be generated by integrating a massive $U(1)$ gauge boson or
a $SU(2)$ triplet.
\medskip
\begin{eqnarray} 
\mathcal{O}_{4} &=&\frac{1}{M_{\ast }^{2}}\int d^{4}\theta \,\,\mathcal{Z}%
_{4}(S,S^{\dagger })\,\,(H_{2}\,H_{1})\,(H_{2}\,H_{1})^{\dagger },\quad 
\qquad \qquad   \nonumber \\
\mathcal{O}_{5} &=&\frac{1}{M_{\ast }^{2}}\int d^{4}\theta \,\,\mathcal{Z}%
_{5}(S,S^{\dagger })\,\,(H_{1}^{\dagger 
}\,e^{V_{1}}\,H_{1})\,(H_{2}\,H_{1}+h.c.)  \nonumber \\
\mathcal{O}_{6} &=&\frac{1}{M_{\ast }^{2}}\int d^{4}\theta \,\,\mathcal{Z}%
_{6}(S,S^{\dagger })\,\,(H_{2}^{\dagger 
}\,e^{V_{2}}\,H_{2})\,(H_{2}\,H_{1}+h.c.)  \nonumber \\ 
\mathcal{O}_{7} &=&\frac{1}{M_{\ast }^{2}}\int d^{2}\theta \,\,\mathcal{Z}%
_{7}(S,0)\,\,W^{\alpha }\,W_{\alpha }\,(H_{2}\,H_{1})+h.c.,  \nonumber 
\\
\mathcal{O}_{8} &=&\frac{1}{M_{\ast }^{2}}\int d^{4}\theta
 \,\,\Big[\mathcal{Z}_{8}(0,S^{\dagger })\,\,(H_{2}\,H_{1})^{2}+h.c.\Big] 
\end{eqnarray}%

\medskip\noindent
where $W_{\alpha }$ is the supersymmetric field strength of a vector 
superfield of the SM gauge group. As an example, $\cO_4$ can be generated by integrating a gauge singlet. 
\medskip  
\begin{eqnarray} 
\mathcal{O}_{9} &=&\frac{1}{M_{\ast }^{2}}\int d^{4}\theta \,\,\mathcal{Z} 
_{9}(S,S^{\dagger })\,\,H_{1}^{\dagger }\,\overline{\nabla }
^{2}\,e^{V_{1}}\,\nabla ^{2}\,H_{1}  \nonumber \\ 
\mathcal{O}_{10} &=&\frac{1}{M_{\ast }^{2}}\int d^{4}\theta \,\,\mathcal{Z}
_{10}(S,S^{\dagger })\,\,H_{2}^{\dagger }\,\overline{\nabla } 
^{2}\,e^{V_{2}}\,\nabla ^{2}\,H_{2}  \nonumber \\ 
\mathcal{O}_{11} &=&\frac{1}{M_{\ast }^{2}}\int d^{4}\theta \,\,\mathcal{Z}
_{11}(S,S^{\dagger })\,\,H_{1}^{\dagger }\,e^{V_{1}}\,\nabla^a\,W^a\,H_{1}  \nonumber \\ 
\mathcal{O}_{12} &=&\frac{1}{M_{\ast }^{2}}\int d^{4}\theta \,\,\mathcal{Z} 
_{12}(S,S^{\dagger })\,\,H_{2}^{\dagger }\,e^{V_{2}}\,\nabla ^{\alpha 
}\,W_{\alpha }\,H_{2}  \label{der} 
\end{eqnarray}

\medskip\noindent
where $\nabla^a$ acts on everything to the right and $\nabla 
_{\alpha }\,H_{i}=e^{-V_{i}}\,D_{\alpha }\,e^{V_{i}} H_i$. In addition to 
the spurion dependence in the wavefunctions $\mathcal{Z}_{i}(S,S^{\dagger })$, extra $(S,S^{\dagger })$ dependence (not shown) can be present under each 
derivative $\nabla _{\alpha }$ in eq.(\ref{der}), in order to ensure the most 
general supersymmetry breaking contribution associated to these operators. 
One may use the equations of motion to replace the operators 
involving  extra derivatives by non-derivative ones\footnote{
Setting higher derivative operators onshell is a subtle issue in this case.
One can also use general 
spurion-dependent field redefinitions to ``gauge 
away'' (some of) these operators,  using the method
 of~\cite{Antoniadis:2008es,Antoniadis:2009rn}.}.
 Note that when computing the fine tuning 
measure eliminating a particular operator will lead to correlations 
between the remaining operators that, strictly, should be taken into 
account. 

As the determination of fine tuning depends on using RGEs between the gauge unification scale and the electroweak scale, one cannot consistently consider all dimension six operators when $M_\ast^{}$ is between these scales. However, this is the case of interest in order to find significant effects on electroweak physics. This limitation arises because the physical degrees of freedom that have been integrated out to generate these operators must be accounted for in the running when above their threshold scales. For this reason, numerical results of fine tuning in cases where dimension six operators are important are not presented. As states that do not significantly affect the running of MSSM states, such as gauge singlets, can generate dimension five operators, reliable fine tuning results in the presence of dimension five operators can still be determined.

\subsection{Validity of effective theory} 
 
The operator analysis used here has a limited range of validity 
because it corresponds to integrating out new heavy degrees of freedom. If 
the mass of these degrees of freedom is not much above the energies being 
probed, the operator analysis breaks down and one must deal with the new 
degrees of freedom directly. The mass, $M_{\ast },$ at which this 
happens corresponds to the point where high dimension operators are not 
suppressed relative to low dimension operators. A measure of this may be 
obtained by dimensional analysis in which the operator matrix elements are 
taken to be determined by the energy scale being probed. Applied here, this 
implies that the operator analysis is reliable provided 
$m_{0},\mu \ll M_{\ast }^{}$. 
 
A potential fault in this  estimate of  the range of 
convergence occurs because higher dimension operators 
may have anomalously large 
matrix elements. An example of this occurs for the dimension-six operators 
listed in Section~\ref{s1}. Consider the first dimension-six operator 
in eq.(\ref{dim6}) 
\medskip
\begin{equation} 
\mathcal{O}_{2}\supset \frac{c_{1}}{M_{\ast }^{2}}\Big[\,S^{\dagger 
}S(H_{2}^{\dagger }\,e^V\,H_{2})^{2}\Big]_{D}\supset 
\frac{c_{1}\,m_{0}^{2}}{M_{\ast }^{2}}\,|h_{2}|^{4}
\label{six} 
\end{equation}
for $c_1^{} \sim O(1)$.
This should be compared to the leading quartic Higgs term coming 
from the dimension-five operators in eq~(\ref{dimensionfive})
 that contributes at 
 $\cO\left( \frac{2\mu _{0}}{M_{\ast }}\left\vert h_{2}\right\vert 
 ^{2}h_{1}h_{2}\right)$.
One may see that the relative magnitude of 
the dimension-six to dimension-five contributions is 
$\cO\left( \frac{m_{0}^{2}}{2 \mu _{0}M_{\ast }}\tan \beta \right) .$
 Thus, strictly, 
the region of validity of the dimension-five operator analysis is 
$\frac{m_{0}^{2}}{2\mu _{0}M_{\ast }}\tan \beta \ll 1.$
However, as discussed in the next section, the new physics
 generating this dimension-six 
operator is different from that generating the dimension-five operator and 
so their coefficients should be uncorrelated. In this case the addition of 
higher dimension operators can reduce the fine tuning for some region in 
parameter space
 so that the analysis with dimension-five 
operators only will provide a useful upper bound even in regions where 
dimension-six contributions are significant.
For this reason the region of validity of the dimension-five operators 
analysis is better described by the original $m_0/M_*\ll 1$ and
$\mu_0/M_*\ll 1$ condition.
This keeps the corrections coming from operators with correlated 
coefficients small. 
 
In the following quantitative analysis, only the effects of
dimension-five operators are included. The convergence criterion found above
gave $m_0/M_*\ll 1$ and $\mu_0/M_*\ll 1$. These bounds are comfortably satisfied in the following
when the restriction, $m_0/M_*, \,\mu_0/M_*\leq 0.035$ is made,
giving upper values $\zeta_{1,2}\leq 0.07$.

\subsection{Fine tuning results from parameter space scan}\label{numerics}

The fine tuning of the electroweak scale is considered here, using the analytic formulae of Chapter~2 for the fine tuning measure, Higgs potential and associated spectrum. The dark matter physics in the presence of higher dimension operators is not considered as the increased set of relevant dimension five operators for LSP annihilation and LSP-nucleon scattering significantly modifies the allowed parameter space for which an analysis is beyond the scope of this discussion. One of the key results of this section is a demonstration of how the dependence of fine tuning tuning on Higgs mass is modified by non-minimal physics. In the mSUGRA case, Higgs masses of greater than 121\,GeV required $\Delta > 100$. This value is quite close to the current Higgs mass bound. The following results indicate that non-minimal physics permit larger Higgs masses with smaller fine tunings.

\begin{figure}[t] 
\begin{tabular}{cc|cr|} 
\parbox{7cm}{\psfig{figure=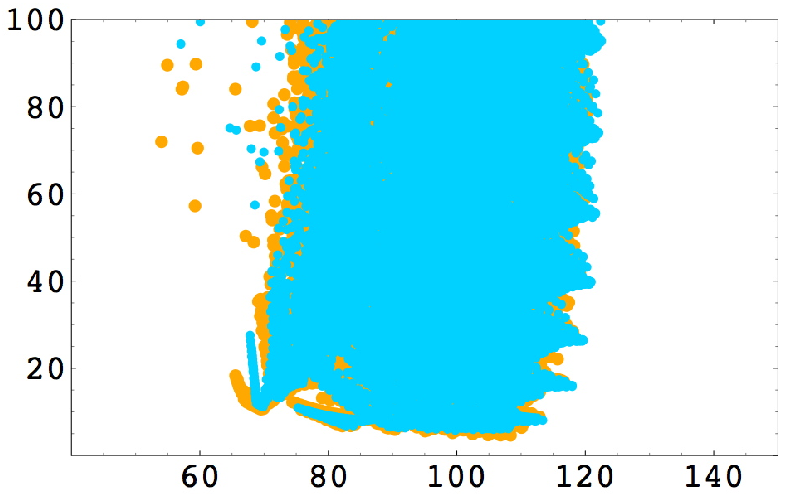,  
height=5.cm,width=6.7cm}} \hspace{0.5cm}  
\parbox{7cm}{\psfig{figure=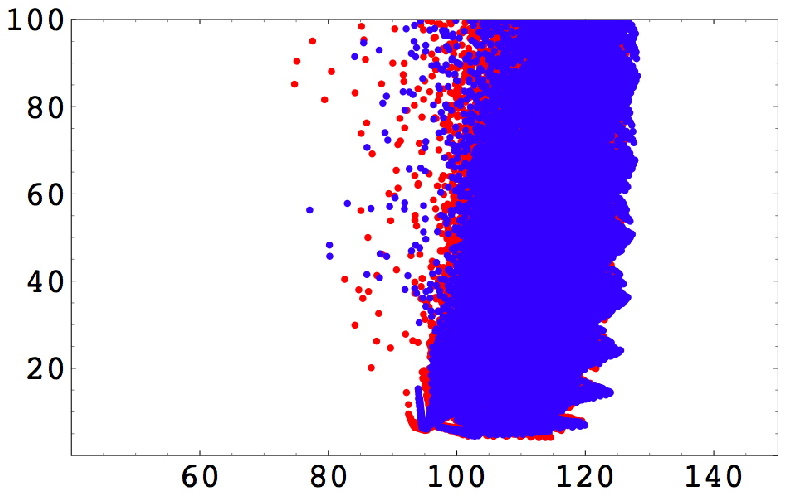, 
height=5.cm,width=6.7cm}}
\end{tabular}%
\def\baselinestretch{1.1}
\caption{{Left figure (a): the MSSM fine tuning
    $\Delta$ as a function of $m_{h}$;
 Right figure (b): the fine tuning in the MSSM with $d=5$ 
operators in terms of $m_{h}$, with 
$\zeta _{1}\!=\zeta_{2}\!=\!0.03$.
 In both figures, the top pole mass considered is $m_{t}=174$ GeV 
for blue (dark blue) 
areas and $m_{t}=171.2$ for yellow (red) areas, respectively. 
Larger $m_{t}$ 
input (blue) shifts the plots towards higher $m_{h}$ by 2-5 GeV. In both 
figures the parameters space scanned 
is: $1.5\leq \tan \protect\beta \leq 10$,
 $50\,\mathrm{GeV}\!\leq m_{0},m_{12}\leq\! 1$ TeV, $-10\leq A_{0}^{} / m_{0}^{} \leq 10$, without application of the experimental constraints. }} 
\label{fig1ab} 
\end{figure} 
\begin{figure}[th] 
\begin{tabular}{cc|cr|} 
\parbox{7.cm}{\psfig{figure=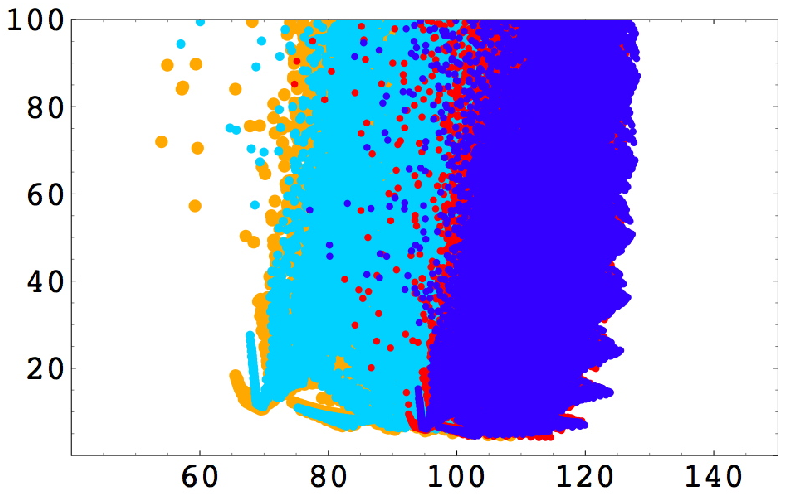, 
height=5.cm,width=6.7cm}} \hspace{0.5cm}  
\parbox{7.cm}{\psfig{figure=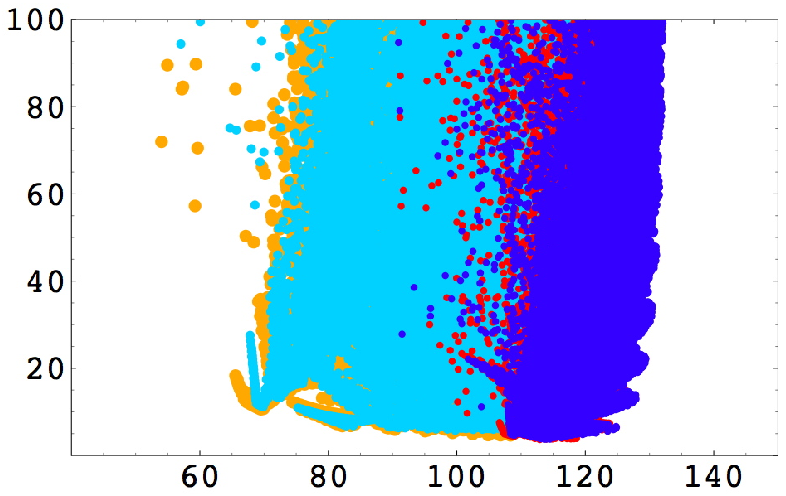, 
    height=5.cm,width=6.7cm}} 
\end{tabular}
\def\baselinestretch{1.1}
\caption{{Left figure (a): the fine tuning $\Delta $ as a
function of $m_{h}$. 
$\Delta $ of MSSM is plotted in light blue ($m_{t}=174$ GeV) with an orange 
edge ($m_{t}=171.2$); $\Delta $ of MSSM with d=5 operators with $\protect%
\zeta _{1}=\protect\zeta _{2}=0.03$ is plotted in dark blue ($m_{t}=174$ 
GeV) with a red edge ($m_{t}=171.2$). Right figure (b): similar to figure 
(a) but with $\protect\zeta _{1}=\protect\zeta _{2}=0.05$. Non-zero or 
larger $\protect\zeta _{i}$ (dark blue and red areas) shift the plots to 
higher $m_{h}$ to allow a reduced $\Delta $ for higher $m_{h}$. In both 
figures $1.5\leq \tan \protect\beta \leq 10$; $50\,\mathrm{GeV}\leq 
m_{0},m_{12}\leq 1$ TeV,  
$-10\leq A_{0}^{} / m_{0}^{} \leq 10$, without application of the experimental constraints.}} 
\label{fig2ab} 
\end{figure} 

In a similar approach to the previous results, a parameter space scan is performed and the resulting fine tuning values plotted for the points tested. Due to the interest in identifying relative changes and general features introduced by higher dimension operators, the experimental constraints are ignored here. This also avoids complications that extra higher dimension operators would affect the experimental constraints.
 
To ensure that the calculations are appropriate (ie neglected contributions are insignificant at the level of precision considered), the range of $\tan\beta$ is restricted to low values between 1.5 and 10. The Higgs potential is also only evaluated at one loop leading log order. The parameter space scan is further defined by, $50\,\mathrm{GeV}\leq m_{0},m_{12}\leq 1$ TeV, and the trilinear coupling is varied according to, $-10\leq A_{0}^{} / m_0^{} \leq 10$. The signs of $\zeta_{1,2}^{}$ are always chosen such that fine tuning is reduced with respect to the mSUGRA case.

The results are shown in Figures~\ref{fig1ab} to \ref{fig4ab}. Note that in these
 figures the structure apparent at small $\Delta$ and large $m_h$  is a scanning artefact due to discrete steps in parameter space. As a benchmark,
 Figure~\ref{fig1ab}(a) shows the fine tuning of the electroweak scale, $\Delta$, in mSUGRA given the method used in this section. It was observed in Chapter~2 that the right hand edge of this plot is insensitive to the experimental constraints. The same dependence of fine tuning on Higgs mass is then reproduced.
 
Figure~\ref{fig1ab}(b) shows $\Delta $ for the case 
of the MSSM with dimension-five operators added, with $\zeta _{1}=\zeta 
_{2}=0.03$. The dominant effects in Figure~\ref{fig1ab}(b) are mostly due to 
the effect of non-zero $\zeta _{1}$, which comes from the 
supersymmetric part of the higher dimensional operator. One may see a 
systematic shift of the allowed region to higher $m_{h}$  which 
(for positive $\zeta _{i})$  is driven by an increase in the 
quartic Higgs coupling which appears in the denominator of the fine tuning 
measure. 
The overall result is that the 
minimum amount of fine-tuning $\Delta $ in the 
presence of $d=5$ effective operators 
is small, of order $\Delta \approx 6$, for $m_{h}$ from 
$95$ to $119$ GeV. Therefore non-zero $\zeta _{i}$ can accommodate larger $%
m_{h}$ while keeping a $\Delta $ significantly smaller than in the MSSM, for a similar value of $m_h$. To 
illustrate the change more directly both plots are superposed in 
Figure~\ref{fig2ab}(a). The effect is enhanced for larger operator
coefficients, as 
may be seen in Figure~\ref{fig2ab}(b), where $\Delta $ is presented for
$\zeta_{1}=\zeta _{2}=0.05$, again shown relative to the MSSM case. One can see 
that in this case, values of $m_{h}>114.4$ GeV can have a low $\Delta \approx 
6$. Therefore $\Delta$ can be significantly reduced from 
the MSSM case, for a similar $m_{h}$. This conclusion is further supported 
by the plots in Figure~\ref{fig4ab} where other values for $%
\zeta _{i}$ are considered. From all plots shown, one sees that $\Delta <10$ 
is easily satisfied for values of the Higgs mass that can be as large
as $130$ GeV, depending on the exact values of $\zeta_{i}$. 

\begin{figure}[t] 
\begin{tabular}{cc|cr|} 
\parbox{7.cm}{ 
\psfig{figure=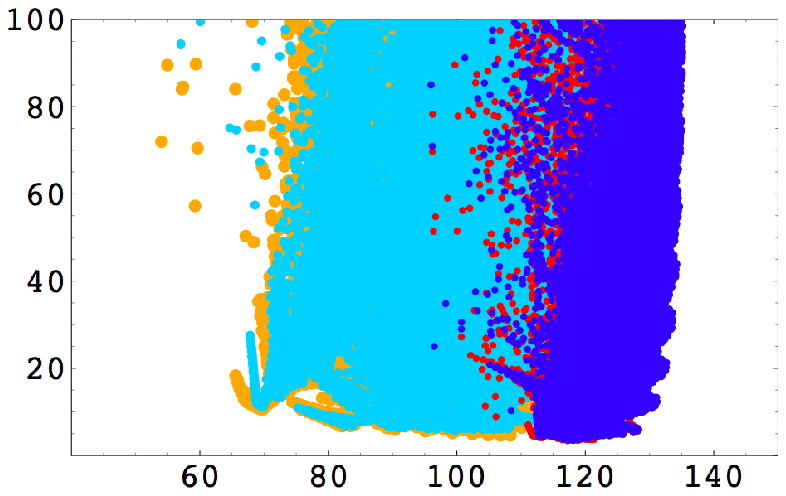,  
height=5.cm,width=6.7cm}} \hspace{0.5cm}  
\parbox{7.cm}{ 
\psfig{figure=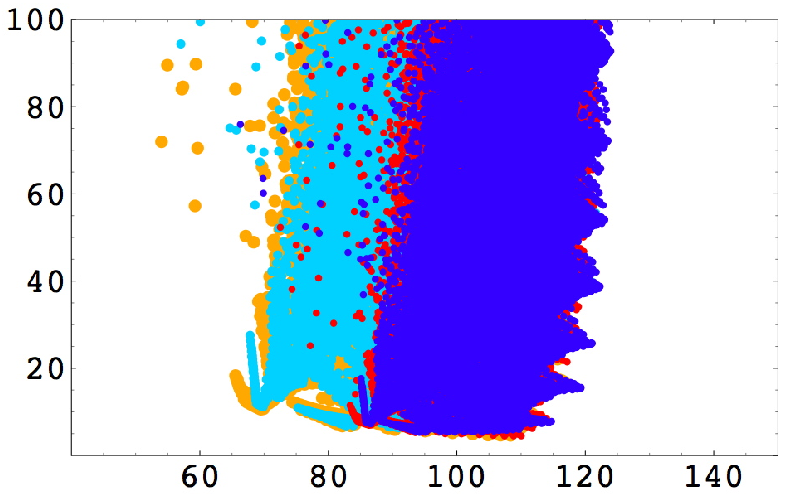, 
height=5.cm,width=6.7cm}} 
\end{tabular}
\def\baselinestretch{1.1}
\caption{{As in Fig.\protect\ref{fig2ab} but with: left 
figure (a): $\zeta_1=0.07$, $\zeta_2=0$; right figure (b):
 $\zeta_1=0$, $\zeta_2=0.1$}} 
\label{fig4ab} 
\end{figure} 
 
Note that, in the MSSM, 
$\Delta$ increases for low $\tan\beta$\,\, ($\ll 10$)
 and $m_h$ above the LEPII bound.
However, 
the effect of the $d=5$ operators is important 
for low $\tan\beta$ and  in their presence
$\Delta$ actually decreases for low $\tan\beta$. 
Thus the reduction in the fine tuning at very low
$\tan\beta$ relative to the MSSM case
is much more marked than that shown.

The lower amount of fine tuning in the presence of effective 
operators is due to two effects. The first, already mentioned, is the 
presence of additional quartic Higgs couplings enhancing the denominator 
which determines the Higgs via $v^{2}=-m^{2}/\lambda $  thus 
allowing for a smaller electroweak breaking scale. The second is the fact 
that higher dimensional operators add a tree level contribution to the Higgs 
mass, which reduces the need for large quantum contributions, and therefore 
the fine tuning. 
 
The scale of new physics that is generating this increase in Higgs mass and reduction of fine tuning is given by,
\begin{equation}
M_*\approx 2\mu_0/\zeta_1 \approx 
(40\,\,\,\mathrm{to}\,\,\,65) \times \mu_0,\qquad 
\zeta_{1,2}=0.05\,\,\,\mathrm{to}\,\,\,0.03
\end{equation}
For $\mu_0$  between the electroweak scale and 1 TeV, 
this shows that large values of $M_*$ are allowed: $M_*\approx (5.2\,\,%
\mathrm{to} \,\,8.45)$ TeV for $\mu_0=130$ GeV and $M_*\approx (8\,\,\mathrm{%
to}\,\,13)$ TeV for $\mu_0=200$ GeV. For larger $\mu_0$ one obtains
values of $M_*$ above the LHC reach. Finally, for $\zeta_1=0.07$ but with
$\zeta_2=0$, 
one has $M_*\approx 30\times \mu_0$ and  $\Delta<10$ for $m_h\approx 130$ GeV.
Thus, the
 EW fine tuning is small $\Delta<10$ for $114\leq m_h\leq 130$ GeV,
for rather conservative values of $\zeta_{1,2}$.
To relax these values, a rough rule is that an increase of $\zeta_1$
by $0.01$ increases  $m_h$ by $2$ to $4$ GeV for the same $\Delta$.

\section{The origin of ``new physics''}\label{origin}

The presence of a higher dimension operator signals new physics and 
it is important to ask what this new physics can be. In the context of new 
renormalisable interactions it may come from the effects of new chiral 
superfields or from new gauge vector superfields. Consider chiral 
superfields first. One may readily obtain the $d=5$  operator of 
eq.(\ref{dimensionfive}) by integrating out a gauge singlet or a
triplet 
\cite{Dine}. Consider the case of a massive gauge singlet 
$X$ with Lagrangian 
\medskip
\[ 
\mathcal{L}_{X}=\int d^{4}\theta \,X^{\dagger }X+\bigg\{\int d^{2}\theta \,%
\Big[\mu H_{1}H_{2}+ \lambda _{x}\,X\,H_{1}H_{2}+\frac{1}{2}\,M_{\ast }\,X^{2}%
\Big]+h.c.\bigg\}. 
\] 

\bigskip\noindent
For $M_{\ast }\gg \mu ,$ $m_{0}$, one may use the equation of 
motion to integrate out $X$, giving, 
\begin{eqnarray}
- \frac14 \, \bar{D}^2 X^\dagger + M X + W^\prime  + X \, W^{\prime \prime}  &=& 0
\end{eqnarray}
where $W = \lambda_x^{} X H_1^{} H_2^{} + \mu H_1^{} H_2^{}$ and $W^\prime = \partial W / \partial X$, and so on for $W^{\prime \prime}$ (which is zero in the above example). A series expansion of this equation of motion generates,
\begin{eqnarray}
X &=& - \frac{ W^\prime}{M} - \frac{\left( \bar{D}^2 \, W^{\prime \dagger} \right)}{4  \left| M  \right|^2} \,  + O\left( \frac{\Box}{M^3} \right)
\end{eqnarray}
which can be used to substitute into the original Lagrangian. This generates dimension five and six operators up to $O(M_\ast^{-2} )$ terms,
\begin{equation} 
\mathcal{L}_{X}^{effective}= \frac{\lambda_x^2}{M_\ast^2} \int d^4 \theta \left| H_1^{} H_2^{} \right|^2  +\left[  \frac{-\lambda _{x}^2}{2\,M_{\ast }}\int 
d^{2}\theta \,(H_{1}^{} H_{2}^{} )^{2}+h.c. \right] \label{ops} 
\end{equation} 
 
\medskip\noindent
The supersymmetry breaking terms associated with this 
operator are obtained by replacing $\lambda \rightarrow \lambda (S)$
giving the $d=5$ operator of interest.
Note that $\mathcal{L}_{X}$ has a similar form to that of 
the NMSSM. However, in the NMSSM, the singlet field has mass of order the 
electroweak breaking scale and cannot be integrated out whereas here, the singlet mass is assumed to be much larger than the EW scale. 
 
However, the origin of the $d=5$ operator cannot be 
uniquely ascribed to a gauge singlet field. Indeed it may equally well point 
to the existence of $SU(2)$ triplets \cite{Espinosa:1998re,Espinosa:1991gr,Espinosa:1991wt},
$T_{1,2,3}$, of hypercharge $\pm 1,0$. In this case a 
Lagrangian of the form 
\medskip  
\[ 
\mathcal{L}_{T}=\!\!\int \!d^{4}\theta \,\Big[T_{1}^{\dagger 
}e^{V}T_{1}+T_{2}^{\dagger }e^{V}T_{2}\Big]+\!\int d^{2}\theta \,\Big[\mu 
H_{1}H_{2}\!+\!M_{\ast }T_{1}T_{2}\!+\!\lambda 
_{1}H_{1}T_{1}H_{1}\!+\!\lambda _{2}H_{2}T_{2}H_{2}\Big]\!+\!h.c  
\]

\medskip\noindent
which generates an equivalent dimension five operator as in eq~(\ref{ops}) except that $\lambda _{x}^2$ is replaced by $\lambda _{1}\lambda _{2}$.
 More generally, one can generate the $d=5$
operator through a combination of both gauge singlets and triplets. 
However  note that the pure singlet $X$  case has the advantage of not 
affecting the gauge couplings unification (at one-loop), 
which is not true for the $SU(2)$ triplet. 
 
What about additional, massive, $SU(2)$ doublets that 
couple to the MSSM Higgs sector? One may readily show that integrating them 
out does not generate, to lowest order in $1/M_{\ast }$, a dimension five operator of the type in eq~(\ref{ops}).
 
There remains the possibility that the new physics is due to the 
effect of new massive vector gauge superfields. The simplest example is the 
case where there is a new $U(1)^{\prime }$ gauge symmetry under which 
the Higgs sector is charged. This brings extra quartic contributions to the 
scalar potential that are expected to reduce the fine-tuning 
\cite{Batra:2003nj,kaplan,Bellazzini:2009ix}. Assuming the
$U(1)^{\prime }$ 
 is broken at $M_{\ast }$  one obtains the effective 
Lagrangian to leading order in inverse powers of $M_{\ast }$
 given by
\medskip
\[ 
\mathcal{L}_{U(1)}^{effective}=-\frac{g^{^{\prime }2}}{M_{\ast }^{2}}\int 
d^{4}\theta \,\,\Big[q_{1}H_{1}^{\dagger }e^{V}H_{1}+q_{2}H_{2}^{\dagger 
}\,e^{V}H_{2}\Big]^{2} 
\]

\medskip\noindent
where $g^{\prime }$ is the $U(1)^{\prime }$
coupling and $q_{1,2}$ are the charges of the Higgses under 
$U(1)^{\prime }$ ($q_{1}+q_{2}=0$). Note that, after 
including the associated supersymmetry breaking operators, this corresponds 
to the $d=6$ effective operators \cite{Dine} of eq.(\ref{dim6}) and 
that no $d=5$ operators are generated. 

\medskip  
In summary, the requirement that the SUSY extension of the MSSM 
should not have significant fine tuning may indicate the presence of
the 
$d=5$  operator of eq.(\ref{dimensionfive}) which, in turn, suggests 
the presence of a massive gauge singlet and/or a $SU(2)$  triplet. 
This is the simplest interpretation based on new renormalisable interactions 
but other, more complicated possibilities to generate the 
$d=5$ operator may be 
possible.

\section{Further remarks on fine tuning}

Effective field theory approaches to the fine tuning 
of the electroweak scale were used before in models of low SUSY breaking
scale scenarios \cite{Casas:2003jx} where both $d=5$ and $d=6$
operators were included. The model in \cite{Casas:2003jx} introduces
supersymmetry breaking through coupling of MSSM states to a 
SM singlet field responsible for supersymmetry breaking.
After integrating this field out, in addition to the $d=5$ operator 
considered here, there are correlated contributions from the 
$d=6$ operators. Using this, the authors find the fine tuning can be
very small even for an arbitrarily high Higgs mass, provided the
scale of supersymmetry breaking is less than $500$ GeV.

How does this analysis relate to the one presented here?
 The examples given in  \cite{Casas:2003jx} are found varying the ratio $\tilde
 m/M$ in the range $0.05$ to $0.8$ where $\tilde m$ is the
 supersymmetry breaking scale and $M$ is the messenger mass.
For $\tilde m/M$ small, the fine tuning is close to that in the MSSM
 but reduces rapidly for $\tilde m/M$ large; in this latter case the
 fine tuning actually reduces as the Higgs mass increases. This range
of values for $\tilde m/M$ corresponds to a 
choice of our $m_0/M_*$ and $\mu_0/M_*$ in a similar range.
The upper value strongly violates the criterion for applicability of
the operator analysis argued here and is a factor of $\approx 10$ larger than the value
 chosen in Figure~\ref{fig4ab}(a).
Ignoring, for the moment, the fact that the contributions of
 higher dimension operators 
are expected to be large for this choice of mediator mass, it can be asked 
what this choice of mediator mass in this analysis would give for 
the Higgs mass consistent with small 
$\Delta$. Since the change in the upper bound on the Higgs mass
 roughly scales with the coefficient of the $d=5$ operator, this would allow a Higgs mass in the region of
 $276$ GeV, much larger than the earlier conservative estimates presented here.
 However, as has been stressed, for this value of the messenger mass the operator
analysis breaks down and one should do the analysis including the
 messenger fields explicitly.

\section{Summary} 
 \label{conclusions} 
 
The LEPII lower bound on the Higgs mass places MSSM Higgs physics at
the forefront  of supersymmetry phenomenology. While this bound can be 
satisfied by including the MSSM quantum corrections, it 
(re)introduces some amount of fine tuning in the model.
To reduce the fine tuning may require new physics beyond the MSSM 
which can be parametrised by higher dimensional operators.
In this chapter, an effective field 
theory framework with $d=5, 6$ operators is considered in the MSSM Higgs sector, 
giving a  model independent approach to the fine tuning problem.

Fine tuning proves to be  very sensitive to the addition of higher
dimensional operators and this is  mostly due to extra
corrections to the quartic couplings of the Higgs field.
For the case of dimension-five operators 
it was showed that one can maintain a reduced fine-tuning
 $\Delta<10$ for a Higgs mass above the LEPII bound and
 as large as $m_h\approx 130$ GeV, 
for  the parameter space considered, with low $\tan\beta$ ($\tan\beta<10$). 
 The scale of new physics $M_*$  responsible for the reduction in fine 
tuning can be rather large, for example $M_*\approx 2\mu_0/\zeta_1 \approx 
(40\,\mathrm{to}\,65) \times \mu_0$, for $\zeta_{1,2}=0.05\,\,\mathrm{to}
\,\,0.03 $, and $M_*\approx 30 \times \mu_0$ for $\zeta_1=0.07$, $\zeta_2=0$.
For values of $\mu_0$  between the electroweak  scale and 1 TeV, 
these results show that large values of $M_*$ are allowed; in the former case
$M_*\approx~(5.2\,\,\mathrm{to} \,\,8.45)$ TeV for $\mu_0=130$
 GeV and $M_*\approx (8\,\,\mathrm{to}\,\,13)$
 TeV for $\mu_0=200$ GeV. For larger $\mu_0$, larger
 values of $M_*$ are possible, even above the LHC reach.
These results follow from  rather conservative choices for
 the coefficients of the quartic couplings  induced by 
the dimension-five  operators, to ensure the convergence of the 
effective operator expansion.

The numerical analysis included the effect of dimension-five operators
only. These  give the leading corrections at low $\tan\beta$, being 
proportional to $1/M_*$. However, dimension-six operators, suppressed
by $m_0^2/M_*^2$ or $\mu_0^2/M_*^2$, give contributions that can be
 enhanced by large $\tan\beta$;
 for $(\tan\beta\,m_0)>M_*$ or $(\tan\beta\,\mu_0)>M_*$
these will be the leading terms. However, due to what would be an inconsistent application of renormalisation group equations, general dimension six operators cannot be reliably tested for fine tuning with this methodology. Depending on the physics that generates the dimension five operators, the results determined here may also not be appropriate.

Of course the crucial question is what is the origin of the physics
beyond the MSSM giving rise to these operators?
The dimension-five operator can be generated by a gauge singlet
superfield or a $SU(2)$ triplet superfield of mass of $\cO(M_*)$
 coupling to the Higgs sector. The dimension-six operators can be
generated, for example, by an extra gauge symmetry with a massive gauge
supermultiplet or additional (Higgs-like) $SU(2)$ doublet supermultiplets
of mass $\cO(M_*)$.

Returning to alternative methods for reducing naturalness, in addition to non-minimal particle content affecting the Higgs potential, experimental constraints may be evaded. Alternatively, the pattern of high energy soft SUSY breaking parameters may generate special features such as focus points at the electroweak scale. These are the main mechanisms for improving naturalness. If the natural parameter space of a minimal model is ruled out, such mechanisms should then seriously be considered, having initally motivated low energy supersymmetry using a naturalness principle.

\chapter{Conclusions}

Supersymmetry has been argued to be the most general spacetime symmetry consistent with a selection of assumed physical principles. It is then appealing to embed a model of nature with such a symmetry. Furthermore, there are theoretical motivations that lead to expecting the physics of supersymmetry to be realised around the TeV scale which is at our current experimental frontier. The motivations include providing a natural solution to the hierarchy problem of the electroweak and Planck scales and a dark matter candidate which may saturate the inferred non-baryonic matter energy density. A thorough understanding of these models may lead to imminent discoveries.

Having motivated supersymmetry by invoking a principle of naturalness, it is then vital to test the allowed regions of parameter space for the presence of a satisfactory degree of fine tuning. The parameter space can also have limits placed upon it for remaining natural. The phenomenology of the natural regions of parameter space can be identified, and search strategies optimised for testing the existence of natural theories.

The work presented here has identified the natural regions of minimal supergravity, provided naturalness limits on the parameter space and spectrum and discussed the phenomenology associated with the natural realisations. The analysis was completed using the latest experimental constraints, and tested the fine tuning of the electroweak scale at the 2 loop order for a scan over all parameters of the theory. The fine tuning of the thermal relic density, controlled by the electroweak interactions, were similarly tested and found to compete with the former fine tuning in the regions of lowest fine tuning.

A fine tuning of 1 part in 15 was found to be consistent with all current experimental constraints for the mSUGRA model. This is much smaller than the commonly repeated value of 1 part in 100 for MSSM theories. The low fine tuning occurs in the region around the scalar focus point, which is a special property obtained for a family of structures of the UV soft scalar masses, for which the universality condition within mSUGRA belongs. The MSSM spectrum around this low fine tuned region is similar to the benchmark point SPS2, for which the phenomenology has been widely investigated.

Methods for improving the naturalness found in mSUGRA through non-minimal models was also discussed. These included special patterns in the fundamental input parameters that lead to an insensitivity of the electroweak scale to these parameters, or non-minimal physics that allows evasion of the experimental constraints.

Non-minimal physics may also increase the effective quartic coupling of the Higgs potential simultaneously increasing the Higgs mass and reducing fine tuning. A model independent analysis was presented using an effective theory formalism to demonstrate that new physics beyond mSUGRA with an energy scale beyond the design reach of the LHC can still allow a Higgs mass of 130\,GeV with fine tuning less than 1 part in 10.

To conclude, a large group of natural supersymmetric theories remain experimentally viable which includes minimal supergravity. If the natural parameter space of a given model is fully excluded, this will motivate introducing further new physics or different UV completions. The phenomenology of natural minimal supergravity suggests that experiments will soon be in a position to discover its physics, allowing a new Standard Model to be written.




\appendix
\chapter{Renormalisation group equations}\label{appendix1}


Using the notation from eqs~(\ref{superpot}) and (\ref{lsoft}), the 1-loop renormalisation group equations (RGEs) are given by \cite{Martin:1993zk}:
\begin{eqnarray}
\frac{dg_a^{}}{dt} &=& \frac{g_a^3}{16\pi^2} \, \left[ \sum_R^{} S_a^{} (R) - 3\, C(G_a^{})  \right] \\
\frac{dM_a^{}}{dt} &=& \frac{g_a^2 M_a^{}}{8\pi^2} \left[ \sum_R^{} S_a^{} (R) - 3\, C(G_a^{})  \right] \\
\frac{dY^{ijk}}{dt} &=& \frac{Y^{ijp} }{16\pi^2} \left[ Y_{pmn}^{} Y^{kmn} - 2\, \delta_p^k \sum_a^{} g_a^2 \, C_a^{}(p)  \right]  + (k \leftrightarrow i) + (k \leftrightarrow j) \\
\frac{d\mu^{ij}}{dt} &=& \frac{\mu^{ip} }{16\pi^2} \left[ Y_{pmn}^{} Y^{jmn} - 2\, \delta_p^j \sum_a^{} g_a^2 \, C_a^{}(p)  \right]  + (j \leftrightarrow i) \\[6pt]
\frac{d \tilde{A}^{ijk}}{dt} &=& \frac{1}{16 \pi^2} \Big[ Y_{lmn}^{} \left( \frac12 \, \tilde{A}^{ijl} \, Y^{mnk} + Y_{ijl} \, \tilde{A}^{mnk}\right)
\nonumber\\
&& ~~
-\, 2 \sum_a^{} \left( \tilde{A}^{ijk} - 2 M_a^{} Y^{ijk} \right) g_a^2\,  C_a^{} (k) \Big]
+ (k \leftrightarrow i) + (k \leftrightarrow j) \\
\frac{d \tilde{B}^{ij}}{dt} &=& \frac{1}{16 \pi^2} \Big[  Y_{lmn}^{} \left( \frac12 \, \tilde{B}^{ij} \, Y^{mnk} + \frac12 \, Y^{ijl} \, \tilde{B}_{mn}^{} + \mu^{il} \tilde{A}^{mnj} \right)
\nonumber\\
&& 
~~
-\, 2 \sum_a^{} \left( \tilde{B}^{ij} - 2 M_a^{} \mu^{ijk} \right) g_a^2\,  C_a^{} (k) \Big]
+\, (j \leftrightarrow i) \\
\frac{d (m^2)_i^j}{dt} &=& \frac{1}{16\pi^2} \Big[ \frac12 \, Y^{pqn}  Y_{ipq}^{}  (m^2)_n^j + \frac12\, Y_{pqn}^{} Y^{jpq} (m^2)_i^n  + 2\, Y_{ipq}^{} Y^{jpr} (m^2)_r^q + \tilde{A}_{ipq}^{} \tilde{A}^{jpq} \nonumber\\
&&\hspace{10mm}
- 8\, \delta_i^j \sum_a^{} M_a^{} M_a^\dagger \, g_a^2 \, C_a^{}(i) + 2 \sum_a^{} g_a^2 \, (T_a^A)_i^j ~ \mbox{Tr} \left[ T_a^A m^2 \right] \Big]
\end{eqnarray}
where the MSSM state masses are assumed to be negligible, and a $SU(5)$ gauge normalisation is chosen. The $SU(5)$ normalised hypercharge, $Y^\prime$, is related to the hypercharge in Table~\ref{mssmRep}, $Y$, by $Y^\prime = Y \sqrt{\frac{3}{5}}$. To keep the Lagrangian invariant, the hypercharge gauge coupling must then also be rescaled, $g_{Y^\prime}^{} = g_{Y}^{} \sqrt{\frac53}$.

The group theoretic factors are given by:
\begin{table}[h]
\center
\begin{tabular}[c]{|rl|rl|}\hline
\multicolumn{2}{|c}{$SU(n)$} & \multicolumn{2}{|c|}{$U(1)_{Y}^{}$} \\[2pt] \hline
$C(G)~~=$ & $n$  \hspace{5mm} &  $C(G)~~=$ & $0$ \\[2pt]
$C_n^{}(\mathbf{n}~\mbox{or}~\mathbf{\bar{n}})~=$ & $\frac{n^2-1}{2n}$  & $C_1^{}\left( Y \right)~=$ & $Y^{2} $ \\[4pt]
$S_n^{}(\mathbf{n}~\mbox{or}~\mathbf{\bar{n}})~=$ & $\frac{1}{2n}$  & $S_1^{}\left( Y \right)~=$ & $Y^{2}$ \\[2pt]
\hline
\end{tabular}
\caption{Group theoretic factors for SM symmetries}
\label{groupfactors1}
\end{table}
and for $SU(n)$, the $S_n^{} \left( \mathbf{1} \right), C_n^{} \left( \mathbf{1} \right) = 0$. 









\end{document}